\overfullrule=0pt

\newcount\mgnf

\mgnf=1
\ifnum\mgnf=0
      \def\openone{\leavevmode\hbox{\ninerm 1\kern-3.3pt\tenrm1}}%
      \def\*{\vglue0.3truecm} 
      \hsize=15truecm
      \vsize=23.truecm
      \parindent=4.pt
      \baselineskip=0.45cm
      \font\titolo=cmbx12 
      \font\titolone=cmbx10 scaled\magstep 2
      \font\cs=cmcsc10 
       
      \font\ottorm=cmr8
       
      \font\euftw=eufm10 
      \font\msytw=msbm10
       
      \font\msytwww=msbm7 
      \font\indbf=cmbx10 scaled\magstep1 
      \fi

\ifnum\mgnf=1
      \def\openone{\leavevmode\hbox{\ninerm 1\kern-3.63pt\tenrm1}}%
      \def\*{\vglue0.5truecm}
      \magnification=1100
      \hoffset=0.truecm
      \hsize=16truecm
      \vsize=24.truecm
      \baselineskip=1.4em  plus 0.05em minus0.1em 
      \parindent=1em
      \lineskip=0.1em\lineskiplimit=0.1em      
      \parskip=0.1pt plus1pt
      \font\titolo=cmbx12 scaled\magstep 1 
      \font\titolone=cmbx10 scaled\magstep 3 
      \font\cs=cmcsc10 scaled\magstep 1
      \font\ottorm=cmr8 scaled\magstep 1 
      \font\euftw=eufm10 scaled\magstep1
      \font\msytw=msbm10 scaled\magstep1 
      \font\msytwww=msbm7 scaled\magstep1
       
      \font\indbf=cmbx10 scaled\magstep2 \fi

\global\newcount\numsec
\global\newcount\numapp
\global\newcount\numfor
\global\newcount\numfig
\global\newcount\numsub
\global\newcount\numlemma
\global\newcount\numtheorem
\global\newcount\numdef
\global\newcount\appflag 

\numsec=0\numapp=0\numfig=1

\def\veroparagrafo{\number\numsec}
\def\veraformula{\number\numfor}
\def\veraappendice{\number\numapp}
\def\verasub{\number\numsub}
\def\verafigura{\number\numfig}
\def\verolemma{\number\numlemma}

\def\section(#1,#2)
    {\advance\numsec by 1\numfor=1\numsub=1%
    \numlemma=1\numtheorem=1\numdef=1\appflag=0%
    \SIA p,#1,{\veroparagrafo} %
    \write15{\string\Fp (#1){\secc(#1)}}%
    \write16{ sec. #1 ==> \secc(#1)}%
    \hbox to \hsize{\titolo\hfill\number\numsec. #2\hfill%
    \expandafter{\alato(sec. #1)}}\*}

\def\appendix(#1,#2)
    {\advance\numapp by 1\numfor=1\numsub=1%
    \numlemma=1\numtheorem=1\numdef=1\appflag=1%
    \SIA p,#1,{A\veraappendice} %
    \write15{\string\Fp (#1){\secc(#1)}}%
    \write16{ app. #1 ==> \secc(#1)  }%
    \hbox to \hsize{\titolo\hfill Appendix A\number\numapp. #2\hfill%
    \expandafter{\alato(app. #1)}}\*}

\def\senondefinito#1{\expandafter\ifx\csname#1\endcsname\relax}

\def\SIA #1,#2,#3 {\senondefinito{#1#2}%
\expandafter\xdef\csname #1#2\endcsname{#3}\else \write16{???? ma
#1#2 e' gia' stato definito !!!!} \fi}

\def \Fe(#1)#2{\SIA fe,#1,#2 }
\def \Fp(#1)#2{\SIA fp,#1,#2 }
\def \Fg(#1)#2{\SIA fg,#1,#2 }
\def \Fl(#1)#2{\SIA fl,#1,#2 }
\def \Ft(#1)#2{\SIA ft,#1,#2 }
\def \Fd(#1)#2{\SIA fd,#1,#2 }

\def\etichetta(#1){(\veroparagrafo.\veraformula)%
\SIA e,#1,(\veroparagrafo.\veraformula) %
\global\advance\numfor by 1%
\write15{\string\Fe (#1){\equ(#1)}}%
\write16{ EQ #1 ==> \equ(#1)  }}

\def\etichettaa(#1){(A\veraappendice.\veraformula)%
\SIA e,#1,(A\veraappendice.\veraformula) %
\global\advance\numfor by 1%
\write15{\string\Fe (#1){\equ(#1)}}%
\write16{ EQ #1 ==> \equ(#1) }}

\def\getichetta(#1){
\SIA g,#1,{\verafigura} %
\global\advance\numfig by 1%
\write15{\string\Fg (#1){\graf(#1)}}%
\write16{ Fig. #1 ==> \graf(#1) }}

\def\etichettap(#1){\veroparagrafo.\verasub%
\SIA p,#1,{\veroparagrafo.\verasub} %
\global\advance\numsub by 1%
\write15{\string\Fp (#1){\secc(#1)}}%
\write16{ par #1 ==> \secc(#1)  }}

\def\etichettapa(#1){A\veraappendice.\verasub%
\SIA p,#1,{A\veraappendice.\verasub} %
\global\advance\numsub by 1%
\write15{\string\Fp (#1){\secc(#1)}}%
\write16{ par #1 ==> \secc(#1)  }}

\def\Eq(#1){\eqno{\etichetta(#1)\alato(#1)}}
\def\eq(#1){\etichetta(#1)\alato(#1)}
\def\Eqa(#1){\eqno{\etichettaa(#1)\alato(#1)}}
\def\eqa(#1){\etichettaa(#1)\alato(#1)}
\def\eqg(#1){\getichetta(#1)\alato(fig. #1)}
\def\sub(#1)#2{\0\palato(p. #1){\bf \etichettap(#1).}{\it #2}}
\def\asub(#1)#2{\0\palato(p. #1){\bf \etichettapa(#1).}{\it #2}}

\def\lemma{\0{\bf Lemma.\hskip.1truecm}}
\def\alemma{\0{\bf Lemma.\hskip.1truecm}}
\def\corollary{\0{\bf Corollary.\hskip.1truecm}}
\def\theorem{\0{\bf Theorem.\hskip.1truecm}}
\def\atheorem{\0{\bf Theorem.\hskip.1truecm}}

\def\definition(#1){\0{\bf Definition #1\hskip.1truecm}}
\def\proof{\0{\bf Proof\hskip.1truecm}}

\def\equv(#1){\senondefinito{fe#1}$\clubsuit$#1%
\write16{eq. #1 non e' (ancora) definita}%
\else\csname fe#1\endcsname\fi}

\def\grafv(#1){\senondefinito{fg#1}$\clubsuit$#1%
\write16{fig. #1 non e' (ancora) definito}%
\else\csname fg#1\endcsname\fi}

\def\secv(#1){\senondefinito{fp#1}$\clubsuit$#1%
\write16{par. #1 non e' (ancora) definito}%
\else\csname fp#1\endcsname\fi}

\def\lmv(#1){\senondefinito{fl#1}$\clubsuit$#1%
\write16{lemma #1 non e' (ancora) definito}%
\else\csname fl#1\endcsname\fi}

\def\almv(#1){\senondefinito{fal#1}$\clubsuit$#1%
\write16{lemma A #1 non e' (ancora) definito}%
\else\csname fal#1\endcsname\fi}

\def\thmv(#1){\senondefinito{ft#1}$\clubsuit$#1%
\write16{th. #1 non e' (ancora) definito}%
\else\csname ft#1\endcsname\fi}

\def\athmv(#1){\senondefinito{fat#1}$\clubsuit$#1%
\write16{ath. #1 non e' (ancora) definito}%
\else\csname fat#1\endcsname\fi}

\def\defzv(#1){\senondefinito{fd#1}$\clubsuit$#1%
\write16{def. #1 non e' (ancora) definito}%
\else\csname fd#1\endcsname\fi}

\def\equ(#1){\senondefinito{e#1}\equv(#1)\else\csname e#1\endcsname\fi}
\def\graf(#1){\senondefinito{g#1}\grafv(#1)\else\csname g#1\endcsname\fi}
\def\secc(#1){\senondefinito{p#1}\secv(#1)\else\csname p#1\endcsname\fi}
\def\lm(#1){\senondefinito{l#1}\lmv(#1)\else\csname l#1\endcsname\fi}
\def\thm(#1){\senondefinito{t#1}\thmv(#1)\else\csname t#1\endcsname\fi}
\def\defz(#1){\senondefinito{d#1}\defzv(#1)\else\csname d#1\endcsname\fi}
\def\sec(#1){{\S\secc(#1)}}

\def\BOZZA{\bz=1
          \def\alato(##1){\rlap{\kern-\hsize\kern-3em{$\scriptstyle##1$}}}
          \def\palato(##1){\rlap{\kern-4em{$\scriptstyle##1$}}}
           }

\def\alato(#1){}
\def\galato(#1){}
\def\palato(#1){}

{\count255=\time\divide\count255 by 60
\xdef\hourmin{\number\count255}
        \multiply\count255 by-60\advance\count255 by\time
   \xdef\hourmin{\hourmin:\ifnum\count255<10 0\fi\the\count255}}

\def\oramin{\hourmin }

\def\data{\number\day/\ifcase\month\or gennaio \or febbraio \or marzo \or
aprile \or maggio \or giugno \or luglio \or agosto \or settembre
\or ottobre \or novembre \or dicembre \fi/\number\year;\ \oramin}
\setbox200\hbox{$\scriptscriptstyle \data $}
\footline={\rlap{\hbox{\copy200}}\tenrm\hss \number\pageno\hss}

\let\a=\alpha \let\b=\beta  \let\g=\gamma     \let\d=\delta  \let\e=\varepsilon
\let\z=\zeta  \let\h=\eta    \let\th=\vartheta
\let\k=\kappa   \let\l=\lambda
\let\m=\mu    \let\n=\nu    \let\x=\xi        \let\p=\pi      \let\r=\rho
\let\s=\sigma \let\t=\tau   \let\f=\varphi     \let\c=\chi
\let\ps=\psi   \let\o=\omega 
\let\G=\Gamma \let\D=\Delta  \let\Th=\Theta   \let\L=\Lambda

\def\EE{{\cal E}}\def\VV{{\cal V}}
\def\CC{{\cal C}}\def\FF{{\cal F}}\def\WW{{\cal W}}
\def\TT{{\cal T}}\def\NN{{\cal N}}\def\BB{{\cal B}}\def\ZZ{{\cal Z}}
\def\RR{{\cal R}}\def\LL{{\cal L}}\def\JJ{{\cal J}}\def\QQ{{\cal Q}}
\def\DD{{\cal D}}\def\GG{{\cal G}}\def\SS{{\cal S}}
\def\AAA{{\cal A}}

\def\xx{{\bf x}}

\def\tt{{\bf t}}

       \def\oo{{\underline \omega}}

          \def\ux{{\underline x}}
\def\ue{{\underline \e}}           \def\uy{{\underline y}}
\def\uz{{\underline z}}            \def\uo{{\underline \o}} 
\def\us{{\underline \s}}

\def\MMM{\hbox{\euftw M}}          
\def\RRR{\hbox{\msytw R}}          
\def\rrr{\hbox{\msytwww R}}        \def\CCC{\hbox{\msytw C}}         
        
\def\NNN{\hbox{\msytw N}}          
        \def\ZZZ{\hbox{\msytw Z}}
        \def\zzz{\hbox{\msytwww Z}}
\def\TTT{\hbox{\msytw T}}          
        \def\EE{\hbox{\msytw E}}

\let\dpr=\partial 

\let\bs=\backslash
\let\==\equiv

\let\io=\infty
\let\0=\noindent

\def\\{\hfill\break}

\def\lft{\left}
\def\rgt{\right}

\def\ket#1{{|#1\rangle}}

\def\norm#1{{\left|\hskip-.05em\left|#1\right|\hskip-.05em\right|}}

\def\eg{\hbox{\it e.g.\ }}
\def\der{\hbox{\rm d}}
\def\tgl#1{\not\!#1}

\def\tende#1{\,\vtop{\ialign{##\crcr\rightarrowfill\crcr
             \noalign{\kern-1pt\nointerlineskip}
             \hskip3.pt${\scriptstyle #1}$\hskip3.pt\crcr}}\,}
\def\otto{\,{\kern-1.truept\leftarrow\kern-5.truept\to\kern-1.truept}\,}

\def\der{{\rm d}}

\def\T#1{{#1_{\kern-3pt\lower7pt\hbox{$\widetilde{}$}}\kern3pt}}
\def\VVV#1{{\underline #1}_{\kern-3pt\lower7pt\hbox{$\widetilde{}$}}\kern3pt\,}
\def\W#1{#1_{\kern-3pt\lower7.5pt\hbox{$\widetilde{}$}}\kern2pt\,}

\def\indica{\leaders \hbox to 0.5cm{\hss.\hss}\hfill}
\def\guida{\leaders\hbox to 1em{\hss.\hss}\hfill}\mathchardef\oo= "0521

\def\Halmos{\hfill\vrule height6pt width4pt depth2pt \par\hbox to \hsize{}}

\def\qed{\raise1pt\hbox{\vrule height5pt width5pt depth0pt}}
\def\hf#1{{\hat \f_{#1}}}

\def\indic{\hbox{\raise-2pt \hbox{\indbf 1}}}

%
%
%
\def\ins#1#2#3{\vbox to0pt{\kern-#2 \hbox{\kern#1 #3}\vss}\nointerlineskip}
%
%
%
\newdimen\xshift \newdimen\xwidth \newdimen\yshift

\def\insertplot#1#2#3#4#5{\par%
\xwidth=#1 \xshift=\hsize \advance\xshift by-\xwidth \divide\xshift by 2%
\yshift=#2 \divide\yshift by 2%
\line{\hskip\xshift \vbox to #2{\vfil%
#3 \includegraphics{#4.pst}}\hfill \raise\yshift\hbox{#5} }}

\openin14=\jobname.aux \ifeof14 \relax \else
\input \jobname.aux \closein14 \fi
\openout15=\jobname.aux

\font\tenmib=cmmib10
\font\sevenmib=cmmib10 scaled 800
\font\titolo=cmbx12
\font\titolone=cmbx10 scaled\magstep 2

\font\book=cmcsc9

\font\journal=cmti10
\font\pagine=cmti10

\font\cs=cmcsc10

\font\ninerm=cmr9
\font\ottorm=cmr8
\textfont5=\tenmib
\scriptfont5=\sevenmib
\scriptscriptfont5=\fivei

\font\euftw=eufm10 scaled\magstep1
\font\msytw=msbm9 scaled\magstep1

\font\msytwww=msbm7 scaled\magstep1
\font\indbf=cmbx10 scaled\magstep2

\newskip\ttglue
\font\ottorm=cmr8\font\ottoi=cmmi8\font\ottosy=cmsy7
\font\ottobf=cmbx7\font\ottott=cmtt8\font\ottosl=cmsl8\font\ottoit=cmti7
\font\sixrm=cmr6\font\sixbf=cmbx7\font\sixi=cmmi7\font\sixsy=cmsy7
\font\fiverm=cmr5\font\fivesy=cmsy5\font\fivei=cmmi5\font\fivebf=cmbx5

\def\ottopunti{\def\rm{\fam0\ottorm}\textfont0=\ottorm%
\scriptfont0=\sixrm\scriptscriptfont0=\fiverm\textfont1=\ottoi%
\scriptfont1=\sixi\scriptscriptfont1=\fivei\textfont2=\ottosy%
\scriptfont2=\sixsy\scriptscriptfont2=\fivesy\textfont3=\tenex%
\scriptfont3=\tenex\scriptscriptfont3=\tenex\textfont\itfam=\ottoit%
\def\it{\fam\itfam\ottoit}\textfont\slfam=\ottosl%
\def\sl{\fam\slfam\ottosl}\textfont\ttfam=\ottott%
\def\tt{\fam\ttfam\ottott}\textfont\bffam=\ottobf%
\scriptfont\bffam=\sixbf\scriptscriptfont\bffam=\fivebf%
\def\bf{\fam\bffam\ottobf}\tt\ttglue=.5em plus.25em minus.15em%
\setbox\strutbox=\hbox{\vrule height7pt depth2pt width0pt}%
\normalbaselineskip=9pt\let\sc=\sixrm\normalbaselines\rm}

%
%
%
%
%
%
%


{\count255=\time\divide\count255 by 60
\xdef\hourmin{\number\count255}
        \multiply\count255 by-60\advance\count255 by\time
   \xdef\hourmin{\hourmin:\ifnum\count255<10 0\fi\the\count255}}

\def\oramin{\hourmin }

\def\data{\number\day/\ifcase\month\or gennaio \or febbraio \or marzo \or
aprile \or maggio \or giugno \or luglio \or agosto \or settembre
\or ottobre \or novembre \or dicembre \fi/\number\year;\ \oramin}
\setbox200\hbox{$\scriptscriptstyle \data $}
\footline={\rlap{\hbox{\copy200}}\tenrm\hss \number\pageno\hss}

%
%
%
%
%
%
\newdimen\xshift \newdimen\xwidth \newdimen\yshift

\def\insertplot#1#2#3#4#5{\par%
\xwidth=#1 \xshift=\hsize \advance\xshift by-\xwidth \divide\xshift by 2%
\yshift=#2 \divide\yshift by 2%
\line{\hskip\xshift \vbox to #2{\vfil%
#3 \includegraphics{#4.ps}}\hfill \raise\yshift\hbox{#5}}}

\def\eqfig#1#2#3#4#5{ \par\xwidth=#1
\xshift=\hsize \advance\xshift by-\xwidth \divide\xshift by 2
\yshift=#2 \divide\yshift by 2 \line{\hglue\xshift \vbox to #2{\vfil #3
\includegraphics{#4.ps} }\hfill\raise\yshift\hbox{#5}}}

\def\8{\write13}

\def\bar#1{{\overline#1}}

\def\qed{\raise1pt\hbox{\vrule height5pt width5pt depth0pt}}

\def\ha{{\widehat \a}}

\def\hb{{\widehat \b}}

\def\hf{{\widehat \f}}
\def\hW{{\widehat W}}
\def\hH{{\widehat H}}
\def\hK{{\widehat K}}
\def\hW{{\widehat W}}

\def\hp{{\widehat \ps}}     

\def\jm{{\jmath}}
\def\hj{{\widehat \jmath}}
\def\hg{{\widehat g}}

\def\hS{{\widehat S}}

\def\hM{{\widehat M}}
\def\hT{{\widehat T}}

\def\indic{\hbox{\raise-2pt \hbox{\indbf 1}}}

\def\RRR{\hbox{\msytw R}} 
\def\rrr{\hbox{\msytwww R}} \def\CCC{\hbox{\msytw C}}
 
\def\NNN{\hbox{\msytw N}} 
 \def\ZZZ{\hbox{\msytw Z}}
 \def\zzz{\hbox{\msytwww Z}}
\def\TTT{\hbox{\msytw T}}

\newcount\mgnf  
\mgnf=0

\ifnum\mgnf=0
\def\openone{\leavevmode\hbox{\ninerm 1\kern-3.3pt\tenrm1}}%
\def\*{\vglue0.3truecm}\fi
\ifnum\mgnf=1
\def\openone{\leavevmode\hbox{\ninerm 1\kern-3.63pt\tenrm1}}%
\def\*{\vglue0.5truecm}\fi

\newcount\tipobib\newcount\bz\bz=0\newcount\aux\aux=1
\newdimen\bibskip\newdimen\maxit\maxit=0pt


\tipobib=0
\def\9#1{\ifnum\aux=1#1\else\relax\fi}

\newwrite\bib
\immediate\openout\bib=\jobname.bib
\global\newcount\bibex
\bibex=0
\def\verabib{\number\bibex}

\ifnum\tipobib=0
\def\cita#1{\expandafter\ifx\csname c#1\endcsname\relax
\hbox{$\clubsuit$}#1\write16{Manca #1 !}%
\else\csname c#1\endcsname\fi}
\def\rife#1#2#3{\immediate\write\bib{\string\raf{#2}{#3}{#1}}
\immediate\write15{\string\C(#1){[#2]}}
\setbox199=\hbox{#2}\ifnum\maxit < \wd199 \maxit=\wd199\fi}
\else
\def\cita#1{%
\expandafter\ifx\csname d#1\endcsname\relax%
\expandafter\ifx\csname c#1\endcsname\relax%
\hbox{$\clubsuit$}#1\write16{Manca #1 !}%
\else\probib(ref. numero )(#1)%
\csname c#1\endcsname%
\fi\else\csname d#1\endcsname\fi}%
\def\rife#1#2#3{\immediate\write15{\string\Cp(#1){%
\string\immediate\string\write\string\bib{\string\string\string\raf%
{\string\verabib}{#3}{#1}}%
\string\Cn(#1){[\string\verabib]}%
\string\CCc(#1)%
}}}%
\fi

\def\Cn(#1)#2{\expandafter\xdef\csname d#1\endcsname{#2}}
\def\CCc(#1){\csname d#1\endcsname}
\def\probib(#1)(#2){\global\advance\bibex+1%
\9{\immediate\write16{#1\verabib => #2}}%
}

\def\C(#1)#2{\SIA c,#1,{#2}}
\def\Cp(#1)#2{\SIAnx c,#1,{#2}}

\def\SIAnx #1,#2,#3 {\senondefinito{#1#2}%
\expandafter\def\csname#1#2\endcsname{#3}\else%
\write16{???? ma #1,#2 e' gia' stato definito !!!!}\fi}

\bibskip=10truept
\def\hboxto{\hbox to}

\catcode`\{=12\catcode`\}=12
\catcode`\<=1\catcode`\>=2
\immediate\write\bib<
        \string\halign{\string\hboxto \string\maxit%
        {\string #\string\hfill}&%
        \string\vtop{\string\parindent=0pt\string\advance\string\hsize%
        by -1.55truecm%
        \string#\string\vskip \bibskip
        }\string\cr%
>
\catcode`\{=1\catcode`\}=2
\catcode`\<=12\catcode`\>=12

\def\raf#1#2#3{\ifnum \bz=0 [#1]&#2 \cr\else
\llap{${}_{\rm #3}$}[#1]&#2\cr\fi}

\newread\bibin

\catcode`\{=12\catcode`\}=12
\catcode`\<=1\catcode`\>=2
\def\chiudibib<
\catcode`\{=12\catcode`\}=12
\catcode`\<=1\catcode`\>=2
\immediate\write\bib<}>
\catcode`\{=1\catcode`\}=2
\catcode`\<=12\catcode`\>=12
>
\catcode`\{=1\catcode`\}=2
\catcode`\<=12\catcode`\>=12

\def\makebiblio{
\ifnum\tipobib=0
\advance \maxit by 10pt
\else
\maxit=1.truecm
\fi
\chiudibib
\immediate \closeout\bib
\openin\bibin=\jobname.bib
\ifeof\bibin\relax\else
\raggedbottom
\input \jobname.bib
\fi
}

\openin13=#1.aux \ifeof13 \relax \else
\input #1.aux \closein13\fi
\openin14=\jobname.aux \ifeof14 \relax \else
\input \jobname.aux \closein14 \fi
\immediate\openout15=\jobname.aux

\def\defi{{\buildrel def \over =}}
\def\wdg{\wedge}

\def\Tr{{\rm Tr}}

\def\wt#1{\widetilde{#1}}
\def\wh#1{\widehat{#1}}
\def\sqt[#1]#2{\root #1\of {#2}}

\def\sde{{\cs SDe}}
\def\wti{{\cs WTi}}
\def\osa{{\cs OSa}}
\def\ce{{\cs CE}}
\def\rg{{\cs RG}}
\def\chapter(#1,#2){{}}
\def\intro(#1,#2){{}}
\frenchspacing

\font\titolocap=cmbx10 scaled\magstep 4
\font\grande=cmbx10 scaled\magstep 5

\global\newcount\numcap
\global\newcount\numtheo
\global\newcount\numatheo
\global\newcount\numcorol
\global\newcount\numlem
\global\newcount\numalem
\numcap=0\numtheo=1\numatheo=1\numcorol=1\numlem=1\numalem=1

\global\newcount\dispari
\dispari=0

\global\newcount\paginacap
\def\asterisco{*}
\paginacap=1

\global\newcount\articolo
\def\art{\item{\bf \number\articolo.}\advance\articolo by 1}
\def\elenco#1{{\articolo=1
           \vskip1em\parindent2.8em #1\vskip1em\par}}
\def\subelenco#1{{\vskip.5em\parindent3.8em #1\vskip.5em\par}}

\def\leaderfill{\leaders\hbox to 1em{\hss.\hss}\hfill}
\def\indice{\null
    \ifodd\pageno
          {\null\vfill\eject\mark{*}}
    \fi
    \null\vfill\eject
    \null\mark{a}
    \paginacap=\pageno
    \numcap=0\numfor=1\numsub=1\numtheo=1%
    \numlemma=1\numtheorem=1\numdef=1\numsec=0\appflag=0%
    \SIA ch,toc,{\verocapitolo}%
    \vskip 5truecm
    \hfill\break
    \vskip 1.8em
    \vtop{\hsize=16truecm\noindent\baselineskip 2.2em 
    \grande Contents\hfill}%
    \hfil\break
    \vskip 4em
    \headline={\if\botmark\asterisco
               {\hfill}
	       \else
	       {\ifnum\pageno=\paginacap
                     {\hfill}
               \else
                     {\noindent
                      \ifodd\pageno
                            {{\bf Contents \hskip0.1em\hrulefill\hskip0.1em \folio\hskip-.5em}}
                      \else
                            {{\bf{\folio}\hskip0.1em \hrulefill\hskip-.5em}}
                      \fi}
                \fi}\fi}
\vskip 2em \hbox to\hsize {{\fam \bffam \tenbf Introduction}\leaders \hbox to 1em{\hss .\hss }\hfill {\fam \bffam \tenbf 7}}
\vskip 2em \hbox to\hsize {{\fam \bffam \tenbf 1\hskip 1em Definitions and Main Results}\leaders \hbox to 1em{\hss .\hss }\hfill {\fam \bffam \tenbf 11}}
\hbox to\hsize {\hskip 2em{1.1}\hskip 1em Euclidean Thirring Model\leaders \hbox to 1em{\hss .\hss }\hfill 11}
\hbox to\hsize {\hskip 4em{1.1.1}\hskip 1emWeyl formalism.\leaders \hbox to 1em{\hss .\hss }\hfill 11}
\hbox to\hsize {\hskip 4em{1.1.2}\hskip 1emSpacetime Lattice.\leaders \hbox to 1em{\hss .\hss }\hfill 12}
\hbox to\hsize {\hskip 4em{1.1.3}\hskip 1emGrassmann Algebra.\leaders \hbox to 1em{\hss .\hss }\hfill 12}
\hbox to\hsize {\hskip 4em{1.1.4}\hskip 1emSchwinger functions.\leaders \hbox to 1em{\hss .\hss }\hfill 13}
\hbox to\hsize {\hskip 4em{1.1.5}\hskip 1emRemarks.\leaders \hbox to 1em{\hss .\hss }\hfill 14}
\hbox to\hsize {\hskip 4em Theorem 1.1\leaders \hbox to 1em{\hss .\hss }\hfill 15}
\hbox to\hsize {\hskip 4em{1.1.6}\hskip 1emWard-Takahashi identities: first anomaly.\leaders \hbox to 1em{\hss .\hss }\hfill 16}
\hbox to\hsize {\hskip 4em Theorem 1.2\leaders \hbox to 1em{\hss .\hss }\hfill 16}
\hbox to\hsize {\hskip 4em{1.1.7}\hskip 1emClosed equation: new anomaly\leaders \hbox to 1em{\hss .\hss }\hfill 17}
\hbox to\hsize {\hskip 4em Theorem 1.3\leaders \hbox to 1em{\hss .\hss }\hfill 18}
\vskip 2em \hbox to\hsize {{\fam \bffam \tenbf 2\hskip 1em Hamiltonian Regularization}\leaders \hbox to 1em{\hss .\hss }\hfill {\fam \bffam \tenbf 21}}
\hbox to\hsize {\hskip 2em{2.1}\hskip 1em  Hamiltonian Thirring Model\leaders \hbox to 1em{\hss .\hss }\hfill 21}
\hbox to\hsize {\hskip 4em{2.1.1}\hskip 1emHamiltonian.\leaders \hbox to 1em{\hss .\hss }\hfill 22}
\hbox to\hsize {\hskip 4em{2.1.2}\hskip 1emCorrelations.\leaders \hbox to 1em{\hss .\hss }\hfill 23}
\hbox to\hsize {\hskip 4em{2.1.3}\hskip 1emPropagator.\leaders \hbox to 1em{\hss .\hss }\hfill 23}
\hbox to\hsize {\hskip 4em{2.1.4}\hskip 1emSchwinger functions.\leaders \hbox to 1em{\hss .\hss }\hfill 24}
\hbox to\hsize {\hskip 4em Theorem 2.1\leaders \hbox to 1em{\hss .\hss }\hfill 25}
\vskip 2em \hbox to\hsize {{\fam \bffam \tenbf 3\hskip 1em Renormalization Group Analysis}\leaders \hbox to 1em{\hss .\hss }\hfill {\fam \bffam \tenbf 27}}
\hbox to\hsize {\hskip 2em{3.1}\hskip 1em  Renormalization Group Analysis for Hard Fermions\leaders \hbox to 1em{\hss .\hss }\hfill 27}
\hbox to\hsize {\hskip 4em{3.1.1}\hskip 1emMomenta slicing.\leaders \hbox to 1em{\hss .\hss }\hfill 27}
\hbox to\hsize {\hskip 4em{3.1.2}\hskip 1emMultiscale integration.\leaders \hbox to 1em{\hss .\hss }\hfill 28}
\hbox to\hsize {\hskip 4em{3.1.3}\hskip 1emDimensional bounds.\leaders \hbox to 1em{\hss .\hss }\hfill 29}
\hbox to\hsize {\hskip 4em Lemma 3.1 \leaders \hbox to 1em{\hss .\hss }\hfill 30}
\hbox to\hsize {\hskip 4em{3}\hskip 1emRemark: superrinormalizability.\leaders \hbox to 1em{\hss .\hss }\hfill 30}
\hbox to\hsize {\hskip 2em{3.2}\hskip 1em Renormalization Group Analysis for Double Fermions\leaders \hbox to 1em{\hss .\hss }\hfill 30}
\hbox to\hsize {\hskip 4em{3.2.1}\hskip 1emMomenta slicing.\leaders \hbox to 1em{\hss .\hss }\hfill 30}
\hbox to\hsize {\hskip 4em{3.2.2}\hskip 1emDimensional bounds.\leaders \hbox to 1em{\hss .\hss }\hfill 31}
\hbox to\hsize {\hskip 2em{3.3}\hskip 1em Renormalization Group Analysis for Soft Fermions\leaders \hbox to 1em{\hss .\hss }\hfill 31}
\hbox to\hsize {\hskip 4em{3.3.1}\hskip 1emMomenta slicing.\leaders \hbox to 1em{\hss .\hss }\hfill 31}
\hbox to\hsize {\hskip 4em{3.3.2}\hskip 1emMultiscale integration.\leaders \hbox to 1em{\hss .\hss }\hfill 31}
\hbox to\hsize {\hskip 4em{3.3.3}\hskip 1emDimensional bounds.\leaders \hbox to 1em{\hss .\hss }\hfill 33}
\hbox to\hsize {\hskip 4em{3.3.4}\hskip 1emLocalization.\leaders \hbox to 1em{\hss .\hss }\hfill 33}
\hbox to\hsize {\hskip 4em Theorem 3.1\leaders \hbox to 1em{\hss .\hss }\hfill 37}
\hbox to\hsize {\hskip 2em{3.4}\hskip 1em  Flows of the Running Coupling Constants\leaders \hbox to 1em{\hss .\hss }\hfill 38}
\hbox to\hsize {\hskip 4em Theorem 3.2\leaders \hbox to 1em{\hss .\hss }\hfill 38}
\hbox to\hsize {\hskip 2em{3.5}\hskip 1em  Equivalence of the Euclidean and Hamiltonian Regularization\leaders \hbox to 1em{\hss .\hss }\hfill 39}
\vskip 2em \hbox to\hsize {{\fam \bffam \tenbf 4\hskip 1em Phase and Chiral Symmetries}\leaders \hbox to 1em{\hss .\hss }\hfill {\fam \bffam \tenbf 41}}
\hbox to\hsize {\hskip 2em{4.1}\hskip 1em  Ward-Takahashi Identities\leaders \hbox to 1em{\hss .\hss }\hfill 41}
\hbox to\hsize {\hskip 4em{4.1.1}\hskip 1em{\cs WTi}{} for the Schwinger functions.\leaders \hbox to 1em{\hss .\hss }\hfill 41}
\hbox to\hsize {\hskip 4em{4.1.2}\hskip 1emFlows of $\n ^{(+)}_N$ and $\n ^{(-)}_N$. \leaders \hbox to 1em{\hss .\hss }\hfill 44}
\hbox to\hsize {\hskip 4em Lemma 4.2 \leaders \hbox to 1em{\hss .\hss }\hfill 44}
\hbox to\hsize {\hskip 4em{4.1.3}\hskip 1emImproved localization I.\leaders \hbox to 1em{\hss .\hss }\hfill 44}
\hbox to\hsize {\hskip 4em Theorem 4.1\leaders \hbox to 1em{\hss .\hss }\hfill 47}
\hbox to\hsize {\hskip 4em Theorem 4.2\leaders \hbox to 1em{\hss .\hss }\hfill 47}
\hbox to\hsize {\hskip 4em Theorem 4.3\leaders \hbox to 1em{\hss .\hss }\hfill 48}
\hbox to\hsize {\hskip 4em{4.1.4}\hskip 1emRemark: anomaly and anomalous exponents.\leaders \hbox to 1em{\hss .\hss }\hfill 49}
\hbox to\hsize {\hskip 2em{4.2}\hskip 1em  Closed Equations\leaders \hbox to 1em{\hss .\hss }\hfill 49}
\hbox to\hsize {\hskip 4em{4.2.1}\hskip 1emSchwinger-Dyson equation\leaders \hbox to 1em{\hss .\hss }\hfill 49}
\hbox to\hsize {\hskip 4em{4.2.2}\hskip 1emClosed equations.\leaders \hbox to 1em{\hss .\hss }\hfill 49}
\hbox to\hsize {\hskip 4em{4.2.3}\hskip 1emFlows of $\mathaccent "0365 {z}^{(\m )}_N$ and $\mathaccent "0365 {\l }^{(\m )}_N$.\leaders \hbox to 1em{\hss .\hss }\hfill 50}
\hbox to\hsize {\hskip 4em{4.2.4}\hskip 1emImproved localization II.\leaders \hbox to 1em{\hss .\hss }\hfill 52}
\hbox to\hsize {\hskip 4em Theorem 4.4\leaders \hbox to 1em{\hss .\hss }\hfill 58}
\hbox to\hsize {\hskip 4em Theorem 4.5\leaders \hbox to 1em{\hss .\hss }\hfill 59}
\hbox to\hsize {\hskip 4em Theorem 4.6\leaders \hbox to 1em{\hss .\hss }\hfill 59}
\hbox to\hsize {\hskip 4em{4.2.5}\hskip 1emVanishing of the Beta function.\leaders \hbox to 1em{\hss .\hss }\hfill 62}
\hbox to\hsize {\hskip 2em{4.3}\hskip 1em  Solution of the closed equation\leaders \hbox to 1em{\hss .\hss }\hfill 62}
\vskip 2em\hbox to\hsize {{\fam \bffam \tenbf A.1\hskip 1em Simple Analytical Properties}\leaders \hbox to 1em{\hss .\hss }\hfill {\fam \bffam \tenbf 65}}
\hbox to\hsize {\hskip 4em {A1.1}\hskip 1emPartial-fraction expansion.\leaders \hbox to 1em{\hss .\hss }\hfill 65}
\hbox to\hsize {\hskip 4em {A1.2}\hskip 1emGevrey compact-support functions\leaders \hbox to 1em{\hss .\hss }\hfill 66}
\hbox to\hsize {\hskip 4em {A1.3}\hskip 1emBounds for the propagators.\leaders \hbox to 1em{\hss .\hss }\hfill 66}
\vskip 2em\hbox to\hsize {{\fam \bffam \tenbf A.2\hskip 1em OS axioms}\leaders \hbox to 1em{\hss .\hss }\hfill {\fam \bffam \tenbf 67}}
\hbox to\hsize {\hskip 4em {A2.1}\hskip 1emTest functions.\leaders \hbox to 1em{\hss .\hss }\hfill 67}
\hbox to\hsize {\hskip 4em Lemma A.2.1 \leaders \hbox to 1em{\hss .\hss }\hfill 68}
\hbox to\hsize {\hskip 4em {A2.2}\hskip 1emReflection Positivity for the Hamiltonian Regularization\leaders \hbox to 1em{\hss .\hss }\hfill 69}
\vskip 2em\hbox to\hsize {{\fam \bffam \tenbf A.3\hskip 1em Tree Expansion and Convergence of the Schwinger functions}\leaders \hbox to 1em{\hss .\hss }\hfill {\fam \bffam \tenbf 71}}
\hbox to\hsize {\hskip 4em {A3.1}\hskip 1emTree structure.\leaders \hbox to 1em{\hss .\hss }\hfill 71}
\hbox to\hsize {\hskip 4em {A3.2}\hskip 1emCluster expansion.\leaders \hbox to 1em{\hss .\hss }\hfill 74}
\hbox to\hsize {\hskip 4em {A3.3}\hskip 1emBounds for the kernels.\leaders \hbox to 1em{\hss .\hss }\hfill 75}
\hbox to\hsize {\hskip 4em Lemma A.3.1 \leaders \hbox to 1em{\hss .\hss }\hfill 75}
\hbox to\hsize {\hskip 4em Lemma A.3.2 \leaders \hbox to 1em{\hss .\hss }\hfill 78}
\hbox to\hsize {\hskip 4em {A3.4}\hskip 1emRemark.\leaders \hbox to 1em{\hss .\hss }\hfill 80}
\hbox to\hsize {\hskip 4em Lemma A.3.3 \leaders \hbox to 1em{\hss .\hss }\hfill 80}
\hbox to\hsize {\hskip 4em {A3.5}\hskip 1emShort memory property.\leaders \hbox to 1em{\hss .\hss }\hfill 84}
\hbox to\hsize {\hskip 4em Lemma A.3.4 \leaders \hbox to 1em{\hss .\hss }\hfill 85}
\hbox to\hsize {\hskip 4em {A3.6}\hskip 1emCompletion of the proof of Theorem {1.1}\leaders \hbox to 1em{\hss .\hss }\hfill 85}
\hbox to\hsize {\hskip 4em {A3.7}\hskip 1emCompletion of the proof of Theorem {1.2}\leaders \hbox to 1em{\hss .\hss }\hfill 85}
\vskip 2em\hbox to\hsize {{\fam \bffam \tenbf A.4\hskip 1em Exact symmetries}\leaders \hbox to 1em{\hss .\hss }\hfill {\fam \bffam \tenbf 87}}
\hbox to\hsize {\hskip 4em {A4.1}\hskip 1emReflection.\leaders \hbox to 1em{\hss .\hss }\hfill 87}
\hbox to\hsize {\hskip 4em {A4.2}\hskip 1emSpace reflection.\leaders \hbox to 1em{\hss .\hss }\hfill 88}
\hbox to\hsize {\hskip 4em {A4.3}\hskip 1emRotation.\leaders \hbox to 1em{\hss .\hss }\hfill 88}
\vskip 2em\hbox to\hsize {{\fam \bffam \tenbf A.5\hskip 1em Proof of Theorem {3.2 }}\leaders \hbox to 1em{\hss .\hss }\hfill {\fam \bffam \tenbf 89}}
\hbox to\hsize {\hskip 4em {A5.1}\hskip 1emBeta and Gamma functions.\leaders \hbox to 1em{\hss .\hss }\hfill 89}
\hbox to\hsize {\hskip 4em Lemma A.5.1 \leaders \hbox to 1em{\hss .\hss }\hfill 89}
\hbox to\hsize {\hskip 4em {A5.2}\hskip 1emFlows of the couplings\leaders \hbox to 1em{\hss .\hss }\hfill 90}
\hbox to\hsize {\hskip 4em Lemma A.5.2 \leaders \hbox to 1em{\hss .\hss }\hfill 90}
\hbox to\hsize {\hskip 4em Lemma A.5.3 \leaders \hbox to 1em{\hss .\hss }\hfill 94}
\hbox to\hsize {\hskip 4em {A5.3}\hskip 1emFurther properties of the Gamma functions.\leaders \hbox to 1em{\hss .\hss }\hfill 95}
\vskip 2em\hbox to\hsize {{\fam \bffam \tenbf A.6\hskip 1em Proof of Lemma {4.2}}\leaders \hbox to 1em{\hss .\hss }\hfill {\fam \bffam \tenbf 97}}
\vskip 2em\hbox to\hsize {{\fam \bffam \tenbf A.7\hskip 1em Proof of Theorems {4.1} and {4.4}}\leaders \hbox to 1em{\hss .\hss }\hfill {\fam \bffam \tenbf 101}}
\vskip 2em\hbox to\hsize {{\fam \bffam \tenbf A.8\hskip 1em Schwinger-Dyson equation}\leaders \hbox to 1em{\hss .\hss }\hfill {\fam \bffam \tenbf 105}}
\hbox to\hsize {\hskip 4em {A8.1}\hskip 1emFunctional derivation.\leaders \hbox to 1em{\hss .\hss }\hfill 105}
\vskip 2em\hbox to\hsize {{\fam \bffam \tenbf A.9\hskip 1em Lowest Order Computations}\leaders \hbox to 1em{\hss .\hss }\hfill {\fam \bffam \tenbf 107}}
\hbox to\hsize {\hskip 4em {A9.1}\hskip 1em{\cs WTi}{} anomaly\leaders \hbox to 1em{\hss .\hss }\hfill 107}
\hbox to\hsize {\hskip 4em {A9.2}\hskip 1em{\cs CE}{} anomaly.\leaders \hbox to 1em{\hss .\hss }\hfill 109}
\hbox to\hsize {\hskip 4em {A9.3}\hskip 1emExplicit computation.\leaders \hbox to 1em{\hss .\hss }\hfill 113}
\vskip 2em \hbox to\hsize {{\fam \bffam \tenbf References}\leaders \hbox to 1em{\hss .\hss }\hfill {\fam \bffam \tenbf 119}}
\null}


\def\verocapitolo{\number\numcap}
\def\veroteorema{\number\numtheo}
\def\veroateorema{\number\numatheo}
\def\verocorollario{\number\numcorol}
\def\verolemma{\number\numlem}
\def\veroalemma{\number\numalem}

\def\intro(#1,#2)
    {\vfill
     \ifodd\pageno
           {\null\vfill\eject\mark{*}}
    \fi
    \null\vfill\eject
    \null\mark{a}
    \paginacap=\pageno
    \numcap=0\numfor=1\numsub=1\numtheo=1%
    \numlemma=1\numtheorem=1\numdef=1\numsec=0\appflag=0%
    \SIA ch,#1,{\verocapitolo} %
    \write15{\string\Fp (#1){\secc(#1)}}%
    \write16{ sec. #1 ==> \secc(#1)}%
    \vskip 5truecm
    {\expandafter{\alato(sec. #1)}}\hfill\break
    \vskip 1.8em
    \vtop{\hsize=16truecm\noindent\baselineskip 2.2em 
    \grande #2\hfill}%
    \hfil\break
    \vskip 4em
    \headline={\if\botmark\asterisco
               {\hfill}
	       \else
	       {\ifnum\pageno=\paginacap
                     {\hfill}
               \else
                     {\noindent
                      \ifodd\pageno
                            {{\bf #2\hskip0.1em\hrulefill\hskip0.1em \folio\hskip-.5em}}
                      \else
                            {{\bf{\folio}\hskip0.1em \hrulefill\hskip-.5em}}
                      \fi}
                \fi}\fi}}

\def\chapter(#1,#2)
    {\vfill
     \ifodd\pageno
          {\null\vfill\eject\mark{*}}
    \fi 
    \null\vfill\eject
    \null\mark{a}
    \paginacap=\pageno
    \advance\numcap by 1\numfor=1\numsub=1\numtheo=1%
    \numlemma=1\numtheorem=1\numdef=1\numsec=0\appflag=0%
    \SIA p,#1,{\verocapitolo} %
    \write15{\string\Fp (#1){\secc(#1)}}%
    \write16{ sec. #1 ==> \secc(#1)}%
    \vskip 5truecm
    {\noindent \titolocap Chapter \number\numcap:
    \expandafter{\alato(sec. #1)}}\hfill\break
    \vskip 1.8em
    \vtop{\hsize=16truecm\noindent\baselineskip 2.2em 
    \grande#2\hfill}%
    \hfil\break
    \vskip 4em
    \headline={\if\botmark\asterisco
                     {\hfill}
               \else
              {\ifnum\pageno=\paginacap
                     {\hfill}
               \else
                     {\noindent
                      \ifodd\pageno
                            {{\bf #2\hskip0.1em\hrulefill\hskip0.1em \folio\hskip-.5em}}
                      \else
                            {{\bf{\folio}\hskip0.1em \hrulefill 
                            \hskip0.1em Chapter \number\numcap.\hskip-.5em}}
                      \fi}
                \fi}\fi}}

\def\appendix(#1,#2)
    {\vfill
     \ifodd\pageno
           {\null\vfill\eject\mark{*}}
     \fi
    \null\vfill\eject
    \null\mark{a}
    \paginacap=\pageno
    \advance\numapp by 1\numfor=1\numsub=1\numtheo=1%
    \numalem=1\numlemma=1\numtheorem=1\numdef=1\appflag=1%
    \SIA p,#1,{A\veraappendice} %
    \write1{\vskip 2em\line{{\bf A.\veraappendice\hskip 1em #2}\leaderfill{\bf \folio}}}%
    \write15{\string\Fp (#1){\secc(#1)}}%
    \write16{ app. #1 ==> \secc(#1)  }%
    \vskip 5truecm
    {\noindent \titolocap Appendix \number\numapp:
    \expandafter{\alato(app. #1)}}\hfill\break
    \vskip 1.8em
    \vtop{\hsize=16truecm\noindent\baselineskip 2.6em 
    \grande#2\hfill}%
    \hfil\break
    \vskip 4em 
    \headline={\if\botmark\asterisco
              {\hfill}
              \else
              {\ifnum\pageno=\paginacap
                     {\hfill}
               \else
                     {\noindent
                      \ifodd\pageno
                            {{\bf #2\hskip0.1em\hrulefill\hskip0.1em \folio\hskip-.5em}}
                      \else
                            {{\bf{\folio}\hskip0.1em \hrulefill 
                            \hskip0.1em Appendix \number\numapp.\hskip-.5em}}
                      \fi}
               \fi}\fi}}

\def\section(#1,#2)
    {\advance\numsec by 1\numfor=1\numsub=1%
    \numlemma=1\numtheorem=1\numdef=1\appflag=0%
    \SIA p,#1,{\verocapitolo.\veroparagrafo} %
    \write1{\line{\hskip 2em{\secc(#1)}\hskip 1em #2\leaderfill \folio}}%
    \write15{\string\Fp (#1){\secc(#1)}}%
    \write16{ sec. #1 ==> \secc(#1)}%
    \vskip 2em
    \hbox to \hsize{\titolo\number\numcap.\number\numsec\hskip 1em #2\hfill%
    \expandafter{\alato(sec. #1)}}\*}

\def\senondefinito#1{\expandafter\ifx\csname#1\endcsname\relax}

\def\SIA #1,#2,#3 {\senondefinito{#1#2}%
\expandafter\xdef\csname #1#2\endcsname{#3}\else \write16{???? ma
#1#2 e' gia' stato definito !!!!} \fi}

\def \Fe(#1)#2{\SIA fe,#1,#2 }
\def \Fp(#1)#2{\SIA fp,#1,#2 }
\def \Fg(#1)#2{\SIA fg,#1,#2 }
\def \Fl(#1)#2{\SIA fl,#1,#2 }
\def \Ft(#1)#2{\SIA ft,#1,#2 }
\def \Fd(#1)#2{\SIA fd,#1,#2 }
   
\def\equv(#1){\senondefinito{fe#1}$\clubsuit$#1%
\write16{eq. #1 non e' (ancora) definita}%
\else\csname fe#1\endcsname\fi}

\def\secv(#1){\senondefinito{fp#1}$\clubsuit$#1%
\write16{par. #1 non e' (ancora) definito}%
\else\csname fp#1\endcsname\fi}

\def\lmv(#1){\senondefinito{fl#1}$\clubsuit$#1%
\write16{lemma #1 non e' (ancora) definito}%
\else\csname fl#1\endcsname\fi}

\def\thmv(#1){\senondefinito{ft#1}$\clubsuit$#1%
\write16{th. #1 non e' (ancora) definito}%
\else\csname ft#1\endcsname\fi}

\def\equ(#1){\senondefinito{e#1}\equv(#1)\else\csname e#1\endcsname\fi}
\def\secc(#1){\senondefinito{p#1}\secv(#1)\else\csname p#1\endcsname\fi}
\def\thm(#1){\senondefinito{t#1}\thmv(#1)\else\csname t#1\endcsname\fi}
\def\lm(#1){\senondefinito{l#1}\lmv(#1)\else\csname l#1\endcsname\fi}

\def\etichetta(#1){(\verocapitolo.\veroparagrafo.\veraformula)%
\SIA e,#1,(\verocapitolo.\veroparagrafo.\veraformula) %
\global\advance\numfor by 1%
\write15{\string\Fe (#1){\equ(#1)}}%
\write16{ eq. #1 ==> \equ(#1)  }}

\def\etichettap(#1,#2){\verocapitolo.\veroparagrafo.\verasub%
\SIA p,#1,{\verocapitolo.\veroparagrafo.\verasub} %
\global\advance\numsub by 1%
\write1{\line{\hskip 4em{\secc(#1)}\hskip1em#2\leaderfill\folio}}%
\write15{\string\Fp (#1){\secc(#1)}}%
\write16{ par. #1 ==> \secc(#1)  }}

\def\etichettapa(#1,#2){A\veraappendice.\verasub%
\SIA p,#1,{A\veraappendice.\verasub} %
\global\advance\numsub by 1%
\write1{\line{\hskip 4em {\secc(#1)}\hskip1em#2\leaderfill\folio}}%
\write15{\string\Fp (#1){\secc(#1)}}%
\write16{ par #1 ==> \secc(#1)  }}

\def\etichettaT(#1){\verocapitolo.\veroteorema%
\SIA t,#1,{\verocapitolo.\veroteorema} %
\global\advance\numtheo by 1%
\global\advance\numcorol by -\numcorol
\global\advance\numcorol by 1
\write1{\line{\hskip 4em  Theorem \thm(#1)\leaderfill\folio}}%
\write15{\string\Ft (#1){\thm(#1)}}%
\write16{ teor. #1 ==> \thm(#1)}}

\def\etichettaC(#1,#2){\thm(#2).\verocorollario%
\SIA t,#1,{\thm(#2).\verocorollario} %
\global\advance\numcorol by 1%
\write1{\line{\hskip 4em  Corollary \thm(#1). \leaderfill\folio}}%
\write15{\string\Ft (#1){\thm(#1)}}%
\write16{ teor. #1 ==> \thm(#1)}}

\def\etichettaaT(#1){\veraappendice.\veroateorema%
\SIA t,#1,{A.\veraappendice.\veroateorema} %
\global\advance\numatheo by 1%
\write1{\line{\hskip 4em Theorem \thm(#1)  \leaderfill\folio}}%
\write15{\string\Ft (#1){\thm(#1)}}%
\write16{ teor. A.#1 ==> \thm(#1)  }}

\def\etichettaL(#1){\verocapitolo.\verolemma%
\SIA l,#1,{\verocapitolo.\verolemma} %
\global\advance\numlem by 1%
\write1{\line{\hskip 4em Lemma \lm(#1) \leaderfill\folio}}%
\write15{\string\Fl (#1){\lm(#1)}}%
\write16{ lem. #1 ==> \lm(#1)  }}

\def\etichettaaL(#1){\veraappendice.\veroalemma%
\SIA l,#1,{A.\veraappendice.\veroalemma} %
\global\advance\numalem by 1%
\write1{\line{\hskip 4em  Lemma \lm(#1) \leaderfill\folio}}%
\write15{\string\Fl (#1){\lm(#1)}}%
\write16{ lem. A.#1 ==> \lm(#1)  }}

\def\Eq(#1){\eqno{\etichetta(#1)\alato({equ(#1)})}}
\def\Eqa(#1){\eqno{\etichettaa(#1)\alato({equ(#1)})}}
\def\theorem(#1)#2{\0\palato({thm(#1)}){\bf Theorem \etichettaT(#1). {\it #2}}}
\def\corollary(#1,#2)#3{\0\palato({thm(#1)}){\bf Corollary \etichettaC(#1,#2). {\it #3}}}
\def\atheorem(#1)#2{\0\palato({thm(#1)}){\bf Theorem A.\etichettaaT(#1). {\it #2}}}
\def\lemma(#1)#2{\0\palato({lm.(#1)}){\bf Lemma \etichettaL(#1). {\it #2}}}
\def\alemma(#1)#2{\0\palato({lm(#1)}){\bf Lemma A.\etichettaaL(#1). {\it #2}}}
\def\sub(#1)#2{\0\palato({secc(#1)}){\bf\etichettap(#1,#2) #2}}
\def\asub(#1)#2{\0\palato({secc(#1)}){\bf\etichettapa(#1,#2) #2}}
\footline={}
\mgnf=1

\font\it=cmsl10
\voffset=1truecm
\vsize=22.5truecm

\def\*{\vskip2em}

\pageno=1

\centerline  {UNIVERSIT\`{A} DEGLI STUDI DI ROMA  ``LA SAPIENZA''}

\vskip.1truecm
\vbox
{\insertplot{90pt}{150pt}%
{}%
{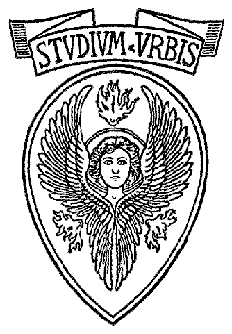}{}
}
\vskip1em

\centerline {FACOLT\`{A} DI SCIENZE MATEMATICHE FISICHE E NATURALI}
\vskip1.5truecm
\centerline {TESI DI DOTTORATO IN MATEMATICA}
\vskip2.7truecm
\centerline{\titolone Rigorous construction of the Thirring model:} 
\vskip1mm 
\centerline{\titolone Ward-Takahashi Identities, Schwinger-Dyson Equations} 
\vskip1mm
\centerline{\titolone and New Anomalies} \vskip1cm

\vskip3truecm

\hbox to \hsize
{\hskip2em \vtop{
            \hbox{\hskip1em Relatori}
            \hbox{Prof. G.Benfatto}
            \hbox{Prof. V.Mastropietro}}
\hfill
      \vtop{
            \hbox{\hskip1em Dottorando}
            \hbox{Dott. P.Falco}}
\hskip2em\null}

\vskip2truecm

\centerline  {Anno Accademico 2004-2005}
\null
\null

\indice

\intro(I,Introduction)
{\bf Historical outlook.} The Thirring
model was proposed in [T58]. It describes Dirac fermions in
$d=1+1$ spacetime dimensions with local current-current
interaction. With summation over repeated indices, the classical
Action for mass $\m$ and coupling $\l$ reads:
$$
 \int\!\der^2x\
 \bar\psi_x\left(i\tgl{\partial}+\m\right)\psi_x-
 {\l\over 2} \int\!\der^2x\
 \r_\n(x) \r^\n(x)\;,
\Eq(1)
$$
where $\psi$ and $\bar\psi\defi\ps^\dagger \g^0$ are 2-spinors;
$x\defi(x_0,x_1)$; $\r^\m(x)\defi \bar\ps_x\g^\m\ps_x$ is the
current; and the $\g$'s matrices are a realization of the Clifford
algebra
$$
 \g^0\defi
 \pmatrix{0&1\cr
          1&0\cr}\;,
 \qquad
 \g^1\defi
 \pmatrix{0&-1\cr
          1&0\cr}\;,
 \qquad
 \g^5\defi i\g^0\g^1
 =
 \pmatrix{i&0\cr
          0&-i\cr}\;,
\Eq(eca)$$
which, for $\h^{\m,\n}\defi \d_{\m,\n}(1-2\d_{\m,1})$, do satisfy the properties
$$\eqalign{
 \{\g^\m,\g^\n\}=2\h^{\m,\n}\;,
&\qquad
 \left(\g^\m\right)^\dagger=-\g^0_M\g^\m_M\g^0_M\;,\cr
 \{\g^5,\g^\m\}=0\;,
 \qquad
&\left(\g^5\right)^2=-1\;,
 \qquad
 \left(\g^5\right)^\dagger=-\g^0\g^5\g^0=-\g^5\;.
}$$
This model is enough simple to be analysed in full details; and
yet it contains many of the typical features of the quantization
of relativistic {\it quantum field theories} (QFT), such as the
{\it anomalous scaling} -- as conjectured in QED, [JZ];
and the {\it anomalous  phase and chiral symmetries} -- like the
anomalous chiral symmetry of QED or Standard Model.

As peculiarity of the $1+1$ spacetime dimension, since there are
only two independent component of the current, the invariance of
the classical massless Lagrangian under phase transformation
$\psi_x\to e^{i\a} \psi_x$ and under chiral transformation
$\psi_x\to e^{i\g^5\a_{x}} \psi_x$ led to the hope to find an
exact solution also for the quantum massless model.

First, Thirring, [T58], derived many matrix elements of the
interacting field; then, Glaser, [G58], gave an explicit formula
for such a field operator, arising the criticism of Pradhan and
Scarf. The breakthrough had place with Johnson, [J61], who first
found the expression for the two point Schwinger functions which,
until nowadays, has been accepted as the {\it exact solution}. In
the end, Klaiber, [K64], with a slightly different technique,
wrote out the general formula for all the Schwinger functions. All
this story is commented upon in [W64]; here it is worthwhile to
stress that {\it all above papers} were plagued by the typical
infinities of relativistic QFT: the virtue of Johnson's
development merely was a greater solidity of the final result.

A remarkable feature in [J61] is the presence of {\it anomalies}
in the
\hbox{Ward-Takahashi} identities (\wti): they
occur -- some years {\it before} the discovery of Adler, [A70] --
as a modification of the field-current commutation relations,
simply guessed in order to avoid triviality of the identities.

Remarkable as well is the procedure of joining of the
Schwinger-Dyson equation (\sde) together to the phase and chiral
\wti, in order to obtain a Closed Equation (CE) for the two point
Schwinger function which can be solved straightforwardly.

In order to clear the result of all the surreptitious calculations
with infinities, Wightman, [W64], stressed that the set of
Schwinger function of Johnson and Klaiber, no matter how they were
derived, only represent {\it good candidates}: if they verified
the requirements of an axiomatic program, they would define a QFT
to be called ``Thirring model" essentially by definition. But no
kind of {\it positive definiteness} of inner product of physical
Hilbert space has ever been possible to prove; up to recent years,
when in [M93] the {\it reflection positivity} was obtained as
consequence of the Hamiltonian formulation of a many particle
model, the Luttinger model, exactly soluble as showed in [ML65]
and in a sense close to the massless Thirring model.

The massive theory is  much less analysed, [GL72]. In such a case
no ``exact solution'' was ever found; as well as no physical
positive metric.

Now, a different point of view can be considered, the
Renormalization Group (\rg) approach {\it \`a la Wilson}. Such a
technique has been revealed very profitable for certain QFT, like
the Yukawa$_2$ model, [S75] and  [MS76], or the ultraviolet part
of Gross-Neveu model, [GK85] and [FMRS85]; the subtle point being
that all such models are superrinormalizable, or were studied in
asymptotically free regimes.

The Thirring model, instead, is renormalizable, but not
superrinormalizable; and no regime is asymptotically free, since
the effective coupling remains essentially constant
over every regime. This property, called {\it vanishing of Beta
function}, was already used in [BoM97] to point out the critical
behavior of the infrared regime of Yukawa$_2$ model; and 
it is a
consequence of the phase and chiral \wti{} -- in agreement with
the general belief that, without the aid of symmetries, \rg{} can
be effective only in constructing {\it trivial} theories.

As matter of fact, there is a basic conflict between the
regularization of the theory and the phase and chiral symmetries.
The situation is very similar to the scaling transformation: the
classical theory is scale invariant; the theory regularized with a
cutoff is no longer; removing the cutoff, scale invariance is
recovered, but with a different exponent, called anomalous. In the
same way, removing the cutoff, the \wti{} are recovered, but  a
change in the factor in front of the currents makes such
identities anomalous.

In recent times, Benfatto and Mastropietro, [BM01],[BM02],[BM04],[BM05],
have developed 
a technique to complete construction of Luttinger liquids 
without any reference to the exact solution of the Luttinger model.
As byproduct of their developments, the anomaly of the 
\wti{} arose.

The aim of this thesis is to use such a technique to
construct, by  a {\it self-consistent \rg{} approach},
uniform in the mass,
the Thirring model at imaginary time.  
And then to make a continuation to Minkowskian spacetime by
verifying the Osterwalder and Schrader axioms, (\osa). 
The occurrence of the phenomenon of {\it fermion doubling} --
peculiar of the discretization  on a lattice --
has been solved introducing a  momentum dependent mass term, as suggested 
in [W76], but also a mass counterterm which avoids the generation of mass in 
the massless theory. 

As main applications,  the anomalous \wti{} stated by Johnson are
derived and, as consequence, the current operator is proved not to
need any renormalization. Anyway, the explicit value of
the anomaly obtained by Johnson are wrong by lowest order calculation, 
and this is {\it in violation of  the Adler-Bardeen's
theorem}, [A69]. Also the rigorous implementation of the
Johnson's closure of the \sde{} is proved: it will be showed,
anyway, the arising of a {\it new anomaly}, missed in the formal
developments, which have driven Johnson to a {\it wrong anomalous
exponent}.
\chapter(RR,Definitions and Main Results)

\section(ETM,Euclidean Thirring Model)

Many properties of a quantum field theory
can be obtained from the {\it Schwinger functions}, 
the ``cumulants'', or the ``truncated expectations''
of a statistical measure which correspond to the 
{\it imaginary-time version} 
of the model. Such a measure 
can be conveniently formulated in terms of
a ``path integral'' on a lattice spacetime.
Since the fields dealt with are  
{\it fermions} -- namely only the case of anticommuting
fields is considered -- they are represented in 
the path integral formulation 
by Grassmannian variables.

\*
\sub(SWF){Weyl formalism.} While in Dirac 
notation of \equ(1) the independent fields are 
the \hbox{2-spinor} $\bar\ps$ and $\ps$, in Weyl notation
they are
$\hp_{k}\defi (\hp^-_{k,+},\hp^-_{k,-})^T,\hp^\dagger_{k}\defi
(\hp^+_{k,+},\hp^+_{k,-})$.
The Euclidean Clifford Algebra is defined to be:
$$\eqalign{
 \{\g^\m,\g^\n\}=2\d^{\m,\n}\;,
&\qquad
 \left(\g^\m\right)^\dagger=\g^\m\;,\cr
 \{\g^5,\g^\m\}=0\;,
 \qquad
&\left(\g^5\right)^2=1\;,
 \qquad
 \left(\g^5\right)^\dagger=\g^5.
}$$
Such requirements are fulfilled by the same $\g'$s matrices
in \equ(eca), by multiplying $\g^1$ and $\g^5$ by the 
imaginary unity; namely, from now on 
the definitions in \equ(eca) are turned into:
$$
 \g^0\defi
 \pmatrix{0&1\cr
          1&0\cr}\;,
 \qquad
 \g^1\defi
 \pmatrix{0&-i\cr
          i&0\cr}\;,
 \qquad
 \g^5\defi
 -i\g^0\g^1
 =
 \pmatrix{1&0\cr
          0&-1\cr}\;.
$$
Accordingly, 
the Euclidean Action, for mass $\m$ and coupling $\l$,
 is defined to be:
$$\eqalign{
&\sum_{\o,\s=\pm}
 \int\!
 {\der^2k\over (2\p)}\
 \hp^+_{k,\o}T_{\o,\s}(k)
 \hp^-_{k,\s}\cr
&\phantom{***********}
 -
 {\l\over 2}\sum_{\o=\pm}
 \int\!{\der^2p\over (2\p)^2}{\der^2q\over (2\p)^2}
 {\der^2k\over (2\p)^2}\
 \hp^+_{p,\o}\hp^-_{q,\o}\hp^+_{k,-\o}\hp^-_{p+k-q,-\o}\;,}
\Eq(el)$$
where the coefficients of the quadratic part are
$$ T_{\o,\s}(k)
 \defi
 \pmatrix{ D_+(k) & -\m \cr
           -\m & D_-(k)\cr}_{\o,\s}\;,
 \qquad
 {\rm with}\quad D_\o(k)\defi -ik_0+\o k_1\;.$$

\*
\sub(STL){Spacetime Lattice.}
Let $a$ and $L$ be respectively the  spacing 
and the side length of the lattice to be constructed, 
such that $L/a$ is an integer. 
Then, in correspondence of such parameters, let the quotient set $Q$ 
be defined as
$$Q\defi\left\{(n_0,n_1)\in \ZZZ^2
\Big|\ n\sim n'\ {\rm if}\ n-n'\in {L\over a}\ZZZ^2 \right\}\;;$$
the spacetime lattice, $\L$, and its reciprocal one, $D$, are defined as
$$\L\defi\{an_0,an_1\left|n\in Q\right.\}\;,
 \qquad
 D
 \defi
 \left\{{2\p\over L}\left.\left(m_0+{1\over 2}\right),{2\p\over L}
 \left(m_1+{1\over 2}\right)
 \right|m\in Q\right\}\;.$$
To shorten the notation, the Riemann sums 
on the lattices are denoted with integrals
$$\int_{\L}\!\der^2x\ 
f(x)\defi a^2\sum_{x\in\L} f(x)\;,
\qquad
\int_{D}\!\der^2k\ 
\wh f(k)\defi 
\left({2\p\over L}\right)^2\sum_{k\in D} 
\wh f(k)\;.
\Eq(rs)$$

\*
\sub(FI){Grassmann Algebra.}
In correspondence of the fields in \equ(el), 
there are  four sets
of Grassmann variables that, {\it with abuse of notation},
are called $\{\hp_{k,\o}^\s\}_{\s,\o=\pm}^{k\in D}$ as well. 
The integration in such a {\it finite algebra} is defined 
so that the integral of a constant is 
zero, while 
$$\int\!\!\der\hp^{\s'}_{k',\o'}\ \hp^\s_{k,\o}=\d_{\s',\s}\d_{k',k}\d_{\o',\o}\;;$$
then the operation is extended by linearity to any polynomial of fields,
considering $\big\{\der \hp_{k,\o}^\s\big\}_{\o,\s}^{k\in D}$ anticommuting 
with themselves and with all the fields.
As consequence, the integration of the monomial $\QQ(\ps)$,
$\int\!\prod_{k\in D}
\prod_{\o=\pm}\der\hp_{k,\o}^+\der\hp_{k,\o}^-\QQ(\ps)$, 
 assigns 1 to
\hbox{$\QQ(\ps)=\prod_{k\in D}\prod_{\o=\pm} \hp_{k,\o}^- \hp_{k,\o}^+$},
and 0 to all the other $\QQ'(\ps)$ which cannot be obtained
as permutation of fields in $\QQ(\ps)$.

The derivative in the Grassmann algebra is defined to be 
equivalent to the integration:
$${\partial \QQ(\ps)\over \partial \hp^+_{k,\o}}
 \defi\int\!\!\der\hp^+_{k,\o}\ \QQ(\ps)\;,
 \qquad
 {\partial \QQ(\ps)\over \partial \hp^-_{k,\o}}
 \defi-\int\!\!\QQ(\ps)\ \der\hp^-_{k,\o}$$
-- hence the derivative in $\hp^-_{k,\o}$ {\it acts from the right}.

\*
\sub(SF){Schwinger functions.}
In order to give a meaning to the path integral formulation of
the Schwinger function, it is necessary 
to introduce a ``cutoff function'', $\c_N(k)$, made as follows.
Let a momentum unity, $\k$, be fixed. Chosen any $\g>1$,
let $N$ be any integer such that 
\hbox{$\k \g^{N+1}\leq 3\p/4a$}.
Then, let $\wh\c_N(t)$ be a $C_0^\io(\RRR)$ function
with compact support $\{t\in \RRR: |t|\le \k\g^{N+1}\}$
and $\wh\c_N(t)\=1$ in \hbox{$\{t\in \RRR:|t|\le \k\g^N\}$.}
Besides, because of technical reason, 
it is convenient to take $\wh\c_N$
in the Gevrey class $\a$: for a positive constant $C$, 
$$\sup_{t\in \rrr}\left|{\der^n\wh\c_N\over \der t^n}(t)\right|
\leq C^n (n!)^\a\;;$$
in particular, $\a=2$ will be good enough.
The possibility of constructing
such a compact support function is discussed in \secc(GCF).
Finally, $\c_N(k)\defi \wh \c_N(k_0)\wh\c_N(k_1)$.
Calling $D_N\subset D$ the support of  $\c_N(k)$,
the {\it Generating Functional} of the Schwinger functions of the
Thirring model is defined to be $\WW(\jm,\f)$:
in correspondence of 
certain parameters $\l_N$, $\m_N$, $Z_N$ and $\z^{(2)}_N$, 
it is such that
$$\eqalign{
 e^{\WW(\jm,\f)}
 \defi
 \int\!\!\der P^{(\leq N)}(\ps)
 \exp
&\left\{ -\l_N  \VV\left(\sqrt{Z_N}\ps\right)+
\z_N^{(2)}\JJ(\jm,\sqrt{Z_N}\ps)
+\FF\left(\f,\ps\right)\right\}\;.}\Eq(gf)$$
The explanation of the above formula is the following. 
The integration 
is done w.r.t. the normalized Gaussian measure given by
$$\eqalign{
 \der P^{(\leq N)}(\ps)
 \defi
&\exp\left\{L^2O_N- Z_N\sum_{\a,\b=\pm}
 \int_{D_N}\!{\der^2k\over (2\p)^2}\
 {T_{\o,\s}(k)\over \c_N(k)}
 \hp_{k,\o}^+\hp_{k,\s}^-\right\}\cr
&\prod_{k\in D_N}\prod_{\o=\pm}\der\hp_{k,\o}^+\der\hp_{k,\o}^-
\;,}\Eq(distr)$$
where  the {\it covariance}
$\hg_{\o,\s}(k)$ is such that:
$$
 \hg^{-1}(k)\defi
 {T(k) \over \c_N(k)}\;,
 \qquad{\rm with}\quad
 T_{\o,\s}(k)
 \defi
 \pmatrix{ D_+(k) & -\m_N \cr
           -\m_N & D_-(k)\cr}_{\o,\s}\;;$$
hence $\hg(k)$ is periodic by the compact support of $\c_N$
and well defined for any $k\in D$, 
also for $\m_N=0$, since the point $(0,0)$
does not belong to $D$. As well as 
$\hg^{-1}(k)$ is well defined in $D_N$, 
since the points in which the cutoff is
zero do not belong to $D_N$. 
The factor $e^{O_N}$ is the normalization of the 
Gaussian measure:
$$O_N\defi\int_{D_N}\!{\der^2k\over (2\p)^2}\
\ln\lft({L^4|k|^2\over \c_N^2(k)}\rgt)\;.$$

The self-interaction is given by the potential
$$\VV(\ps)
 \defi
 {1\over 2}\sum_\o\int_{D}\!{\der^2p\over (2\p)^2}
 {\der^2q\over (2\p)^2}{\der^2k\over (2\p)^2}\
 \hp^+_{p,\o}\hp^-_{q,\o}\hp^+_{k,-\o}\hp^-_{p+k-q,-\o}\;;$$
while the interaction 
with the external sources are
$$\eqalign{
 \JJ_\s(\jm,\ps)
&\defi
 \sum_\o\int_{D}\!{\der^2k\over (2\p)^2}{\der^2p\over (2\p)^2}\
 \hj_{p-k,\o}\hp^+_{k,\s\o}\hp^-_{p,\s\o}\;,\cr
 \FF(\f,\ps)
&\defi
 \sum_\o\int_{D}\!{\der^2k\over (2\p)^2}\
 \left[\hf^+_{k,\o}\hp^-_{k,\o}+\hp^+_{k,\o}\hf^-_{k,\o}\right]\;;}$$
and $\{\hj_{k,\o}\}_{k,\o}$ are a commuting variable, while
$\{\hf^\s_{k,\o}\}_{k,\o,\s}$ are anticommuting.

Finally, w.r.t. the classical Action \equ(1), 
$\l$ has been replaced with $\l_N Z^2_N$, the ``bare coupling'';
$\m$ with $\m_N$, the ``bare mass'';
the free action was multiplied times $Z_N$, the ``field strength'';
and the interaction with the external source $\jm$ 
brings a coupling $Z^{(2)}_N\defi \z^{(2)}_N Z_N$,
the ``density strength'':
such  parameters are essential in order to have a finite
interactive quantum theory, see Theorem \thm(T1). 
Besides, in has to be remarked 
that the introduction of the cutoff has required a reference
momentum, $\k$, absent in the classical action of the
massless theory, which will allow the arising of the anomalous
dimension without violating the scaling symmetry.

The Fourier transform of the fields defines
a Grassmann algebra also in the lattice $\L$.
The conventions are:
$$\ps_{x,\o}^\s\defi\int_D\!{\der^2k\over (2\p)^2}\ 
e^{i\s kx} \hp_{k,\o}^\s\;;
\qquad
\f_{x,\o}^\s\defi\int_D\!{\der^2k\over (2\p)^2}\ 
e^{i\s kx} \hf_{k,\o}^\s\;;$$
$$\jm_{x,\o}\defi\int_D\!{\der^2k\over (2\p)^2}\ 
e^{i kx} \hj_{k,\o}\;.$$
The definition of derivative extends also 
to the fields $\big\{\ps^\s_{x,\o}\big\}_{\o,\s=\pm}^{x\in \L}$,
$\big\{\hf^\s_{k,\o}\big\}_{\o,\s=\pm}^{k\in D}$ and 
$\big\{\f^\s_{x,\o}\big\}_{\o,\s=\pm}^{x\in \L}$;
while the derivative w.r.t. the fields 
$\big\{\hj_{k,\o}\big\}_{\o=\pm}^{k\in D}$ and 
$\big\{\jm_{x,\o}\big\}_{\o=\pm}^{x\in \L}$ is the conventional one. 

Well then, setting $\ux\defi x^{1},\ldots,x^{n}$,
and $\uz\defi z^{1},\ldots,z^{m}$,
collections of points in $\L$,
for any given choice of the labels 
$\us\defi(\s_1\ldots,\s_m)$,
$\uo\defi(\o_1\ldots,\o_n)$
and $\ue\defi(\e_1\ldots,\e_n)$,
the Schwinger functions are defined as
$$
 S^{(m;n)(\ue)}_{\us;\uo}(\uz;\ux)\defi
 {\partial^{n+m}\WW\over\partial
 \jm_{z^{1},\s_1}\cdots\partial\jm_{z^{m},\s_m}
 \partial\f^{\e_1}_{x^{1},\o_1}\cdots\partial\f^{\e_n}_{x^{n},\o_n}}
 (0,0)\;.\Eq(sd11)$$
 In order to
shortening the notations of the Schwinger functions
which will be most used in the following, let
$$S^{(2)}_{\o}(x-y)\defi S^{(0;2)(-,+)}_{\o,\o}(x,y)\;,
\qquad
 S^{(1;2)}_{\s;\o}(z;x-y)\defi S^{(1;2)(-,+)}_{\s;\o,\o}(z;x,y)\;.$$

\*
\sub(c){Remarks.}
The role of the lattice discretization is only to have 
a finite Grassmann algebra: the 
{\it continuous limit}, 
$\k L,(\k a)^{-1}\to\io$
is taken as soon as the Schwinger function are derived; 
it is trivial, since, on the other hand, the use of 
the functional integral suggest, 
but it is not strictly necessary to, the developments.

On the contrary, the function $\c_N$ is an essential cutoff
on the large momenta: the parameters  
$\l_N$, $\m_N$, $Z_N$ and $\z^{(2)}_N$
will be chosen in a way to compensate 
the divergences of 
the {\it limit of removed cutoff}, $N\to +\io$, 
of the Schwinger functions.

\*
\theorem(T1){
There exists $\e>0$ and two positive constant,  
$c$ and $C$, such that, for any $\l:|\l|\le \e$
and  $\m:0\leq \m\leq \k\g^{-1}$,
and for suitable $\l_N$, $\m_N$, $Z_N$ and $Z^{(2)}_N$,
analytic function of $\l$,
the following properties of the Schwinger functions hold.
\elenco{
\art There exist three critical indices, $\h_{\l}$,
$\h^{(2)}_{\l}$, and 
$\bar\h_{\l}$, independent from the cutoff scale $N$ 
and from the mass $\m$, analytic functions of $\l$ and  such that 
$$\h_\l=\h_{2}\l^2+{\rm O}(\l^3)\;,\qquad
\h^{(2)}_\l=\h^{(2)}_{2}\l^2+{\rm O}(\l^3)\;,$$ 
$$\bar\h_\l=-\bar\h_{1}\l+{\rm O}(\l^2)\;,$$ 
with $\h_{2}$, $\h^{(2)}_{2}$ 
and $\bar\h_{1}$ strictly positive; and, for any $N$, 
$$Z_N=\g^{-N\h_\l}\big(1+{\rm O}(\l^2)\big)\;, 
\qquad
Z^{(2)}_N=\g^{-N\h^{(2)}_\l}\big(1+{\rm O}(\l^2)\big)\;,$$
$$\m_N=\m\g^{-N\bar\h_\l}\big(1+{\rm O}(\l))\;,$$
where ${\rm O}(\l)$ are finite in $N$.
\art 
In the limit of removed cutoff, the Schwinger
function are well defined distribution, 
fulfilling the \osa{}.
\art
In the limit of removed cutoff,
the two point Schwinger function verifies 
the bound
$$\lft|S^{(2)}_{\o}(x-y)\rgt|
\leq
{\k C\over \big(\k|x-y|\big)^{1+\h_{\l}}}
e^{-c\sqrt{\lft({\m\over \k}\rgt)^{1+\bar\t}{\k|x-y|}}}\;,$$
for $\bar\t\defi -\bar\h_\l/(1+\bar\h_\l)$.
The same bound holds also for  
$S^{(0;2)(-,+)}_{\o,-\o}(x,y)$.
\art
In the limit of removed cutoff and 
of vanishing mass, i.e. $\m=0$, 
$$S^{(2)}_{\o}(x-y)=
 (1+\l B_{\l})\int\!{\der^2 k\over (2\p)^2}\ e^{-ik(x-y)}
 {1\over D_\o(k)}
 \lft({|k|\over \k}\rgt)^{\h_\l}\;,\Eq(sf)$$
with $B_{\l}$ analytic and O$(1)$ in $\l$. 
While $S^{(0;2)(-,+)}_{\o,-\o}(x,y)\=0$.}}

\0The proof of the first three statements is
obtained by the analysis in Chapter \secc(R), 
the study of the flows of the effective couplings
in \secc(BGF),
the convergence of the Schwinger functions, \lm(SF)
and \secc(P3P1), 
and by the equivalence of the Euclidean and Hamiltonian 
regularization, \secc(PTM). 
The fourth statement is consequence of
symmetries: see \secc(SCE).

 The \osa{} are reported in Appendix \secc(OS). When they hold,
the \hbox{Osterwalder-Schrader} reconstruction theorem guarantees the
possibility of analytically continuing the set of Schwinger
functions to a set of functions obeying the Wightman axioms:
this means the construction of a consistent
relativistic and quantum field theory.

By item 2., the parameters $Z_N$ and $Z^{(2)}_N$ are
 vanishing in the limit of removed cutoff;
whereas $\m_N$ is vanishing or diverging according to the sign 
of $\l$.

\*
\sub(1.3){Ward-Takahashi identities: first anomaly.} In the
massless case, the phase and chiral symmetry makes current
expectations and field expectations strictly related.
By neglecting {\it formally} the presence of the cutoffs, and
performing a combination of the phase and chiral transformation of
the fields, it holds the following identity for the Fourier transform
of such  Schwinger functions:
$$ {D_\s(p)\over \z^{(2)}_N}
 \hS^{(1;2)}_{\s;\o}(p;k)
 =\d_{\s,\o} \left[\hS^{2}_{\o}(k)
-\hS^{2}_{\o}(k+p)\right]\;. \Eq(wil0)
$$
\vbox
{\insertplot{330pt}{90pt}%
{}%
{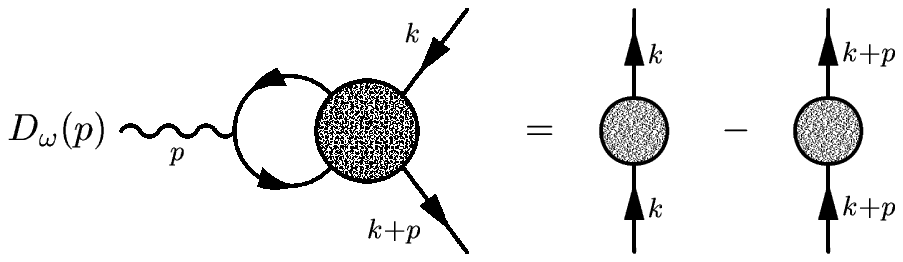}{\eqg(f23)}
\centerline{{\bf Fig \graf(f23)}: Graphical representation of \equ(wil0)}}
\vskip2em
This relation is actually {\it wrong}. Indeed, the presence of the
cutoff -- {\it essential ingredient} of meaningful QFT's -- breaks the
symmetries and generates a correction term
$\hH^{(1;2)}_{\s;\o}$:
$$
 {D_\s(p)\over \z^{(2)}_N} \hS^{(1;2)}_{\s;\o}(p;k)
=\d_{\s,\o}
 \left[\hS^{(2)}_{\o}(k) -\hS^{(2)}_{\o}(k+p)\right]
 +\hH^{(1;2)}_{\s;\o}(p;k)\;.
 \Eq(wi)
$$
What is at first sight surprising is that in the limit of removed
cutoff  {\it the corrections are not vanishing}; and yet  {\it
anomalous \wti{}, strictly different from \equ(wil0)}, are  valid.

\*
\theorem(P2)
{\it There exists $\e>0$ and two positive constants,  
$c$ and $C$,
such that, for any \hbox{$\l:|\l|\le\e$} and $\m:0\le \m\le \k\g^{-1}$,
the following properties hold.
\elenco{
\art
For $\m=0$,
there exists two ``bare parameters'', $\l_b$ and 
$\z^{(2)}_b$, analytic in $\l$, 
such that the coupling $\l_N$ and the 
field strength $\z^{(2)}_N$, as chosen in Theorem \thm(T1),  
are independent form 
the scale of the cutoff, $N$; and are
\hbox{$\l_N=\l_b$},  
\hbox{$\z^{(2)}_N=\z^{(2)}_b$}.
\art
 For $\m=0$, there exist two coefficients, 
$a$ and $\bar a$, analytically
dependent on $\l$, such that 
$${1\over \z^{(2)}_b}\hS^{(1;2)}_{\s;\o}(p,k)
 ={a+\bar{a}\s\o\over2}\
 {\hS^{(2)}_{\o}(k)-\hS^{(2)}_{\o}(k+p)\over  D_\s(p)}\;,
 \Eq(kjk)$$
with $(a+\bar a\s\o)/2\neq \d_{\o,\s}$ whenever $\l\neq 0$.
\art The current-current correlation satisfies the bound
$$\left|S^{(2;0)}_{\s,\o}(x,y)\right|\leq {C\over (\k|x-y|)^2}
 e^{-c\sqrt{\k\left({\m\over\k}\right)^{1+\bar\t_{\l}}|x-y|}}\;,\Eq(dcc)$$
for any allowed value of the mass $\m$.}}

\*
\0The coupling $\l_N$ and the density strength $\z^{(2)}_N$
do not depend on the cutoff scale since,
the mass being zero, the theory is scaling invariant.
The second statement is a sub-case of Theorem \thm(LWTI);
while the third is proved in \secc(P2P2). 

By item 3, the short distance behavior is the same as
in the  free theory: no critical index occurs and changes the 
exponent $2$ of $1/(\k|x-y|)$. 

It is interesting to see how the anomalous \wti{} arises. It is
possible to find two finite counterterms, $\n^{(+)}$ and $\n^{(-)}$,
analytically dependent on $\l$ and independent 
on $N$,  such that  the correction can be
decomposed as
$$\eqalign{
 \hH^{(1;2)}_{\s,\o}(p;k)
=&\n^{(+)} D_\s(p)\hS^{(1;2)}_{\s;\o}(p;k)
 +\n^{(-)} D_{-\s}(p)\hS^{(1;2)}_{-\s;\o}(p;k)\cr
&+\D \hH^{(1;2)}_{\s;\o}(p;k)\;;} \Eq(gg)$$
and, for $p$ and $k$ {\it fixed independently from $N$},
the rest $\D\hH^{(1;2)}_{\s;\o}(p;k)$ is now really 
vanishing. To adhere to the Johnson's notation, let
$$
 a\defi{1\over 1-\left(\n^{(-)}+\n^{(+)}\right)}\;,
 \qquad
 \bar a\defi{1\over 1-\left(\n^{(-)}-\n^{(+)}\right)}\;;
$$
then, replacing \equ(gg) in \equ(wi), and taking the limit of
removed cutoff, gives \equ(kjk). Johnson's \wti{} is precisely given by
\equ(kjk); and his explicit values for $a$ and $\bar a$ 
are in agreement with the Adler-Bardeen theorem on absence
of radiative correction to the anomaly. Anyway, these values are 
{\it wrong}: while Johnson states $\n^{(+)}=0$, by lowest order 
computation, for $\l$ small enough, $\n^{(+)}<0$ (see \secc(LOC)). 

Despite the anomaly, and  despite 
the phase and chiral symmetry hold only in the massless case,
it is possible to prove the finiteness of 
the limit value of $\z^{(2)}_N$,
{\it even in the massive model}; and accordingly 
the finiteness  of  
the current-current Schwinger function,  with no arising
of an anomalous exponent.

\*
\sub(SA){Closed equation: new anomaly}. 
The fields
equation can be turned into an equation for the Schwinger
function, the {\it Dyson-Schwinger equation}. In 
the massless case, the one for  the two point Schwinger function
reads
$$\eqalign{
  {\hS^{(2)}_{\o}(k)\over g_\o(k)}
 =
 {1\over Z_N}
 -{\l_b\over \z^{(2)}_b}
 \int_{D}\!{\der^2 p\over (2\p)^2}\ \hS^{(1;2)}_{-\o;\o}(p;k-p)\;.
}\Eq(DS2)$$
\vbox
{\insertplot{330pt}{60pt}%
{}%
{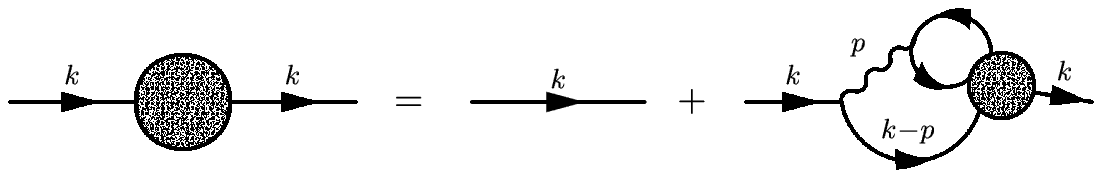}{\eqg(f24)}
\centerline{{\bf Fig \graf(f24)}: Graphical representation of \equ(DS2)}}
\vskip2em
Inserting the \wti{} \equ(wi) and the identity
\equ(gg) in \equ(DS2), since
$\int_{D}\!\der^2 p \ D^{-1}_{-\o}(p)=0$ by oddness,
$$\eqalign{
 {\l_b\over \z^{(2)}_b}
 \int_{D}\!{\der^2 p\over (2\p)^2}\
 \hS^{(1;2)}_{-\o;\o}(p;k-p)
&={a-\bar{a}\over2}\l_b
  \int_{D}\!{\der^2p\over (2\p)^2}\ {\hS^{(2)}_{\o}
  (k-p)\over D_{-\o}(p)}\cr
&+
 \sum_\m {a-\m\o\bar{a}\over2}\l_b
 \int_{D}\!{\der^2p\over (2\p)^2}\
 \D \hH^{(1;2)}_{\m;\o}(p;k-p)\;.
}\Eq(intwti)$$
In the limit of removed cutoff, {\it if} the integral of
$\D\hH^{(1;2)}_{\m;\o}$ had been vanishing, \equ(intwti)
would have been turned into 
$$\eqalign{
 {\l_b\over \z^{(2)}_b}
 \int_{D}\!{\der^2 p\over (2\p)^2}\
 \hS^{(1;2)}_{-\o;\o}(p;k-p)
&={a-\bar{a}\over2}\l_b
  \int_{D}\!{\der^2p\over (2\p)^2}\ {\hS^{(2)}_{\o}
  (k-p)\over D_{-\o}(p)}\;.}\Eq(intwti)$$
Replacing it  into \equ(DS2),
it would have held the equation
$${\hS^{(2)}_{\o}(k)\over g_\o(k)}
 ={1\over Z_N}
 -{a-\bar a \over 2}\l_b 
 \int_{D}\!{\der^2p\over (2\p)^2}\
 {\hS^{(2)}_{\o}(k-p)\over D_{-\o}(p)}\;,\Eq(rene)$$
where $1/Z_N$ is divergent and should compensate 
the divergence of the integral.
The above equation, {\it in a sense stated by Johnson}
-- actually his operations were even more formal; but his final finite solution
is exactly the solution of \equ(rene) -- is {\it wrong}.
Indeed, $\D\hH^{(1;2)}_{\m;\o}$ was said to be vanishing only
for fixed arguments, while here it is integrated over all the
scales allowed by the cutoff. This seems to waste the possibility
of the closure of the \sde; and yet, again, an {\it anomalous \ce}
still holds.

\*
\theorem(P3)
{Under the same assumptions of Theorem \thm(T1):
\elenco{
\art
 The following equation holds,
asymptotically in the limit of removed cutoff
$${\hS^{(2)}_{\o}(k)\over g_\o(k)}
 ={B_N\over Z_N}
 -A \l_b  {a-\bar{a} \over 2}
 \int_{D}\!{\der^2p\over (2\p)^2}\
 {\hS^{(2)}_{\o}(k-p)\over D_{-\o}(p)}\;,\Eq(secan)$$
where $A$, the ``new anomaly'', is analytic and O$(1)$ in $\l$;
while $B$ is $1+{\rm O}(\l)$ and analytic in $\l$ as well.
\art
It holds the following relations between the anomalous
exponent and the coefficients in the first and second anomaly:
$$\h_\l= A {\l_b\over 2\p} {a-\bar a\over 2}\;.\Eq(jj)$$
}}

\0This result is a sub-case of Theorem \thm(LCE),
with the explicit expression of $\h_\l$
is discussed in \secc(SCE).

The name ``new anomaly'' is justified since 
such is an effect of using 
a symmetry, exact only at removed cutoff,
inside an integral which in the same limit is divergent; 
it has been overlooked not only in rigorous works, 
but even in the physical literature.
In particular, $A\neq 1$ would imply a 
{\it striking and net difference 
w.r.t. the Johnson critical index}.

Such a difference could have been 
checked directly by lowest order computation of $\h_\l$ itself;
but, since the fourth  is the first non-trivial order,
the actual computation is almost prohibitive. 
Therefore \equ(jj) is a shortcut, since it gives
$\h_\l$ in terms of
the easier  calculations of $a-\bar{a}$ and $A$. 

Now, by symmetry reasons,  the first order of $A$ is equal to 1,
while $a-\bar a={\rm O}(\l)$: 
this is in agreement with the the fact that 
$\h_\l$ is an even function of $\l$ -- as can 
be easily proved by transformation \hbox{$\hp^{\s}_{k,\o}\to \hp^{\s}_{\s k,\s\o}$}
in the functional-integral 
measure.
But there is no general reason 
for which this result should survive also
 at the second order, at least for a {\it generic}
choice of the cutoff function: 
in \secc(LOC) there is a Montecarlo simulation
which {\it does not prove,} but enforces the clue  that $A\neq 1$.

It is appropriate to disclose here  the developments 
leading to \equ(secan), leaving to the next chapters
the proofs and the generalizations to the 
multi-point Schwinger functions.
For a suitable choice of four counterterms, 
$\{\a^{(\m)}\}_{\m=\pm}$ and $\{\s^{(\m)}\}_{\m=\pm}$,
analytically dependent on $\l$,
$$\eqalign{
&\l_b
 \sum_\m {a-\bar{a}\m\o\over2 }
 \int_{D}\!{\der^2p\over (2\p)^2}\
 \D\hH^{(1;2)}_{\m;\o}(p;k-p)
 =
\left(\sum_\m{a-\bar a\m\o\over 2}\s^{(\m\o)}\l_b\right){\hS^{(2)}_\o(k)\over \hg_\o(k)}\cr
 &+
 \left(\sum_\m{a-\bar a\m\o\over 2}\a^{(\m\o)}\l_b\right)
 {\l_b\over \z^{(2)}_b}
 \int_{D}\!{\der^2p\over (2\p)^2}\
 \hS^{(1;2)}_{-\o;\o}(p;k-p)+\D\hK_{\o}(k)}\;,$$
where,  for $k$ {\it fixed independently from $N$}, the rest $\D\hK_{\o}(k)$ is vanishing.
Putting together the above identity with \equ(intwti) and
\equ(DS2), \equ(secan) holds for
$$\eqalign{
 A&\defi{1\over 1-(\l_b/2)\sum_\m (a-\bar{a}\m)\left(\a^{(\m)}-\s^{(\m)}\right)}\;,\cr
 B&\defi{ 1-(\l_b/2)\sum_\m (a-\bar{a}\m)\a^{(\m)}
   \over 1-(\l_b/2)\sum_\m (a-\bar{a}\m)\left(\a^{(\m)}-\s^{(\m)}\right)}\;.}\Eq(A1)$$ 
\chapter(HR,Hamiltonian Regularization)
Two different {\it regularizations} of the Thirring model can be considered:
the Euclidean one, depicted in the previous Chapter, and the Hamiltonian one,
introduced in the present Chapter.  
As well as, two are the main requirements
of the \osa: 
the {\it Euclidean invariance} and the   
the {\it reflection positivity}.

Well then, the former property is evident 
only in the former regularization -- and even 
false in the latter, if 
the limit of removed cutoff is not taken;
while the latter property is built-in in the latter, 
and not so clear in the former.
  
But it is possible to prove
that, for two (in general) {\it different} choices of the parameters 
of the Lagrangian,
the two regularization, in the limit of 
removed cutoff, are equivalent,
in the sense that the Schwinger function 
derived in the one or in the other 
scheme are {\it exactly the same}. And 
therefore they fulfill both the crucial
properties.

This theorem is a first example of the effectiveness 
of the \rg{} approach, which is introduced in the next Chapter.

\*
\section(HM, Hamiltonian Thirring Model)
 
This time only the space is discretized. 
Then, a finite dimensional Fock space, together to a
many-body Hamiltonian, is built, guaranteeing {\it a priori} the
validity of the {\it reflection positivity} (see \secc(RP)) 
also after taking the continuum space limit.

Other constructions, different from the Hamiltonian formalism and
verifying such positivity property,  would have been possible: \eg
a certain lattice discretization of both space and time 
(different from the one in Chapter 1)
would have turned the quantum field model into a statistical mechanical
lattice model, nearest neighbours interactive, which is reflection
positive by standard proof, [OS77]. Anyway,
despite of the popularity of the latter route, here the former
is preferred, since the consequent integration of the {\it hard
fermions} (see later) was called upon, but never explicitly proved
in [BM01] and in the following papers -- where the setting can only
be Hamiltonian, since they deal with many-body quantum models.
As consequence, space and time are not managed on the same ground,
and the phenomenon  of {\it light velocity modification} occurs
(as first noticed in [M93]):  it is necessary to introduce a {\it
counterterm} to fix the light velocity to 1.

In any case, lattice discretization of  fermionic  QFT -- no matter if
it affects only the space or both space and time -- encounters
the well known problem of the {\it doubling of fermions}. In order
to make the effect of the double fermions to vanish, a
possibility is to use a momentum dependent mass term, [W76]; but
it destroys the symmetries of the propagators and generates a mass
term even in the massless theory: a counterterm also for the mass is
necessary, so that the mass on physical scale can be fixed
to chosen value $\m\geq 0$.

\*
\sub(1.2){Hamiltonian.} A finite dimensional Fock space
is constructed in terms of the periodic spatial lattice, $\L_1$,
as follows. Let $\k$ be fixed. Choosing $\g>1$ and integer, let $a$ and
$L$ be respectively the lattice spacing and the lattice side
length, s.t. $4\k a\defi \g^{-N}$ and $4\k L\defi \g^{-h}$, for
$N,-h$ large positive integers; then, the periodicity of the
lattice is given by the quotient set
$$
  Q_1\defi\left\{n \in \ZZZ
 \ \big|\ n\sim n'\ {\rm if}\ n-n'\in {L\over a}\ZZZ \right\},
$$
so that the lattice $\L_1$ and its reciprocal $D_1$ are
$$
 \L_1\defi\left\{an_1\ \big|\ n_1\in Q_1\right\}\;,
 \qquad
 D_1\defi \left\{{2\p\over L} \lft(m_1+{1\over 2}\rgt)\ \big|\ m_1\in Q_1\right\}\;.
$$
Now, let two couples of fermionic creation and destruction
operators $\{a^\s_{k_1,\o}\}_{k_1\in \L_1}^{\s,\o=\pm}$ be defined
with empty state $\ket{0}$; setting $c(k_1)\defi[1-\cos(k_1a)]/2a$,
$e(k_1)\defi \sin(k_1a)/a$ --
the Fourier transform of the discrete derivative in $x_1$ --
and, for any choice of the mass $\m\geq 0$, 
letting $\m(k_1)\defi \m+c(k_1)$ be the ``momentum dependent
mass term'', the free Hamiltonian is
$$
 H_0
 \defi {1\over L}
 \sum_\o
 \sum_{k_1\in \D_1}
 \o e(k_1)
 a^+_{k_1,\o}a^-_{k_1,\o}
 +
 {1\over L} \sum_\o \sum_{k_1\in \D_1}
 \m(k_1) a^+_{k_1,\o}a^-_{k_1,-\o}\;.
$$
In the limit $a\to 0$, the energy dispersion $e(k_1)$ is
asymptotic to {\it two} linear dispersion: one containing $k_1=0$,
which is the Euclidean Thirring dispersion; another one
containing $k_1=\p/a$, and representing the double fermions: the
role of $\m(k_1)$ is to assign
to the doubles a mass which is diverging with $N$.

The Hamiltonian is made interactive by the term 
$$
 {\l\over 2}
 {1\over L^3}\sum_\o
 \sum_{k_1,p_1,q_1\in D_1}
 a^+_{k_1,\o}a^-_{p_1,\o}a^+_{q_1,-\o}a^-_{k_1+q_1-p_1,-\o}\;.
\Eq(int)$$
As in the Euclidean regularization, 
the parameter of the Lagrangian have to be tuned 
so to have a finite theory.
Then, $\l$ and $\m$ are replaced with $\l_N$ and $\m_N$;
and $H_0$ is multiplied times the {\it field strength} $Z_N$.

Furthermore, to fix the mass to the chosen value 
and to have Schwinger functions with light velocity 
equal to 1 (as in the Euclidean regularization),
it is necessary to introduce two further counterterms  
$d_N$ and $\g^Nn_N$, such that, setting $\n_N\defi n_N/Z_N$ and $\d_N\defi d_N/Z_N$,
the interactive Hamiltonian finally reads
$$\eqalign{
 H
&\defi {1\over L}
 \sum_\o
 \sum_{k_1\in D_1}
 \o e(k_1) Z_N(1 +\d_N)
 a^+_{k_1,\o}a^-_{k_1,\o}\cr
 &+
 {1\over L} \sum_\o \sum_{k_1\in D_1}
 \Big(\m(k_1)+\g^N\n_N\Big)Z_N a^+_{k_1,\o}a^-_{k_1,-\o}\cr
 &+
 {\l_NZ^2_N\over 2}
 {1\over L^3}\sum_\o
 \sum_{k_1,p_1,q_1\in D_1}
 a^+_{k_1,\o}a^-_{p_1,\o}a^+_{q_1,-\o}a^-_{k_1+q_1-p_1,-\o}\;.
}\Eq(hint)$$ 
\sub(GF){Correlations.}
Let the fields and the density be
defined
$$
 \ps^\s_{x,\o}\defi
  e^{-x_0H}\lft({1\over L}\sum_{k_1\in D_1}e^{i\s  k_1 x_1}a^\s_{k_1,\o}\rgt) e^{x_0H}\;,
 \qquad
 \r^{R}_{x,\o}\defi Z^{(2,+)}_N\ps^+_{x,\o}\ps^-_{x,\o}+Z^{(2,-)}_N\ps^+_{x,-\o}\ps^-_{x,-\o}\;,
$$
where $Z^{(2,+)}_N$ and $Z^{(2,-)}_N$
are the density strengths: they are two,
rather than one as in the Euclidean regularization, 
since in this setting space and time are on different
ground and the symmetry which make $Z^{(2,+)}_N=Z^{(2,-)}_N$
is missing. 
 
For any 
$\uz\defi z^{(1)},\ldots,z^{(m)}$
and  
$\ux\defi x^{(1)},\ldots,x^{(n)}$,
fixed set on spacetime points
such that $0<z_0^{(1)}<z_0^{(2)}<\cdots<x_0^{(1)}<\cdots<x_0^{(n)}$,
the {\it correlations} are defined to be,
$$G^{(m;n)(\ue)}_{\us;\uo}(\uz;\ux)\defi
{\Tr\lft[e^{-LH} \r^R_{z^{(1)},\s_1}\cdots \r^R_{z^{(m)},\s_m}
\ps_{x^{(1)},\o_1}^{\e_1}\cdots \ps_{x^{(n)},\o_n}^{\e_n}\rgt]
\over \Tr\big[e^{-L H}\big]}\;,
\Eq(avr)$$
where  $\Tr$ is the
trace over a complete set of states of the quantum lattice model.
\*

\sub(TWT){Propagator.}
Also in this case the Schwinger function can be obtained in terms
of a path integral formula, and a Grassmannian integration.
The free Hamiltonian can
be diagonalized  in terms of a set of new creation and destruction
operators, $\{b^\s_{k_1,\o}\}_{k_1\in\L_1}^{\s,\o=\pm}$, and
energy dispersion $E(k_1)\defi \sqrt{e^2(k_1)+\m^2(k_1)}$:
$$ H_0=
 {1\over L}\sum_\o\sum_{k_1\in \L_1}
 \o E(k_1) b^+_{k_1,\o}b^-_{k_1,\o}\;,$$
where $b_{k_1,\m}^\s\defi\sum_\n a^\s_{k_1,\n}
\Big(C^{-1}(k_1)\Big)_{\n,\m}$ for
$$
 C(k_1)\defi\pmatrix{ \m(k_1)& E(k_1)-e(k_1)\cr
                      e(k_1)-E(k_1)& \m(k_1)\cr}
 {1\over \sqrt{\m^2(k_1)+\big[E(k_1)-e(k_1)\big]^2}}\;.
$$
Calling $T$ the time ordering, it is useful to define the 
{\it propagator}
as
$$\eqalign{
 g_{\a,\b}(x)
&\defi
 {\Tr\lft[e^{-L H_0} T \big(a^+_{k_1,\a}a^-_{k_1,\b}\big)\rgt]
\over \Tr\big[e^{-L H_0}\big]}
 =
 \sum_\o
 {\Tr\lft[e^{-L H_0} T \big(b^+_{k_1,\o}b^-_{k_1,\o}\big)\rgt]
 \over \Tr\big[e^{-L H_0}\big]}
 C(k_1)_{\o,\a}C(k_1)_{\o,\b}
 \cr
&=
 {1\over L} \sum_{k_1\in D_1}
  e^{-ix_1k_1 -x_0 \o E(k_1)}\cr
&\phantom{*******}
 \cdot\sum_\o
 \left\{{\c(x_0>0)\over 1+e^{-\o E(k_1)L}}
 -
 {\c(x_0\leq0)e^{-\o E(k_1)L}\over 1+e^{-\o E(k_1)L}}
 \right\}
 C(k_1)_{\o,\a}C(k_1)_{\o,\b}\;.
}$$
By {\it partial-fraction expansion} of the meromorphic functions in the curl
brackets (see \secc(PFE)), the propagator is
turned into:
$$
 g_{\a,\b}(x)
 =\lim_{M\rightarrow \infty}
 {1\over L\b} \sum_{k\in D}
 e^{-ik\cdot x}
 {\wh\c_M(k_0)
 \over
 \m_N^2(k_1)+k^2_0+e^2(k_1)}
 \pmatrix{ik_0+e(k_1)&\m_N(k_1)\cr
 \m_N(k_1)&ik_0-e(k_1)\cr}_{\a,\b}\;,
\Eq(prop)$$
with $D\defi D_0\times D_1$ and $D_0\defi
\left\{{2\p\over\b}(m+{1\over 2})\right\}_{m\in \zzz}$
(namely $D$ is the product of a periodic lattice
in the space direction times an unbounded one in the time
direction); $\wh\c_M(k_0)$ a non-negative, smooth cutoff,
introduced to give a meaning to the above expression -- which is a
{\it generalized summation} of a series which does not converge in
absolute sense. Specifically, with reference to
the function $\wh\c_N(t)$ defined in  
\secc(SF), the cutoff is defined to be 
$\wh\c_M\defi \wh\c_N\left(\g^{-M+N}t\right)$. 
\*

\sub(SFHR){Schwinger functions.}
As well know consequence of the Trotter formula
for the expansion of the evolution operator, 
$e^{x_0H}$, and the Wick theorem  (see for instance [FW]),
the correlations in \equ(avr)
can be generated from the 
functional $\ZZ(\jm,\f)\defi e^{\WW(\jm,\f)}$,
where, in its turn, $\WW(\jm,\f)$ is defined to be the 
{\it generating functional of the Schwinger function
in the Hamiltonian regularization}:
$$\eqalign{
 e^{\WW(\jm,\f)}
 \defi
 \int\!\der P^{(\leq M)}(\ps)
 \exp
&\left\{ -\l_N  \VV\left(\sqrt{Z_N}\ps\right)
+\g^N\n_N\NN\left(\sqrt{Z_N}\ps\right) +\d_N
\DD\left(\sqrt{Z_N}\ps\right)\right.\cr &\left.+
\sum_{\s}\z_N^{(2,\s)}\JJ_\s(\jm,\sqrt{Z_N}\ps)
+\FF\left(\f,\ps\right)\right\}\;. }\Eq(gf)$$
The settings are the following. 
The Gaussian free measure is given by
$$\eqalign{
 \der P^{(\leq M)}(\ps)
 \defi
&\exp\left\{L^2O_N- Z_N\sum_{\a,\b=\pm}
 \int_{D_M}\!{\der^2k\over (2\p)^2}\
 {T_{\a,\b}(k)\over \wh\c_M(k_0)}
 \hp_{k,\o}^+\hp_{k,\o}^-\right\}\cr
&\prod_{k\in D_M}\prod_{\o=\pm}\der\hp_{k,\o}^+\der\hp_{k,\o}^-\;,}
\Eq(distr)$$
where $O_M$ is the normalization,
$\z_N^{(2,\s)}\defi Z^{(2,\s)}_N/Z_N$ and the covariance
$\hg_{\m,\n}(k)$ is:
$$\hg^{-1}(k)\defi 
  {T(k)\over \wh\c_M(k_0)}\;,
  \qquad
  T(k)\defi
  \pmatrix
  {-ik_0+e(k_1) & \m_N(k_1) \cr
       \m_N(k_1)  & -ik_0-e(k_1)\cr}\;,$$
with
$$e(k_1)\defi{\sin(k_1a)\over a}\;,\qquad
 \m_N(k_1)\defi{1-\cos(k_1a)\over a}+\m_N\;;
\Eq(rr)$$
the lattice $D_M\defi\{k\in D:\wh\c_M(k_0)\neq 0\}$; the
self-interaction is given by the potentials
$$\VV(\ps)
 \defi
 {1\over 2}\sum_\o\int_{D}\!{\der^2p\over (2\p)^2}{\der^2q\over (2\p)^2}{\der^2k\over (2\p)^2}\
 \hp^+_{p,\o}\hp^-_{q,\o}\hp^+_{k,-\o}\hp^-_{p+k-q,-\o}\;,$$
and
$$\DD(\ps)\defi\sum_\o\int_D\!{\der^2 k\over (2\p)^2}\
 \o\e(p_1)\hp^+_{p,\o}\hp^-_{p,\o}\;,
 \qquad
 \NN(\ps)\defi\sum_\o\int_D\!{\der^2 k\over (2\p)^2}\
 \hp^+_{p,\o}\hp^-_{p,-\o}\;.$$
In order to generate the Schwinger functions, there are also
interactions with external sources:
$$\JJ_\s(\jm,\ps)\defi
 \sum_\o\int_{D}\!{\der^2k\over (2\p)^2}{\der^2p\over (2\p)^2}\
 \hj_{p-k,\o}\hp^+_{k,\s\o}\hp^-_{p,\s\o}\;.$$
\theorem(EEH){There exists $\e>0$, a suitable choice 
of the parameters of the Hamiltonian model, 
$\l_N$, $\m_N$, $Z_N$, $Z^{(2,+)}_N$, $Z^{(2,-)}_N$,
$\n_N$, $\d_N$, and a suitable choice of the parameters
of the Euclidean model, $\l_N$, $\m_N$ $Z_N$,  
$Z^{(2)}_N$ -- the analogous parameters 
of the two model being, in general, different -- such that,
in the limit of removed cutoff,  
each Schwinger function in the former regularization
coincides with the analogous Schwinger function in the latter one.}
\*

\0The proof is deferred to the next Chapter: see \secc(PTM).
\chapter(R,Renormalization Group Analysis)
After slicing the momenta in scales, 
the parameters of the generating functional 
are turned into {\it effective parameters}
for each given momentum scale;
in this way obtaining a
sequence, the {\it flow of the running coupling constants}, 
which is controlled by the 
{\it vanishing of the Beta function}.

\section(IN1, Renormalization Group Analysis for Hard Fermions)

\sub(2.1)
{Momenta slicing.} From now on, to be
definite, the {\it scaling parameter} $\g$ is fixed to be equal to
3 -- but any other value would be fine, suitable changing the
following definition of the cutoff. Then, $\k\g^{N+1}=3\p/4a$,
and the {\it cutoff function}
$\wh\c_0(k_0)$ is defined, for $t\in \RRR$,
$$\wh\c_0(t)\defi
 \left\{
 \matrix
 {1\hfill&\hfill{\rm for\ }|t|\leq \k \cr\cr
  0\hfill&\hfill{\rm for\ }3\k \leq \k|t| \leq 4\k \cr\cr
 \in (0,1)\hfill&\hfill{\rm otherwise }\;;}
 \right.
$$
the actual shape in the third domain is here inessential: it will
be chosen in \secc(CD). Accordingly, for $h=N,\ldots,M$, it
is set $\wh\c_h(t)\defi\wh\c_0\left(\g^{-h}t\right)$. With
$\wh\c_0$ it is possible to make a partition of the momenta
scales: for any $h=N,\ldots,M$,
$$
 \wh\c_M(t)
 =
 \wh\c_h(t)
 +
 \sum_{k=h+1}^M \wh{f}_k(t)\;,
 \qquad
 {\rm with}\
 \wh f_k(t)
 \defi
 \wh\c_k(t)-\wh\c_{k-1}(t).
\Eq(cut1)$$
It is worthwhile to remark $\wh f_k$ has compact support 
$\lft\{t:\k\g^{k-1}\leq |t|\leq \k\g^{k+1}\rgt\}$.
\*

\sub(IR) {Multiscale integration.} The decomposition
\equ(cut1) has the purpose to obtain the following scale
integration of $\WW(\f,\jm)$: for any integer $h:N,\ldots,M$,
$$e^{\WW(\jm,\f)}
 =
 e^{E_h}
 \int\!\der \wh P^{(\leq h)}(\ps)\
 e^{\WW^{(h)}\left(\f,\jm,\sqrt{Z_N}\ps\right)}\;,
 \Eq(g1)$$
where the {\it vacuum energy} on scale $h$, $E_h$, do not depend
on the fields; the measure $\der \wh P^{(\leq h)}$ is the same as
\equ(distr), with $\wh\c_M(k_0)$ replaced by $\wh\c_h(k_0)$; the
{\it effective potential} on scale $h$, $\WW^{(h)}$, is a
functional of the fields:
$$\eqalign{
 \WW^{(h)}\left(\f,\jm,\sqrt{Z_N}\ps\right)
 \defi
&-\l_N\VV\left(\sqrt{Z_N}\ps\right)
 +\g^N\n_N\NN\left(\sqrt{Z_N}\ps\right)
 +\d_N\DD\left(\sqrt{Z_N}\ps\right)\cr
&
 +\sum_{\s=\pm}\z^{(2,\s)}_N\JJ_\s\left(\jm,\sqrt{Z_N}\ps\right)
 +\FF\left(\f,\jm\right)
 + \WW^{(h)}_{\rm irr}\left(\f,\jm,\sqrt{Z_N}\ps\right)\;;
}\Eq(g2)$$
namely it has the same expression of the argument of the
exponential in the r.h.s. member of \equ(gf), apart from the {\it
irrelevant contribution} $\WW^{(h)}_{\rm irr}$.

Scale integration \equ(g1) can be verified by induction. Indeed,
it is true for $h=M$, with $E_M=0$ and $\WW^{(M)}_{\rm irr}\equiv
0$; while the procedure to obtain $E_{h-1}$, $\WW^{(h-1)}$ and
$\WW^{(h-1)}_{\rm irr}$ is the following.

The field $\ps$ is decomposed into the sum of fields
$\ps\rightarrow \ps+ \left(Z_N\right)^{-1/2} \x$, both with
Gaussian distribution. The propagator  on scale $h$ of $\x$, the
{\it hard fermion field on scale $h$}, is given by
$$
 g^{(h)}_{\a,\b}(x)
 \defi
 \int_{D}\!{\der^2 k\over (2\p)^2}\
 e^{-ik\cdot x}
 {\wh f_h(k_0)
 \over
 \m^2_N(k_1)+k^2_0+e^2(k_1)}
 \pmatrix{ik_0+e(k_1)&-\m_N(k_1)\cr
 -\m_N(k_1)&ik_0-e(k_1)\cr}_{\a,\b}\;;
\Eq(ph)$$
hence, by decomposition \equ(cut1), $\ps$ is left with propagator
$$
 g^{(\leq h-1)}_{\a,\b}(x)
 \defi
 \int_{D}\!{\der^2 k\over (2\p)^2}\
 e^{-ik\cdot x}
 {\wh\c_{h-1}(k_0)
 \over
 \m^2_N(k_1)+k^2_0+e^2(k_1)}
 \pmatrix{ik_0+e(k_1)&-\m_N(k_1)\cr
 -\m_N(k_1)&ik_0-e(k_1)\cr}_{\a,\b}\;.
\Eq(left)$$
Then, calling $\der\wh P^{(\leq h-1)}(\ps)$ and $\der\wh
P^{(h)}(\x)$ the measure \equ(distr), with propagators \equ(ph)
and \equ (left) respectively, the hard fermion is integrated out:
$$\eqalign{
 \int\!\der \wh P^{(\leq h)}(\ps)\
 e^{\WW^{(h)}\left(\f,\jm,\sqrt{Z_N}\ps\right)}
&=
 \int\!\der  \wh P^{(\leq h-1)}(\ps)\
 \int\!\der  \wh P^{(h)}(\x)\
 e^{\WW^{(h)}\left(\f,\jm,\sqrt{Z_N}\ps+\x\right)}\cr
&\defi e^{\D E_{h-1}}\int\!\der \wh P^{(\leq h-1)}(\ps)\
 e^{\WW^{(h-1)}\left(\f,\jm,\sqrt{Z_N}\ps\right)}\;,
}\Eq(trex1)$$
where $\D E_{h-1}$ is the part of the integration constant  the
fields. Therefore, the vacuum energy on scale $h-1$ is defined to
be:
$$
 E_{h-1}\defi E_{h}+\D E_{h-1}\;;
$$
while,
$$\eqalign{
&\WW_{\rm irr}^{(h-1)}
 \lft(\f,\jm,\sqrt{Z_N}\ps\rgt)\cr
\defi&
\ln\int\!\der  \wh P^{(h)}(\x)\
 e^{\WW^{(h)}\lft(\f,\jm,\sqrt{Z_N}\ps+\x\rgt)}-\D E_{h-1}\cr
=&\sum_{n^\ps,n^\f,n^\jm \geq 1}^{n^\ps+n^\f+n^\jm\neq 0}
 \sum_{\underline\o,\underline\s}
 \int_{\L}\!\der^2\ux\ \der^2\uy\ \der^2\uz\ \cr
&\left(\prod_{i=1}^{n^\ps}\sqrt{Z_N}\ps^{\s_i}_{x^{(i)},\o_i}\right)
 \left(\prod_{i=1}^{n^\f} {\f^{\s'_i}_{y^{(i)},\o'_i}\over \sqrt{Z_N}}\right)
 \left(\prod_{j=1}^{n^\jm} \jm_{z^{(i)},\o''_i}\right)
 W^{(h-1)}_{n^\ps;n^\f;n^\jm,\underline\o,\underline\s}
 (\ux,\uy,\uz)\;,
}\Eq(5.4)$$%
where $\ux$, $\uy$ and $\uz$ are short notations for
$x^{(1)},\ldots,x^{(n^\ps)}$, $y^{(1)},\ldots,y^{(n^\f)}$ and
$z^{(1)},\ldots,z^{(n^\jm)}$ respectively. By the well known
formula for the truncated expectation w.r.t. a Gaussian measure,
the function $W^{(h-1)}_{n^\ps;n^\f;n^\jm,\uo,\us}(\ux,\uy,\uz)$
is a power series in the couplings $\l_N, \n_N, \d_N$, and
coefficient given by all the Feynman graphs with
$n^\ps+n^\f+n^\jm$ external legs of kind $n^\ps,n^\f,n^\jm$
attached respectively to the points $\ux,\uy,\uz$, with eventually
a constraint that some among the point in $\ux$ may coincide: this
is explained in more details in Appendix~\secc(TE). The remarkable
fact is that the number of the Feynman graphs at $n$-the order
expansion is about $n!$; and yet, by {\it cluster expansion} and
anticommutativity of the fermion fields, it is possible to prove a
$C^n$-bound, making the power series defining
$W^{(h-1)}_{n^\ps;n^\f;n^\jm,\uo,\us}(\ux,\uy,\uz)$ {\it
absolutely convergent} for $\l_N, \n_N, \d_N$ small enough (see
\secc(ClE)).

Finally, $\WW^{(h-1)}_{\rm irr}$ is defined by \equ(g2):
in power series expansion, 
it corresponds to the terms in \equ(5.4)
which are at least O$(\l_N)$,
except the terms for $n^\ps=4$, $n^\f=n^\jm=0$
and linear in $\l_N$.
\*

\sub(DB)
{Dimensional bounds.} In order to have a bound for
$W^{(h)}_{n^\ps;n^\f;n^\jm,\uo,\us}$, it is possible to prove the
following decay property of the diagonal and antidiagonal
propagators: there exist two positive constants $c$ and $C$ such that
$$\eqalign{
&\left| g^{(h)}_{\o,\o}(x)\right|
 \leq
 C\g^{N} e^{-c\sqrt{\g^N\k|x|}}e^{-c\sqrt{\g^h\k|x_0|}}\;,\cr 
&\left| g^{(h)}_{\o,-\o}(x)\right|
 \leq
 \g^{-(h-N)}C \g^{N} e^{-c\sqrt{\g^N\k|x|}}e^{-c\sqrt{\g^h\k|x_0|}}\;.}\Eq(prop>N)$$
Since $h>N$, the more factor $\g^{-(h-N)}$ in the 
bound of the antidiagonal propagator
represents a ``gain factors''  w.r.t. the bound of the diagonal one.

In the end of the integration of all hard fermions scales,
\equ(g1) reads
$$e^{\WW(\jm,\f)}
 =
 e^{E_N}
 \int\!\der  \wh P^{(\leq N)}(\hp)\
 e^{\WW^{(N)}\left(\f,\jm,\sqrt{Z_N}\hp\right)}\;,
 \Eq(4.6)$$
which is the starting point of the analysis of the double and
light fermions in the next sections. Let $d(\ux)$ be the {\it tree
distance} of the points $\ux$, namely the length of the shortest
tree path connecting every point in $\ux$.

\*
\lemma(T1bis)
{There exist $\e>0$ and the positive 
constants $c$ and $C$ s.t., for any choice of the couplings
$|\l_N|,|\d_N|, |\n_N|<2\e$, the following bounds hold.
\elenco{
\art If $n^\f+n^\jm\neq 0$,
$$
 \int_{\L}\!\der^2\ux\
 \left|W^{(N)}_{n^\ps;n^\f;n^\jm,\uo,\us}
 (\ux,\uy,\uz)\right|
 \leq
 C{\g^{N\big(2-(1/2)n^\ps-(3/2)n^\f-n^\jm\big)}
  \over e^{{c\over 2(n^\f+n^\jm)}\sqrt{\g^N\k d(\uy,\uz)}}}\;.
$$%
\art If $n^\f+n^\jm= 0$,
$$
 \int_{\L}\!\der^2_*\ux\
 \left|W^{(N)}_{n^\ps;0;0,\uo,\us}
 (\ux)\right|
 \leq
 C\g^{N\big(2-(1/2)n^\ps\big)}\;,
$$
where $\der^2_*\ux$ means that the integration is performed w.r.t.
all but any one variable among $x^{(1)},\ldots,x^{(n)}$. }}
\0The proof is the same of Lemma \lm(SS).
\*

\sub(R) 
{Remark: superrinormalizability.} 
The key feature, here, is the scaling
$\left(Z_N\right)^{-1/2}$ of hard fermion in the decomposition
$\hp\rightarrow \ps+ \left(Z_N\right)^{-1/2} \x$: this factor is
{\it the same for all the scales $h>N$}, so that there is no generation
of anomalous dimension in the hard fermion regime.

\*

\section(DF,Renormalization Group Analysis for Double Fermions)

\sub(CDDF)
{Momenta slicing.}
At this point it is convenient to choose the image in $(0,1)$ of
the cutoff function so that the constant function $I\equiv 1$ on
the periodic lattice $D_1$ is equal to the sum of two
$\wh\c_N$ functions, the former centred in $k_1=0$, and the latter
centred in $k_1=\p/a$:
$$
 \wh\c_N(t)+\wh\c_N\left(t-{\p\over a}\right)\equiv 1\;
\Eq(dec)$$
(and such that $\wh\c_0$ is a Gevrey function: see \secc(GCF)).
After the integration of the hard fermions, it was left the
measure $\der \wh P^{(\leq N)}(\ps)$, with propagator given by
\equ(left) for $h=N$: it is possible now to decompose the fields
$\ps$ into the sum $\ps\rightarrow \ps+ \left(Z_N\right)^{-1/2}
\x$, where the 
{\it double fermion field}, $\x$, has propagator
$$\eqalign{
 g^{({\rm D})}_{\a,\b}(x)
&\defi
 \int_{D}\!{\der^2 k\over (2\p)^2}\
 e^{-ik\cdot x}
 {\wh\c_N(k_0)\wh\c_N\left(k_1-(\p/a)\right)
 \over
 \m^2_N(k_1)+k^2_0+ e^2(k_1)}
 \pmatrix{ik_0+e(k_1)&-\m_N(k_1)\cr
 -\m_N(k_1)&ik_0-e(k_1)\cr}_{\a,\b}\;;
}\Eq(ph2)$$
therefore, because of \equ(dec) and setting
$\c_N(k)\defi\wh\c_N(k_0)\wh\c_N(k_1)$, $\ps$ is left with
propagator
$$
 g^{(\leq N,D)}_{\a,\b}(x)
 \defi
 \int_{D}\!{\der^2 k\over (2\p)^2}\
 e^{-ik\cdot x}
 {\c_{N}(k)
 \over
 \m^2_N(k_1)+k^2_0+e^2(k_1)}
 \pmatrix{ik_0+e(k_1)&-\m_N(k_1)\cr
 -\m_N(k_1)&ik_0-e(k_1)\cr}_{\a,\b}\;.
\Eq(left2)$$

\*
\sub(DBDF){Dimensional bounds.} 
Because of the definition of
$\m_N(k_1)$, the propagator $g^{\rm D}_{\m,\n}(x)$ is massive, and
hence, without decomposition of
$\wh\c_N(k_0)\wh\c_N\left(k_1-(\p/a)\right)$ into scales, it
enjoys the bound,
for $c$ and $C$ two positive constants, 
$$\eqalign{
&\left|g^{({\rm D})}_{\a,\b}(x)\right|
 \leq
 C\g^{N} e^{-c\sqrt{\g^N\k|x|}}\;.
}\Eq(prop=N)$$
Indeed in the
support of $\wh \c_N(k_0)\wh\c_N\left(k_1-(\p/a)\right)$,
it holds $\p/4a\le |k_1|\le \p/4$, while 
$|k_0|$ can be very small: since 
the mass $\m_N$ is supposed non-negative, 
the denominator is not 
lower than $\m^2_N(k_1)\geq c^2(k_1)\ge \lft(\k\g^N(2-\sqrt2)/2\p\rgt)^2$.
And the bound follows by dimensionality argument.  In this
way the effects of the second pole are confined on the scale of the
cutoff, $N$: since it will be proved that the Schwinger functions
do not depend on contribution on such scales, the addition of
$c(k_1)$ to the mass has had the effect to suppress the effects of
the double fermions.

Integrating out the double field now requires 
a localization, which will be explained in the next 
section.

\*
\section(I,Renormalization Group Analysis for Soft Fermions)
 
\sub(CD)
{Momenta slicing.} 
The last, more
involved regime to be studied is the set of momentum scales below
$N$. Let $\c_N(k)$ be decomposed over the scales
$$\c_N(t_0,t_1)=\c_h(t_0,t_1)+
 \sum_{k=h+1}^N f_k(t_0,t_1)\;,\Eq(cut2)$$
where the function $f_k(t_0,t_1)$
is defined to be $\c_k(t_0,t_1)-\c_{k-1}(t_0,t_1)$
and has squared support
\hbox{$\left\{(t_0,t_1):
\k\g^{k-1}\leq \max\{|t_0|,|t_1|\}\leq \k\g^{k+1}\right\}$}.
\*

\sub(j=N)
{Multiscale integration.} As for the hard fermions, the
functional integration of the soft fermions is performed scale by
scale. By induction, for any integer $h:h\leq N$, it holds:
$$e^{\WW(\jm,\f)}
 =
 e^{E_h}
 \int\!\der\wt P^{(\leq h)}(\ps)\
 e^{\WW^{(h)}\left(\f,\jm,\sqrt{Z_h}\ps\right)}\;,
 \Eq(g2)$$
where the effective potential on scale $h$ is
$$\eqalign{
 \WW^{(h)}\left(\f,\jm,\sqrt{Z_h}\ps\right)
 \defi
&-\l_h \VV\left(\sqrt{Z_h}\ps\right)
 +\g^h\n_h\NN\left(\sqrt{Z_h}\ps\right)
 +\d_h\DD\left(\sqrt{Z_h}\ps\right)\cr
&
 +\sum_{\s=\pm}\z^{(2,\s)}_h\JJ_\s\left(\jm,\sqrt{Z_h}\ps\right)
 +\FF\left(\f,\jm\right)
 +\WW^{(h)}_{\rm irr}\left(\f,\jm,\sqrt{Z_h}\ps\right)\; ;
}\Eq(gi3)$$
the measure $\der\wt P^{(\leq h)}$, the couplings $\l_h,\n_h$,
$\d_h$, $\z_h^{(2,\s)}$ and the irrelevant potential
$\WW^{(h)}_{\rm irr}$ are inductively specified by the procedure
to construct $\WW^{(h-1)}$.

The field $\ps$ is decomposed into the sum of two fields,
$\ps\rightarrow \ps+ \left(Z_{h}\right)^{-1/2} \x$, both with
Gaussian distribution. The  propagator of the {\it soft fermion}
field, $\x$ is, {\it for $h\neq N$}:
$$\eqalign{
 g^{(h)}_{\a,\b}(x)
&=
 \int_{D}\!{\der^2 k\over (2\p)^2}\
 e^{-ik\cdot x}
 {\wt{f}^{(h)}(k)
 \over
 \wt\m_h^2(k)+k^2_0+ e^2(k_1)}
 \pmatrix{ik_0+e(k_1)&-\wt\m_h(k)\cr
 -\wt\m_h(k)&ik_0-e(k_1)\cr}_{\a,\b}\cr}\;,\Eq(p1)$$
with 
$$\wt{f}^{(h)}(k)\defi f_{h}(k)\wt C^{(1)}_h(k)\;,
\qquad
\wt c_h(k)\defi
{Z_N\over Z_h}c(k_1) \wt C^{(1)}_h(k)\;,$$
$$\wt\m_h(k)\defi \m_h\wt C^{(2)}_{h}(k)+\wt c_h(k_1)\;,$$
and 
the quantities  $Z_h$, $\m_h$, $\wt C^{(1)}_h(k)$ and $\wt C^{(1)}_h(k)$ will
be constructed in the following {\it localization}. 
For $h=N$, to the above expression for the propagator it has to be 
added the propagator deriving from the the double fermions, 
$g^{(\rm D)}_{\a,\b}(k)$.

Since in presence of $\c_{h-1}(k)$, by simply support compatibility, $\wt
C^{(1)}_h(k)\equiv\wt C^{(2)}_h(k)\equiv1$, by \equ(cut2), $\ps$
is left with propagator:
$$
 g^{(\leq h-1)}_{\a,\b}(x)
 \defi
 \int_{D}\!{\der^2 k\over (2\p)^2}\
 e^{-ik\cdot x}
 {\c_{h-1}(k)
 \over
 \m^2_{h-1}(k_1)+k^2_0+e^2(k_1)}
 \pmatrix{ik_0+e(k_1)&-\m_{h-1}(k_1)\cr
 -\m_{h-1}(k_1)&ik_0-e(k_1)\cr}_{\a,\b}\;,
\Eq(p2)$$
with
$$
 \m_h(k_1)\defi \m_h + {Z_N\over Z_h}c(k_1)\;,
$$
without any residue of $\wt C^{(1)}_h(k)$ or $\wt C^{(2)}_h(k)$.

The soft fermions can be integrated out, scale by scale; this time
this operation does not give directly $\WW^{(h-1)}$, but rather
$\wt\WW^{(h-1)}$. Calling $\der P^{(\leq h-1)}(\ps)$ and $\der
P^{(h)}(\x)$ the measure \equ(distr), with $Z_N$ replaced by $Z_h$
and propagators respectively given by \equ(p2) and \equ(p1)
$$\eqalign{
 \int\!\der\wt P^{(\leq h)}(\ps)\
 e^{\WW^{(h)}\left(\f,\jm,\sqrt{Z_h}\ps\right)}
&=
 \int\!\der P^{(\leq h-1)}(\ps)\
 \int\!\der P^{(h)}(\x)\
 e^{\WW^{(h)}\left(\f,\jm,\sqrt{Z_h}\ps+\x\right)}\cr
&\defi\int\!\der P^{(\leq h-1)}(\ps)\
 e^{\wt\WW^{(h-1)}\left(\f,\jm,\sqrt{Z_h}\ps\right)+\D E_{h-1}}\;,
}\Eq(trex2)$$%
where $\D E_{h-1}$ is the part of the integration constant in the
fields. Again, by the well known formulas of the truncated
expectations:
$$\eqalign{
&\wt\WW^{(h-1)}
 \left(\f,\jm,\sqrt{Z_h}\ps\right)\cr
&=\sum_{n^\ps,n^\f,n^\jm\geq 1}^{n+n^\f+n^\jm\neq 0}
 \sum_{\underline\o,\underline\s}
 \int_{\L}\!\der^2\ux
 \der^2\uy\der^2\uz\ \cr
&\left(\prod_{i=1}^n\sqrt{Z_h}\ps^{\s_i}_{x^{(i)},\o_i}\right)
 \left(\prod_{i=1}^{n^\f} {\f^{\s'_i}_{y^{(i)},\o'_i}\over \sqrt{Z_h}}\right)
 \left(\prod_{j=1}^{n^\jm} \jm_{z^{(i)},\o''_i}\right)
 \wt W^{(h-1)}_{n^\ps;n^\f;n^\jm,\uo,\us}(\ux,\uy,\uz)\;.
}\Eq(g3)$$
For the light fermions a further step is necessary to extract
parts of $\wt\WW^{(h-1)}$ that can be absorbed either into the
free measure $\der P^{(\leq h-1)}$, or in the couplings; this is
the {\it Localization}. In the end of this operation they are
left a potential $\WW^{(h-1)}$ and a measure $\der\wt P^{(\leq
h-1)}$, which fulfil \equ(gi3).
\*

\sub(DBSF)
{Dimensional bounds.} It is convenient to decompose
the propagator $g^{(h)}_{\o,\s}$
into the one of the Euclidean Model, $g^{({\rm E},h)}_{\o,\s}$, 
plus the rest, $g^{({\rm R},h)}_{\o,\s}$, plus the eventual 
contribution of the double fermion, $g^{({\rm D})}_{\o,\s}$; 
in their turn,
let $g^{({\rm E1},h)}_{\o,\s}$, $g^{({\rm R1},h)}_{\o,\s}$ and 
$g^{({\rm D1})}_{\o,\s}$ be respectively the part of  
$g^{({\rm E},h)}_{\o,\s}$, $g^{({\rm R},h)}_{\o,\s}$ 
and $g^{({\rm D})}_{\o,\s}$ 
which is constant or linear in the mass. Finally:
$$\eqalign{
 g^{(h)}_{\o,\s}(x)
&\defi g^{({\rm E1},h)}_{\o,\s}(x)
+g^{({\rm R1},h)}_{\o,\s}(x)+ \d_{h,N}g^{({\rm D1})}_{\o,\s}(x)
 +r^{(1,h)}_{\o,\s}(x)+
 r^{(2,h)}_{\o,\s}(x)}\;,\Eq(decp)$$
with the following definitions
$$\eqalign{
&g^{({\rm E1},h)}_{\o,\o}(x)
 \defi
 \int_{D}\!{\der^2 k\over (2\p)^2}\
 e^{-ik\cdot x}
 { \wt f^{(h)}(k)
 \over D_\o(k)}\;,
\qquad
g^{({\rm E1},h)}_{\o,-\o}(x)
 \defi
 \int_{D}\!{\der^2 k\over (2\p)^2}\
 e^{-ik\cdot x}
 {-\wt\m_h(k)\over
 k_0^2 + k_1^2}\wt f^{(h)}(k)\;,\cr
&g^{({\rm R1},h)}_{\o}(x)
 \defi
 \int_{D}\!{\der^2 k\over (2\p)^2}\
 e^{-ik\cdot x}
 \left[
 {ik_0+\o e(k_1)
 \over
 \wt c^2_h(k)+k^2_0 +e^2(k_1)}
 -
 {-D_{-\o}(k)
 \over
 k_0^2 + k_1^2}\right]\wt f^{(h)}(k)\;,\cr
&g^{({\rm R1},h)}_{\o,-\o}(x)
 \defi
 \int_{D}\!{\der^2 k\over (2\p)^2}\
 e^{-ik\cdot x}
 \left[
 {-\wt \m_h(k)
 \over
 \wt c^2_h(k)+k^2_0 +e^2(k_1)}
 -
 {-\wt \m_h(k)
 \over
 k_0^2 + k_1^2}\right]\wt f^{(h)}(k)\;,\cr
&r^{(1,h)}_{\o,\o}(x)
 \defi
 \int_{D}\!{\der^2 k\over (2\p)^2}\
 e^{-ik\cdot x}\left[
 {-D_{-\o}(k)\over
 \wt\m_h^2(k) + k_0^2 + k_1^2}-
 {-D_{-\o}(k)\over
 k_0^2 + k_1^2}\right]\wt f^{(h)}(k)\;,\cr
&r^{(1,h)}_{\o,-\o}(x)
 \defi
 \int_{D}\!{\der^2 k\over (2\p)^2}\
 e^{-ik\cdot x}\left[
 {-\wt\m_h(k)\over
 \wt\m_h^2 + k_0^2 + k_1^2}-
 {-\wt\m_h(k)\over
 k_0^2 + k_1^2}\right]\wt f^{(h)}(k)\;;}$$
then 
$g^{({\rm D1})}_{\o,\s}$ is given by the sum of 
$g^{({\rm E1},N)}_{\o,\s}$ and  $g^{({\rm R1},N)}_{\o,\s}$,
with the cutoff $\wh f_N(k)$ replaced by 
$\wh \c_N(k_0)\wh\c_N(k_1-(\p/a))$;
and  
$r^{(2,h)}_{\o,\s}(x)$ is defined in consequence of \equ(decp).

For $\e$ small enough, 
(so that, by the inductive hypothesis \equ(ind1) 
$1-c_0\e\ge 3/4$),
there exists two positive constants, $c$ and $C$ s.t.:
$$\eqalign{
\left|g^{({\rm E1},h)}_{\o,\o}(x)\right|
\leq
{C\g^{h} \over e^{c\sqrt{\g^h\k|x|}}}\;,
\phantom{****}\qquad&
 \left|g^{({\rm R1},h)}_{\o,\o}(x)\right|
 \leq
 \g^{-(3/4)(N-h)}
 {C\g^{h}\over e^{c\sqrt{\g^h\k|x|}}}\;,\cr
 \left|g^{({\rm E1},h)}_{\o,-\o}(x)\right|
 \leq
 \left|{\m_h\over \g^h\k}\right|
 {C\g^{h}\over e^{c\sqrt{\g^h\k|x|}}}\;,
 \qquad&
 \left|g^{({\rm R1},h)}_{\o,-\o}(x)\right|
 \leq
 \left|{\m_h\over \g^h\k}\right|
 \g^{-(3/4)(N-h)}
 {C\g^{h}\over e^{c\sqrt{\g^h\k|x|}}}\;,\cr
 \left|g^{({\rm D1})}_{\o,\o}(x)\right|
 \leq
 {C\g^{N}\over e^{c\sqrt{\g^N\k|x|}}}\;,
 \phantom{*****}\qquad&
 \left|g^{({\rm D1})}_{\o,-\o}(x)\right|
 \leq
 \left|{\m_N\over \g^N\k}\right|
 {C\g^{N}\over e^{c\sqrt{\g^N\k|x|}}}\;,\cr
 \left|r^{(1,h)}_{\o,\s}(x)\right|
 \leq
 \left|{\m_h\over \g^h\k}\right|^2
 {C\g^{h}\over e^{c\sqrt{\g^h\k|x|}}}\;,
 \qquad&
 \left|r^{(2,h)}_{\o,\s}(x)\right|
 \leq
 \left|{\m_h\over \g^h\k}\right|^3\g^{-(3/4)(N-h)}
 {C\g^{h}\over e^{c\sqrt{\g^h\k|x|}}}\;.
}\Eq(prop<N)$$
It is remarkable the propagators  $g^{({\rm R1},h)}_{\o}$ 
and  $r^{(2,h)}_{\o,\s}$ have a gain
factor $\g^{-(3/4)(N-h)}$ more than the standard bounds.
Clearly, the above bounds are useful whenever
$\m_h\leq \k\g^h$: when this condition is not
satisfied, then the mass in the propagator is so large that it is
possible to integrate the remaining scales all at once,
as it was done for the double fermion propagator
(see later the definition of the scale $h^*$).

\*
\sub(L) 
{Localization.}
The contribution to $\wt\WW^{(h-1)}$
of certain kinds of Feynman graphs is extracted from the rest by
{\it localization}: it extracts the 0-th or the 1-th order 
Taylor expansion in the momenta and the 0-th or the 1-th order 
expansion in the mass parameters $\{\m_k\}_k$.
Since the space of the momenta, $D$, does not contain $(0,0)$,
and is not continuous, the Taylor expansion should be done 
taking {\it discrete derivatives} in the four 
nearest neighbour lattice site surrounding $0$.
This subtlety cannot be very important, since 
the continuous limit (for the lattice $D$ only),
$L\to\io$, was not taken since the beginning,
not to be involved with  an infinite Grassmannian algebra.
(The analogous argument is not valid also for the lattice $\L$,
since it is essential to make the limit $N\to+\io$ {\it after}
the renormalization has taken place.) 
Therefore, for shake of simplicity, the following developments,
are {\it as if the lattice $D$
were continuous rather than discrete},
leaving the correct technicality to [BM01].

Well then,  it 
is convenient to introduce the directional derivatives
$$\partial_\o^k\defi{1\over 2}
\left[i{\partial\over \partial_{k_0}}+\o{\partial\over \partial_{k_1}}\right]\;,$$
which are orthogonal is the sense that the two relations are true:
$\Big(\partial_\o D_\s\Big)(k) =\d_{\o,\s}$
and \hbox{$\sum_{\o=\pm}D_\o(k)\partial_\o\=k_0\partial_{k_0}+k_1\partial_{k_1}$}.
\elenco{
\art 
Let $\hW_{2,\a,\b}^{(h-1)}(k)$ be considered. 
If $\a=\b$, $\hW_{2,\a,\a}^{(h-1)}(0)$ =0 by \equ(refl); if
$\b=-\a$, independently on $\a$ by \equ(srefl), 
it is possible to define
$$\hW_{2,\a,-\a}^{(h-1)}(0)=s_{h-1}+ \g^{h-1}\D n_{h-1}
+\D s^{(\m)}_{h-1}\;,$$
where,  $\D s^{(\m)}_{h-1}$ is the sum of the graphs in
the expansion of  $\hW_{2,\a,-\a}^{(h-1)}(0)$ which are at 
least quadratic in the masses $\{\m_k\}_k$;
while $s_{h-1}$ is the sum of  all the graphs linear in the 
masses, and therefore made 
with only antidiagonal propagator 
$g^{({\rm E1},k)}_{\o,-\o}$, $g^{({\rm R1},k)}_{\o,-\o}$
or $g^{({\rm D1})}_{\o,-\o}$;
finally, the sum of the graphs which are independent on the
masses is in $\g^{h-1}\D n_{h-1}$.
Then, let $\left(\partial_\s\hW_{2,\a,\b}^{(h-1)}\right)(k)$ be considered. 
By \equ(refl), for $\b=-\a$, $\left(\partial_\s\hW_{2,\a,-\a}^{(h-1)}\right)(0)=0$;
while, for $\a=\b$, it is possible to define, 
independently on $\a$ by \equ(srefl),
$$
\left(
 \partial_\s\hW^{(h-1)}_{2,\a,\a}\right)(0)
 \left\{
 \matrix{\defi d_{h-1}^{(+)}+\D d^{(1,+)}_{h-1} \hfill && \hfill{\rm for}\ \ \s=\a \cr
         \defi d_{h-1}^{(-)}+\D d^{(1,-)}_{h-1}\hfill && \hfill{\rm for}\ \ \s=-\a\;,\cr
}\right.
$$
where $\D d^{(1,\s)}_{h-1}$ is the sum of the graphs
which are at least linear in the masses; while $d_{h-1}^{(\s)}$
is the sum of the masses independent graphs. 
Defining $z_{h-1}\defi d_{h-1}^{(+)}+d_{h-1}^{(-)}$, and 
$\D d_{h-1}\defi -2 d_{h-1}^{(-)}$  and,
accordingly,
$$\D t_{h-1}(k)
 \defi
 \pmatrix{ z_{h-1}\big(-ik_0+e(k_1)\big)
 & s_{h-1}\cr
 s_{h-1}
 & z_{h-1}\big(-ik_0-e(k_1)\big)\cr}\;,
$$
the localization is:
$$\eqalign{\LL
&\left[
 \sum_{\a,\b}\int_{D_{h-1}}\!{\der^2k\over (2\p)^2}\ 
 \hp^+_{k,\a}\hp^-_{k,\b}\hW_{2,\a,\b}^{(h-1)}(k)\right]
 =\g^{h-1}\D n_{h-1}
 \sum_{\o}\int_{D_{h-1}}\!{\der^2k\over (2\p)^2}\ 
 \hp^+_{k,\o}\hp^-_{k,-\o}\cr
&+\D d_{h-1}\sum_\o
 \int_{D_{h-1}}{\der^2k\over (2\p)^2}\ 
 \hp^+_{k,\o}\hp^-_{k,\o} \o e(k)
+\sum_{\a,\b}
 \int_{D_{h-1}}\!{\der^2k\over (2\p)^2}\ 
 \hp^+_{k,\a}\hp^-_{k,\b}\big( \D t_{h-1}\big)_{\a,\b}(k)\;.}$$
Setting $\RR\defi 1-\LL$:
$$\eqalign{\RR
&\left[
 \sum_{\a,\b}\int_{D_{h-1}}\!{\der^2k\over (2\p)^2}\
 \hp^+_{k,\a}\hp^-_{k,\b}\hW_{2,\a,\b}^{(h-1)}(k)\right]\cr
&=\D s^{(\m)}_{h-1}
 \sum_{\o}\int_{D_{h-1}}\!{\der^2k\over (2\p)^2}\ 
 \hp^+_{k,\o}\hp^-_{k,-\o}
 +\sum_{\s,\o} \D d^{(\m,\s)}_{h-1}
 \int_{D_{h-1}}{\der^2k\over (2\p)^2}\ 
 \hp^+_{k,\o}\hp^-_{k,\o} D_{\s\o}(k)\cr
&+z_{h-1}\sum_{\a,\s}
 \int_{D_{h-1}}\!{\der^2k\over (2\p)^2}\
 \hp^+_{k,\a}\hp^-_{k,\a}
 \Big[D_\s(k)- \big(-ik_0+\s e(k_1)\big)\Big]\cr
&+\sum_{\a,\b,\o,\s}
 \int_{D_{h-1}}\!{\der^2k\over (2\p)^2}\
 \hp^+_{k,\a}\hp^-_{k,\b}D_\o(k) D_\s(k)
 \int_0^1\! \der\t\ (1-\t)
 \left(\partial_\o\partial_\s\hW_{2,\a,\b}^{(h-1)}\right)(\t k)\;.
}$$
The local part $\D t_{h-1}$ is absorbed in the free measure.
Calling:
$$
 \wt C^{(1)}_{h-1}(k)\defi
 {1+z_{h-1}+\D z_{h-1}\over
 1+\c_{h-1}(k)z_{h-1}+\c_{h-1}(k)\D z_{h-1}}\;,
$$
$$
 \wt C^{(2)}_{h-1}(k)\defi
 {1+z_{h-1}+\D z_{h-1}\over
 1+\c_{h-1}(k)z_{h-1}+\c_{h-1}(k)\D z_{h-1}}
 {1+\c_h(k)\left(s_{h-1}/ \m_{h-1}\right) \over
 1+\left(s_{h-1}/ \m_{h-1}\right)}\;,
$$
and, since $s_{h-1}$ is linear in the masses, 
$m_{h-1}\defi s_{h-1}/\m_{h-1}$, the {\it effective field
strength} and the {\it effective mass} on scale $h-1$ are:
$$
Z_{h-1}\defi Z_h(1+z_{h-1})\;,
\qquad
\m_{h-1}\defi
\m_h{Z_h\over Z_{h-1}}(1+m_{h-1})\;. \Eq(ga1)$$
Then, in the same way, 
the local parts  $\D n_{h-1}$ and $\D d_{h-1}$ are absorbed
in the {\it effective counterterms} on scale $h-1$, $\n_{h-1}$ and
$\d_{h-1}$:
$$
 \d_{h-1}\defi\left({Z_h\over Z_{h-1}}\right) (\d_h + \D d_{h-1})\;,
 \qquad
 \n_{h-1}\defi\left({Z_h\over Z_{h-1}}\right) \g (\n_h + \D n_{h-1})\;.
\Eq(b1)$$
A remarkable feature is that $Z_{h-1}$, $\n_{h-1}$ and $\d_{h-1}$
are {\it independent from the mass flow,} $\{\m_k\}_k$.
Finally, in changing free measure on scale $h-1$ from $\der
P^{(\leq h-1)}$ to $\der\wt P^{(\leq h-1)}$, it has to be taken
into account the change of the normalization:
$$
 \D \wt E_{h-1}\defi -
 \ln\left\{
 \left({Z_{h-1}\over Z_h}\right)^2
 \int_{D_{h-1}}\!{\der^2 k\over (2\p)^2}\left[{k^2_0+e^2(k_1)+\wt\m^2_{h-1}(k_1)
 \over
  k^2_0+e^2(k_1)+\wt\m^2_{h}(k)}\right]
 \left({1\over \wt C_{h-1}^{(1)}(k)}\right)^2\right\}\;.$$
so that the {\it effective vacuum energy} on scale $h-1$ is
$$
 E_{h-1}\defi E_h+\D E_{h-1} +\D\wt E_{h-1}\;.
$$
\art
Let $\hW_{4,\o,-\o}^{(h-1)}(k,p,q)$ be considered; and let
$$\hW_{4,\o,-\o}^{(h-1)}(0,0,0)\defi\D l_{h-1} + \D l^{(1)}_{h-1}\;,$$
where $\D l^{(1)}_{h-1}$ is the sum of all the graphs
in the expansion of $\hW_{4,\o,-\o}^{(h-1)}(0,0,0)$ 
which are at least linear in the masses.
Then
$$\eqalign{
&\LL
 \left[\sum_\o
 \int_{D_{h-1}}\!{\der^2k\over (2\p)^2}{\der^2p\over (2\p)^2}
 {\der^2q\over (2\p)^2}\
 \ps^+_{k,\o}\ps^-_{k+p-q,\o}\ps^+_{p,-\o}
 \ps^-_{q,-\o}\hW_{4,\o,-\o}^{(h-1)}(k,p,q)\right]\cr
&=\D l_{h-1}
 \sum_\o
 \int_{D_{h-1}}\!{\der^2k\over (2\p)^2}{\der^2p\over (2\p)^2}{\der^2q\over (2\p)^2}\
 \ps^+_{k,\o}\ps^-_{k+p-q,\o}\ps^+_{p,-\o}\ps^-_{q,-\o}\;,\cr
&\RR
 \left[\sum_\o
 \int_{D_{h-1}}\!{\der^2k\over (2\p)^2}{\der^2p\over (2\p)^2}{\der^2q\over (2\p)^2}\
 \ps^+_{k,\o}\ps^-_{k+p-q,\o}\ps^+_{p,-\o}\ps^-_{q,-\o}\hW_{4,\o,-\o}^{(h-1)}(k,p,q)\right]\cr
&=\D l^{(1)}_{h-1}
 \sum_\o
 \int_{D_{h-1}}\!{\der^2k\over (2\p)^2}{\der^2p\over (2\p)^2}{\der^2q\over (2\p)^2}\
 \ps^+_{k,\o}\ps^-_{k+p-q,\o}\ps^+_{p,-\o}\ps^-_{q,-\o}\cr
&+\sum_{\o,\s}\sum_{p'=k,p,q}
 \int_{D_{h-1}}\!{\der^2k\over (2\p)^2}{\der^2p\over (2\p)^2}{\der^2q\over (2\p)^2}\
 \ps^+_{k,\o}\ps^-_{k+p-q,\o}\ps^+_{p,-\o}\ps^-_{q,-\o}
 D_\s(p')\cr
&\phantom{*************************}
 \cdot\int_0^1\!\der\t
 \left(\partial_\s^{p'}\hW_{4,\o,-\o}^{(h-1)}\right)(\t k,\t p,\t q)\;.
}$$
The local part $\D l_{h-1}$ is
absorbed in the effective coupling on scale $h-1$:
$$
\l_{h-1}\defi \left({Z_h\over Z_{h-1}}\right)^2 (\l_h + \D
l_{h-1})\;, \Eq(b2)$$
and also $\l_{h-1}$ is independent from the flow $\{\m_k\}_k$.
\art
Let $\hW^{(h-1)}_{1;2,\m;\n}(0,0)$ be considered;  
since by \equ(srefl1), it does not depend on $\s$,
it is possible to define
$$
 \hW_{1;2,\s;\o}^{(h-1)}(0;0)
 \defi
 \left\{
 \matrix{
 z^{(2)}_{h-1} + \D z^{(2,+)}_{h-1}+ 
 \D d^{(2,+)}_{h-1}\hfill&\hfill{\rm for }\ \s=\o\cr
 \D z^{(2,-)}_{h-1}+\D d^{(2,-)}_{h-1}\hfill&\hfill 
 {\rm for }\ \s=-\o\;;\cr}\right.$$
where $ \D d^{(2,\s)}_{h-1}$ is the sum of the graphs 
at least linear in the masses; then $z^{(2)}_{h-1}$ and 
$\D z^{(2,+)}_{h-1}$ are mass independent: 
the former is the sum of all the graphs made only 
with (diagonal) propagators $\{g^{({\rm E1},k)}_{\o,\o}\}_k$,
and interaction $\VV$ (namely all the mass-independent 
graphs obtained in the case of the Euclidean model for such 
a kernel); while $\D z^{(2,\s)}_{h-1}$ is the sum 
of the graphs made with least one propagator 
$\{g^{({\rm R},k)}_{\o,\s}\}_k$ or an interaction 
$\NN$ or $\DD$. 
Then
$$\eqalign{
&\LL
 \left[\sum_{\s,\o}
 \int_{D_{h-1}}\!{\der^2k\over (2\p)^2}{\der^2p\over (2\p)^2}
 \jm_{p-k,\s}\ps^+_{k,\o}\ps^-_{p,\o}
 \hW^{(h-1)}_{1;2,\s;\o}(k,p)\right]\cr
 &=\left(z^{(2)}+\D z^{(2,+)}_{h-1}\right)
 \sum_{\s}
 \int_{D_{h-1}}\!{\der^2k\over (2\p)^2}{\der^2p\over (2\p)^2}
 \jm_{p-k,\s}\ps^+_{k,\s}\ps^-_{p,\s}\cr
 &+
 \D z^{(2,-)}_{h-1}\sum_{\s}
 \int_{D_{h-1}}\!{\der^2k\over (2\p)^2}{\der^2p\over (2\p)^2}
 \jm_{p-k,\m}\ps^+_{k,-\s}\ps^-_{p,-\s}\;,\cr
&\RR
 \left[\sum_{\s,\o}
 \int_{D_{h-1}}\!{\der^2k\over (2\p)^2}{\der^2p\over (2\p)^2}
 \jm_{p-k,\s}\ps^+_{k,\o}\ps^-_{p,\o}
 \hW^{(h-1)}_{1;2,\s;\o}(k,p)\right]\cr
&=\sum_{\s,\o}\D z^{(2,\o)}_{h-1}\sum_{\o}
 \int_{D_{h-1}}\!{\der^2k\over (2\p)^2}{\der^2p\over (2\p)^2}
 \jm_{p-k,\o}\ps^+_{k,\o\s}\ps^-_{p,\o\s}\cr
&+\sum_{\m,\n,\s}
 \sum_{q=k,p}
 \int_{D_{h-1}}\!{\der^2k\over (2\p)^2}{\der^2p\over (2\p)^2}
 \jm_{p-k,\m}\ps^+_{k,\n}\ps^-_{p,\n}
 D_\s(q)
 \int_0^1\!\der\t
 \left(\partial^q_\s\hW^{(h-1)}_{1;2,\m;\n}\right)(\t k,\t p)\;.
}$$
The local parts are absorbed into the {\it effective density
strength} on scale $h-1$, $\z^{(2,\s)}_{h-1}$:
$$
 \pmatrix{
 \z^{(2,+)}_{h-1}\cr
 \z^{(2,-)}_{h-1}\cr}
 \defi
 \left({Z_h\over Z_{h-1}}\right)
 \pmatrix{
  1+z^{(2)}_{h-1}+\D z^{(2,+)}_{h-1} &  \D z^{(2,-)}_{h-1}\cr
  \D z^{(2,-)}_{h-1}             &  1+z^{(2)}_{h-1}+\D z^{(2,+)}_{h-1}\cr}
 \pmatrix{
 \z^{(2,+)}_{h}\cr
 \z^{(2,-)}_{h}\cr}\;.
\Eq(ga3)$$
}

\0Multiscale integration goes on over all the scales $k$ s.t.
$\m_k\leq \k \g^k$, the first scale for which this is not true
being $k=h^*$. It is simply to verify that , for $h=h^*+1$ the
propagator \equ(p2) has the same dimensional bound of \equ(p1)
$$ \left|g^{(\leq h^*)}_{\o,\s}(x)\right|
 \leq C \g^{h^*}e^{-c\sqrt{\g^{h^*-1}\k|x|}}\;.$$
Finally,  it holds the following theorem.
\*

\theorem(T4)
{Let it be supposed  there exists $\e>0$ and  the constants
$c_0>0$  such that at any RG step
$h:h^*\leq h\leq N$, 
the effective parameters satisfy:
$$\g^{-c_0\e^2}\leq{Z_{h}\over Z_{h+1}}\leq\g^{c_0\e^2}\;,
 \qquad
 \g^{-2c_0\e}\leq{\m_h\over \m_{h+1}}\leq\g^{2c_0\e}\;,
 \qquad
 \g^{-2c_0\e}\leq{\z^{(2,\s)}_h\over \z^{(2,\s)}_{h+1}}\leq\g^{2c_0\e}\;,
\Eq(ind1)$$
$$|\n_h|,|\d_h|,|\l_h|\leq 2\e\;.
\Eq(ind2)$$
Then, for suitable positive constants $C,c$:
\elenco{
\art 
If $n^\f+n^\jm\neq 0$,
$$\int_{\L}\!\der^2\ux\
 \left|W^{(h)}_{n^\ps;n^\f;n^\jm,\uo,\us}
 (\ux,\uy,\uz)\right|
 \leq
 C{\g^{h\big(2-(1/2)n^\ps-(3/2)n^\f-n^\jm\big)}
  \over e^{{c\over 2(n^\f+n^\jm)}\sqrt{\g^h\k d(\uy,\uz)}}}\;;$$
\art if $n^\f+n^\jm= 0$,
$$\int_{\L}\!\der^2_*\ux\
 \left|W^{(h)}_{n^\ps;0;0,\uo,\us}
 (\ux)\right|
 \leq
 C\g^{h\big(2-(1/2)n^\ps\big)}\;;
$$%
}}

\0The proof is follows by simple dimensional analysis, and is
consequence of  the Appendices \secc(TE) and \secc(P).  
Since, by the first item, $\m_h/\g^h$
is strictly decreasing in $h$, for any choice of the mass $0\leq
\m\leq \g^{-1}\k$, the scale $h^*$ is negative; and:
$$
 {\log_\g(\m/\k)\over 1-2c_0\e}-1
 \leq h^*\leq
 {\log_\g(\m/\k)\over 1+2c_0\e}\;;
$$
hence, in the massless case, $h^*=-\io$.

\*

\section(BGF, Flows of the Running Coupling Constants)

A remarkable feature of the Localization is that 
among the flows of the effective parameters, only
the one for the mass is constructed with massive propagator;
the others are constructed with propagators 
$\{g^{({\rm E1},k)}_{\o,\o}\}_k$,
$\{g^{({\rm R1},k)}_{\o,\o}\}_k$
or 
$\{g^{({\rm D1},k)}_{\o,\o}\}_k$,
and therefore are independent on the mass flow.
Since the scale $h^*$ was introduced only to 
avoid bad bound on the massive propagators, 
all the flow, except $\{\m_k\}_k$, 
can be extended  from 
the range of scales $h^*\le k\le N $, 
to the range $k\le N$.

Other features of the flows of the effective parameters are depicted in
the following Theorem.
\*

\theorem(FC)
{Fixed any $\th: 0<\th <1/16$,
there exists $\e>0$ and two positive constants $c$ and $c_2$,
such that in correspondence of any  
parameters $\m$ and $\l$ satisfying $0\leq \m\leq \k\g^{-1}$ 
and $|\l|\leq \e$, there exist the parameters 
$\l_N$, $\m_N$, $Z_N$, $Z_N^{(2,+)}$, 
$Z_N^{(2,-)}$ and $\d_N$, $\n_N$, 
such that the following properties hold.
\elenco{
\art
The flow of  $\l_N$ is such that  
$$\lim_{h\to-\io}\l_h=\l\;;
 \qquad \left|\l_{h-1}-\l_h\right|
 \leq
 c\e^2\g^{-(\th/2)(N-h)}\;.
\Eq(l1)$$
\art 
The flows of $Z_N$ and $\m_N$
are such that 
$\m_0=\m$ and $Z_0=1$; furthermore
there exist $\h_\l$ and $\bar \h_\l$, 
independent from the regularization
used (Euclidean or Hamiltonian)
from the cutoff $N$, and from the mass $\m$, 
such that 
$$Z_h=\g^{-h\h_\l +\D G_h}\;,
 \qquad
 \m_h=\m\g^{-h\bar\h_{\l}+\D\bar G_h}\;, \Eq(z1)$$
with the rests, $\D G_h$ and $\D \bar G_h$,
summable in $h$:
$\left|\D G_h\right|,\left|\D\bar G_h\right|
 \leq c_2\e^2\g^{-(\th/2)\big(N-h\big)}$.
\art
 The flows of $Z_N^{(2,+)}$ and  
$Z_N^{(2,-)}$ are such that 
$Z^{(2,+)}_0=Z^{(2,-)}_0=1$; furthermore
there exist $\h^{(2)}_\l$ independent 
from the regularization, as well as from the 
mass $\m$ and  the cutoff $N$, 
such that 
$$Z^{(2,+)}_h=
 \g^{-h \h_{\l}^{(2)}+\D G^{(2,+)}_h}\;,
 \qquad
 Z^{(2,-)}_h=
 \g^{-h \h_{\l}^{(2)}+\D G^{(2,-)}_h}\;,\Eq(zz1)$$
with the rests $\{\D G^{(2,\s)}_h\}_{\s=\pm}$ 
summable in $h$:
$\left|\D G^{(2,\s)}_h\right|\leq c_2\e^2
\g^{-(\th/2)(N-h)}$.
\art
 The flows of  $\d_N$ and $\n_N$ 
are such that
$|\d_h|,|\n_h| \leq 2\e\g^{-\th(N-h)}$.}}

The proof is given in Appendix \secc(P). It is based on the vanishing of
the Beta function of massless Thirring model.
\*

\section(PTM, Equivalence of the Euclidean and  Hamiltonian Regularization)

\proof {\bf of Theorem \thm(EEH).} 
It is a  corollary of the Theorem~\thm(FC).
It can be obtained in the same way 
as the proof of Lemma \lm(LRC).
Anyway, using the{\it short memory property }
(see \secc(smp)), and the compact support of
the propagators,
a slightly easier proof is available for 
the Fourier transform of the Schwinger functions
with at least one field insertion.
Indeed,  the $(m;n+1)$-Schwinger functions 
calculated at fixed momenta $p_1,\ldots,p_m$, 
$q_1,\ldots, q_n$,  no matter if they are  obtained 
from the Hamiltonian or the Euclidean regularization, 
asymptotically
in the limit of 
removed cutoff are equal to  
the sum of the following  Feynman graphs:
all the graphs found in the 
expansion of the Schwinger functions, excluding those ones having 
an interaction on scale $m\geq N$, or an interaction 
$\DD$ or $\NN$, or a propagator $\{g^{({\rm R},k)}\}_{k}$,
and replacing the parameters 
$\l_k$, $Z_k$, $Z^{(2,\s)}_k$ and $\m_k$, respectively   
with $\l$, $\g^{k\h_\l}$, $\g^{k\h^{(2)}_\l}$
and $\m\g^{k\bar\h_\l}$. Indeed, these graphs
do not depend on the regularization; then, the 
difference between the  sum of such graphs 
and the corresponding Schwinger function
is bounded by the modulus of the sum 
of the graphs with one external 
fermionic propagator 
on the scale of the momentum $q_1$,
called $h_1$ -- fixed $q_1$, by compact support 
function, $h_1$ can be chosen between two adjoining momenta scales --
an effective parameter or propagator on scale $m$,
and falling in one of the following cases.
\elenco{
\item{\bf i.} It is $m\geq N$. Then, by the short memory 
property, the sum of such graphs is bounded, up to a constant, 
by $\g^{-\th(N-h_1)}$.
\item{\bf ii.} It is $m<N$ and the parameter is $\d_m$ or $\n_m$.
By the property of the flows of $\d_N$ and $\n_N$, and
by the short memory property,
the sum of such graphs is bounded, up to a constant, 
by $\g^{-\th|m-h_1|}\g^{-(\th/2)(N-m)}\leq$
$\g^{-(\th/2)(N-h_1)}\g^{-(\th/2)|m-h_1|}$.
\item{\bf iii.} There is a propagator $g^{(R,m)}_\o$ on scale $m<N$. By the 
bound of such a propagator and the short memory property,
the sum of such graphs is bounded 
by $\g^{-\th|m-h_1|}\g^{-(3/4)(N-m)}\leq$
$\g^{-(\th/2)(N-h_1)}\g^{-(\th/2)|m-h_1|}$, for $\th<3/4$.
\item{\bf iv.} It is $m<N$ and effective parameter
$\l_m-\l$, or $Z_m-\g^{m\h_\l}$, 
or $\m_m-\g^{m\bar\h_\l}$, 
or $Z^{(2,\s)}_m-\g^{m\h^{(2)}_\l}$. 
By the property of the flows, and
by the short memory property,
the sum of such graphs is bounded, up to a constant, 
by $\g^{-\th|m-h_1|}\g^{-(\th/2)(N-m)}\leq$
$\g^{-(\th/2)(N-h_1)}$ $\g^{-(\th/2)|m-h_1|}$.}

\0Furthermore the scale $h^*$, in the limit of removed cutoff,
only depends on $\l,\m$. Therefore,  it is possible to perform the sum over $m$ 
and to get for the difference of the Schwinger function
derived in the two different settings 
a bound $\g^{-(\th/2)(N-h_1)}$, for $0<\th<1/16$, up to a constant.
Anyway, in order to have, for  different regularizations,
identical  values of $\l$ and $\m$ (and consequently also of $\h_\l$, 
$\bar\h_\l$ and $\h^{(2)}_\l$), the initial parameters will be  
generally different.\hfill\qed\hskip1em\null
\chapter(WTI,Phase and Chiral Symmetries)

\section(WTI1, Ward-Takahashi Identities)

The classical Lagrangian is invariant under the {\it global}
transformations of the fields:
$$
 \ps^\s_{x,\o}\rightarrow e^{i\s\a_\o}\ps^\s_{x,\o}\;;\Eq(tr)
$$
as the phase, $\{\a_{\o}\}_{\o=\pm}$ does depend on the component
of the fermion fields, $\o$, this transformation is a combination
of the phase and chiral transformations in the Dirac notation.

This symmetry can be implemented in the generating functional of
the Euclidean Thirring model; and in particular, in order to obtaining
the identity $\h_\l=\h^{(2)}_\l$ and the vanishing of the Beta
function it will be useful to consider the generating functional
with infrared cutoff on scale $h$. It has to be performed a real
exponential transformation and to allow a dependence of the
parameter $\{\a_{\o}\}_{\o=\pm}$ on the space points: a new real
field, $\{\a_{x,\o}\}_{\o=\pm}^{x\in\L}$ arises -- this
prescription looks like, but has not to be confused with, the
implementation of a {\it gauge symmetry}.
\*
\sub(WTISF) 
{\wti{} for the  Schwinger functions.} 
An essential condition to get the consequences of the \wti{}
in the functional integration framework is to transform the 
field in {\it every site} of  $\L$. This seems to be forbidden
by the choice of a compact support cutoff function, ad the consequent 
restriction to the momenta in  $D_N$.
Therefore, let $\c^\d_N(k)$ be the cutoff function 
obtained adding to $\c_N(k)$ an exponential decaying 
tail $\d\D \c_N(k)$, alway strictly positive.
 
Hence, let the
following transformation of the integration variables in Fourier
space be considered
$$
 \hp^\s_{k,\o}\longrightarrow
 \hp_{k,\o}^\s-\s\int_{D}\!{\der^2p\over (2\p)^2}\
 \ha_{p,\o}\hp^\s_{k-\s p,\o}\;.
 \Eq(qtr)
$$
Calling $\c^\d_{h,N}(k)\defi \c^\d_N(k)-\c^\d_h(k)$, the \equ(qtr) implies
the following transformation of the kernel of the free measure
$$\eqalign{
 \int_{D}\! {\der^2k\over(2\p)^2}\
&\hp_{k,\o}^+{D_\o(k)\over \c^\d_{h,N}(k)}\hp_{k,\o}^-
 \longrightarrow
 \int_{D}\! {\der^2k\over(2\p)^2}\
 \hp_{k,\o}^+{D_\o(k)\over
 \c^\d_{h,N}(k)}\hp_{k,\o}^-\cr
 +
&\int_{D}\! {\der^2p\over(2\p)^2}{\der^2k\over(2\p)^2}\
\ha_{p,\o}\hp_{k,\o}^+\hp_{k+p,\o}^-
 \left[{D_\o(k)\over \c^\d_{h,N}(k)}-{D_\o(k+p)\over \c^\d_{h,N}(k+p)}\right]\;,}$$
and
$$\eqalign{
 {D_\o(k)\over \c^\d_{h,N}(k)}
-&{D_\o(k+p)\over \c^\d_{h,N}(k+p)}
 \defi
 -D_\o(p) - C^\d_\o(k,k+p)\cr
&=-D_{\o}(p)
 -
 \left[D_{\o}(k)\left(1-\big(\c^\d_{h,N}\big)^{-1}(k)\right)
-D_{\o}(k+p)\left(1-\big(\c^\d_{h,N}\big)^{-1}(k+p)\right)\right]\;.}$$
It is suitable to introduce the interactions with the external
source $\ha_\o$:
$$\eqalign{
 \AAA_0 (\a,\ps)
 &\defi
 \sum_{\o=\pm}\int_{D}\!
 {\der^2q\over (2\p)^2}{\der^2p\over (2\p)^2}\
  C^\d_\o(q,p)\ha_{p-q,\o}\hp^+_{q,\o}\hp^-_{p,\o}\;,\cr
 \AAA_\s(\a,\ps)
 &\defi
 \sum_{\o=\pm}\int_{D}\!
 {\der^2q\over (2\p)^2}{\der^2p\over (2\p)^2}\
 D_{\s\o}(p-q)\ha_{p-q,\o}\hp^+_{q,\s\o}
 \hp^-_{p,\s\o}\;, \quad {\rm for \ }\s=\pm\;,
}$$
\vbox
{\insertplot{300pt}{90pt}%
{}%
{f0}{\eqg(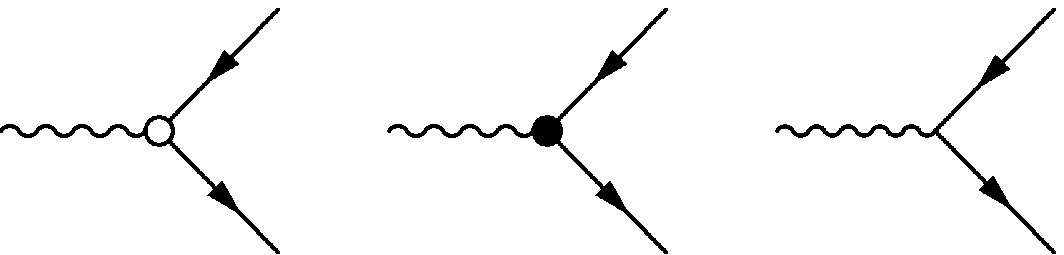)}
\centerline{{\bf Fig \graf(f0.ps)}: Graphical representation of {$\AAA_0$},
{$\AAA_-$} and {$\AAA_+$}}}
\vskip2em
so that, the transformation of $\WW^{(h)}$ reads
$$
 \eqalign{
 e^{\WW^{(h)}(\jm,\f)}
 =\lim_{\d\to 0}
 \int\!
&\der P^{[h,N]}(\ps)
 \exp\left\{
 -l_N\VV(\ps)+ Z_N^{(2)}\JJ(\jm,\ps)
 +\FF(\f,\ps)\right\}\cr
&\cdot\exp\left\{
 Z_N\AAA_+\left(\a,\ps\right)
 +
 Z_N
 \AAA_0\left(\a,\ps\right)
 \right\}\cr
&\cdot\exp\left\{
 \sum_{\o=\pm}
 \int\!{\der^2p\over (2\p)^2}{\der^2k\over (2\p)^2}\
 \ha_{p,\o}
\left[\hf^+_{k,\o}\hp^-_{k+p,\o}-\hp^+_{k,\o}\hf^-_{k+p,\o}\right]\right\}\;.}
\Eq(trW)$$
Being that $\WW^{(h)}$ is independent of $\a$, summing and
subtracting in the argument of the exponential $Z_N\sum_{\m=\pm}
\n^{(\m)}_N\AAA_\m\left(\a,\ps\right)$,
and then taking a derivative in $\ha_{p,\m}$ for $\ha= 0$, it
yields:
$$\eqalign{
 \left({1-\n^{(+)}_N\over \z^{(2)}_N}\right)
&D_\m(p){\partial \WW^{(h)}\over \partial \hj_{p,\m}}(\jm,\f)
 -{\n^{(-)}_N\over \z^{(2)}_N}
 D_{-\m}(p){\partial \WW^{(h)}\over \partial \hj_{p,-\m}}(\jm,\f)\cr
&=
 \int_{D}\!{\der^2k\over (2\p)^2}
 \left[
 {\partial \WW^{(h)}\over \partial \hf^-_{k,\m}}\hf^-_{k+p,\m}-
 \hf^+_{k,\m}
 {
 \partial \WW^{(h)}\over \partial \hf^+_{k+p,\m}}\right]
 -{\partial \WW^{(h)}_\AAA\over \partial \ha_{p,\m}}(0,\jm,\f)\;,
}\Eq(grez)$$
where the last term is given is the derivative of the functional
$$\eqalign{
 e^{\WW^{(h)}_\AAA
 (\a,\jm,\f)}\defi
 \int\!
 \der P^{[h,N]}\
 (\ps)
&\exp\left\{
 -l_N \VV(\ps) + Z^{(2)}_N\JJ(\jm,\ps) + \FF(\f,\ps)\right\}\cr
&\exp\left\{
 Z_N\Big[\AAA_0
 +\sum_{\m=\pm}\n^{(\m)}_N\AAA_{\m}\Big]\left(\a,\ps\right)
 \right\}\;.
}\Eq(WA)$$
Its derivatives are {\it remainders} which will be proved to
vanish in the limit of removed cutoff. Anyway, this holds for
$\{\n^{(\s)}_N\}_{\s=\pm}$ having {\it non-vanishing limit}: w.r.t.
the formal \wti, they represent an {\it anomaly}. Adhering to the
Johnson's notations, let the following definitions be considered:
$$
 a_N\defi{1\over 1-\left(\n^{(-)}_N+\n^{(+)}_N\right)}\;,
 \qquad
 \bar{a}_N\defi{1\over 1+\left(\n^{(-)}_N-\n^{(+)}_N\right)}\;;
$$
now, the \wti{} due to the {\it phase symmetry} (to be compared
with formula (16) of [J61]) is obtained summing \equ(grez) over
$\m$:
$$\eqalign{
 \sum_\m D_{\m}(p){1\over \z^{(2)}_N}
 {\partial \WW^{(h)}\over \partial \hj_{p,\m}}(\jm,\f)
&=
  a_N\sum_\m
 \int_{D}\!{\der^2k\over (2\p)^2}
 \left[
 {\partial \WW^{(h)}\over \partial \hf_{k,\m}}\hf^-_{k+p,\m}
 -
 \hf^+_{k,\m}
 {\partial \WW^{(h)}\over \partial \hf_{k+p,\m}}\right]\cr
&-a_N\sum_\m {\partial \WW^{(h)}_\AAA\over \partial
\ha_{p,\m}}(0,\jm,\f)\;; }$$
whereas the one due to the {\it chiral symmetry} (to be compared
with formula (17) of [J61]) is obtained multiplying both members
of \equ(grez) times $\m$ and summing over $\m$:
$$\eqalign{
 \sum_\m \m D_{\m}(p){1\over \z^{(2)}_N}
 {\partial \WW^{(h)}\over \partial \hj_{p,\m}}(\jm,\f)
&=\bar{a}_N\sum_\m \m
 \int_{D}\!{\der^2k\over (2\p)^2}
 \left[
 {\partial \WW^{(h)}\over \partial \hf^-_{k,\m}}\hf^-_{k+p,\m}-
 \hf^+_{k,\m}
 {\partial \WW^{(h)}\over \partial \hf^+_{k+p,\m}}\right]\cr
&-\bar{a}_N\sum_\m \m{\partial \WW^{(h)}_\AAA\over \partial
\ha_{p,\m}}(0,\jm,\f)\;. }$$
Finally, being that $(1+\s\m)/2=\d_{\s,\m}$, summing the two above
equations, the final expression for the \wti{} reads:
$$\eqalign{
 D_{\s}(p){1\over \z^{(2)}_N}
 {\partial \WW^{(h)}\over \partial \hj_{p,\s}}(\jm,\f)
&=\sum_\m {a_N+\bar{a}_N\s\m \over 2}
 \int_{D}\! {\der^2k\over (2\p)^2}\
 \left[
 {\partial \WW^{(h)}\over \partial \hf^-_{k,\m}}\hf^-_{k+p,\m}-
 \hf^+_{k,\m}
 {\partial \WW^{(h)}\over \partial \hf^+_{k+p,\m}}\right]\cr
&-\sum_\m {a_N+\bar{a}_N\s\m \over 2}\ {\partial
\WW^{(h)}_\AAA\over \partial \ha_{p,\m}}(0,\jm,\f)\;. }\Eq(WT1)$$
By taking suitable derivatives w.r.t. the field $\hf$ for
$\jm=\f=0$, \equ(WT1) generates all the \wti{} involving one
density insertion: for instance, by taking  derivatives w.r.t.
$\hf^+_{k,\o}$ and $\hf^-_{k+p,\o}$, \equ(WT1) gives \equ(wi) and
\equ(gg), for
$$\D\hH^{(1;2)}_{\s,\o}(p;k)\defi
{\partial\WW^{(h)}_\AAA\over
\partial \ha_{p,\m}\partial\hf^+_{k,\o}\partial\hf^-_{k+p,\o}}
(0,0,0)\;.$$
\*
\sub(F+-)
{Flows of $\n^{(+)}_N$ and $\n^{(-)}_N$. } 
The
remainder of the above \wti{} are the Schwinger
 functions generated from the functional
$\WW^{(h)}_\AAA$ with one -- and only one -- derivation in the
field $\ha$, and various number of derivation in the fields $\f$'s.
Therefore it is necessary to study the renormalization of the
contraction of the vertices $\{\AAA_{a}\}_{a=0,\pm}$, up to linear
order in $\ha$, which lead to the flows of $\n^{(+)}_N$ and
$\n^{(-)}_N$.

By induction, having integrated the scale from the $N$-th below to
the $j$-th, it is possible to prove that, up to the
renormalization of the coupling constants already present in
functional $\WW^{(h)}$, the functional $\WW^{(h)}_\AAA$ reads:
$$\eqalign{
 e^{\WW^{(h)}_\AAA
 (\a,\jm,\f)}\defi
 \int\!
 \der P^{[h,j]}\
 (\ps)
&\exp\left\{\WW^{(j)}\left(\f,\jm,\sqrt{Z_j}\ps\right)
 +\WW_{\AAA,{\rm irr}}^{(j)}\left(\a,\f,\jm,\sqrt{Z_j}\ps\right)
 \right\}\cr
&\exp\left\{\Big[\left({Z_N\over Z_j}\right)\AAA_0
 +\sum_{\m=\pm}\n^{(\m)}_j\AAA_\m\Big]\left(\a,\sqrt{Z_j}\ps\right)
 \right\}\;,
}$$
where $\WW^{(j)}$ and $\WW_{\AAA,{\rm irr}}^{(j)}$ are defined as
in formula \equ(g3), but with propagators and couplings obtained
for the Euclidean massless Thirring model; besides in the monomials of
the fields of $\WW_{\AAA,{\rm irr}}^{(j)}$ there is also one
$\a$-field and either $n^\ps+n^\f\geq 2$ or $n^\jm\geq 1$.

From this section to the end, since all the developments will be
about the Euclidean Massless Thirring model,  let $\hg^{({\rm E1},h)}_{\o,\s}$
be called, with abuse of notation,  $\hg^{(h)}_{\o}$.
\*
\lemma(L2)
{Let the kernel $U_{\e;\o}^{(i,j)}(k,p)\defi
C^\d_\o(k,p)\hg^{(j)}_\o(k)\hg^{(i)}_\o(p)$ be considered. It can be
decomposed into
$$U_{\e;\o}^{(i,j)}(k,p)
 \defi
 \sum_\s D_\s(p-k)S_{\e;\o,\s}^{(i,j)}(k,p)\;,$$
and $S_{\o,s}^{(i,j)}$, the limit $\e\to 0$ of  
$S_{\e;\o,s}^{(i,j)}$, satisfies the bound  
$$|\partial^{s_i}_k\partial^{s_j}_p S_{\o,\s}^{(i,j)}(k,p)|
 \leq
 \left\{
 \matrix{
 C\g^{-i(1+s_i)-j(1+s_j)}\hfill&&\hfill {\rm if\ } i{\rm \ or\  }j=h, N\cr
 0\hfill&&\hfill {\rm otherwise}\;.\cr
 }\right.
$$
}
\0The proof of the bound is given in appendix \secc(BD). It means
that {\it formally} $C^\d_\o$ can be thought as a 1-dimensional
kernel: since the monomial $\a\ps\ps$ has dimension 1, the power
counting for the graphs with insertion of the vertex $\AAA_0$ will
be found to be always satisfied.
\*
\sub(IL)
{Improved localization I.} 
As for the effective
potential, also the multiscale integration of $\WW_\AAA$ is
companied by a localization and absorption in the effective
parameters the graphs which are divergent according to the
dimensional analysis. At the $j-1$-th scale, with the inductive
hypothesis the previous scales were integrated and the local terms
were extracted, they holds the following cases.
\elenco{ 
\art
One field $\hp$ of the interaction $\AAA_0$, contracted with a
kernel $\hW^{(j)}_{2,\o}(k)$, has vanishing local part since
$\hW^{(j)}_{2,\o}(0)=0$ by symmetries; furthermore, for compact
support arguments, such a contraction can only occur at
scale $j$:
$$\eqalign{
&\LL\left[\int_{D}\!{\der^2k\over (2\p)^2}{\der^2q\over (2\p)^2}\
\ha_{k-q,\o}\hp_{q,\o}^+\hp_{k,\o}^-\
 C^\d_\o(q,k)\hg^{(j)}_\o(k) \hW^{(j)}_{2,\o}(k)\right]=0\;,\cr
&\RR\left[\int_{D}\!{\der^2k\over (2\p)^2}{\der^2q\over (2\p)^2}\
\ha_{k-q,\o}\hp_{q,\o}^+\hp_{k,\o}^-\
 C^\d_\o(q,k)\hg^{(j)}_\o(k)
 \hW^{(j)}_{2,\o}(k)\right]\cr
&=
 \sum_{\m=\pm}\int_{D}\!{\der^2k\over (2\p)^2}{\der^2q\over (2\p)^2}\
 \ha_{k-q,\o}\hp_{q,\o}^+\hp_{k,\o}^-\cr
&\phantom{************}
 D_\m(k)
 \left[C^\d_\o(q,k)\hg^{(j)}_{\o}(k)
 \int_0^1\!\der\t\
 \left(\partial_\s\hW^{(j)}_{2,\o}\right)(\t k)\right]\;;
}$$
the derivative clearly improves the bound on the kernel
$\hW^{(j)}_{2,\o}$ of one negative dimension, at a loss of the
bound on the kernel that will be obtained contracting the field
$\hp^-_{k,\o}$ in a scale lower than $j-1$.\\
This automatic dimensional gain is due to the fact that this
situation cannot occur in more than one node $v$ in the tree
expansion, and in its first preceding $v'$; hence an alternative way to
cure it is to multiply by $\g^{-2}\g^{2}$: the former factor makes
negative the dimension of such a graph, the latter worsen the
bound of a constant.
\art 
As in the previous point, one $\hp$-field of the vertex
$\sum_\s\n_{j}^{(\s)}\AAA_\s$, contracted with a kernel
$\hW^{(j)}_{2,\s\o}(k)$ has vanishing local part; since
$\hp_{k,\o}^+$ has to be contracted on scale $j$:
$$\eqalign{
&\LL\left[\int_{D}\!{\der^2k\over(2\p)^2}{\der^2q\over (2\p)^2}\ 
\ha_{k-q,\o}\hp_{q,\s\o}^+\hp_{k,\s\o}^-\
 D_\o(k-q)\hg^{(j)}_{\s\o}(k) \hW^{(j)}_{2,\s\o}(k)\right]=0\;,\cr
&\RR\left[\int_{D}\!{\der^2k\over(2\p)^2}{\der^2q\over (2\p)^2}\ 
\ha_{k-q,\o}\hp_{q,\s\o}^+\hp_{k,\s\o}^-\
 D_\o(k-q)\hg^{(j)}_{\s\o}(k)
 \hW^{(j)}_{2,\s\o}(k)\right]\cr
&=
 \sum_{\m=\pm}
 \int_{D}\!{\der^2k\over (2\p)^2}{\der^2q\over (2\p)^2}\
 \ha_{k-q,\o}\hp_{q,\s\o}^+\hp_{k,\s\o}^-\
 D_\m(k)\cr
&\phantom{***************}
 \cdot\left[D_\o(k-q)\hg^{(j)}_{\s\o}(k)
 \int_0^1\!\der\t\
 \left(\partial_\s\hW^{(j)}_{2,\s\o}\right)(\t k)\right]\;.
}$$
\vbox
{\insertplot{400pt}{90pt}%
{}%
{f2}{\eqg(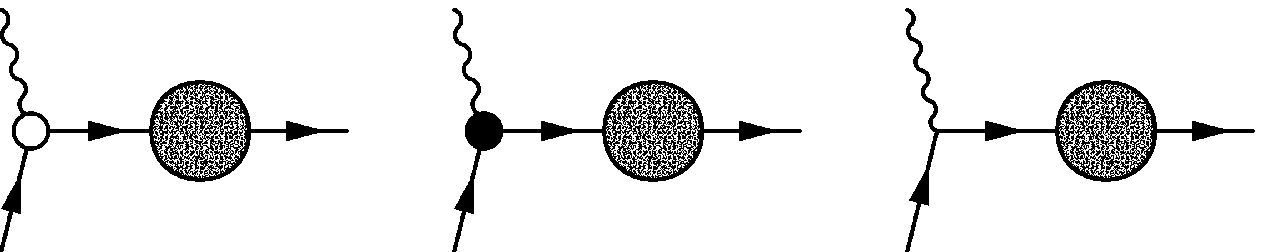)}
\centerline{{\bf Fig \graf(f2.ps)}: Graphical representation of items 1. and 2.}}
\vskip2em
\art
Both $\hp$-field of the interaction $\AAA_0$, contracted
with a graph $\hW^{(j)}_{4,\o,\m}$, is identically vanishing,
except if at least one of the two propagators is on scale $N$, or
$h$. It is convenient to define:
$$
 \hM_{\o,\s\o,\m}^{(r,s),(4)}(p,q)\defi
 \int_{D}\!{\der^2k\over (2\p)^2} \
 S_{\o,\s\o}^{(r,s)}(q+k,p+k)
 \hW^{(j)}_{4,\o,\m}(q,p,k)\;.
$$
By symmetry under rotation and under space reflection
(\secc(SR) and \secc(Rot)), it holds:
$$
 \hM_{\o,\s\o,\m}^{(r,s),(4)}(0,0)
 \left\{
 \matrix
 {=0\hfill&\hfill {\rm \ for\ }\m=-\s\o\cr
  \defi  \D n_j^{(0,\s)} \hfill&\hfill {\rm \ for\ }\m=\s\o\;.}
 \right.
$$
Hence the localization of such graphs gives:
$$
 \eqalign{
 \LL
&\left[ \sum_{\s}\int_{D}\!
 {\der^2q\over (2\p)^2}{\der^2p\over
 (2\p)^2}\
 D_{\s\o}(p-q)
 \ha_{p-q,\o}\hp_{q,\m}^+\hp_{p,\m}^-
  \hM_{\o,\s\o,\m}^{(r,s),(4)}(p,q)\right]\cr
&=\sum_{\s}\D n_j^{(0,\s)}
 \int_{D}\!{\der^2q\over (2\p)^2}{\der^2p\over (2\p)^2}\
 D_{\s\o}(p-q)
 \ha_{p-q,\o}\hp_{q,\s\o}^+\hp_{p,\s\o}^-\;,\cr
\RR &\left[ \sum_{\s}\int_{D}\!{\der^2q\over
(2\p)^2}{\der^2p\over (2\p)^2}\
 D_{\s\o}(p-q)
 \ha_{p-q,\o}\hp_{q,\m}^+\hp_{p,\m}^-
 \hM_{\o,\s\o,\m}^{(r,s),(4)}(p,q)\right]\cr
&=
 \sum_{k=p,q}\sum_{\s,\n=\pm}
 \int_{D}\!{\der^2q\over (2\p)^2}{\der^2p\over (2\p)^2}\
  D_{\s\o}(p-q)
 \ha_{p-q,\o}\hp_{q,\m}^+\hp_{p,\m}^-D_\n(k)\cr
&\phantom{********************}
 \cdot\int_0^1\!\der\t
 \left(\partial_\n^{k} \hM_{\o,\s\o,\m}^{(r,s),(4)}\right)
 (\t p,\t q)\;.
}$$
\art
For the contraction of both $\hp$-field of the vertex
$\sum_\s\n_i^{(\s)}\AAA_\s$ with a graph
$\hW^{(j)}_{4,\o\s,\m}$ it is
convenient to define:
$$
 \hM_{\s\o,\m}^{(i,r,s),(4)}(p,q)\defi
 \n^{(\s)}_i\int_{D}\!{\der^2k\over (2\p)^2} \
 \hg^{(r)}_{\s\o}(q+k)\hg^{(s)}_{\s\o}(p+k)
 \hW^{(j)}_{4,\s\o,\m}(q,p,k)\;.$$
As in the previous item, by symmetries it holds:
$$
 \hM_{\s\o,\m}^{(i,r,s),(4)}(0,0)
 \left\{
 \matrix
 {=0\hfill&\hfill {\rm \ for\ }\m=-\s\o\cr
  \defi  \D n_j^{(\s)}  \hfill&\hfill {\rm \ for\ }\m=\s\o\;.}
 \right.
$$
hence, the localization of such graphs gives:
$$
 \eqalign{
 \LL
&\left[
 \int_{D}\!{\der^2q\over (2\p)^2}{\der^2p\over (2\p)^2}\
 D_{\s\o}(p-q)
 \ha_{p-q,\o}\hp_{q,\m}^+\hp_{p,\m}^-
 \hM_{\s\o,\m}^{(i,r,s),(4)}(p,q)\right]\cr
&=\D n_j^{(\s)}
 \int_{D}\!{\der^2q\over (2\p)^2}{\der^2p\over (2\p)^2}\
 D_{\s\o}(p-q)
 \ha_{p-q,\o}\hp_{q,\s\o}^+\hp_{p,\s\o}^-\;,\cr
\RR
&\left[
 \int_{D}\!{\der^2q\over (2\p)^2}{\der^2p\over (2\p)^2}\
 D_{\s\o}(p-q)
 \ha_{p-q,\o}\hp_{q,\m}^+\hp_{p,\m}^-
 \hM_{\o,\s\o,\m}^{(i,r,s),(4)}(p,q)\right]\cr
&=
 \sum_{k=p,q}\sum_{\n=\pm}
 \int_{D}\!{\der^2q\over (2\p)^2}{\der^2p\over (2\p)^2}\
  D_{\s\o}(p-q)
 \ha_{p-q,\o}\hp_{q,\m}^+\hp_{p,\m}^-D_\n(k)\cr
&\phantom{********************}
 \cdot\int_0^1\!\der\t
 \left(\partial_\n^{k} \hM_{\s\o,\m}^{(i,r,s),(4)}\right)(\t p,\t q),
}$$
\vbox
{\insertplot{400pt}{90pt}%
{}%
{f1}{\eqg(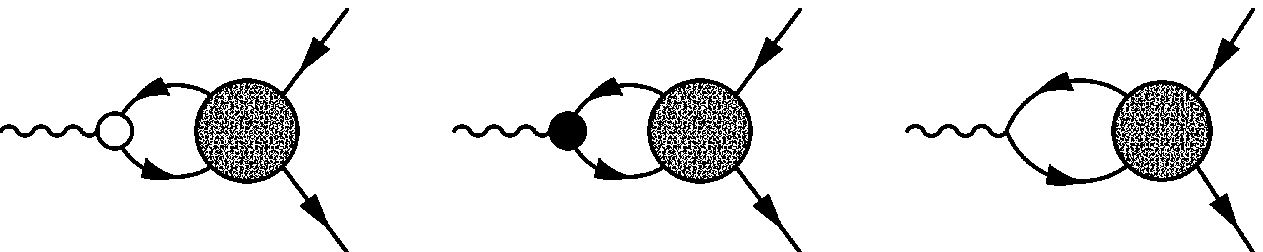)}
\centerline{{\bf Fig 13:} Graphical representation of items 3. and 4.}}
\vskip2em
\art
The self-contraction of the interactions $\AAA_0$ 
would give divergences because of $C_\o^\e$.
Anyway, such a self-contraction, either for $\AAA_0$,
and for $\{\AAA_\s\}_\s$, cannot occur in the expansion of the Schwinger
function: in such expansions they cannot occur subgraphs 
with no external fields of type $\ps$ or  $\f$. }

\0The local parts are absorbed into the effective parameter on scale
$h-1$:
$$\n^{(\s)}_{j-1}\defi{Z_j\over Z_{j-1}}
 \left(\n^{(\s)}_{j} + \D n_j^{(\s)} +\D n_j^{(0,\s)}\right)\;.$$

\theorem(TWTI)
{Fixed any $\th :0<\th<1/16$, 
there exists $\e>0$, a positive constant $c_4$  and two counterterms
$\n^{(+)}$ and $\n^{(-)}$, analytically dependent on $\l$,
such that, for any fixed  cutoff scale, $N$, and choosing 
$\n^{(\s)}_N=\n^{(\s)}$, it holds
$$\left|\n^{(\s)}_j\right|\leq c_4\e \g^{-(\th/2) (N-j)}\;.\Eq(TWTI)$$ 
}

The proof is in appendix \secc(PL). It is a simple application 
of the  fixed point theorem; once two counterterms 
$\{\n^{(\s)}_N\}_{\s=\pm}$ with the required property 
are found, it is easy to
verify they are sum of scaling invariant graphs, and therefore they 
are independent on the scale of the cutoff, $N$.
Accordingly, it is natural to
define:
$$a\defi{1\over 1-\left(\n^{(-)}+\n^{(+)}\right)}\;,
 \qquad
 \bar{a}\defi{1\over 1+\left(\n^{(-)}-\n^{(+)}\right)}\;.$$
Now it is possible to prove that, even removing the cutoff, the
\wti{} are not equal to the formal one because of the non-vanishing
anomaly $a-\bar{a}$.
\*

\theorem(LWTI) 
{In the same hypothesis of theorem 
\thm(TWTI), all the anomalous \wti{}
for Schwinger functions, with only one 
density insertion and calculated at fixed momenta w.r.t.
the cutoff scales, $h$ and $N$,  in the limit $-h,N\to \io$
are generated by suitable derivatives of the following identity:
$$D_{\s}(p){\partial \WW\over \partial \hj_{p,\s}}(\jm,\f)
 =\z^{(2)}_b\sum_\m {a+\bar{a}\s\m \over 2}
 \int_{D}\!{\der^2k\over (2\p)^2}
 \left[
 {\partial \WW\over \partial \hf^-_{k,\m}}\hf^-_{k+p,\m}
 -
 \hf^+_{k,\m}
 {\partial \WW\over \partial \hf^+_{k+p,\m}}\right]\;.\Eq(finale)$$
In particular, $(a+\bar a\s\m)/2 = \d_{\s,\m}+ \d_{\s,-\m}{\l/4\p} +{\rm O}(\l^2)$.}
\*

\0The essence of the anomaly is that  $(a+\bar a\s\m)/2\neq \d_{\s,\m}$,
which implies, in spite of the formal result, the non-vanishing of
$\hS^{(1;2)}_{-\o;\o}$. A celebrated consequence of the \wti, not
wasted by the anomaly, is the following.
\*

\theorem(Co){\it The anomalous exponent of the field
strength and the anomalous exponent of the density strength
coincide: $\h^{(2)}_\l=\h_\l$.}
\*

\0This is what in formal language is stated as $Z^{(2)}=Z$.
\*

\proof{\bf of Theorem \thm(LWTI)}.
With reference to \equ(WT1), 
it is only required to prove that the derivatives
of $\WW_\AAA^{(h)}$, made w.r.t. 
one field $\a$ and various fermionic fields
at fixed momenta, fulfil the same 
bound of the derivatives 
of $\WW^{(h)}$, with $\a$ replaced by $\jm$, 
with a more factor which is vanishing in the limit of
removed cutoff. 
Hence, let any integer $n\in \NNN$, any 
set of labels
$\e_1,\ldots,\e_n$ and $\o_1,\ldots,\o_n$,
and any momenta $p,k_1,\ldots,k_n$,
chosen independently from $h,N$, 
be considered.  It holds the
bound
$$\eqalign{
 {1\over |p|}
&\left|{\partial^{1+n} \WW^{(h)}_\AAA\over
 \partial \ha_{p,\m}
 \partial \hf^{\e_1}_{k_1,\o_1}
 \ldots \partial \hf^{\e_n}_{k_n,\o_n}}\right|
 _{\jm\equiv\f\equiv0}
 \leq{ C_{n;p,h_1,\ldots,h_n}\over \prod_{j=1}^n   \sqrt{Z_{h_j}}}
 \left(\g^{-(\th/2)(N-h_1)}
 +\g^{-(\th/2)(h_1-h)}\right)}\;,
\Eq(pbound)$$
where $\{h_j\}_{j=1}^n$ are the scales of $\{k_j\}_j$:
$\k\g^{h_j-1}\le |k_j|\le \k \g^{h_j}$ and
$C_{n;p,h_1,\ldots,h_n}/ \prod_{j=1}^n   \sqrt{Z_{h_j}}$ is 
the bound for the same derivatives of the functional $\WW^{(h)}$.
Such a  bound  can be obtained by the following argument.
The graphs in the expansion of the l.h.s. member of \equ(pbound)
has to have an external propagator on scale $h_1$
-- besides external propagators  on scales $h_2,\ldots,h_n$;
and they fall in one of the following cases.
\elenco{
\art 
An interaction $\AAA_0$ is contracted: this can happen 
only on scale $m=N,h$. 
By the short memory property (see \secc(smp)),
the sum of all such graphs is bounded, up to a constant, 
with $\g^{-(\th/2)(N-h_1)}$ or $\g^{-(\th/2)(h_1-h)}$.
\art
An interaction $\AAA_\s$ is first contracted on scale $m$,
and hence brings a coupling $\n^{(\s)}_m$.
By the short memory property 
and the bound in theorem \thm(TWTI),
the sum of  such graphs is bounded, up to a constant, 
with $\g^{-\th|m-h_1|}$ $\g^{-(\th/2)(N-m)}$
$\le \g^{-(\th/2)(N-h_1)}$$\g^{-(\th/2)|m-h_1|}$.}
\0Hence it is possible to take the sum over 
$m$, obtaining \equ(pbound).\hfill\qed\hskip1em\null

\*
\proof{\bf of Theorem \thm(Co).} It simply follows from lowest
order expansion of \equ(wi),  and from 
the proof of Theorem \thm(FC) -- in particular from the 
features of the anomalous exponents depicted in \secc(FP).
Indeed, since
$$
 \left|\h_\l-\G_h\right|\leq c\e\g^{-\th(N-h)}\;,
 \qquad
 \left|\h_\l^{(2)}-\G^{(2)}_h\right|\leq c\e\g^{-\th(N-h)}\;,$$
then
$$
 \log_\g\left({\z^{(2)}_h\over \z^{(2)}_N}\right)
 =
 (N-h)\left(\h_\l-\h_\l^{(2)}\right)+{\rm O}(\l^2)\;,
 \Eq(ette)$$
where ${\rm O}(\l^2)$ is a term of the order of $\l^2$ and bounded
for every $h$. Calling $\bar k$ any momentum $\k\g^{h}\leq |\bar k|\leq
\g^{h+1}$, by the lowest order  graph expansion
in Appendix \secc(TE), it holds,
$$\hS^{(1;2)}_{\o,\o}(2\bar k;\bar k)
={\z^{(2)}_h\over Z_h}{1+{\rm O}(\l^2)\over D^2_\o(\bar k)}\;, 
\qquad
\hS^{(2)}_{\o}(\bar k) ={1\over Z_h}{1+{\rm O}(\l^2)\over
D_\o(\bar k)}\;,$$
$$\D\hH^{(1;2)}_{\o,\o}(2\bar k;\bar k)
={1\over Z_h}{{\rm O}(\l^2)\over D_\o(\bar k)}\;, 
\qquad 
{a_N+\bar
a_N\over 2} =1+{\rm O}(\l^2)\;.$$
Replacing the above identities into  \equ(wi) and \equ(gg), 
the bound \hbox{$\log_\g(\z^{(2)}_h/ \z^{(2)}_N)={\rm O}(\l^2)$}
holds for any $h\leq N$: to be consistent with \equ(ette), it
cannot be but $\h_\l-\h_\l^{(2)}=0$.\hfill\qed\hskip1em\null

\*
\sub(R3){Remark: anomaly and anomalous exponents.}
Formally, by the phase and chiral symmetry, it is possible to
prove the identity of the field and density strength, $Z_N=Z^{(2)}_N$,
so that the renormalization $\z^{(2)}_N\=1$. But in a
rigorous setting, \wti{} are seen to break this identity. 
Anyway, since the anomaly only changes a factor in front 
of the current, the identity 
between the exponents with which  
$Z_N$ and $Z^{(2)}_N$ diverge {\it remains true};
therefore $\z_N^{(2)}$, although no longer constant,
is anyway bounded.
\*
\section(CE, Closed Equations)
\sub(SDE){Schwinger-Dyson equation}. The fermionic fields
satisfy an evolution equation which can be turned into a set of
equations for the Schwinger functions: see Appendix \secc(a3).
Such equations relate the $n$-points Schwinger functions to the
$m$-points Schwinger function with $m\le n$ and one density insertion. 
Using the \wti{} to write the latter in terms of $m$-point Schwinger functions,
the \ce's arise.
\*

\sub(CE2)
{Closed equations.} In Appendix \secc(a3), the
following equation, generator of all the \sde, is proved for any
$k:\g^h\k\leq |k|\leq \g^N\k$ -- where the cutoff $\c_{h,N}(k)\=1$:
$$\eqalign{
D_\o(k){\partial e^{\WW^{(h)}} \over\partial \hf^+_{k,\o}}
 ={\hf^-_{k,\o}e^{\WW^{(h)}}\over Z_N}
 -
 {\l_N\over \z^{(2)}_N}
 \int_{D}\!{\der^2p\over(2\p)^2}\
 {\partial^2 e^{\WW^{(h)}}\over \partial\hj_{p,-\o}
 \partial\hf^+_{k-p,\o}}\;,}\Eq(DSE2)$$
for $\jm\=0$ -- since here only the \ce{} for Schwinger function
{\it without} density insertion are studied.
Since it is  possible to prove the convergence of the last integral 
for small $p$; and since $|p|\leq 2\g^{N}\k$, 
it is convenient make in the argument of
the integral the following replacement:
$$ 1\=\c_{N+2}(p)\=\c_{h+2,N+2}(p)+\c_{h+2}(p)\defi\bar\c_{h,N}(p)+\bar\c_{h}(p) \;.$$
where $\c_{h+2,N+2}(p)\defi \c_{N+2}(p)-\c_{h+2}(p)$.
Then, from the generator of the \wti, \equ(WT1), it holds
the following integral identity:
$$\eqalign{
&{1\over \z^{(2)}_N}
 \int_{D}\! {\der^2p\over (2\p)^2}\
 \bar \c_{h,N}(p){\partial^2 e^{\WW^{(h)}}\over \partial \hj_{p,-\o}
 \partial\hf^+_{k-p,\o}}\cr
=&\sum_\m {a_N+\bar{a}_N\s\m \over 2}
 \int_{D}\! {\der^2p\over (2\p)^2}{\der^2q\over (2\p)^2}\
 {\bar\c_{h,N}(p)\over D_{-\o}(p)}
 \left[
 {\partial^2 e^{\WW^{(h)}}\over \partial\hf^+_{k-p,\o}
 \partial \hf^-_{q,\m}}\hf^-_{q+p,\m}
 -\hf^+_{q,\m}
 {\partial^2 e^{\WW^{(h)}}\over
 \partial \hf^+_{q+p,\m}\partial\hf^+_{p-k,\o}}\right]\cr
 &-\sum_\m {a_N+\bar{a}_N\s\m \over 2}
 \int_{D}\! {\der^2p\over (2\p)^2}\
 {\bar\c_{h,N}(p)\over D_{-\o}(p)}
 {\partial^2e^{\WW^{(h)}_\AAA}\over
 \partial\ha_{p,\m}\partial\hf^+_{k-p,\o}}(0,\jm,\f)\;.
 }\Eq(intwt11)$$
Taking a derivative in $\hf^-_{k,\o}$, and putting $\f=0$,
gives \equ(intwti) -- apart from the function $\bar\c_{h,N}(p)$
that had been skipped for reproducing the Johnson's argument. 
By the general analysis of the previous
section, the term proportional to the derivatives of the
functional $\WW_\AAA$ would have been vanishing in the limit of removed
cutoff {\it if the external momenta had been fixed}. But in this case
the external momenta are integrated over, and there is no reason
that this term is vanishing -- differently from what implicitly
stated in [J61].

\*
\sub(FWY){Flows of $\wt z^{(\m)}_N$ and $\wt \l^{(\m)}_N$.}
To overcome the problem of not having, 
neither in the limit, a real closed equation, 
it is possible  to write such a rest as addends that are
already present in the \sde. 
To this purpose, let the functionals
$\WW^{(h)}_{\TT,\m}$, for $\m=\pm$ be defined as
$$\eqalign{
 e^{\WW^{(h)}_{\TT,\m} (\b,\jm,\f)}
 \defi\int\!
&\der P^{[h,N]}(\ps)\
 \exp\Bigg\{
 -\l_{N} \VV\left(\sqrt{Z_N}\ps\right)
 + \z^{(2)}_{N}\JJ\left(\jm,\sqrt{Z_N}\ps\right)
 +\FF(\f,\ps)\Bigg\}\cr
&\exp\left\{
 \left[\TT^{(\m)}_0+\sum_{\s=\pm}\n_N^{(\s)} \TT^{(\m)}_{\s}
 -\a^{(\m\o)}\l_N\BB^{(3)}-\s^{(\m\o)}\BB^{(1)}\right]
 \left(\sqrt{Z_N}\ps,\sqrt{Z_N}\b\right)\right\}\;,
}$$
with $\{\a^\m\}_{\m=\pm}$ and  $\{\s^\m\}_{\m=\pm}$ 
four real parameters later fixed;
and
$$\eqalign{
 \TT^{(\m)}_0(\ps,\b)
&\defi
 \sum_{\o=\pm}
 \int\! {\der^2k\over (2\p)^2}{\der^2p\over (2\p)^2}{\der^2q\over (2\p)^2}\
 \bar\c_{h,N}(p){C_\m(q,p+q)\over D_{-\o}(p)}
 \hb_{k,\o}\hp^-_{k-p,\o}\hp^+_{q,\m}\hp^-_{p+q,\m}\;,\cr
 \TT^{(\m)}_{\s}(\ps,\b)
&\defi
 \sum_{\o=\pm}
 \int\! {\der^2k\over (2\p)^2}{\der^2p\over (2\p)^2}{\der^2q\over (2\p)^2}\
 \bar\c_{h,N}(p){ D_{\s\m}(p)\over D_{-\o}(p)}\
 \hb_{k,\o}\hp^-_{k-p,\o}\hp^+_{q,\s\m}\hp^-_{p+q,\s\m}\;;}$$
\vbox
{\insertplot{350pt}{90pt}%
{}%
{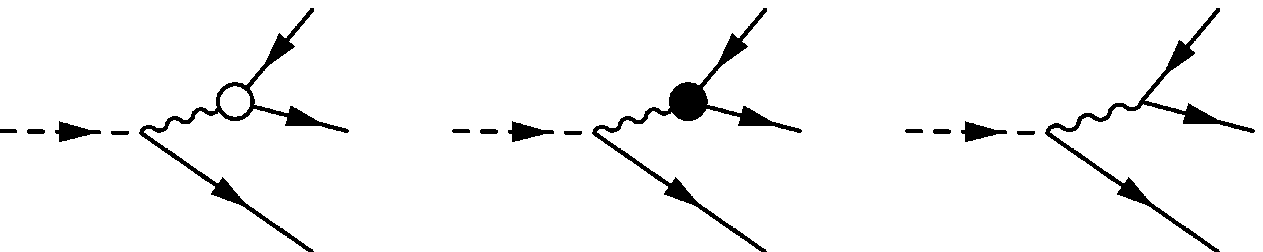}{\eqg(f15)}
\centerline{{\bf Fig \graf(f15)}: Graphical representation of the interactions {$\TT_0^{(\m)}$},  
{$T_-^{(\m)}$} and {$T_+^{(\m)}$}}}
\vskip1em
$$\eqalign{
 \BB^{(3)}(\ps,\b)
&\defi
 \sum_{\o=\pm}\int\!{ \der^2k\over (2\p)^2}{ \der^2p\over (2\p)^2}{ \der^2q\over (2\p)^2}\
 \hb_{p+k-q,\o}\hp^-_{p,\o}\hp^+_{q,-\o}\hp^-_{k,-\o}\;,\cr
 \BB^{(1)}(\b,\ps)
&\defi
 \sum_{\o=\pm}\int\!{\der^2k\over (2\p)^2}\
 \hb_{k,\o} D_\o(k)\hp^-_{k,\o}\;.
}$$
Because of the identity
$$\eqalign{
 \int\!{\der^2 p\over (2\p)^2}\
 {\bar\c_{h,N}(p)\over D_{-\o}(p)}
 {\partial^2e^{\WW_\AAA}\over \partial \ha_{p,\m}
 \partial \hf^+_{k-p,\o}}
 =
 {1\over Z_N}
 {\partial e^{\WW^{(h)}_{\TT,\m}}\over \partial \hb_{k,\o}}
&+
 \a^{(\m\o)}{\l_N\over \z^{(2)}_N}
 \int\! {\der^2p\over (2\p)^2}\
 {\partial^2 e^{\WW}\over
 \partial \hj_{p,-\o} \partial \hf^+_{k-p,\o}}\cr
&+
 \s^{(\m\o)} D_{\o}(k)
 {\partial e^{\WW}\over
 \partial \hf^+_{k,\o}} \;,
}\Eq(ce1)$$
it is possible to turn equation \equ(intwt11) into:
$$\eqalign{
&\left(1-\sum_\m {a_N-\bar{a}_N\m \over 2}
 \a^{(\m)}\l_N\right)
 {1\over \z^{(2)}_N}
 \int_{D}\! {\der^2p\over (2\p)^2}\
 \bar\c_{h,N}(p)
 {\partial e^{\WW^{(h)}}\over \partial \hj_{p,-\o}
 \partial\hf^+_{k-p,\o}}\cr
&=
 \left(\sum_\m {a_N-\bar{a}_N\m \over 2}
 \s^{(\m)}\right)D_\o(k)
 {\partial e^{\WW^{(h)}}\over 
 \partial\hf^+_{k,\o}}
 \cr
&+\sum_\m {a_N-\bar{a}_N\o\m \over 2}
 \int_{D}\! {\der^2p\over (2\p)^2}{\der^2q\over (2\p)^2}\
 {\bar\c_{h,N}(p) \over  D_{-\o}(p)}\left[
 {\partial e^{\WW^{(h)}}\over \partial\hf^+_{k-p,\o}
 \partial \hf^-_{q,\m}}\hf^-_{q+p,\m}
 -\hf^+_{q,\m}
 {\partial e^{\WW^{(h)}}\over
 \partial \hf^+_{q+p,\m}\partial\hf^+_{p-k,\o}}\right]\cr
 &-
 {1\over Z_N}\sum_\m {a_N-\bar{a}_N\o\m \over 2}
 {\partial e^{\WW^{(h)}_{\TT,\m}}\over \partial \hb_{k,\o}}\cr
 &-\lft(\sum_\m {a_N-\bar{a}_N\m \over 2}
 \a^{(\m)}\l_N\rgt)
 {1\over \z^{(2)}_N}
 \int_{D}\! {\der^2p\over (2\p)^2}\
 \bar\c_{h}(p)
 {\partial e^{\WW^{(h)}}\over \partial \hj_{p,-\o}
 \partial\hf^+_{k-p,\o}}\;.
}\Eq(CE111)$$
The term proportional to the derivatives of
$\WW^{(h)}_{T,\m}$ {\it does vanish} for a suitable choice of the
counterterms; as well as the second term in the last line vanishes, 
at least in some important cases -- the \ce{}
for $S^{(2)}$ and for $S^{(4)}$. As consequence, it is suitable to 
replace \equ(CE111) in \equ(DSE2), obtaining:
$$\eqalign{
&D_\o(k)
 {\partial e^{\WW^{(h)}} \over\partial \hf^+_{k,\o}}
 ={B_N\over Z_N}\hf^-_{k,\o}e^{\WW^{(h)}}\cr
&-\l_N A_N
 \sum_\m {a_N-\bar{a}_N\o\m \over 2}
 \int_{D}\! {\der^2p\over (2\p)^2}{\der^2q\over (2\p)^2}\
 {\bar\c_{h,N}(p)\over D_{-\o}(p)}
 \left[
 {\partial e^{\WW^{(h)}}\over \partial\hf^+_{k-p,\o}
 \partial \hf^-_{q,\m}}\hf^-_{q+p,\m}
 \right.\cr
 &\phantom{************************************}
  \left.
 -\hf^+_{q,\m}
 {\partial e^{\WW^{(h)}}\over
 \partial \hf^+_{q+p,\m}\partial\hf^+_{p-k,\o}}\right]\cr
 &
 -
 {\l_NA_N\over Z_N}\sum_\m {a_N-\bar{a}_N\o\m \over 2}
 {\partial e^{\WW^{(h)}_{\TT,\m}}\over \partial \hb_{k,\o}}
 -{\l_N A_N\over \z^{(2)}_N}
 \int_{D}\! {\der^2p\over (2\p)^2}\
 \bar\c_{h}(p)
 {\partial e^{\WW^{(h)}}\over \partial \hj_{p,-\o}
 \partial\hf^+_{k-p,\o}}\;,
}\Eq(CECE)$$ %
where it was set
$$\eqalign{
&A_N\defi{1\over 1-(\l_N/2)\sum_\m
(a_N-\bar{a}_N\m)\left(\a^{(\m)}-\s^{(\m)}\right)}\;,\cr
&B_N\defi{1-(\l_N/2)\sum_\m(a_N-\bar{a}_N\m)\a^{(\m)}\over 
1-(1/2)\sum_\m(a_N-\bar{a}_N\m)\left(\a^{(\m)}-\s^{(\m)}\right)}\;.
}\Eq(A)$$
Deriving \equ(CECE) w.r.t. $\hf^-_{k,\o}$, for $\f\equiv0$; 
since by the tree expansion, see \secc(TE),
$$\lft|\int_{D}\! {\der^2p\over (2\p)^2}\
 \bar\c_{h}(p)
 \hS^{(1;2)}_{-\o;\o}(p;k)\rgt|\leq C\g^{(1-\th)(h_1-h)}\;,
\Eq(ale)$$
for any $\th:0<\th<1$ and for $h_1$ the scale of the momentum $k$;
and supposing  the derivatives of $\WW_{\TT,\m}$ are vanishing,
in the limit of removed cutoff, it holds 
the asymptotic formula \equ(secan).

More in general, in order to prove the derivatives of
$\WW_{\TT,\m}$ are vanishing in the limit of removed cutoff, it is
necessary a multiscale expansion.
\*

\sub(IL2){Improved localization II.}
After the multiscale integration, down to the $j$-th scale, it
holds:
$$\eqalign{
&e^{\WW_{\TT,\m}^{(h)}(\b,\jm,\f)}
 \defi
 \int\!
 \der P^{[h,j]}(\ps)\
 \exp\left\{
 \WW^{(j)}\left(\f,\jm,\sqrt{Z_j}\ps\right)
+\WW^{(j)}_{\TT,{\rm
irr}}\left(\b,\f,\jm,\sqrt{Z_j}\ps\right)\right\}\cr
&\cdot\exp\left\{
 \Big[\left({Z_N\over Z_j}\right)^2\TT^{(\m)}_0
 +{Z_N\over Z_j}\sum_{\s=\pm}\n_j^{(\s)} 
 \TT^{(\m)}_{\s} \Big]
 \left(\sqrt{Z_j}\ps,\sqrt{Z_j}\b\right)\right\}\cr
&\cdot\exp\left\{
 \Bigg[
 \wt\z^{(3,\m\o)}_j\BB^{(3)}
 +\sum_{k=j}^N{Z_k\over Z_j}\wt\z^{(1,\m\o)}_k\BB^{(1)}\Bigg]
 \left(\sqrt{Z_j}\ps,\sqrt{Z_j}\b\right)\right\}\;,}\Eq(FWY)$$
where $\wt\z^{(3,\m)}_N\defi
-\a^{(\m)}\l_N$, while, for $j\le N-1$,
$\wt\z^{(3,\m)}_j\defi\left(\wt\l^{(\m)}_j-\a^{(\m)}\l_j\right)$;
and, $\wt\z^{(1,\m)}_N\defi \s^{(\m)}$, while,
for $j\le N-1$,  $\wt\z^{(1,\m)}_j\defi\left(\wt z^{(\m)}_{j}-\a^{(\m)} z_{j}\right)$.
Indeed, these are the following possible contractions
of the interactions in $\WW_{\TT,\m}^{(h)}$.
\elenco{
\art
 The contraction of  the interactions 
$\TT^{(\m)}_0$, $\TT^{(\m)}_\s$, $\BB^{(3)}$ and $\BB^{(1)}$
through only one external field 
$\hp^-_{k,\o}$ with a kernel $\hW_{2,\o}^{(j)}$ are apparently marginal; 
instead the localization is
proportional to $\hW^{(j)}_{2,\o}(0)$, vanishing by symmetries;
for instance, in the case of the occurring of the interaction $\TT^{(\m)}_0$,
it holds:
$$\eqalign{
&\LL\left[
 \int\!{\der^2k\over (2\p)^2}{\der^2p\over (2\p)^2}{\der^2q\over (2\p)^2}\
 \bar\c_{h,N}(p-q)\hb_{k+p-q}\hp^-_{k,\o}\hp^+_{q,\m}\hp^-_{p,\m}
 {C_\o(q,p)\over D_{-\o}(p-q)}\hg^{(s)}_\o(k)\hW^{(j)}_{2,\o}(k)\right]=0\;,\cr
&\RR\left[
 \int\!{\der^2k\over (2\p)^2}{\der^2p\over (2\p)^2}{\der^2q\over (2\p)^2}\
 \bar\c_{h,N}(p-q)\hb_{k+p-q}\hp^-_{k,\o}\hp^+_{q,\m}\hp^-_{p,\m}
 {C_\o(q,p)\over D_{-\o}(p-q)}\hg^{(s)}_\o(k)\hW^{(j)}_{2,\o}(k)\right]\cr
&=\sum_\s\int{\der^2k\over (2\p)^2}{\der^2p\over
(2\p)^2}{\der^2q\over (2\p)^2}\
\bar\c_{h,N}(p-q) \hb_{k+p-q}\hp^-_{k,\o}D_\s(k)
 \hp^+_{q,\m}\hp^-_{p,\m}
 {C_\o(q,p)\over D_{-\o}(p-q)}\hg^{(s)}_\o(k)\cr
&\phantom{******************************}
 \cdot\int_0^1 \!\der\t\ 
 \big(\partial_\s\hW^{(j)}_{2,\o}\big)(\t k)\;.}$$
\*
The above case is given by one-particle reducible graphs,
therefore an alternative argument is the one similar to item 1. and
2. of the previous section.\*
\vbox
{\insertplot{400pt}{70pt}%
{}%
{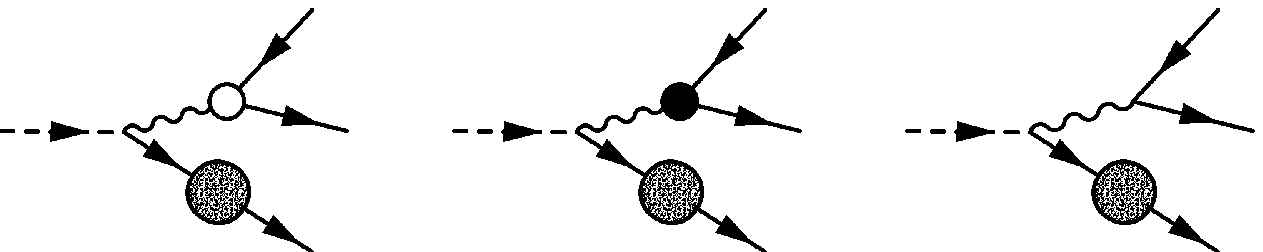}{\eqg(f3)}
\centerline{{\bf Fig \graf(f3)}: Graphical representation of items 1.}}
\vskip2em
\art 
The fields $\hp^-_{k,\o}$ and $\hp^-_{p,\m}$ of the vertex
$\TT_0^{(\m)}$ with the kernel $\hW^{(j)}_{4,\m,\o}$, is non-irrelevant; 
by the explicit expression of $C^\d_\o$, it holds:
$$\eqalign{
&\int\!{\der^2k'\over (2\p)^2}\
\bar\c_{h,N}(p+k'-q)\hg^{(r)}_\o(k-k'){C^\d_{\m}(q,p+k')\over
D_{-\o}(p+k'-q)}\hg^{(s)}_\m(p+k')
 \hW^{(j)}_{4,\m,\o}(k',p,k)\cr
&=
 \int {\der^2k'\over (2\p)^2}\
 \bar\c_{h,N}(p+k'-q)\left[
 \hg^{(r)}_\o(k-k'){D_\m(q)\left(1-(\c^\d_{h,N})^{-1}(q)\right)\over D_{-\o}(p+k'-q)}
 {f_s(p+k')\over D_{\m}(p+k')}\right.\cr
&\phantom{******************}
 +
 \left.\hg^{(r)}_\o(k-k'){\big(\d_{r,N}+\d_{s,h}\big)u_s(p+k')\over D_{-\o}(p-q+k')}\right]
 \hW^{(j)}_{4,\m,\o}(k',p,k)\;;
}$$
only the second term has a non-irrelevant part; indeed, for $j\geq
h+2$, because of $f_s(p+k')$, with $s\geq j$, and because of
$\left(1-(\c^\d_{h,N})^{-1}(q)\right)$, which,
for $q\to 0$
compels $q$ to
be contracted on scale $h$,
$$\eqalign{
 |D_{\m}(p+k'-q)|
&\geq |D_{\m}(p+k')|-|D_{\m}(q)|
 \geq\g^{j-1}-\g^{h+1}\cr
&\geq(1- \g^{-1})\g^{j-1}\;; }$$
this means that the bound of such a 
kernel, w.r.t. the standard bound,
has a more factor $\g^{-(j-h)}$
which gives a gain of one unity 
in the dimension of the kernel, making it strictly negative
down to scale $h$, where the third field
of the interaction, $\hp_{q,\m}$, {\it is compelled
to be contracted} by $\left(1-(\c^\d_{h,N})^{-1}(q)\right)$.
On the contrary, the other term
$$\eqalign{
&\hM^{(r,s),(4)}_{\m,\o}(p,k,q)\cr
&\defi
 \int\!{\der^2k'\over (2\p)^2}\
 \bar\c_{h,N}(p-q+k')\hg^{(r)}_\o(k-k'){\big(\d_{r,N}+\d_{s,h}\big)u_s(p+k')\over D_{-\o}(p-q+k')}
 \hW^{(4)}_{\m,\o}(k',p,k)\;,}$$
can occur only if $r$ in on scale 
$N$, or $s$ is
on scale $h$; and requires the extraction of the coefficient:
$$
 \hM^{(r,s),(4)}_{\m,\o}(0,0,0)
 \left\{
 \matrix{
 =0\hfill&&\hfill{\rm \ if\ }\o\m=1\cr
 \defi \D\wt l_j^{\;(-,0)}\hfill&&\hfill{\rm \ if\ }\o\m=-1\;,
 }\right.
$$
so that the above contraction is equal to
$$\eqalign{
 \LL
&\left[
 \int\!{\der^2k\over (2\p)^2}{\der^2p\over (2\p)^2}{\der^2q\over (2\p)^2}
 \hb_{k+p-q,\o} \hp^-_{k,\o}\hp^+_{q,\m}\hp^-_{p,\m}\right.\cr
&\cdot
 \left.\int {\der^2k'\over (2\p)^2}\
 \bar\c_{h,N}(p+k'-q)\hg^{(r)}_\o(k-k'){C^\d_{\m}(q,p+k')\over D_{-\o}(p+k'-q)}\hg^{(s)}_\m(p+k')
 \hW^{(j)}_{4,\m,\o}(k',p,k)\right]\cr
=&\D\wt l_j^{\;(-,0)}\int {\der^2k\over (2\p)^2}{\der^2p\over
(2\p)^2}{\der^2q\over (2\p)^2}
 \hb_{k+p-q,\o} \hp^-_{k,\o}\hp^+_{q,-\o}\hp^-_{p,-\o}\;,\cr
 \RR
&\left[
 \int {\der^2k\over (2\p)^2}{\der^2p\over (2\p)^2}{\der^2q\over (2\p)^2}
 \hb_{k+p-q,\o} \hp^-_{k,\o}\hp^+_{q,\m}\hp^-_{p,\m}\right.\cr
&\cdot
 \left.\int {\der^2k'\over (2\p)^2}\
 \bar\c_{h,N}(p+k'-q)
 \hg^{(r)}_\o(k-k'){C^\d_{\m}(q,p+k')\over D_{-\o}(p+k'-q)}\hg^{(s)}_\m(p+k')
 \hW^{(j)}_{4,\m,\o}(k',p,k)\right]\cr
=& \int\!{\der^2k\over (2\p)^2}{\der^2p\over (2\p)^2}{\der^2q\over
(2\p)^2}
 \hb_{k+p-q,\o} \hp^-_{k,\o}\hp^+_{q,\m}\hp^-_{p,\m}\cr
&\cdot\int {\der^2k'\over (2\p)^2}\
 \left[
 \bar\c_{h,N}(p+k'-q)
 \hg^{(r)}_\o(k-k'){D_\m(q)\left(1-(\c^\d_{h,N})^{-1}(q)\right)\over D_{-\o}(p+k'-q)}
 {f_s(p+k')\over D_{\m}(p+k')}\right]\cr
&+
 \sum_{p'=k,p}
 \sum_{\n}
 \hb_{k+p-q,\o} \hp^-_{k,\o}\hp^+_{q,\m}\hp^-_{p,\m}D_\n(p')
 \int_0^1d\t\
 \left(\partial_\n^{p'}\hM^{(r,s),(4)}_{\m,\o}\right)(\t p,\t k)\;.
}$$
With similar developments it is extracted $\D \wt\l_j^{(+,0)}$,
the local part of the graphs with 
the fields $\hp^-_{k,\o}$ and $\hp^+_{q,\m}$ of the interaction
$\TT_0^{(\m)}$ contracted with the kernel $\hW^{(j)}_{4,\m,\o}$.
\art
The contraction of the 
fields $\hp^-_{k,\o}$ and $\hp^-_{p,\s\m}$ of the interaction
$\TT_\s^{(\m)}$ with the kernel $\hW^{(j)}_{4,\m,\o}$ is non-irrelevant. 
Setting:
$$\eqalign{
&\wh N_{\m,\s,\o}^{(r,s),(4)}(p,k,q)\cr
&\defi
 \int\!{\der^2k'\over (2\p)^2}\
 \hg^{(r)}_\o(k-k')\bar\c_{h,N}(p+k'-q)
 {D_{\s\m}(p+k'-q)\over
 D_{-\o}(p+k'-q)}\hg^{(s)}_{\s\m}(p+k')
 \hW^{(j)}_{4,\m,\o}(k',p,k)}\;,$$
such a contraction  requires the extraction of the coefficient:
$$
 \wh N^{(r,s),(4)}_{\m,\s,\o}(0,0,0)
 \left\{
 \matrix{
 =0\hfill&&\hfill{\rm \ if\ }\o\m=0\cr
 \defi \D\wt l_j^{\;(-,\s)}\hfill&&\hfill{\rm \ if\ }\o\m=-1\;,\cr
 }\right.
$$
so that the above contraction is equal to
$$\eqalign{
 \LL
&\left[
 \int\!{\der^2k\over (2\p)^2}{\der^2p\over (2\p)^2}{\der^2q\over (2\p)^2}
 \hb_{k+p-q,\o} \hp^-_{k,\o}\hp^+_{q,\m}\hp^-_{p,\m}
 \wh N^{(r,s),(4)}_{\m,\s,\o}(p,k,q)\right]\cr
&=\D\wt l_j^{\;(-,\s)}\int {\der^2k\over (2\p)^2}{\der^2p\over
(2\p)^2}{\der^2q\over (2\p)^2}
 \hb_{k+p-q,\o} \hp^-_{k,\o}\hp^+_{q,-\o}\hp^-_{p,-\o}\;,\cr
 \RR
&\left[
 \int {\der^2k\over (2\p)^2}{\der^2p\over (2\p)^2}{\der^2q\over (2\p)^2}
 \hb_{k+p-q,\o} \hp^-_{k,\o}\hp^+_{q,\m}\hp^-_{p,\m}
 \wh N^{(r,s),(4)}_{\m,\s,\o}(p,k,q)\right]\cr
&= \int\!{\der^2k\over (2\p)^2}{\der^2p\over (2\p)^2}{\der^2q\over(2\p)^2}
 \sum_{p'=k,p,q}
 \sum_{\n}
 \hb_{k+p-q,\o} \hp^-_{k,\o}\hp^+_{q,\m}\hp^-_{p,\m}D_\n(p')\cr
&\phantom{******************************}\cdot
 \int_0^1d\t\
 \left(\partial_\n^{p'}\wh N^{(r,s),(4)}_{\m,\s,\o}\right)(\t p,\t k,\t q)\;.
}$$
With similar developments it is extracted $\D \wt\l_j^{(+,\s)}$,
the local part of the graphs with 
the fields $\hp^-_{k,\o}$ and $\hp^+_{q,\m}$ of the interaction
$\TT_\s^{(\m)}$ contracted with the kernel $\hW^{(j)}_{4,\m,\o}$.
\*
\vbox
{\insertplot{350pt}{100pt}%
{}%
{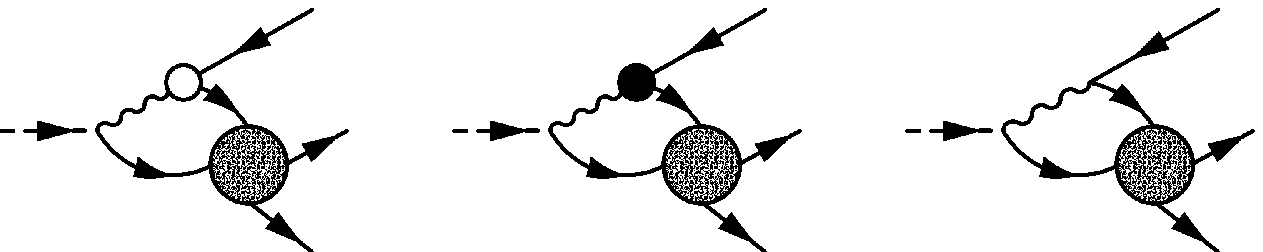}{\eqg(f4)}
\centerline{{\bf Fig \graf(f4)}: Graphical representation of items 2. and 3.}}
\vskip2em
\art
The contraction of all and three $\ps$-field of $\TT^{(\m)}_0$
with the kernel $\hW^{(j)}_{6,\m,\o,\n}$ is non-vanishing if at
least one between the two  propagators $g_\m$, has scale $N$
or $h$, otherwise the product
$C_{\m}(p,k)\hg_{\m}(k)\hg_{\m}(p+k)$ vanish; it generates non-irrelevant
operators.
Let the contraction be:
$$\eqalign{
&\hM^{(r,s,t),(6)}_{\o,\s,\m,\r}(k,q,p)\defi
 \int {\der^2k'\over (2\p)^2}{\der^2q'\over (2\p)^2}\
 \hg^{(r)}_{\o}(k-k')
 \bar\c_{h,N}(p+k'-q)
 { D_{\s\m}(p+k'-q)\over D_{-\o}(p+k'-q)}\cr
&\phantom{*****************}\cdot 
 S^{(s,t)}_{\m,\s\m}(q+q',p+k'+q')
 \hW^{(6)}_{\m,\o,\r}(q,p,k,k',q')\;,
}$$
and let the following coefficient be considered:
$$\sum_{\s}\hM^{(r,s,t),(6)}_{\o,\s,\m,\r}(0,0,0)
 \left\{
 \matrix{
 =0\hfill&&\hfill{\rm\ if \ }\r=\o\cr
 \defi\D \wt \l^{(0,0,\m\o)}_j \hfill&&\hfill{\rm\ if \ }\r=-\o\;.
 }
 \right.
$$
Then, the decomposition into marginal operator plus irrelevant one is:
$$\eqalign{
&\LL\left[
 \int\!{\der^2k\over (2\p)^2}{\der^2p\over (2\p)^2}{\der^2q\over (2\p)^2}
 \hb_{p+k-q,\o} \hp^-_{k,\o}\hp^+_{q,\r}\hp^-_{p,\r}
\sum_{\s}\hM^{(r,s,t),(6)}_{\o,\s,\m,\r}(k,q,p)\right]\cr
&=\D \wt\l^{(0,0,\m\o)}_j
 \int{\der^2k\over (2\p)^2}{\der^2p\over (2\p)^2}{\der^2q\over (2\p)^2}\
 \hb_{p+k-q,\o} \hp^-_{k,\o}\hp^+_{q,-\o}\hp^-_{p,-\o}\;,\cr
&\RR\left[
 \int\!{\der^2k\over (2\p)^2}{\der^2p\over (2\p)^2}{\der^2q\over (2\p)^2}
 \hb_{p+k-q,\o} \hp^-_{k,\o}\hp^+_{q,\r}\hp^-_{p,\r}
\sum_{\s}\hM^{(r,s,t),(6)}_{\o,\s,\m,\r}(k,q,p)\right]\cr
&=\sum_{p'=q,p,k}
 \sum_{\s,\s'}\hb_{p+q-k,\o} \hp^-_{q,\o}\hp^+_{k,\n}\hp^-_{p,\n}D_{\s'}(p')\cr
&\phantom{****************}
 \cdot\int\!{\der^2k'\over (2\p)^2}{\der^2q'\over (2\p)^2}
 \int_0^1\!\der\t \
\left( \partial_{\s'}^{p'}\hM^{(r,s,t),(6)}_{\o,\s,\m,\r}\right)
(\t k,\t q,\t p)\;.}$$
Besides, similar decomposition is done when 
$\TT_0^{(\m)}$ is replaced by $\TT_\s^{(\m)}$, with 
the replacements of $S^{(s,t)}_{\m,\s\m}$  with 1,
and of $\D \wt\l^{(0,0,\m\o)}_j$ with $\D \wt\l^{(0,\s,\m\o)}_j$.
\*
\insertplot{350pt}{100pt}%
{}%
{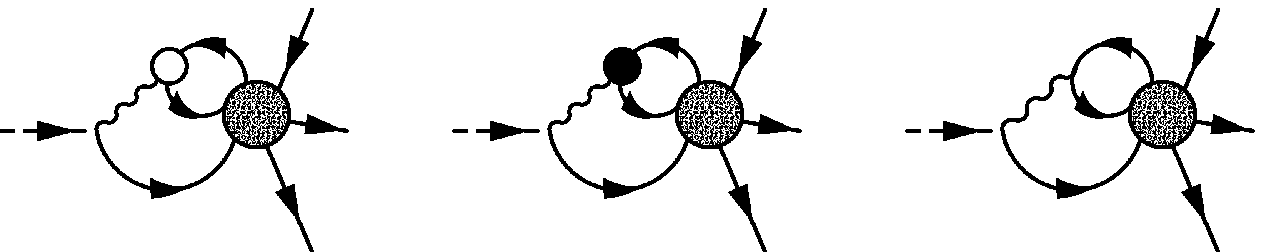}{\eqg(f5)}
\*
\centerline{{\bf Fig \graf(f5)}: Graphical representation of item 4}
\*
\art
The contraction of all and three $\ps$-fields of
$\TT^{(\m)}_0$ with the kernel $\hW^{(j)}_{4,\m,\n}$ is non-vanishing 
if at least one between the two above propagators $g_\m$,
has scale $N$ or $h$.
Let the contraction 
$$\eqalign{
&\hM^{(r,s,t),(4)}_{\o,\r,\m}(p)\cr
&\defi
 \int\!{\der^2k\over (2\p)^2}{\der^2q\over (2\p)^2}\
 \hg^{(r)}_{\o}(k)
 \bar\c_{h,N}(p+k'-q){ D_{\r\m}(p-k)\over D_{-\o}(p-k)}
 S^{(s,t)}_{\m,\r\m}(q,p-k+q)
 \hW^{(j)}_{4,\m,\o}(q,p,k)\;;
}$$
then $\hM^{(r,s,t),(4)}_{\o,\r,\m}(0)=0$
by transformation under rotation; while
$$
 \sum_\r\left(\partial_\s \hM^{(r,s,t),(4)}_{\o,\r,\m}\right)(0)
 \left\{
 \matrix{
 =0\hfill&&\hfill{\rm\ if \ }\s=-\o\cr
 \defi \D \wt z^{(0,\m\o)}_j\hfill&&\hfill{\rm\ if \ }\s=\o\;.}
 \right.
$$
Finally:
$$\eqalign{
 \LL
&\left[
 \sum_{\r}
 \int {\der^2p\over (2\p)^2}
 \hb_{p,\o}\hp^-_{p,\r}\hM^{(r,s,t),(4)}_{\o,\r,\m}(p)\right]
 =\D \wt z^{(0,\m\o)}_j
 \int {\der^2p\over (2\p)^2}\
 \hb_{p,\o}D_\o(p) \hp^-_{p,\o}\;,\cr
 \RR
&\left[
 \sum_{\r}
 \int {\der^2p\over (2\p)^2}
 \hb_{p,\o} \hp^-_{p,\r}
 \hM^{(r,s,t),(4)}_{\o,\m,\r}(p)\right]\cr
&=
 \sum_{\s,\s'}\hb_{p,\o} \hp^-_{p,\r}D_{\s}(p)D_{\s'}(p)
 \int_0^1d\t \ (1-\t)
\left( \partial_{\s'}^{p}\partial_{\s'}^{p}\hM^{(r,s,t),(4}_{\o,\r,\m}\right)(\t p)\;.
}$$
Besides, similar decomposition is done when 
$\TT_0^{(\m)}$ is replaced by $\TT_\s^{(\m)}$, with 
the replacements of $S^{(s,t)}_{\m,\r\m}$  with 1,
and of $\D \wt z^{(0,\m\o)}_j$ with $\D \wt z^{(\s,\m\o)}_j$.
\*
\vbox{
\insertplot{350pt}{60pt}%
{}%
{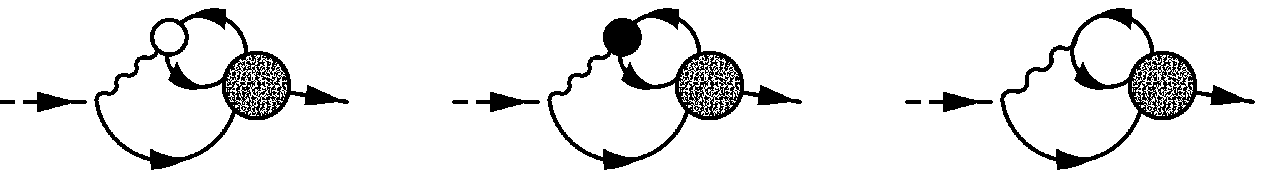}{\eqg(f6)}
\centerline{{\bf Fig \graf(f6)}: Graphical representation of item 5}}
\*
\art
The self-contraction of the 
fields $\hp^+_{q,\m}$ and $\hp^-_{k-p,\o}$
of the interactions $\TT_0^{(\m)}$,   is non-vanishing 
for $\o=\m$ and $q=k-p$.
The kernel is
$$\eqalign{
&\int\!{\der^2p\over (2\p)^2}\
 \hg^{(s)}_{\o}(k-p)
 \bar\c_{h,N}(p){ C^\d_\o(k-p,k)\over D_{-\o}(p)}\cr
&=\int\!{\der^2p\over (2\p)^2}\
 \lft[
 \bar\c_{h,N}(p){(\d_{s,N}+\d_{s,h})u_s(k-p)\over D_{-\o}(p)D_{\o}(k-p)}\rgt.\cr
&\phantom{*********}
 \lft.-{f_s(k-p)\over D_{\o}(k-p)}
 \bar\c_{h,N}(p){D_\o(k)\Big(1-(\c^\d_{h,N})^{-1}(k)\Big)\over D_{-\o}(p)}\rgt]\;;
}$$
only the former addend has non-irrelevant part.
Indeed, in the latter one, for $j\geq h+2$, because of $f_s(k-p)$, 
with $s\geq j$, and because of $\Big(1-(\c^\d_{h,N})^{-1}(k)\Big)$,
which compels the momentum $k$ to stay on scale $h$, 
$$\eqalign{
&|D_{-\o}(p)|\geq |D_{-\o}(k-p)| - |D_{-\o}(k)|\cr
&\geq \g^{j-1}-\g^{h+1}\geq \lft(1-\g^{-1}\rgt)\g^{j-1} 
\;;}$$
hence, as in item 2, there is a more factor $\g^{-(j-h)}$
in the bound of such a kernel, which  gives it 
negative dimension down to scale $h$,
where the field $\hp^-_{k,\m}$
is compelled to be contracted by 
$\Big(1-(\c^\d_{h,N})^{-1}(k)\Big)$
in the limit $\d\to 0$.
Then, let
the former addend be 
$$\hT^{(s),(0)}_{\o}(k)\defi
 \int\!{\der^2p\over (2\p)^2}\
 \bar\c_{h,N}(p){(\d_{s,N}+\d_{s,h})u_s(k-p)\over D_{-\o}(p)D_{\o}(k-p)}\;.$$
It is $ \hT^{(s),(0)}_{\o}(0)=0$ by  
transformation under rotation; while
$$
 \left(\partial_\s \hT^{(s),(0)}_{\o}\right)(0)
 \left\{
 \matrix{
 =0\hfill&&\hfill{\rm\ if \ }\s=-\o\cr
 \defi \D \wt z^{(T,0)}_j\hfill&&\hfill{\rm\ if \ }\s=\o\;.
 }
 \right.
$$
Finally:
$$\eqalign{
 \LL
&\left[
 \int {\der^2p\over (2\p)^2}
 \hb_{p,\o}\hp^-_{p,\o}
 \hT^{(s),(0)}_{\o}(p)\right]
 =\D \wt z^{(T,0)}_j
 \int {\der^2p\over (2\p)^2}\
 \hb_{p,\o}D_\o(p) \hp^-_{p,\o}\;,\cr
 \RR
&\left[
 \int {\der^2p\over (2\p)^2}
 \hb_{p,\o} \hp^-_{p,\o}
 \hT^{(s),(0)}_{\o}(p)\right]\cr
&=
 \sum_{\s,\s'}\hb_{p,\o} \hp^-_{p,\r}D_{\s}(p)D_{\s'}(p)
 \int_0^1d\t \ (1-\t)
\left( \partial_{\s'}^{p}\partial_{\s'}^{p}\hT^{(s),(0)}_{\o}\right)(\t p)\;.}$$
\art
The self-contraction of the 
fields $\hp^-_{k-p,\o}$ and $\hp^+_{q,\s\m}$ of the interaction
$\TT_\s^{(\m)}$ is non-irrelevant. Setting:
$$\wh T_{\o}^{(s)}(k)
\defi
 \int\!{\der^2p\over (2\p)^2}\
 \hg^{(s)}_\o(k-p)\bar\c_{h,N}(p)
 {D_{\o}(p)\over
 D_{-\o}(p)}\;,$$
since $\wh T_{\o}^{(s),(\s)}(0)$,
such a contraction  requires the extraction of the coefficient:
$$
 \lft(\partial_\n^p\wh T^{(s)}_{\o}\rgt)(0)
 \left\{
 \matrix{
 =0\hfill&&\hfill{\rm \ if\ }\o\n=-1\cr
 \defi \D\wt z_j^{(T)}\hfill&&\hfill{\rm \ if\ }\o\n=1\;,\cr
 }\right.
$$
so that the above contraction is equal to
$$\eqalign{
 \LL
&\left[
 \int\!{\der^2k\over (2\p)^2}
 \hb_{k,\o} \hp^-_{k,\o}
 \wh T^{(s)}_{\o}(k)\right]
 =\D\wt z_j^{(T)}\int {\der^2k\over (2\p)^2}
 \hb_{k,\o} \hp^-_{k,\o}D_\o(k)\;,\cr
 \RR
&\left[
 \int {\der^2k\over (2\p)^2}
 \hb_{k,\o} \hp^-_{k,\o}
 \wh T^{(s)}_{\o}(k)\right]\cr
&= \int\!{\der^2k\over (2\p)^2}
 \sum_{\n,\n'}
 \hb_{k,\o} \hp^-_{k,\o}
 D_{\n}(k)D_{\n'}(k)
 \int_0^1d\t\
 \left(\partial_{\n}^k\partial_{\n'}^k\wh T^{(s)}_{\o}\right)(\t k)\;.
}$$
\*
\vbox
{\insertplot{350pt}{50pt}%
{}%
{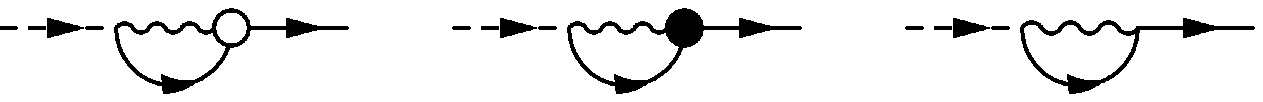}{\eqg(f14)}
\centerline{{\bf Fig \graf(f14)}: Graphical representation of items 6. and 7.}}
\vskip2em
\art
The self-contraction of the 
fields $\hp^+_{q,\m}$ and $\hp^-_{p+q,\m}$
of the interaction  $\TT_0^{(\m)}$, or 
  the 
fields $\hp^+_{q,\s\m}$ and $\hp^-_{p+q,\s\m}$
of the interactions $\{\TT_\s^{(\m)}\}_{\s=\pm}$,
would give problems;  but it arises
only for $p=0$ and it is forbidden 
by the cutoff function $\bar\c_{h,N}(p)$.
\art
The contraction of one of or both the fields $\hp^+_{q,\m}$ and
$\hp^-_{p,\m}$ was already discussed in the previous section, and
give rise to the flow of $\{\n^{(\s)}_j\}^{\s=\pm}_{j=h,...\ N}$.
} 
\0Finally, the same above developments can be done 
for the contractions of the interactions $\BB^{(3)}$:
the localization containing the couplings $\wt \l^{(\m\o)}_j$ and $\wt z^{(\m\o)}_j$
are $\D \wt\l_{j-1}^{(\m\o)}$ and $\D \wt z_{j-1}^{(\m\o)}$; while 
the localization containing $\a^{(\m\o)}$ are exactly the same of the 
flows of $\l_N$ and $Z_N$.
Then:
$$\eqalign{\wt\l_{j-1}^{(\m\o)}
&\defi
 \left({ Z_{j}\over Z_{j-1}}\right)^2
 \left(\wt\l_j^{(\m\o)}+\D \wt\l_{j-1}^{(\m\o)}+
 \d_{\o\m,-1}
 \sum_{a=\pm}\D \wt l_{j-1}^{\;(a,0)}
 \right.\cr
&\left.\phantom{*********}
+\d_{\o\m,-1}\sum_{\s,a=\pm}{Z_N\over Z_{j}}\n^{(\s)}_j
 \D \wt l_{j-1}^{\;(a,\s)}
 +\D \wt l_{j-1}^{\;(0,0,\m\o)}
 +\sum_{\s=\pm}{Z_N\over Z_{j}}
\n^{(\s)}_j\D \wt l_{j-1}^{\;(0,\s,\m\o)}\right)\;,}$$
$$\wt z_{j-1}^{(\m\o)}
 \defi
 \left(\D \wt z_{j-1}^{(\m\o)}+
 \D \wt z_{j-1}^{\;(0,\m\o)}
 +\d_{\o\m,1}\D \wt z_{j-1}^{\;(T,0)}
 +\sum_{\s=\pm}{Z_N\over Z_{j}}\n^{(\s)}_j
 \lft(\D \wt z_{j-1}^{\;(\s,\m\o)}+\d_{\o\m,1}
 \D \wt z_{j-1}^{\;(T)}\rgt)\right)\;.$$
The remarkable point is that the following theorem holds.
\*

\theorem(TT)
{For any fixed $\th:0<\th<1$, 
there exist $\e>0$, a constant $c$ and two  counterterms
$\{\a^{(\m)}\}$ , analytically dependent on $\l$,
such that, for any fixed cutoff scale, $N$, and choosing 
$\a^{(\m)}_N=\a^{(\m)}$, it holds:
$$
 \left|\wt\z^{(3,\m)}_j\right|\leq c\e\g^{-(\th/2)(N-j)}
 \qquad
 \left|\wt \z^{(1,\m)}_j\right|\leq c\e^2\g^{-(\th/2)(N-j)}\;.\Eq(TT)$$
}

\0The proof is given in appendix \secc(PL). It is a simple application
of the fixed point theorem in Banach spaces. Once 
two $\{\a^{(\m)}_N\}_\m$ are found with the required properties,
it is simply to verify that they are actually independent from $N$.
\*

\theorem(LCE)
{In the same hypothesis of 
theorem \thm(TT) and choosing 
$$\s^{(\m)}=
-\sum_{k\leq N-1}{Z_k\over Z_N}\wt\z^{(1,\m)}_k\;;$$
in the limit of removed cutoff, 
the following asymptotic  identity 
$$\eqalign{
&D_\o(k)
 {\partial e^{\WW} \over\partial \hf^+_{k,\o}}
 ={B_N\over Z_N}\hf^-_{k,\o}e^{\WW}\cr
&-\l_N A_N
 \sum_\m{a_N-\bar{a}_N\o\m \over 2} 
 \int_{D}\! {\der^2p\over (2\p)^2}{\der^2q\over (2\p)^2}\
 {1\over D_{-\o}(p)}
 \left[
 {\partial e^{\WW}\over \partial\hf^+_{k-p,\o}
 \partial \hf^-_{q,\m}}\hf^-_{q+p,\m}
\right.\cr&\phantom{***********************************}\left.
 -\hf^+_{q,\m}
 {\partial e^{\WW}\over
 \partial \hf^+_{q+p,\m}\partial\hf^+_{p-k,\o}}\right]\;,
}\Eq(CEth)$$
generates the anomalous \ce{} of those Schwinger functions
which have no density insertion and 
the addend relative to which generated by
\hbox{$\int_{D}\! {\der^2p\over (2\p)^2}\
 \bar\c_{h}(p)
 {\partial e^{\WW^{(h)}}\over \partial \hj_{p,-\o}
 \partial\hf^+_{k-p,\o}}$}
is vanishing.}

\*
The last requirement is fulfilled, as already stated,
for the $S^{(2)}$ Schwinger function, see \equ(ale).
A similar bound is valid also for $S^{(4)}$.

\*
\theorem(VBF)
{For $\e$ small enough, for $\th:0<\th<1/16$, 
and for any scale
$h\leq N$, the effective coupling is almost 
constant:
$$\l_h-\l_N={\rm O}(\l^2)\;.\Eq(vbf)$$ 
where ${\rm O}(\l^2)$ is bounded uniformly in $h$.
}

\*
\proof{\bf of Theorem \thm(LCE).} The choice of $\s^{(\m)}$ makes sense:
by \equ(TT) and \equ(ind1), for $c_0\e^2\le \th/4$
it is finite:
$$\left|\sum_{k\leq N-1}{Z_k\over Z_N}\wt\z^{(1,\m)}_k\right|
\le  c \e^2\left(1-\g^{-(\th/4)}\right)^{-1}\;.$$
With reference to \equ(CECE), the theorem is proved
once it is shown 
the bound for the derivatives of $\WW_{\TT,\m}^{(h)}$
has a vanishing factor more than the bound 
of the derivatives of $\WW^{(h)}$.
Hence, let any integer $n\in \NNN$, any choice of the label
$\ue\defi(\e_1,\ldots,\e_n)$ 
and $\uo\defi(\o,\o_1,\ldots,\o_n)$, and
any momenta $\underline k\defi(k,k_1,\ldots,k_n)$
be considered. The 
\ce{} equation for the Schwinger function 
$\hS^{(0;n+1)(\ue)}_{\uo}(\underline q;\underline k)$
is obtained by suitable derivatives of the above 
functional, plus the limit $-h,N\to\io$ of the following rest:
$$\eqalign{
 {1\over Z_N}
&\left|{\partial^{1+n} \WW^{(h)}_{\TT,\m}\over
 \partial \hb_{k,\o}
 \partial \hf^{\e_1}_{k_1,\o_1}
 \ldots \partial \hf^{\e_n}_{k_n,\o_n}}\right|
 _{\jm\equiv\f\equiv0}\cr
&\leq {C_{n;h_0,h_1\ldots, h_n}\over \sqrt{Z_{h_0}}\prod_{j=1}^n \sqrt{Z_{h_j}}}
 \left(\g^{-(\th/4)(N-h_0)}
 +\g^{(\th/4)(h_0-h)}\right)\;,}\Eq(CEbound)$$
where $\{h_j\}_{j=0}^n$ are the scales of the momenta $(k,\underline k)$:
$\k\g^{h_j-1} \le|k_j|\le \k\g^{h_j}$, with $k\defi k_0$;
and $C_{n;h_0,h_1\ldots, h_n}/ \prod_{j=0}^n \sqrt{Z_{h_j}}$
is the bound for the derivatives of $\WW^{(h)}$.
The bound derives  from the following  arguments.
By the explicit 
choice of $\s^{(\m)}$, and by \equ(TT), \equ(ind1),
for $c_0\e^2$ smaller than $\th/4$, it holds:
$$\eqalign{
&\left|\sum_{k=m}^N{Z_k\over Z_m}\wt\z^{(1,\m)}_k\right|
=
\left|\sum_{k\le m-1}{Z_k\over Z_m}\wt\z^{(1,\m)}_k\right|\cr
&\le c \e^2\sum_{k\le m-1}\g^{c_0\e^2(m-k)}\g^{-(\th/2)(N-k)}
\le  \wt c \e^2\g^{-(\th/2)(N-m)}\;.
}\Eq(sos)$$ 
for $\wt c\geq c \left(1-\g^{-(\th/4)}\right)^{-1}$.
Then, the graphs in the expansion 
of the r.h.s. member of \equ(CEbound)
has  one external 
propagator on scale $h_0$, and fall in one of the 
following classes.
\elenco{
\art
An interaction   $\TT^{(\m)}_0$ is first
contracted on scale $j$; there also 
has to be one propagator on scale $m=h,N$. 
The factor $1/Z_N$ in the r.h.s. member
of \equ(CEbound), times  factors coming form the multiscale integration
(see \equ(FWY))
gives $(Z_N/Z_j)^2(\sqrt{Z_j}/Z_N)\le 
(Z_N/Z_j)(1/\sqrt{Z_{h_0}})\g^{\e^2c_0|j-h_0|}$.
Now it is $Z_N/Z_m<1$; 
while $\g^{\e^2c_0|j-h_0|}$  is turned into
$\g^{-(\th/2)|m-h_0|}$ by a short memory factor between
the scales $m$ and $\min\{j,h_0\}$.
\art 
An interaction 
$\TT^{(\m)}_\s$ is first contracted
on scale $j$.
The factor to be studied is  now
$(Z_N/Z_j)\lft|\n^{(\s)}_j\rgt|(\sqrt{Z_j}/Z_N)\leq 
(1/\sqrt{Z_{h_0}})\g^{\e^2c_0|h_0-j|}\g^{-(\th/2)(N-j)}$;
and, as in the previous item, 
extracting  a short memory factor,
$\g^{-(\th/2)(N-j)}\g^{\e^2c_0|h_0-j|}$ is turned into the factor
$\g^{-(\th/2)(N-h_0)}\g^{-(\th/2)|h_0-j|}$.
\art
An interaction $\BB^{(3)}$ first contracted
on scale $j$.
In this case the factor to be studied is
$(\sqrt{Z_j}/Z_N)\lft|\z^{(3,\o\m)}_j\rgt|\leq 
(1/\sqrt{Z_{h_0}})\g^{\e^2c_0(N-j)}\g^{\e^2c_0|j-h_0|}\g^{-(\th/2)(N-j)}$;
then $\g^{\e^2c_0|j-h_0|}$ is changed by the short memory factor
into $\g^{-(\th/2)|m-h_0|}$;
and then, for $\e$ small, it holds 
$\g^{-(\th/2)|m-h_0|}\g^{-(\th/2-\e^2c_0)(N-j)}\le \g^{-(\th/4)|m-h_0|}
\g^{-(\th/4)(N-h_0)}$.
\art
The contraction of the interaction $\BB^{(1)}$
can only occur in a scale compatible with 
the momentum $k$ (hence two possible contiguous scales):
let it be $h_0$.
Then there is a factor 
$\sqrt{Z_{h_0}}/Z_N\lft|\sum_{j=h_0}^N(Z_j/Z_{h_0})\wt\z^{(1)}_j\rgt|\le
(1/Z_{h_0})\g^{-(\th/4)(N-h_0)}$.
} 

\0Besides the decay factor, in the first three
items there is also $\g^{-(\th/2)|j-h_0|}$, 
controlling the summation over $j$.
This proves \equ(CEbound). Hence,
 keeping $k$ fixed and non-zero, in the limit of removed cutoff,
such derivatives are vanishing.
\hfill\qed\hskip1em\null

\*
\proof{\bf of Theorem \thm(VBF).}
Taking in \equ(CECE) the derivatives $\partial\f^-_{k+q-s,\o}
 \partial\f^+_{q,-\o}\partial\f^-_{s,-\o}$,
for $\f\equiv0$,
it holds the following 
\ce{}  for $S^{(4)}$
$$\eqalign{
&{\hS^{(4)}_{\o,-\o}(k,q,s)\over \hg_{\o}(k)}\cr
&=
 -\l_N A_N{a_N+\bar{a}_N\over2}
  {\hS^{(2)}_{-\o}(s)-\hS^{(2)}_{-\o}(q)\over D_{-\o}(s-q)}
  \hS^{(2)}_{\o}(k+q-s)\cr
&+
 \l_N A_N{a_N-\bar{a}_N\over2}
 \int_{D}\!{\der^2p\over (2\p)^2}\
 {\bar\c_{h,N}(p)\over D_{-\o}(p)}\hS^{(4)}_{\o,-\o}(k-p,q,s)\cr
&+
 \l_N A_N{a_N+\bar{a}_N\over2}
 \int_{D}\!{\der^2p\over (2\p)^2}\
 {\bar\c_{h,N}(p)\over D_{-\o}(p)}
 \lft[\hS^{(4)}_{\o,-\o}(k-p,q,s-p)-\hS^{(4)}_{\o,-\o}(k-p,q+p,s)\rgt]\cr
&-\l_N A_N\sum_\m {a_N-\bar{a}_N\o\m\over2 }
 {1\over Z_N}
 {\partial^4 \WW_\TT^{(\m)}\over 
\partial \hb_{k,\o}\partial \hf^-_{k+q-s,\o}
 \partial \hf^+_{q,-\o}\partial \hf^-_{s,-\o}}\cr
&-{\l_N A_N\over \z^{(2)}_N}
 \int_{D}\! {\der^2p\over (2\p)^2}\
 \bar\c_{h}(p)
 \lft[\hS^{(1;4)}_{-\o;\o,-\o}(p;k-p,q,s)
-\d(q-s)\hS^{(2)}_{-\o}(q)\hS^{(1;2)}_{-\o;\o}(p;k-p)\rgt]\;,
}\Eq(ce4)$$
where $A_N$ was defined in \equ(A). Now, fixing $-q=s=k=\bar k$,
for any $\bar k:\k\g^{h}\leq |\bar k|\le \k\g^{h+1}$,
by lowest order computation it holds:
$${\hS^{(4)}_{\o,-\o}(\bar k,-\bar k,\bar k)\over \hg_{\o}(\bar k)}
 =
 {1\over  Z_h^2}{\l_h+{\rm O}(\l^2)\over \bar k^2 D_{-\o}(\bar k)}\;,$$
$$ \l_N A_N{a_N+\bar{a}_N\over2}
 {\hS^{(2)}_{-\o}(\bar k)\over D_{-\o}(\bar k)}
 \hS^{(2)}_{\o}(\bar k)
 ={1\over Z^2_h}{-\l_N+{\rm O}(\l^2)\over \bar k^2 D_{-\o}(\bar k)}\;;$$
while (see also [BM04] for more details)
$$
 \left|\l_N A_N{a_N-\bar{a}_N\over2}
 \int_D\!{\der^2p\over (2\p)^2}\
 {\bar\c_{h,N}(p)\over D_{-\o}(p)}\hS^{(4)}_{\o,-\o}(\bar k-p,-\bar k,\bar k)\right|
 \leq {\g^{-3h}\over Z_h^2}{\rm O}(\l^2)\;,$$
and identical bound for 
$$\left|\l_N A_N{a_N-\bar{a}_N\over2}
 \int\!{\der^2p\over (2\p)^2}\
 {\bar\c_{h,N}(p)\over D_{-\o}(p)}
 \lft[\hS^{(4)}_{\o,-\o}(\bar k-p,-\bar k,\bar k-p)-
 \hS^{(4)}_{\o,-\o}(\bar k-p,p-\bar k,\bar k)\rgt]\right|\;,$$
and
$$\lft|{\l_N A_N\over \z^{(2)}_N}
 \int_{D}\! {\der^2p\over (2\p)^2}\
 \bar\c_{h}(p)
 \hS^{(1;4)}_{-\o;\o,-\o}(p;\bar k-p,-\bar k,\bar k)\rgt|\;.
$$
Finally, by the study of the flow of $\WW_\TT$,
it also hold
$$\lft|\l_N A_N\sum_\m {a_N-\bar{a}_N\o\m\over2 }
 {1\over Z_N}
 {\partial^4 \WW_\TT^{(\m)}\over \partial \hb_{\bar k,\o}
\partial \hf^-_{-\bar k,\o}
 \partial \hf^+_{-\bar k,-\o}\partial \hf^-_{\bar k,-\o}}\rgt|
\le{\g^{-3h}\over Z_h^2}{\rm O}(\l^2) $$
(namely, in this case, since $k$ is on the infrared cutoff scale,
the rest is not vanishing; but it diverges in $h\to -\io$
with the same exponent, $3-2\h_\l$, of the other
terms in \equ(ce4)).
Considering together the above bound with \equ(ce4),
it holds the theorem.
\hfill\qed\hskip1em\null

\*
\sub(VVBBFF){Vanishing of the Beta function.}
In the end, it is remarkable how \equ(vbf)
is read in terms of the {\it Beta function}
for the effective couplings.
In agreement with  \equ(d2),
the Beta function for the massless Thirring
model, in Euclidean regularization is
such that 
$$\l_{h-1}-\l_h\defi  \b^{({\rm T})}_h(\l_h)+
\sum_{m=h}^N\b^{({\rm T},\l)}_{h,m}(\l_m-\l_h)\Eq(vbf2)$$
(see \secc(cf) for the explanation of the addends).
As done for the anomalous exponent, 
by scaling invariance of the graphs in the expansion 
of $\{\b^{({\rm T})}_h\}_h$, it is possible to prove that 
there exist a real function  $B$ such that 
$$|\b^{({\rm T})}_h(\l_h)-B(\l_h)|\leq c\e^2\g^{-\th(N-h)}\;.\Eq(vbf3)$$
Well then, as consequence of \equ(vbf), $B\= 0$. Otherwise, 
if the coefficient of the $m$-th order expansion of $B(\l)$,
$B^{(m)}$,
where non-zero,  then replacing the expansion
 $\l_h\defi\sum_{n>0}c_h^{(n)}\l^n$ in \equ(vbf3), 
it would be possible to prove -- by an iterative procedure similar to 
the one in \secc(cf) -- that for any $h$, and for any $n<m$:
$$
 \lft|\sum_{m=h}^N\b^{({\rm T},\l)(n)}_{h,m}(\l_m-\l_h)\rgt|\le C^n\g^{-(\th/2)(N-h)}\;,
 \qquad
 \lft|c^{(n)}_{h-1}-c^{(n)}_{h}\rgt|\le C^n\g^{-(\th/2)(N-h)}\;;$$
while, for $n=m$, 
$$c^{(m)}_{h-1}=c^{(m)}_{h}+B^{(m)}+{\rm O}(\g^{-(\th/2)(N-h)})\;.$$
Therefore $\{c^{(m)}_h\}_{h\le N}$ would be a diverging sequence,
in contradiction with \equ(vbf).

\*
\section(SCE, Solution of the closed equation)

With simple symmetry considerations and  multiscale
integration, it possible to prove the following general 
expression for the two point Schwinger function:
$$
 \hS_{\o}^{(2)}(k)
 =
 {1\over D_\o(k)}
 \left({|k|\over \k}\right)^{\h_\l}
 F_{h,N}
 \left({|k|\over \k}\right)\;,\Eq(gex)$$
where $F_{h,N}$ is finite, uniformly in $h,N$, and
 such that, for a suitable real constant $F$,
$$
 \sup_{\g^{(h/2)}\k\leq|p|\leq \g^{(N/2)}\k}
 \left| F_{h,N}
 \left({|p|\over \k}\right)-F \right|
 =C\left(\g^{-(\th/4)N}+\g^{(\th/4)h}\right)\;.
\Eq(esdecay)$$
Indeed,
once the factor $1/(D_\o(k)Z_{h_0})$  is extracted
(with $h_0$ the scale of $k$),  the expansion of 
$\hS_{h,N;\o}^{(2)}\left(k\right)$ 
is given by scaling invariant graphs.
Calling $F$ the limit of $ F_{h,N}$, with
all the couplings $\{\l_j\}_j$ replaced by $\l$,
all the ratios  $\{Z_{j-1} /Z_j\}_j$ replaced by 
$\g^{\h_\l}$ and the factor $(|k|/\k)^{\h_\l}(1/Z_{h_0})$ 
with 1, the difference between $F_{h,N}$
and  $F$ is the sum of all the graphs with an external propagator on scale 
$h_0$ and falling in one of the following cases.
\elenco{
\art
There is an interaction on scale 
$m>N$ or $m<h$. By the short memory property, 
given any $\th:0<\th<1/16$, 
the sum of 
all such graphs is bounded with $\g^{-\th(N-h_0)}+\g^{-\th(h_0-h)}$, 
up to a constant.
\art
There is a coupling
$\left[(|k|/\k)^{\h_\l}(1/Z_{h_0})-1\right]$. 
By the feature 
of the flow of the field strength --
namely the analogous for the Euclidean regularization
of \equ(z1) --  the sum of 
all such graphs is bounded with $\g^{-(\th/2)(N-h_0)}$, 
up to a constant. 
\art
There is an interaction  $\l_m-\l$, or $(Z_{m-1}/Z_m)-\g^{\h_\l}$
on scale $m:h\leq m\leq N$ .
By the short memory factor an features of the flows
-- analogous for the Euclidean regularization 
of \equ(l1) and \equ(z1) --
the sum of  such graphs is bounded by $\g^{-\th|m-h_0|}$
$\g^{-(\th/2)(N-m)}\leq$  
$\g^{-(\th/2)(N-h_0)}$
$\g^{-(\th/2)|m-h_0|}$,
up to a constant.}

\0Hence, after summing over $m$,  \equ(esdecay) holds.
\*

Now, replacing \equ(gex) in the \ce{} for the two point Schwinger
function, and taking the limit $h\to -\io$, it holds:
$$\eqalign{
 \left|{k\over \k}\right|^{\h_\l}
 F_{N}(k)
=&
 {B_N\over Z_N}
 -\l_NA_N {a_N-\bar{a}_N \over 2}
 \int_{D}\!{\der^2p\over (2\p)^2}\
 \left|{p\over \k}\right|^{\h_\l}
 {F_{N}(p)\over D_{-\o}(k-p)D_{\o}(p)}\cr
&+{\D \wh K_{N,\o}(k)}\;,}$$
where, by \equ(CEbound),
$$\sup_{|k|\leq \g^{(N/2)}\k}
\left|\D\wh K_{N,\o}(k)\right|\leq {C\over Z_{h_0}}
\g^{-(\th/8)N}\;.
$$
The equation for $k=0$ -- then $Z_{h_0}=+\io$ -- gives 
$${B_N\over Z_N}=\l_NA_N {a_N-\bar{a}_N \over 2}
 \int_{D}\!{\der^2p\over (2\p)^2}\
 \left|{p\over \k}\right|^{\h_\l}
 {F_{N}(p)\over p^2}\;;$$
therefore:
$$\eqalign{
 \left|{k\over \k}\right|^{\h_\l}
 F_{h,N}(k)
&=
 \l_NA_N {a_N-\bar{a}_N \over 2}
 \int_{D}\!{\der^2p\over (2\p)^2}\
 \left|{p\over \k}\right|^{\h_\l}
 F_{N}(p){k^2+D_{-\o}(p)D_\o(k)\over(k-p)^2 p^2}\cr
&+\D\wh K_{N;\o}(k)\;.
}$$
Now it is possible to take the limit $N\to+\io$, for 
$k$ fixed: since the rest is vanishing, by 
finiteness of $F_N$ uniformly in $N$ and 
by \equ(esdecay), the limit can be exchanged with the integral in 
$\der^2 p$, it holds:
$$
 \left|k\right|^{\h_\l}
 =
 \l_b A {a-\bar{a} \over 2}
 \int\!{\der^2p\over (2\p)^2}\
 \left|p\right|^{\h_\l}
 {k^2+D_{-\o}(p)D_\o(k)\over(k-p)^2 p^2}\;.$$
The integral can be elementarily computed:
the pure imaginary part is zero by symmetries, 
while for the real one it holds, for $\th$ 
the angle between the vector $p$ and the vector $k$,
for $t\defi \tan(\th/2)$,
and calling, with abuse of notation, $k$ and $p$ the moduli 
of the vectors $k$ and $p$ themselves,
$$\eqalign{
&{1\over (2\p)^2}\int_0^\io\!\!\!\der p\
 p^{\h_\l-1}
\int_{-\p}^\p\!\!\!\der \th\
 {k^2-pk\cos(\th)\over k^2+p^2-2pk\cos(\th)}\cr
=&
{1\over (2\p)^2}\int_0^\io\!\!\!\der p\
 p^{\h_\l-1}
 \int_{-\io}^\io\!\!\!\der t\
 {2k\over 1+t^2}
 {\big(k+p\big)t^2+\big(k-p\big)\over 
\big(k+p\big)^2t^2+\big(k-p\big)^2}\cr
=&
{1\over (2\p)^2}\int_0^\io\!\!\!\der p\
 p^{\h_\l-1}
\int_{-\io}^\io\!\!\!\der t\
 {2k\over k+p}
 \lft[{1\over 1+t^2}-{k-p\over 2k}
 \lft({1\over 1+t^2}-{(k+p)^2\over (k+p)^2t^2+(k-p)^2}\rgt)\rgt]\cr
=&
{2\over (2\p)^2}\int_0^k\!\!\!\der p\
 p^{\h_\l-1}
\int_{-\io}^\io\!\!\!\der t\
 {1\over 1+t^2}={1\over 2\p\h_\l}k^{\h_\l}\;.}$$
This gives  the following
expression for the critical index $\h_\l$:
$$\h_\l=A{\l_b\over 2\p}  {a-\bar{a} \over 2}\;,$$
to be compared with the formula for the half value
of $\h_\l$  given  in [J61] just after (36)
-- with the following identification: Johnson's $\a$
is here $\h_\l/2$; Johnson's $\l$
is $\l_b/2$; while $a-\bar a$ is, according to Johnson,
equal to $2{\l/2\p\over1-(\l/2\p)^2}$.
\appendix(SAR,Simple Analytical Properties)

\asub(PFE){Partial-fraction expansion.}
The functions
$$
 f^-_L(z)\defi{ e^{-(x_0+L) z }\over 1+e^{-L z}}
 \quad{\rm for\ } -L<x_0 < 0\;,
 \qquad f^+_L(z)\defi{ e^{-x_0 z }\over 1+e^{-L z}}
 \quad{\rm for\ } 0< x_0 < L\;,$$
are both meromorphic, since in any  circles, $\CC_R$,
of radius $R$ and centre the origin, their only singularities
are a finite number of poles.
In particular, setting 
$D_0\defi \big\{{2\p \over L}(m+{1\over 2})\big\}_{m\in \zzz}$, 
they are  on the imaginary axis, in 
$\{ik_0:k_0\in D_0\}$.
Therefore, by the Cauchy theorem, 
for any $e\in \RRR$, 
$R>|e|$ and $\s=\pm$,
$$
 f^\s_L(e)=
 \oint_{\CC_R}\!{\der z\over 2\p i}\ 
 {f^\s_L(z)\over z-e}
 +\s
 {1\over L} \sum_{k_0\in D_0}^{|k_0|\le R}
 {e^{-ix_0k_0} \over -ik_0+e}\;. \Eqa(pfe)$$
Since, for $0\leq \th \leq \p/2$, 
$\cos\th\geq 1-2\th/\p$, then
it holds the following bound:
$$\left|\oint_{\CC_R}\!{\der z\over 2\p i}\ 
 {f^+_L(z)\over z-e} \right|
 \leq {2 R\over R-|e|}
 \int_0^{\p}\!\der \th {e^{-x_0R\cos\th}\over 1+ e^{-LR\cos\th}}
 \leq  {2 \over R-|e|}\lft[{\p\over 2x_0}+{\p\over 2(L-x_0)}\rgt]\;,$$
and similarly for $f^-_L$.
Hence the first addend in the r.h.s. member of \equ(pfe)
vanish for $R\to\io$, and than, for any $x_0\neq 0: |x_0|< L$,
$$
 f^+_L(e)\c(x_0>0)- f^-_L(e)\c(x_0<0)=\lim_{R\to+\io}
 {1\over L} \sum_{k_0\in D_0}^{|k_0|\le R}
 {e^{-ix_0k_0} \over -ik_0+e}\;.$$
Such a series, not absolutely convergent,
can be written as $\sin^{-1}(\p x_0/L)$,
times an absolutely convergent series -- 
and border terms vanishing for large $R$
-- so that it is clear the possibility of
replacing the sharp constraint  $|k_0|\leq R$
with a smooth cutoff function.

\*
\asub(GCF){Gevrey compact-support functions}
It is easy to construct a compact support-function 
which also fulfil the Gevrey constraint
on the derivatives.

Indeed, let the following $C^\io$ 
function be considered for any number $p>0$:
$$
 \th(t)\defi\left\{
 \matrix{
 0\hfill&\hfill{\rm for\ }t<0\cr
 e^{1-(1/t^p)}\hfill&\hfill{\rm for\ }0\le t\le 1\cr
 1\hfill&\hfill{\rm for\ }t>1\;.}\right.
$$
For $t\leq 0$ and $t\geq 1$ all the  derivatives 
are identically zero.
For $t:0<t<1$, it is possible to find
a bound for the derivatives using the analyticity of $\th(t)$ 
in the half-plane \hbox{$\CCC_+\defi\{z\in \CCC:{\rm Re}(z)>0\}$}.
For any $t:0<t<1$, 
let the disc \hbox{$D_t\defi\{z\in \CCC:|z-t|\leq t\sin(\p/4p)\}$}
be considered.
By the Cauchy theorem:
$$
 |\th^{(n)}(t)|\leq {n!\over 2\p\Big(t\sin(\p/4p)\Big)^n}
\max_{z\in D_t}|\th(z)|\;.
$$
For any $z\defi r e^{i\f}\in D_t$, since
the lines passing through $z=0$ and tangent to $D_t$
have angular parameter $\pm\p/4p$, then 
${\rm Re}\left(z^{-p}\right)\geq r^{-p}\cos(\f p)
\ge (2t)^{-p}\cos(\p/4)$.
Hence, since for any $x\geq 0$, and any constant $c>0$,
it holds \hbox{$x^n e^{-cx^p}\leq C^n (n!)^{(1/p)}$},
then for a certain constant $C>1$,
$$
 |\th^{(n)}(t)|\leq C^n (n!)^{1+(1/p)}\;;
$$
namely $\th(t)$ is a Gevrey function of order $\a=1+(1/p)$.
Finally, if  $\wh\c_0(t)\defi 1-\th\left({t-1\over \g-1}\right)$,
then $\wh f_j(t)\defi \wh\c_0(t\g^{-j})-\wh\c_0(t\g^{-j+1})$
is a compact-support Gevrey function for any \hbox{integer $j$.}

\*
\asub(DP){Bounds for the propagators.}
If $K$ is the compact support of $f_0(k)$, the $n$-th derivatives of
$1/D_\o(k)$ are bounded in $K$ by $C_Kc_K^n n!$, for suitable $K$-dependent
constants $C_K$ and $c_K$. Therefore, by Leibniz formula it
follows that it $f(k)$ is a Gevrey, compact-support function of
class $\a\geq1$, also  $f(k)/D_\o(k)$ is.
Therefore, for any $n_0, n_1\in \NNN$, by partial derivation
and Stirling formula, 
$$\eqalign{
 |g^{(0)}_\o(x)|
&\le 
 {1\over |x_0|^{n_0}|x_1|^{n_1}}
 \sup_{k\in K}
 \left|\partial_0^{n_0}\partial_1^{n_1} {f_0(k)\over D_\o(k)}\right|\cr
&\leq C \left(\left|{c\over x_0}\right|^{1/\a}{n_0\over e}\right)^{\a n_0}
 \left(\left|{c\over x_1}\right|^{1/\a}{n_1\over e}\right)^{\a n_1}\;.
}$$
Therefore, choosing for $n_j$ such that $(|x_j|/c)^{1/\a}-1\leq n_j \le (|x_j|/c)^{1/\a}$,
it holds:
$$
 |g^{(0)}_\o(x)|
 \le C e^{-\a (|x_0|/c)^{1/\a}} e^{-\a (|x_1|/c)^{1/\a}}\;.
 $$
Finally, with similar argument, it is possible to 
obtain the same bounds for lattice-spacetime 
propagators. 
\appendix(OS,OS axioms)

\asub(TF){Test functions.}
For any $n\in \NNN$, setting $\ux\defi (x^{(1)},\ldots,x^{(n)})$,
let $\SS\big(\RRR^{2n}\big)$ be the space of the complex test 
functions
on $\RRR^{2n}$, with labels, $\uo\defi
(\o_1,\ldots,\o_n)$, $\ue\defi (\e_1,\ldots,\e_n)$, s.t., for any
integer $m$, and any $f^{(\ue)}_{n,\uo}(\ux)\in
\SS\big(\RRR^{2n}\big)$, the Schwartz norm
$$
 ||f^{(\ue)}_{n,\uo}||_m\defi \max_{\underline r:\sum_jr_j\leq m}
\sup_{x^{(j)}\in \rrr^2}
 \left|
 \left(1+\sum_{i=1}^n|x^{(i)}|^{m}\right)\dpr_1^{r_1}\cdots\dpr_n^{r_n} f^{(\ue)}_{n,\uo}(\ux)
 \right|
$$
is finite. Let $\SS_{\neq}\big(\RRR^{2n}\big)$ be the space of the
functions in $\SS\big(\RRR^{2n}\big)$ which vanish, together with
all their partial derivatives, if $x^{(i)}=x^{(j)}$ for some $1\leq
i<j\leq n$; and let $\SS_<\big(\RRR^{2n}\big)$ be the space of the
functions in $\SS_{\neq}\big(\RRR^{2n}\big)$ which vanish, together
with all their partial derivatives, if the ordering of the times
$x_0^{(1)},\dots,x_0^{(n)}$ is different from
\hbox{ $0<x_0^{(1)}<x_0^{(2)}<\dots<x_0^{(n)}$}.

Let the ``space translation'', $\t_y$, for $y=(0,y_1)$, be defined
as
$$\big(\t_yf\big)^{(\ue)}_{n,\uo}(\ux)
 \defi
 f^{(\ue)}_{n,\uo}(\t_y\ux)\;,$$
with $\t_y\ux\defi (x^{(1)}+y,\ldots,x^{(n)}+y)$.

Let the ``time reflection'' be defined as
$$
 \left(\Th f\right)^{(\ue)}_{n,\uo}(\ux)
 \defi
 \lft(f^{(\ue^*)}_{n,\underline{\o}^*}\rgt)^*(\th_0\ux)\;,$$
with $\th_0\ux\defi (\th_0 x^{(1)},\ldots,\th_0 x^{(n)})$, 
where $\th_0(x_0,x_1)\defi (-x_0,x_1)$; 
$f^*(x^{(1)},\ldots,x^{(n)})$ is the
complex conjugate of $f(x^{(n)},\ldots,x^{(1)})$; 
and the labels $\uo^*$ and $\ue^*$ 
are defined respectively to be $\o_n,\ldots,\o_1$
and $-\e_n,\ldots,-\e_1$ (see [OS72], formula
(6.2)).

In the end, it has to be noticed the following fact: for 
$\WW$ being the generating functional of the Schwinger functions, then 
$e^{\WW}$ is the generating functional of the correlations.
Hence, each Schwinger function -- also called ``truncated correlation'' --
can be written as finite linear combination of correlations,
in term of which the \osa{} are now listed --
with the simplification in the notation that 
$G^{(0,n)(\ue)}_{\us;\uo}(\uz,\ux)\defi G^{(n)(\ue)}_{\uo}(\ux)$.

\*
\alemma(OSII)
{Given $\e$ small enough, for any  $\l: |\l|<\e$ 
and $\m:0\le \m\le \k\g^{-1}$, 
the correlations satisfy the Osterwalder-Schrader axioms:
\elenco{
\item{\bf E1.} 
$G^{(n)(\ue)}_\uo(\ux)$ is a
distribution on $\SS_{<}(\RRR^{(2n)})$; and there exists an  integer
$m$ and  two constants $c_m, C_m>0$ s.t., for any $n$
$$\norm{G^{(n)(\ue)}_\uo}_{m\cdot n}
 \defi
 \sup_{f\in \SS_{<}\left(\rrr^{(2n)}\right)}
 {\lft|\left(G^{(n)(\ue)}_\uo,f\right)\rgt|
  \over
 \norm{f}_{m\cdot n}} \leq C_m(n!)^{c_m}\;.$$
\item{\bf E2.}
 $G^{(n)(\ue)}_\uo$ is  covariant
under the Euclidean group of translation and rotation of all the
coordinates.
\item{\bf E3.}
 $ G^{(n)(\ue)}_\uo$ is
antisymmetric under the exchange of the $x^{(i)}, \o_i,\e_i$
respectively with $x^{(j)}, \o_j,\e_j$, for any $1\leq i < j\leq
n$.
\item{\bf E4.}
 For any finite sequence of ``time ordered'' test functions,
$\left\{f^{(\ue)}_{n,\underline{\o}}(\ux)\in
\SS_{<}(\RRR^{(2n)})\right\}_{n\geq 0,\underline{\o},\ue}$,
the correlations are ``reflection invariant'':
$$\lft[G^{(n)(\ue)}_\uo\Big((\Th f)^{(\ue)}_{n,\uo}\Big)\rgt]^*=
  G^{(n)(\ue)}_\uo(f^{(\ue)}_{n,\underline{\o}})$$
and ``reflection positive'':
$$\sum_{m,\uo',\ue'}\sum_{n,\uo,\ue}
 G^{(m+n)(\ue',\ue)}_{\uo',\uo}
 \Big((\Th f)^{(\e')}_{m,\underline{\o}'}
 \otimes  f^{(\e)}_{n,\underline{\o}}\Big)
 \geq 0\;.\Eqa(rp)$$
\item{\bf E5.} 
For any $f^{(\e)}_{n,\uo}\in
\SS_{<}(\RRR^{(2n)})$ and $g^{(\e')}_{m,\uo'}\in
\SS_{<}(\RRR^{(2m)})$, decorrelation  holds: 
$$\eqalign{
&\lim_{|y|\rightarrow \io}
 G^{(m+n)(\ue',\ue)}_{\uo',\uo}
 \lft((\Th g)^{(\ue')}_{m,\uo'}
 \otimes  (\t_yf)^{(\ue)}_{n,\uo}\rgt)\cr
&\phantom{*************}
=G^{(m)(\ue')}_{\uo'}
 \Big((\Th g)^{(\ue')}_{m,\uo'}\Big)
  G^{(n)(\ue)}_{\uo}
 \Big(f^{(\ue')}_{n,\uo}\Big)\;.}$$
}}
The last property, called {\it cluster decomposition},
in terms of the Schwinger function reads:
$$\lim_{|y|\rightarrow \io}
 S^{(m+n)(\ue',\ue)}_{\uo',\uo}
 \lft((\Th g)^{(\ue')}_{m,\uo'}
 \otimes  (\t_yf)^{(\ue)}_{n,\uo}\rgt)
=0\;.\Eqa(cldec)$$

From the \osa, it is possible to derive the theory in
Minkowskian spacetime, from the Euclidean one. The main difficulty,
here, is to prove the validity of E2 and E4: a
regularization that makes clear the one, usually makes obscure the
other.

\*
\asub(RP){Reflection Positivity for the Hamiltonian Regularization}

The Euclidean fields operator in Heisemberg picture are:
$$\ps^\s_{x,\o}
 \defi
 e^{-x_0H}
 \left({1\over L} \sum_{k\in D} e^{\s ikx_1} a^\s_{k,\o}\right)e^{x_0H}\;,
 \qquad
 x\defi(x_0,x_1)\in \RRR\times\TTT\;;$$
therefore $\ps^\s_{x,\o}$ is {\it not} the Hermitian conjugate
of $\ps^{-\s}_{x,\o}$ -- as it were in the Minkowskian picture: 
it is therefore suitable to 
define the operator $\th$ ``time reflection'' s.t.
$\th x=(-x_0,x_1)$, so that $\ps^\s_{x,\o}$ is the Hermitian of 
$\ps^{-\s}_{\th x,\o}$.

Let now the space $\FF$ of the linear functionals
of the operator-valued fields: namely the operators on the Fock space 
of the form:
$$
 F(\ps)=
 \sum_{n\geq 0} \sum_{\uo,\us}
 \int d^2x^{(1)}\ \cdots d^2x^{(n)}\  
 f_{n,\uo,\us}\left(x^{(1)},\ \dots,x^{(n)}\right)
 \ps^{\s_1}_{x^{(1)},\o_1}\ \cdots\ps^{\s_n}_{x^{(n)},\o_n}
$$
for any choice of the
test functions $f_{\uo,\us}\in\SS_{<}\Big((\RRR\times\TTT)^n\Big)$.
Then,
it is simply to verify that $\Th$ on the space $\FF$ is the 
Hermitian conjugation. 
Hence, for any real $L$,
the following quantity is non-negative:
$$\Tr\lft[e^{-LH}(\Th F) F\rgt]\geq 0\;.$$
Such an inequality, by the definition of the correlations, 
reads as in \equ(rp).
\appendix(TE,Tree Expansion and Convergence of
the Schwinger functions)

The renormalization procedure used here is slightly 
different from the classical one, the BPHZ scheme.

As noticed in the early works on the renormalization, 
the localization is necessary and effective in extracting
the divergent contribution of the subgraphs whenever 
the momenta flowing in the internal propagators of the subgraphs
are in some sense higher than the momenta flowing on the external ones
({\it Hepp's sectors}). Anyway, the localization 
has a further complication in the massless case: while it improves the
convergence at large momenta, it 
worsen consequently the convergence at small ones. 

Accordingly, in the BPHZ scheme, the propagators of the graphs  
are decomposed {\it a posteriori}
in scales, and the subgraphs, selected by the Hepp
procedure,  are localized: this is done 
by extracting the first orders of the Taylor 
expansion around zero external momenta, if 
the theory is massive;  around any fixed non-zero
value, if the theory is massless: in the latter case
some discrete symmetries are broken, and
more ``relevant'' and ``marginal'' terms, even a mass term,
are generated.

In the scheme here depicted, instead, 
the multiscale integration 
not only produces directly only subgraphs
satisfying the Hepp's property; 
but it makes clear the possibility of localizing at 
zero external momenta {\it even the subgraphs with
massless propagators}, since such a localization
is naturally stopped below the scales of the 
momenta of the Schwinger function at hand.

\*
\asub(3.2){Tree structure.}
By expanding iteratively the truncated expectations \equ(trex1) and
\equ(trex2), starting from $\WW^{(M)}$, it is possible to write the
effective potential on scale $\WW^{(h)}$, for $h\le M$, in terms of
a {\it tree expansion}, quite similar to that described, for
example, in [BGPS].
\elenco{
\art
Let a tree, $\t$, be a tree graph with the 
following features:  if there are $n+1$ points with incidence 
number equal to 1,  one of such points is the 
{\it root}; the other $n$ points are  the {\it endpoints}; 
the integer $n$ is the {\it order} of the tree. 
All the points of the tree graphs, except the root and the endpoint,
are  called {\it nodes}. The only
node paired to the root by the tree graph is the {\it first node}:
it is required not to be an endpoint.  
\art
The nodes, the root and
the endpoints are partially
ordered in the natural way by the tree structure, 
so that the root is lower than the endpoints: 
$v<v'$ means $v$ is lower that $v'$.
In correspondence of any node $v$, the integer $s_v$ is the number
of minimal nodes or endpoints greater than $v$: 
such nodes or endpoints are 
also said to be {\it first followers of $v$}, and  
are denoted $v_1,\ldots,v_{s_v}$. 
If $s_v>1$, then $v$ is a {\it branching node}. 
In correspondence of  a node or an end point $v$, the unique
maximal node lower than it is the {\it first preceding of $v$}, and is
denoted $v'$.
\art
Let the {\it topological trees} be the
quotient set of the above depicted trees, 
in which any two of them are identified if,  
by a suitable continuous deformation 
of the length of the links 
and of the angled between them,
-- included permutation 
of the links coming out of the same branching node -- 
 they can be superposed.
It is then easy to verify that,
since the number of the branching nodes 
of a tree with $n$ endpoints is not larger 
than $n-1$, then the number of all the 
topological tree with $n$ endpoints 
is bounded by $4^{2n-1}<16^n$.
\art 
With each node $v$ of the tree, a scale $h_v:h\leq h_v\leq M$
is assigned, with the compatibility condition that $v'<v$ imply
$h_{v'}<h_{v}$: therefore it is possible to draw the trees as lying
vertically along a family of horizontal parallel lines, each one marking a
scale \hbox{$j:h-1\leq j\leq M+1$}, so that the each node $v$ is contained in
the horizontal line with index $h_v$.  The scale  $h_u$
of the endpoint $u$ ranges from $h+1$ to $M+1$; if
$v$ is the first preceding of such an endpoint, $h_u=h_v+1$.  
The scale of the first node is $h$: because of the distinction that 
will be done between the nodes in correspondence  of the hard fermion
regime and the soft fermion regime, 
$h$ is allowed to be $\le N+1$; the scale of the root is 
$h_r=h-1$.
\art 
There are two kinds of endpoints, {\it normal} and {\it special}.
With each normal endpoint $u$, it is associated one of the three
self-interactions $\l_{h_u-1}\VV$, $\g^{h_u-1}\n_{h_u-1}\NN$ or
$\d_{h_v-1}\DD$, if $h_u-1\leq N$; otherwise 
the interactions $\l_N\VV$, $\g^N\n_N\NN$ or
$\d_N\DD$. They are called the endpoints of type $\l$, $\n$,
$\d$, with an obvious correspondence.
With each special endpoint $u$ it is associated one of the three
interactions with the external sources, $\z^{(2,+)}_{h_u-1}\JJ_+$,
$\z^{(2,-)}_{h_u-1}\JJ_-$ or $\FF$, if $h_u-1\leq N$; otherwise
the interactions  $\z^{(2,+)}_N\JJ_+$,
$\z^{(2,-)}_N\JJ_-$ or $\FF$. They are called the endpoints
of type $\f$, $\jm_+$ and $\jm_-$.
The endpoints of type $\jm$ are the union of the
ones of type $\jm_+$ and $\jm_-$.
\art
Given a node $v$, $n^\f_v$ and $n^\jm_v$ are respectively the
number of endpoints of type $\f$, and of type $\jm$
greater than $v$; $n^{(4)}_v, n^{(2)}_v$ are respectively the
number of normal endpoint of type $\l$ and of type $\n$ or $\d$
greater than $v$; $n_v\defi n^{(4)}_v+n^{(2)}_v$. Analogously,
given a tree $\t$, the integers $n^\f_\t,n^\jm_\t, n^{(4)}_\t,
n^{(2)}_\t$ and $n_\t$ are respectively the number of endpoints of
type $\f$, of type $\jm$, of type $\l$, of type $\n$ or $\d$ and
the total number of normal endpoints of the tree.
\art
 For any node $v$, the {\it cluster } $L_v$ with frequency $h_v$
is the set of endpoints greater than the node $v$; if $v$ is an
endpoint, it is itself a ({\it trivial}) cluster. The tree provides
an organization of endpoints into a hierarchy of clusters:
$L_{w}<L_{v}$ if $L_{w}\subset L_{v}$
\art
A {\it field label} $f$ distinguishes a field
involved in the interactions. If $v$ is an endpoint,
$I_v$ is the
the set of all the fields $\ps$, $\f$ and $\jm$ involved in the
interaction in $v$. If $v$ is a node, $I_v$ is defined as the union
of the sets $I_u$, for any endpoint $u:u>v$; $x(f)$, $\s(f)$ and
$\o(f)$ denote the spacetime point, the (eventual) $\s$ index and
the $\o$ index, respectively, of the field $f$.
If $h_v< N$, one of the field variables belonging to $I_v$ may also
carry a derivative. It is associated with each field label $f$ an
integer $m(f)\in\{0,1,2\}$, denoting the order of the derivative.
\art
In correspondence of any node or endpoint $v$,
let $P_v\subset I_v$, the {\it external
fields} of $v$, be constructed as follows. In each endpoint $u$ all
the fields are external: $P_u\defi I_v$. If $v$ is a node, and
$v_1,\ldots,v_{s_v}$ are its first followers, then $P_v$ can be 
any set s.t. $P_v\subset \left(\cup_i P_{v_i}\right)$. Let
$Q_{v_i}\defi P_v\cap P_{v_i}$: the union of the complementary ones,
$\cup_i P_{v_i}\bs Q_{v_i}$, is the set of the {\it internal
fields} of $v$
-- or the fields {\it contracted} in correspondence of the node $v$ --
and have not to be an empty at least
\subelenco{
\item{$\bullet$}
in the first node, except if its scale is $h=N+1$.
\item{$\bullet$}
in the branching points;
\item{$\bullet$}
in the first preceding nodes of the endpoints.} 
Hence, the endpoints are attached to nodes
where some of their  external fields are 
actually contracted; while the first point is 
the lowest node in correspondence of which 
some contraction actually occur, 
except in the case of trees  lying 
only on the scales $\ge N+1$, for which 
the first point has been set to be on scale $N+1$. 
Among the fields in $P_v$, the set of all the 
fields of type $\f$ and $\jm$ will be called $S_v$,
the set of the ``special fields''.
Finally, $|P_v|=n^\ps_v+n^\f_v+n^\jm_v$, where $n^\ps_v$ is
the number of external fields of type $\ps$, while 
$n^\f_v,$ $n^\jm_v$,
as already  defined, are
the the number of
external fields $\f$ and $\jm$ -- 
indeed there is only one source field in the special 
endpoint.
\art
Let $\TT_{w;h;n}^{n^\ps,n^\f,n^\jm}$ be the set of all 
topological trees, with all the above depicted constraints, 
with root on scale $h$, first node $w$ on scale $h+1$, 
and with $n$ normal endpoints, 
$n^\ps$ external fields of type $\ps$, 
$n^\f$ endpoints of type $\f$ and $n^\jm$ 
endpoints of type $\jm$. To each such tree
 it corresponds a sequence of
instructions to built a class of Feynman graphs. 
\art
Let $\GG$ one of the Feynman graphs
corresponding to the tree
$\t\in\TT_{w;h;n}^{n^\ps,n^\f,n^\jm}$.
The endpoints of $\t$ represents the vertices of
$\GG$, with the specified couplings. Any node $v$ is in 
correspondence with a subgraph $\GG_v\subset
\GG\=\GG_w$, in which the external legs are the external
fields of $v$. Specifically, if $v_1,\ldots,v_{s_v}$ ($s_v\geq 1$)
are the first followers of $v$, the Feynman graph $\GG_v$ is constructed
by pairing the internal fields of $v$ with propagators $g^{(h)}$,
in a way that the subgraphs $\GG(v_1),\ldots,\GG(v_{s_v})$ remains
connected. 
There are many possible way to chose $\{P_v\}_{v}$,
or equivalently  many possible ways of selecting the internal fields
to be involved in the contractions;
and there are many possible 
connecting contractions: that is why to
each $\t$ is associated a family of many different Feynman graphs.
\art
Let the set of the nodes of $\t$ --
hence considering neither the root, nor
the endpoints -- 
be denoted,
with abuse of notation, $\t$ as well.
For each node $v$, the integer $l_v$ is the number of lines of
the Feynman graph $\GG_v$; while $l_{o,v}$ is the number of lines
in $\GG_v$, which are not in $\cup_{i=1}^{s_v}\GG_{v_i}$. 
Similarly, $l^{\rm anti}_{v}$ and $l^{\rm anti}_{o,v}$
count the number of lines of the graph which 
correspond to antidiagonal propagators.
Two fundamental relations are
$$\eqalign{
&\sum_{u\in \t}^{u\geq v}(s_u-1)
=n_v+n^\f_v+n^\jm_v-1\;,\cr
&\sum_{u\in \t}^{u\geq v
}l_{o,v}=l_v=2n^{(4)}_v+n^{(2)}_v+(1/2)n^\f_v+n^\jm_v-(1/2)n^\ps_v
\;.}\Eqa(trel)$$ 
For instance, from them,  by telescopic decomposition
of the differences of the scales,
$h_u-h_v=\sum_{w\in \t}^{v<w\leq u} h_w- h_{w'}$,
 other two identities descend:
$$\eqalign{
 \sum_{u\in \t}^{u\geq v}(h_u-h_v)(s_u-1)
&=
 \sum_{u\in \t}^{u\geq v}(h_u-h_{u'})(n_v+n^\f_v+n^\jm_v-1)\;,\cr
 \sum_{u\in \t}^{u\geq v}(h_u-h_v)l_{o,v}
&=
 \sum_{u\in \t}^{u\geq v}(h_u-h_{u'})l_u\;.
}\Eqa(treeid)$$
The above formulas are stated as they are for
shake of clarity;
but sometimes it will be used that, by definition,
$h_w-h_{w'}=1$. 
\art 
It is natural to consider the following decomposition.
Given any $\t\in \TT_{w;h;n}^{n^\ps,n^\f,n^\jm}$, let
the ``auxiliary tree'', $\t^a\subset \t$,
be the union of the paths in $\t$ 
which connects the  special endpoint with the root $r$;
for any $w\in \t^a$, let $s_w^*$,
the number of the nodes first followers of 
$v$ and in $\t^a$.
Besides,
if $w$ is one of the maximal nodes in $\t^a$, 
let the integers $n^\jm_{*,w}$,  
$n^\f_{*,w}$, be the number of 
the external fields of type $\jm$ or 
of type $\f$ which are in the cluster $L_w$;
otherwise, for $w\in \t^a$ but not maximal,
let them be the number of 
the external fields of type $\jm$ or 
of type $\f$ which are in the cluster $L_w$,
but not in the following clusters $L_{w_1},\ldots,L_{w_{s_w}}$. 
Finally, the ``main tree'', $\t^*\subset \t^a$,
is given by the auxiliary tree,
deprived of the nodes above the maximal nodes
with $s_w^*\geq 2$;
for $w\in \t^*$, 
let the integer $b^*_w$ be the number of 
nodes of $\t^*$ first followers of $w$: 
hence $s_w^*= b^*_w+n^\f_{*,w}+n^\jm_{*,w}$.
\art
Given any set of fields $M$,  let $x(M)\defi\cup_{f\in M}x(f)$. 
Let $D_v$ be the {\it tree distance} among
$x\left(I_{v_1}\right),\ldots, x\left(I_{v_{s_v}}\right)$
the sets of the spacetime points
of the clusters $L_{v_1}\ldots L_{v_{s_v}}$:
namely $D_v\defi\min_{g\in
\CC}\sum_{l\in g}|l|$, where $\CC$ the set 
of all the possible tree graphs $g$
connecting the spacetime points in  
$x\left(I_{v_1}\right),\ldots, x\left(I_{v_{s_v}}\right)$,
and $l$ are the links.
Similarly, $D_{0,w}$ and $D_{1,w}$ are respectively 
the ``time'' and ``space'' 
tree distance
and are defined as the tree distance 
among the time component and the space component 
of the spacetime points in 
$x\left(I_{v_1}\right),\ldots, x\left(I_{v_{s_v}}\right)$.
}

\*
\asub(ClE){Cluster expansion.} A standard tool in the
fermionic Renormalization Group -- first introduced in [Le87] --
is the cluster expansion of the truncated expectations (see
[B84]). It explains why in the bounds it is better to consider 
altogether all Feynman graphs corresponding  to
one tree, rather than one Feynman graph singly.

Let $P_1,\ldots,P_s$ be disjoint sets of $\ps$ fields s.t.
$\left|\cup_i P_i\right|=2n$; and let $P_j^{\s}\defi\{f\in
P_j:\s(f)=\s\}$. A pairing $l$ is the couple of a field $f^+_l$ in
$\cup_j P^+_j$ and a field $f^-_l$ in $\cup_j P^-_j$: let
$x(f^+_l)-x(f^-_l)
\defi x_l$; and
$\big(\o(f^+_l),\o(f^-_l)\big)\defi \underline\o_l$. Then, the
truncated expectation w.r.t. the Gaussian measure of propagator
$g^{(h)}$ is given, up to a global sign, by:
$$
 \EE^T_{h}\big[\psi(P_1),\ldots,\psi(P_s)\big]=
 \sum_{T}\left(\prod_{l\in T}  g^{(h)}_{\underline\o_l}(x_l)\right)
 \int\!\der P_{T}(t)\  \det G^{h,T}(t)\;,
\Eqa(ce)$$
where $T$ is a set of pairings of elements of $\cup_i P_i$, which
would be a connected tree graph if all the points in the same set $P_i$ 
where identified; the parameters $t=\big\{t_{i,j}\in [0,1] :
i,j=1,\ldots,s\big\}$ have a certain normalized distribution $\der
P_{T}(t)$; finally $G^{h,T}(t)$ is a $(n-s+1)\times (n-s+1)$
matrix, the entries of which
 are given by
$G^{h,T}_{f^-_l,f^+_l}= g^{(h)}_{\underline\o_l}(\underline
x_l)t_{\underline i_l}$, where $\underline i\defi (i^+_l, i^-_l)$
s.t. $f^-_{l}\in P^-_{i^-_l}$ and $f^+_{l}\in P^+_{i^+_l}$, for any
possible pair $l$ of elements of $\cup_i P_i$, s.t. $l\notin T$.

The importance of this formula is that, if all the entries
$M_{i,j}$ of an $n\times n$ matrix $M$ are give by scalar products,
$M_{i,j}=(v^{(i)},w^{(j)})$, where $v^{(1)},\ldots,v^{(n)}$ and
$w^{(1)},\ldots,w^{(n)}$ are vectors, bounded in norm by a constant
$C_0$, the sum of $n!$ monomials that gives the determinant of $M$
can be bounded with $C_0^n$, by a simple application of the volume
inequality. In this way factorial bounds are avoided.
\*

\asub(BST){Bounds for  the kernels.} 
Setting $(h\wdg N)\defi \min\{h,N\}$,
the effective potential on scale $h$ 
is a polynomial of the fields 
with coefficients  given by the kernels:
$$\eqalign{
&\WW^{(h)}
 \left(\f,\jm,\ps\right)\cr
=&
 \sum_{n>0}\sum_{n^\ps,n^\f,n^\jm\geq 0}
 \sum_{\t_v\in \TT_{v;(h\wdg N);n}^{n^\ps,n^\f,n^\jm}}
 \sum_{P_{v}\subset I_{v}}^{|P_v|=n^\ps+n^\f+n^\jm}
 \int\!\der^2 x(P_{v})\
 f(P_{v}) W^{(h)}\Big(\xx(P_{v});\t_v; P_{v}\Big)}\;,$$ 
where, $ f(P_{v})$ denotes the product of
every external field in $P_v$.
In its turn,
the kernel is a sum over the Feynman graphs 
of the product of a propagator for each line of the graphs,
$K^{(h)}\Big(\xx(I_{v});\t_v;P_{v}\Big)$
integrated w.r.t. all the internal points 
of the cluster $L_v$:
$$W^{(h)}\Big(\xx(P_{v});\t_v;P_{v}\Big)
 =
 \int\!\der^2x(I_{v}\bs P_{v})\
 K^{(h)}\Big(\xx(I_{v});\t_v;P_{v}\Big)\;.$$
A useful norm to bound the kernels is obtained by 
integrating the product of the propagators 
w.r.t. all the spacetime points $x\left(I_v\right)$,
except the ``fixed points'', 
$x(F_v)$: they are, if $S_v$ is not empty, 
the points in $F_v\defi S_v$; otherwise 
the point in $F_v\defi\{x_v\}$,
for any choice of $x_v\in P_v$. It holds the following
lemma.
\*

\alemma(SS){If $h>N$,
there exists a constant $C_2\geq C$ such that,
for  any choice of the  
tree $\t_v\in \TT_{v;N;n}^{n^\ps,n^\f,n^\jm}$, 
with root $r$,
$$\eqalign{
&\int\!
 \der^2 x\big(I_{v}\bs F_v\big)\
\left| K^{(h)}
 \Big(x(I_{v});\t_v;P_{v}\Big)\right|\cr
 \leq&
 (C_2\e)^n C_2^{n^\f+n^\jm}
 \g^{Nd_r}
 \left(\sum_{\{P_{w}\}_{w> r}}
 \prod_{w\in \t_v}
 \g^{d_w+r_w}\right)\cr
&\cdot
 \left(
 \prod_{w\in \t^*_v}^{s_v^*\geq 2}
 {\g^{(N+h_w)(s^*_w-1)}\over
 e^{{c\over 2(n^\f+n^\jm)}\left(\sqrt{\g^{N} D_{w}}+\sqrt{\g^{h_w} D_{0,w}}\right)}}\right)
 \left(\prod_{i=1}^{n^\f}{1\over \sqrt{Z_{N}}}\right)
 \left(\prod_{i=1}^{n^\jm}{Z^{(2)}_{N}\over Z_{N}}\right)\;,
}\Eqa(l2)$$
with 
$$d_w\defi \left\{
\matrix{1-n_w-n^\jm_w-n^\f_w \hfill&\hfill{\rm for\ }h_w\ge N+1\cr
        2-(1/2)n^\ps-(3/2)n^\f-n^\jm\hfill&\hfill{\rm for\ }w=r\;,}\right.$$
and $r_w$ such that $d_w+r_w\le -1/2 -(1/8)n^\ps_w$.
}
\*

\proof.
Let $\t_{v_1},\ldots,\t_{v_{s_v}}$ be the subtrees of $\t_v$
branching from $v$ -- namely with root in $v$, and first nodes
$v_1,\ldots,v_{s_v}$; the product of propagators
$K^{(h_v)}\Big(\xx(I_{v});\t_v;P_{v}\Big)$ is obtained 
as
$$\eqalign{
 K^{(h_v)}
 \Big(\xx(I_v);\t_v;P_{v}\Big)
=&{1\over s_{v} !}\sum_{P_{v_1},\ldots, P_{v_{s_{v}}}}
 \left(\prod_{i=1}^{s_{v}}
 K^{(h_v+1)}\Big(\xx(I_{v_i});\t_{v_i};P_{v_i}\Big)\right)\cdot\cr
&
 \cdot\EE^T_{h}
 \left[\psi(P_{v_1}\bs Q_{v_1}),\ldots,\psi(P_{v_{s_{v}}}\bs Q_{v_{s_{v}}})
 \right]\;.
}\Eqa(3.38)$$ 
Applying \equ(ce), and iterating till the endpoints, it
holds:
$$\eqalign{
 K^{(h_v)}
&\Big(\xx(I_{v});\t_v;P_{v}\Big)
=
 \left(\prod_{u}^{\rm{e.p.}}\r_u\right)
 \cdot\cr&\cdot
 \prod_{w\in\t_v}
 \sum_{P_{w}}
 \sum_{T_{w}}{1\over s_{w} !}
 \left(\prod_{l\in T_w}  g^{(h_w)}_{\underline\o_l}(x_l)\right)
 \int\!\der P_{T_w}(\tt)\  \det G^{h_w,T_w}(\tt)\;,
}\Eqa(iter)$$
where $\r_u$ denotes the
coupling in the endpoints: $\l_N$, $\g^N\n_N$ or $\d_N$,
if $u$ is a normal 
endpoint; $\z^{(2,\s)}_N$ if  $u$ is an endpoint 
of type $\jm^{(\s)}$; $1$  if $u$ is an endpoint 
of type $\f$.
Then, a bound for   
the integral of \equ(iter) can be obtained as follows.
\elenco{
\art 
Calling $b_{h}(x-y)\defi  
e^{-(c/2)\left(\sqrt{\g^N| x_{l}|}+\sqrt{\g^{h}| x_{0,l}|}\right)}$,
by  \equ(prop>N) 
each
of the $s_w-1$ propagators in a tree $T_w$ 
is bounded with $ C \g^N b^2_{h_w}(x-y)$; while 
$\left|\det G^{h_w,T_w}(\tt)\right|$ is bounded
with a factor $C_0C\g^{N}$ for each of the 
$l_{o,w}-(s_w-1)$ rows of the matrix 
$G^{h_w,T_w}(\tt)$: globally, 
the product of the propagators can be bounded with 
$$\left(C_0C\g^N\right)^{l_{o,w}}\prod_{l\in T_w} 
b^2_{h_w}(x_l)\;.$$
\art 
Collecting the products over $b_{h_w}(x_l)$
for any node of the tree $\t_v$,
since the branching nodes of the main tree 
are not more than the special 
endpoints $n^\f+n^\jm$,
$$\eqalign{
 \prod_{w\in \t_v}
 \prod_{l\in T_w} b^2_{h_w}(x_l)
 \leq
& \prod_{w\in \t_v}
 \prod_{l\in T_w}  b_{h_w}(x_l)\cr
&\prod_{w\in \t^*_v}
 e^{-{c\over 2(n^\f+n^\jm)}\left(\sqrt{\g^N  D_{w}}+\sqrt{\g^{h_w} D_{0,w}}\right)}
\;. }\Eqa(tdis)$$
\art 
The integrations in $\der^2 x(I_v/F_v)$ are performed,
the left integrand  being  the product 
of the $ b_{h_w}$'s, increased by replacing in them 
$\g^N|x_l|$ with $\g^N|x_{1,l}|$, times constant factors.
It holds
$$\int\!\der^2 x(I_v/F_v)\prod_{w\in \t_v}
 \prod_{l\in T_w} b_{h_w}(x_l)\le
 \prod_{w\in \t}\left(C_1\g^{-(N+h_w)}\right)^{(s_w-s^*_w)}\;.
\Eqa(cas)$$
Indeed, the above formula is obtained 
iteratively starting from the first node, $v$. 
Let the labels 
$w_1,\ldots,w_{s_w}$ be assigned to the nodes 
following $w$ so that: 
for $j=1,\ldots,s^*_w$ the cluster $L_{w_j}$
contains at least a special endpoint, 
$S_{w_i}\neq \emptyset$, 
and is called ``special cluster'';
for $j=s^*_w+1,\ldots,s_w$, the cluster $L_{w_j}$
contains no special endpoints, $S_{w_i}= \emptyset$ -- eventually it
may be $s^*_w=0,s_w$.
Now, the graph $T_w$ can be thought as a tree graph: 
the cluster $L_{w_1}$ is its root,  
$L_{w_2},\ldots,L_{w_{s_w}}$ are its nodes,
while the factors $ b_{h_w}$'s are its links.
Then, considering the first node $v$, and starting from
the endpoints of $T_v$, let 
$L_{v_j}$  be the first followers of  $L_{v_{j'}}$, and 
let  $b_{h_v}$ be the link connecting them.
If $L_{v_j}$ is a special cluster,
than $b_{h_v}$ is simply bounded with its maximum, 
$\norm{b_{h_v}}_\io$;
whereas, if  $L_{v_j}$ is a normal cluster, 
the link $b_{h_v}$ is bounded with $\norm{b_{h_v}}_1$,
the integral being taken w.r.t. the point in $F_{v_j}$.
Since $\norm{b_{h_v}}_\io\le1$, while 
$\norm{b_{h_v}}_1\leq C_1\g^{-(N+h_v)}$,
this gives the factor in \equ(cas) for 
$w=v$.
Iterating  to all the nodes following
the first, the complete bound is found.
\art
 The sum over $T_w$ is bounded
by the number of the topological graphs 
with $s_w$ nodes, $4^{s_w}$, times the number 
of the possible permutations of such nodes,
$s_w!$ .
\art 
Each factor $\r_u$ are bounded, by \equ(ind2),
with $2\e$ if $u$ is a normal endpoint; otherwise
$\r_u=1/\sqrt{Z_N}$ or $Z^{(2)}_N/Z_N$ if respectively 
$u$ is of type $\f$ or $\jm$.}
\*

\0In the end, the factorial in item 4. is compensated 
by the one in the denominator of \equ(iter); while 
the powers of $2\e$, $C$, $C_0$, $C_1$ and $4^{s_w}$ is
all together bounded with

$$\prod_{w\in \t_v}(4C_1)^{s_w}(C_0C)^{l_{o,w}}(2\e)^{n_{o,w}}
 \leq (C_2\e)^n C_2^{n^\f+n^\jm}\;,$$
for $C_2\geq (4CC_0C_1)^2$.
And the rest of 
the bound is reduced to simple dimensional analysis. 
For each of the
$l_{o,w}$ propagators there is a factor
$\g^{N}$; for each of the  $s_w-s^*_w$ integrals
there is a factor $\g^{-(N+h_w)}$ more. 
Furthermore, not yet counted in the above items, by \equ(prop>N)
there is a factor $\g^{-(h_w-N)}$ more for any antidiagonal propagator.
Finally, in correspondence 
of each endpoint of type $\d$ and $\n$ there 
is a factor  $\g^{N}$. Therefore the collection of 
all such factors gives
$$\eqalign{
 &\left(\prod_{w\in \t^*_w}\g^{(N+h_w)(s^*_w-1)}\right)
 \prod_{w\in \t_v}\g^{h_w\big(1-s_w-l_{o,w}^{\rm anti}\big)}
 \g^{N\big(l_{o,w}-(s_w-1)+l_{o,w}^{\rm anti}+ n_{o,w}^{(2)}\big)}
 \cr\le &
 \left(\prod_{w\in \t^*_w}\g^{(N+h_w)(s^*_w-1)}\right)
 \g^{Nd_r}
 \prod_{w\in \t_v}\g^{d_w+r_w}\;, }\Eqa(coll21)$$
where $r_w\defi-l_w^{\rm anti}$ for $n_w=1$, $n_w^\ps=n_w^\jm=0$,
and $r_w\defi 0$ otherwise.
Now it is possible to prove that 
$d_w+r_w\leq-(1/2) -(1/16)n_w^\ps\;.$
Indeed, there are the following possibilities. 
\elenco{
\art
The number of normal endpoints is zero.
Then, since in the nodes of the tree
there has to be at least a contraction, 
and since the self-contraction of 
the fields in  the endpoint of type $\jm$
is zero by oddness of the diagonal propagator,
$n^\f_w+n^\jm_w\geq2$. Then, since
in such graphs the external fields 
of type $\ps$ cannot be more than 
$2(n_w^\f+n_w^\jm)$,
it holds $d_w\leq -(1/2)(n^\f_w+n^\jm_w)\leq -(1/2)-(1/8)n^\ps_w$.
\art
 The number of the normal endpoints is 1, while 
$n^\f_w+n^\jm_w=0$.
Then $d_w+r_w\leq -l_w^{\rm anti}$. By explicit 
inspection, such graphs, made of self-contractions,
either are zero by oddness of the diagonal 
propagator, or
have at least one antidiagonal propagator;
furthermore the number of external $\ps$ fields
cannot be more than two. Therefore
\hbox{$d_w+r_w\leq -(1/2)-(1/4) n^\ps_w$}. 
\art 
The number of the total endpoints,
$n_w+n^\f_w+n^\jm_w$, is greater or equal to 2.
Since in such graphs the external  fields $\ps$
cannot be more than
$4(n_w+n^\f_w+n^\jm_w)$,
and $r_w=0$, then
\hbox{$d_w+r_w\leq -(1/2)(n_w+n^\f_w+n^\jm_w)\leq 
-(1/2)-(1/16)n^\ps_w$}. }
{\0The proof is complete. \hfill\qed\hskip1em}

\*
\alemma(DBST){\it If $h\leq N-1$,  
and for $\e$ small enough,
there exists
a constant $C_2\geq C$ such that 
$$\eqalign{
&\int\!\der^2\xx\big(I_{v}\bs F_v\big)\
 \left|K^{(h)}
 \Big(\xx(I_{v});\t_v;P_{v}\Big)\right|\cr
 \leq&  (C_3\e)^n C_3^{n^\f+n^\jm}
 \g^{h d_r}
 \left(\sum_{\{P_{w}\}_{w> r}}\prod_{w\in \t_v}
 \g^{d_w+r_w}\right)
 \cdot\cr
&\left(
 \prod_{w\in \t^*_v}^{s_v^*\geq 2}
 {\g^{\big((h_w\wedge N)+h_w\big)(s^*_w-1)}\over
 e^{{c\over 2(n^\f+n^\jm)}\left(\sqrt{\g^{(h_w\wedge N)} D_{w}}+\sqrt{\g^{h_w} D_{0,w}}\right)}}\right)
 \left(\prod_{i=1}^{n^\f}{1\over \sqrt{Z_{(h_i\wedge N)}}}\right)
 \left(\prod_{i=1}^{n^\jm}{Z^{(2)}_{(k_i\wedge N)}\over Z_{(k_i \wedge N)}}\right)\;,
}\Eqa(lemma3)$$
where
$$d_w\defi\left\{
 \matrix{1-n_w-n_w^\f-n_w^\jm\hfill&\hfill {\rm for\ }h_w\geq N+1\cr
  2-(1/2)n_w^\ps-(3/2)n_w^\f-n_w^\jm\hfill&\hfill {\rm for\ }h_w\le N}\right.\;,
$$
and $r_w$ is such that $d_w+r_w\le -1/4 -(1/12)n^\ps_w$.
}
\*

\proof.
Neglecting the effects of the localization, 
with argument similar to the proof 
of the previous lemma, the bound 
is reduced to simple dimensional analysis:
for each of the $l_{o,w}$ propagators there is a factor
$\g^{h_w}$; for each of the the  $s_w-s^*$ integrals
there is a factor
$\g^{-2h_w}$. 
Finally,  regarding the endpoints, there is a factor $2\e$ for each
endpoint of type $\l$; $2\e \g^{h_w}$ for each endpoint of type
$\d$ or $\n$. Therefore,
collecting only the factors coming from the dimensional analysis,
$$\eqalign{
 &\left(\prod_{w\in \t^*_w}\g^{2h_w(s^*_w-1)}\right)
 \prod_{w\in \t_v}\g^{h_w\big(l_{o,w}-2(s_w-1)+ n^{(2)}_{o,w}\big)}
 \cr=&
 \left(\prod_{w\in \t^*_w}\g^{2h_w(s^*_w-1)}\right)
 \g^{h d_v}
 \prod_{w\in \t_v}\g^{d_w}\;,
}\Eqa(coll2)$$
with $d_w\defi 2-(1/2)n^\ps_w-(3/2)n^\f_w-n^\jm_w$.
Now the point is that they can occur nodes with non-negative
dimension:
 here comes the role of the localization, which improves their
dimension by  absorbing the localized part of the graphs into the
coupling constants. Indeed, for the kernel bringing an
$\RR$-operator, 
with reference to the items at point \secc(L), 
the following facts have to be considered.
\elenco{
\art
The local part $z_{h_w} D_\s$,
occurring in a certain node $w$,
is bounded, up to a constant, by 
\hbox{$\g^{h_w}\g^{-(h_w-h_{w_0})}$},
if $w_0$ is the node, lower than 
$w$, in correspondence of
which one of the field of momenta $k$ 
is contracted. While the local part
$z_{h_w}\big|-ik_0+\o e(k_1)-D_\o(k)\big|$
is instead bounded, up to a constant, 
with $\g^{h_w}\g^{-(h_w-h_{w_0})}\g^{-(N-h_{w_0})}\leq 
\g^{h_w}\g^{-(N-h_{w_0})}\g^{-2(h_w-h_{w_0})}$:
the standard power counting, 
as it were using only the factor
$\g^{h_w}$, because of $\g^{-2({h_w}-h_{w_0})}$,
is improved in  all the nodes $u$ along the path 
connecting $w$ with $w_0$ by  $r_u=2$.
Furthermore, with reference to the proof 
of the equivalence of the Euclidean and the 
Hamiltonian regularization, 
the factor $\g^{-(N-h_w)}$ 
makes such a kernel -- generated only 
in the latter regularization --
vanishing in the limit of removed cutoff.
\art
One or two increments 
$D_\o$, and respectively one or two derivatives in the companying kernels
-- the kernel occurring  at node $w$, 
the increment having the same momenta of
a $\ps$-field contracted on a lower node, $w_0$ -- gives
a gain w.r.t. the standard power counting:
each derivative gives a factor $\g^{-h_w}$ more,
while each increment gives a factor $\g^{h_{w_0}}$ more.
Since
$$\g^{-(h_w-h_{w_0})r}=\prod_u^{w_0\leq u\leq w} 
\g^{-r}\;,\qquad {\rm for }\  r=1,2\;,$$
all the nodes $u$ in the path connecting 
the node $w$ with the node $w_0$
have a gain $r_u=1$ or 2.
\art
The local terms which are 
linear or quadratic in the factors $\{\m_k/\g^k\}_k$
gives a gain in the bounds since, if they occur
in the node $w$ on scale $h$, $k$ has to be
greater or equal to $h$, and,
by \equ(ind1) and the definition of $h^*$:
$$\left({\m_k\over \k\g^k}\right)^r\leq 
\left({\m_{h^*}\over \k\g^{h^*}}\right)^r
 \g^{-r(1-2c_0\e)(k-h^*)}\leq 
 \prod_{u\leq w}\g^{-r(1-2c_0\e)}\;,$$
and therefore, 
for $\e$ small enough,
the dimension of every node $u$ occurring
along the path connecting the node $w$
with the root is improved by $r_u=r3/4$.
\art 
In the kernels corresponding to 
nodes $w$ with  $n^\jm_w=0$, and $n^\ps_w=n_w^\f=1$,
the dimension is zero. It is possible to 
obtain a gain $r_w=1$ at the price
of worsening the final constant $C_3$
of a factor $\g^2$. Indeed, because of the
compact support of the propagators, it is clear 
that such nodes can be both among  
the preceding ones of the $n^\f$
special endpoints of type $\f$,
let them be  $w_1,\ldots,w_q$,
and among the ones preceding 
$w_1,\ldots,w_q$ themselves: namely 
no more than  
$2 n^\f$ nodes.}
\*

\0Therefore, with developments similar to
the ones in the previous proof, 
it is possible to prove 
that $d_w+r_w\le -(1/4)-(1/12)n^\ps_w $.
But since  the 
localization produces the flows of the
field and densities strengths, 
\equ(coll2) has to be replaced
with 
$$\g^{h d_v}\left(\prod_{i=1}^{n^\f}{1\over \sqrt{Z_{h_i}}}\right)
 \left(\prod_{i=1}^{n^\jm}{Z^{(2)}_{k_i}\over Z_{k_i}}\right)
  \left(\prod_{w\in \t^*_w}\g^{2h_w(s^*_w-1)}\right)
 \left(\prod_{w\in \t_v}
 \left({Z_{h_w}\over Z_{h_w'}}\right)^{(n^\ps_w/ 2)}\g^{d_w+r_w}\right).
$$
This completes the proof. \hfill\qed\hskip1em\null
\*

\asub(R2){Remark.}
The argument in the last item does not apply in  the case $n^\jm=1$ and
$n^\ps=2$. This is the main difference of the external sources
$\jm$ and $\f$: while the former requires a coupling constant for
absorbing divergences due to interaction with the source, the latter
need not, since it in
interacts only by {\it one particle reducible
graphs}.
\*

\alemma(SF){For $\e$ small enough, the perturbative expansion 
for the $(n^\jm;n^\f)$-Schwinger
functions is absolutely convergent to a distribution
fulfilling property E1 and E5. of the \osa{}. }
\*

\proof.
The expansion for the Schwinger function is given by the expansion
for the effective potential in the case $P_v=S_v$ and
for any scale of the first node  $h:h^*-1\le h\le N+1$.

Since the case $h^*$ finite is much more easier 
of the case $h^*=-\io$, the following development 
will concern only the latter.

Calling
$\TT_{v,h;\underline k;\underline h;n}^{0,n^\f,n^\jm}$ the set of
trees $\t\in \TT_{v,h, n}^{0,n^\f,n^\jm}$ having the $n^\f$ external
fields of type $\f$ on scales $h_1,\ldots, h_{n^\f}$, and the
$n^\jm$ external fields of type $\jm$ on scales $k_1,\ldots,
k_{n^\jm}$, it holds
$$
 S_{\us;\uo}^{(n^\jm;n^\f)(\ue)}
 \left(\underline z;\underline x\right)
 \defi
 \sum_{n\leq 0}\sum_{h\leq M}
 \sum_{\underline k}^{h< k_j\leq M}
 \sum_{\underline h}^{h< h_j\leq M}
 \sum_{\t_v\in \TT_{v,(h\wdg N);\underline h;\underline k;n}^{0,n^\f,n^\jm}}
 W^{(h)}\Big(\xx(S_v);\t_v;S_v\Big)
\Eqa(sch)$$
and, by the just proved bound on the kernels,
$$\eqalign{
&\left|W^{(h)}\Big(\xx(S_v);\t_v;S_v\Big)\right|
 \leq
 (C_2\e)^n C_2^{n^\f+n^\jm}
 \g^{h d_r}\prod_{w\in \t^*_v}^{h_w\geq N+1}
 e^{-{c\over 2(n^\f+n^\jm)}\sqrt{\g^{h_w} D_{0,w}}}\cr
&\cdot\left(
 \prod_{w\in \t^*_v}^{s_v^*\geq 2}
 {\g^{\Big((h_w\wdg N)+h_w\Big)(s^*_w-1)}\over
 e^{{c\over 2(n^\f+n^\jm)}\sqrt{\g^{(h_w\wdg N)} D_{w}}}}\right)
 \left(\prod_{i=1}^{n^\f}{1\over \sqrt{Z_{(h_i\wdg N)}}}\right)
 \left(\prod_{i=1}^{n^\jm}{Z^{(2)}_{(k_i\wdg N)}\over Z_{(k_i\wdg N)}}\right)
 \cdot\cr&\cdot
 \left(
 \sum_{\{P_{w}\}_{w> r}}\prod_{w\in \t_v}
 \left({Z_{(h_w\wdg N)}\over Z_{(h_{w'})\wdg N}}\right)^{n^\ps_w\over 2}\g^{d_w+r_w}\right)\;.
}\Eqa(lemma4)$$
Let the following facts be considered.
\elenco{
\art 
For the main tree it holds an identity 
similar  to \equ(treeid), with $s_v$ replaced by $s^*_v$,
and with $n_v$ removed from the r.h.s. member; so that:
$$\eqalign{
&\sum_{w\in \t^*_v}\Big((h_w\wdg N)+h_w\Big)(s^*_w-1)
=2h(n^\f +n^\jm-1)\cr
&\phantom{****}
+\sum_{w\in \t^*_v}^{h_w\le N}(h_w-h_{w'})2(n_w^\f+n_w^\jm-1)
+\sum_{w\in \t^*_v}^{h_w\ge N+1}(h_w-h_{w'})(n_w^\f+n_w^\jm-1)\cr
&\defi h\D d_v+\sum_{w\in \t_v^*}(h_w-h_{w'})\D d_w\;.}$$
These factors can be absorbed into 
the dimension of  any node $w$ of the main tree,
changing it from  $d_w$ to 
$$d_w + \D d_w=
\left\{\matrix{
n^\jm_w+(1/2)n^\f_w -(1/2)n_w^\ps \hfill&\hfill{\rm for\ }h_w\le N\cr
-n_w\hfill&\hfill{\rm otherwise.}}\right.$$
\art
Since $n^f_w=\sum_{v\geq w}n^f_{*,v}$ for $f=\f,\jm$,
then
$$\eqalign{
&\sum_{w\geq v}^{h_w<N}(h_w-h)\big(n_{*,w}^\jm+(1/2)n_{*,w}^\f\big)
+\sum_{w\geq v}^{h_w=N+1}(N-h)\big(n_{w}^\jm+(1/2)n_{w}^\f\big)\cr
&=\sum_{w\geq v}^{h_w\le N}(h_w-h_{w'})\big(n_{w}^\jm+(1/2)n_{w}^\f\big)\;,
}$$
which formula  gives:
$$\eqalign{
&\g^{h \big(n^\jm + (1/2)n^\f\big)}
\prod_{w\in \t^*_v}^{h_w\le N}
\g^{(h_w-h_{w'})\big(n^\jm_w + (1/2)n^\f_w -(1/2)n^\ps_w\big)}\cr
&\phantom{*********}
\cdot\prod_{w\in \t^*_v}^{h_w\ge N+1}
\g^{-(h_w-h_{w'})n_w}\cr
=&\prod_{w\in \t^*_v}^{h_w\le N}
\g^{h_w\big(n^\jm_{*,w} + (1/2)n^\f_{*,w}\big)}
\prod_{w\in \t^*_v}^{h_w\le N}\g^{-(h_w-h_{w'})(1/2)n^\ps_w}\cr
&\cdot\prod_{w\in \t^*_v}^{h_w= N+1}
\g^{N\big(n^\jm_{w} + (1/2)n^\f_w\big)}
\prod_{w\in \t^*_v}^{h_w\ge N+1}\g^{-(h_w-h_{w'})n_w}\;.}\Eqa(ddee)$$
\art
In view of the proof  of cluster decomposition,
since it can be, for $w={v^*_0}$, the  lowest 
branching point of $\t^*$, 
\hbox{$ n^\jm_{*,w}+(1/2)n^\f_{*,w}=0$},
a further modification of the above decomposition is
performed.
Setting 
$m\defi n^\jm+(1/2)n^\f$, 
$m_w\defi n^\jm_w+(1/2)n^\f_w$ and 
$m_{*,w}\defi n^\jm_{*,w}+(1/2)n^\f_{*,w}$;
and letting $h_0$ be the scale of $v_0^*$,
the following identity 
$$1=\g^{-\big(h_w-h_0\big){1\over 8}{m_{*,w}\over m}}
 \g^{\big(h_w-h_0\big){1\over 8}{m_{*,w}\over m}}\;,$$
for each node $w\in\t^*:h_w\le N$ 
turns \equ(ddee) into
$$\eqalign{
&\g^{h_0(1/8)}\prod_{w\in \t^*_v}^{h_w\le N}
\g^{h_wm_{*,w}\big(1-(1/8m)\big)}
\prod_{w\in \t_v}^{h_w\le N}\g^{(h_w-h_{w'})\big((m_w/8 m)-(1/2)n^\ps_w\big)}\cr
&\cdot\prod_{w\in \t^*_v}^{h_w= N+1}
\g^{Nm_{w}\big(1-(1/8m)\big)}
\prod_{w\in \t_v}^{h_w\ge N+1}\g^{-(h_w-h_{w'})n_w}\;.}\Eqa(mvm)$$
\art
Let each  factor $1/\sqrt{Z_{h_i\wdg N}}$ be considered
for $h_i\le N$: 
if the $w\in t_v^*$, is the 
highest branching point in $\t$  
lower than the $i$-th endpoint of type $\f$, $u_i$,
by \equ(ind1), such a factor can be moved to the 
node $w$,
\hbox{$1/\sqrt{Z_{h_i}}\leq 1/\sqrt{Z_{h_w}}\g^{(c_0/2)\e^2(h_i-h_w)}$},
at the price of the factor 
$\g^{(c_0/2)\e^2(h_i-h_w)}=\prod_{w'}^{w\leq w'\le u_i}\g^{(c_0/2)\e^2}$:
it  is absorbed in the 
dimension
of the nodes along the path 
connecting  $u_i$ with the node 
$w$ -- by definition such nodes are not in the main tree -- 
changing it, for $\e$ small enough, from 
\hbox{$d_w+r_w\le -1/4-(1/12)n_w^\ps$} to 
the new dimension
$\wh d_w\le-1/8 -(1/12)n^\ps_v$. 
Similar decomposition is done in case 
$h_i\ge N+1$: the lost in the dimension is
only in the nodes on scales $h_w\le N$.
\art
Similar procedure is executed
 for  each  factor $Z^{(2)}_{k_i\wdg N}/Z_{k_i\wdg N}$,
for $h_w\le N$: 
if $w\in\t^*$,  
is the highest branching point in $\t$ lower than 
the $i-$th endpoint of type $\jm$, $u_i$ 
by \equ(ind1),
\hbox{$Z^{(2)}_{k_i}/Z_{k_i}\leq Z^{(2)}_{h_w}/Z_{h_w}
\g^{2c_0\e(k_i-h_w)}$}; 
the factor $\g^{2c_0\e(k_i-h_w)}$ is absorbed in the 
dimension of the nodes 
along the path  connecting $u_i$ with 
$w$, again changing it from 
\hbox{$d_w+r_w=-1/4-(1/12)n^\ps_w$}
to the new dimension  $\wh d_w\le-1/8-(1/12)n^\ps_v$, for 
$\e$ small enough. Similar decomposition is done in case 
$k_i\ge N+1$.
\art
The exponent $(m_w/8m)-(1/2)n^\ps_w$ of 
the factors 
in the second product in formula \equ(mvm)
can be bounded with $-1/8-(1/12)n^\ps_w$.
\art
Since in every node $w:h_w\leq N$, both in the main tree and in the
rest of the tree, the dimension has been left to be
$\wh d_w=-1/8-(1/12)n^\ps_w$, and since for $\e$ small enough, 
$(Z_{h_w}/Z_{h_w'})\leq \g^{c_0\e^2}\leq \g^{(1/12)}$,
it is possible to absorb all the factors 
$(Z_{h_w}/Z_{h_w'})^{(1/2){n^\ps_w}}$ into the dimension 
$\wh d_w$, turning it into $d'_w\le  -1/8 -(1/24)n^\ps_w$.
\art
Regarding the nodes $w:h_w\geq N$,
if $n_w>0$, by inspection of
the graphs -- eventually involving the interaction of type $\jm$
and $\f$ -- it can be $n_w\neq 0$, 
and than $-n_w\le -(1/4)n^\ps_w\le -1/8 -(1/24)n^\ps_w$;
otherwise $n_w=0$: this can happen only on the highest 
node, in the sense that a node with $n_w=0$
cannot be lower than any node $v$ with
$n_v\neq 0$ -- since $n_w$ is a cumulative counter --
then the graphs corresponding to this latter case are  
contractions of special vertices only, 
and  $n_w^\ps\le 2$.
Hence in the region of the tree
where $n_w=0$ there  can be no 
more than $n^\ps+n^\jm$ branching points:
it is in any case possible, multiplying  $C_2$ by a factor
$\g^{2/24}$, to extract a factor $\g^{-(1/24)n^\ps_w}$
for every node $w:h_w\geq N$ such that $P_w\neq P_{w'}$,
namely where some contraction really occur.
\art
The product over the nodes where at least 
a contraction of internal fields does occur, 
$\prod_{w\in \t_v}^{\rm b.p.}\g^{-(\th/24)n^\ps_w}$,
allows to control the summation in $P_w$ -- which,
fixed the tree $\t_v$, is actually only a summation in 
$P_w\bs S_w$:
$$\eqalign{
&\prod_{w\in \t_v}^{\rm b.p.}
 \sum_{P_w}\g^{-(1/24)n^\ps_w}
 \leq 
 \prod_{w\in \t_v}^{\rm b.p.}\sum_{n_w^\ps}\g^{-(1/24)n^\ps_w}
 \left(\matrix {n_{w_1}^\ps+\cdots+n^\ps_{w_{s_w}}\cr n_w^\ps}\right)
 \cr&\leq
 \prod_{u\in \t_v}^{\rm e.p.}
 \left(1-\g^{-(1/24)}\right)^{-n^\ps_u}\le \left(1-\g^{-(1/24)}\right)^{-4(n+n^\f+n^\jm)}\;,
}$$
where the last-but-one  inequality 
can be easily proved by induction
by thinking the endpoints $u$ as the node at 
which are attached one or more further branches;
while the last simply follows from the fact that 
$n_u^\ps\leq 4$.}

\*
\0Finally, once $C_3$ is 
taken greater or equal to $C_2\g^{2/24}(1-\g^{-(1/24)})^{-4}$,
the bound for the Schwinger function has become
$$\eqalign{
&\left|\WW^{(h)}\Big(\xx(S_v);\t_v;S_v\Big)\right|\cr
 \leq&
 (C_3\e)^n C_3^{n^\f+n^\jm}
 {\g^{h_0(m+1/8)}\over
 e^{{c\over 2(n^\f+n^\jm)}\sqrt{\g^{h_0}D_{v_0^*}}}}
 \left(
 \prod_{w\in \t^*_v}^{h_w= N}{\g^{h_wm_{w}(1-(1/8m))}\over
 e^{{c\over 2(n^\f+n^\jm)}\sqrt{\g^{N} D_{w}}}}\right)\cr
&\cdot\left(
 \prod_{w\in \t^*_v}^{h_w\le N}
 {\g^{h_wm_{*,w}(1-(1/8m))}\over
 e^{{c\over 2(n^\f+n^\jm)}\sqrt{\g^{h_w} D_{w}}}}
 \left({Z^{(2)}_{h_w}\over Z_{h_w}}\right)^{n^\jm_{*,w}}
 \left({1\over Z_{h_w}}\right)^{(1/2)n_{*,w}^\f}\right)\cr
&\cdot
 \left(\prod_{w\in \t_v}^{**}\g^{-1/8}\right)
 \left(\prod_{w\in \t_v}^{***}\g^{h_w\over 2(n^\f+n^\jm)}
e^{-{c\over 2(n^\f+n^\jm)}\sqrt{\g^{h_w} D_{0,w}}}\right)\;.
}\Eqa(sbound)$$ 
The product $\prod_{w\in \t_v}^{**}$ is over all the nodes in the tree,
except the ones higher than the branching points $w$
with $h_w\geq N+1$ and  $n_w=0$.
The product $\prod_{w\in \t_v}^{***}$ is over all the 
branching points $w$ with $h_w\geq N+1$ and  $n_w=0$;
and the factors $\g^{h_w\over 2(n^\f+n^\jm)}$ --
strictly greater than 1 -- are added for later purposes.

This bound is enough to prove the convergence of the 
Schwinger function.
Indeed,   for any $m>(1/4)$ and $d,\b,z>0$, 
the two inequalities hold:
$$z^{m}e^{-(\b/m)\sqrt{z}}\leq C_\b^{2m} (4m)!\;,\Eqa(in1)$$
$$\eqalign{
\sum_{h=-\io}^{+\io}\left(\g^h d\right)^me^{-(\b/m)\sqrt{\g^hd}}
&\leq \sum_{h\leq 0}\g^{hm}+\sum_{h>0}\g^{hm}e^{-(\b/m)\sqrt{\g^h}}\cr
&\le C_\b^{4m}(8m)!(1-\g^{-{(1/8)}})^{-1}\;.}\Eqa(in2)$$
Then \equ(in1) allows to bound 
each factor of the product $\prod_{w\in \t^*_v}^{h_w\le N}$,
as:
$$\eqalign{
&{\g^{h_w m_{*,w}\big(1-(1/8m)\big)}\over
 e^{{c\over2( n^\f+n^\jm)}\sqrt{\g^{h_w} D_{w}}}}
 \left({Z^{(2)}_{h_w}\over Z_{h_w}}\right)^{n^\jm_{*,w}}
 \left({1\over \sqrt{Z_{h_w}}}\right)^{n^\f_{*,w}}\cr
&\phantom{**********}
 \leq C_w
 \left({1\over D_w}\right)^{m_{*,w}(1-(1/8m)-\h_\l)+ n^\jm_{*,w}\h^{(2)}_\l}\;,}$$
for $C_w\sim (m!)^p$, for some positive integer $p$; 
and $\prod_{w\in \t^*_v}^{h_w\le N}D_w^{-m_{*,w}(1-(1/8m)-\h_\l)- n^\jm_{*,w}\h^{(2)}_\l}$
is integrable against test functions which vanish with all their derivatives 
for each $D_w=0$.   
Furthermore,  \equ(in2) allows in a similar manner to control the summation over
the scales of the branching points with $n_w=0$ of the factors in the 
product $\prod_{w\in \t_v}^{***}$: apart a constant,
it gives a factor $\prod_{w\in \t_v}^{***}D_{0,w}^{-[1/2(n^\f+n^\jm)]}$,
which is integrable against a test function, even if it 
does not vanish for $D_{0,w}$, since the number of the factor 
is not larger than $n^\f+n^\jm$.

The summation over the scales
$\underline h, \underline k$, 
taking fixed the lowest, $h$,  
and also over the scales of all the 
remaining  branching point in the tree $\t_v$
is clearly controlled by the factors 
$\prod_{w\in \t_v}\g^{-1/8}$
and, since the number of the branches in a 
tree is no more than twice the number of 
the endpoints, it  is bounded by $(1-\g^{-{1/8}})^{-2(n+n^\f+n^\jm)}$.

Then it is possible 
to take the summation also over 
$-\io<h_0\le N$, which is convergent by \equ(in2),
and gives a further factor $D_v^{-(m+1/8)}$, which,
besides not to waste the integrability  against the test function,
guarantees the cluster decomposition, 
namely that the Schwinger function 
vanish if the distance of {\it any two points} is
sent to infinity. 
 
The summation over the topology of the trees,
is bounded by $16^{(n+n^\f+n^\jm)}$. 
Finally  the summation over 
$n$ is convergent 
for any  $\e\leq \Big(16C_3 (1-\g^{-{1/8}})^{-2})\Big)^{-1}$.

{The lemma is proved. \hfill\qed\hskip1em}

\*
\asub(smp)
{Short memory property.}
Before performing the summation 
over the scales  in the product 
$\prod_{w\in \t_v}\g^{-1/8}$,
it is possible to extract a factor
$\g^{-(1/16)\big((h_{\rm max}\wdg N)-\h_{\rm min}\big)}$,
for $h_{\rm max}$ and $h_{\rm min}$ respectively the scale of the 
one of the maximal nodes and of the minimal node
of the tree, leaving $\prod_{w\in \t_v}\g^{-1/16}$ to 
control such a summation.

Many consequences derives from such a factor. 
An example is the following lemma.
\*

\alemma(LRC)
{In the limit of removed cutoff,
the trees with unbounded maximal scale
gives vanishing contribution 
to the  integration of the Schwinger function
against the test functions.}
\*

\proof. Before removing the cutoff, 
let $M_N\defi h_{\rm max}\wdg N$; then $M_N\to +\io$.
With reference to the summation over $-\io< h\leq N$
of the factor $ \g^{h (m+1/8)}
 e^{-{c\over 2(n^\f+n^\jm)}\sqrt{\g^{h_0} D_{v_0^*}}}$,
the following facts hold.
\elenco{
\art
Since the integration against test functions
over all the space time  
is finite, the integration in the region  
 $\k|D_{v_0^*}|\leq \g^{-(M_N/4)}$ is vanishing.
\art
In the domain  $\k|D_{v_0^*}|\leq \g^{-(M_N/4)}$,
the summation for  $h\ge (M_N/2)$ is vanishing
faster than $e^{-{c\over 4(n^\f+n^\jm)}\g^{M_n/8}}$.
\art
Trees with first node on scale $h\le (M_N/2)$
have a short memory factor 
$\le \g^{-(1/16)(M_N/2)}$,
which is vanishing too.}
\*

\asub(P3P1){Completion of the proof of Theorem {\thm(T1)}}
The bound for the two point Schwinger function is, accordingly to 
\equ(sbound), for $\e$ small enough,
$$\left|S^{(2)}_{\o}(x-y)\right|\leq
  C\sum_{h=h^*}^N {\g^{h}\over e^{(c/4)\sqrt{\g^h\k|x-y|}}}
 {1\over Z_h}\;.\Eq(tpbound)$$
Setting $h_o$ s.t. $\g^{-h_o}\leq k|x-y|< \g^{-h_o+1}$,
if $h_o<h^*$, then
$$\sum_{h=h^*}^N {\g^{h}\over e^{(c/4)\sqrt{\g^h\k|x-y|}}}
{1\over Z_h}\leq K 
{\g^{h^*}\over e^{(c/8)\sqrt{\g^{h^*}\k|x-y|}}}{1\over Z_{h^*}}\;;$$ 
while, if $h_o > h^*$, then
$$\sum_{h=h^*}^N {\g^{h}\over e^{(c/4)\sqrt{\g^h\k|x-y|}}}
{1\over Z_h}\leq K 
\g^{h_o}{1\over Z_{h_o}}\;.$$ 
Since $\m_{h^*}$ is proportional to $\k \g^{h^*}$, then: 
$\m_{h^*}$  is proportional to $\k (\m/\k)^{(1/1+\bar\h_{\l})}$;
$Z_{h^*}$ is proportional to $ (\m/\k)^{-(\h_{\l}/1+\bar\h_{\l})}$;
for $h_o\leq N/2$, in the limit $N\to +\io$, $Z_{h_o}$
is proportional to $(\k|x-y|)^{\h_\l}$.
Hence the item is proved for
$1+\bar\t_{\l}\defi (1/1+\bar\h_{\l})$. \hfill\qed\hskip1em\null

\*
\asub(P2P2){Completion of the proof of Theorem {\thm(P2)}}
The bound for the current-current Schwinger function
is the same of \equ(tpbound), with the replacement of
$\g^h/Z_h$ with $\g^{2h}(Z^{(2)}_h/Z_h)^2$. Therefore,
with the same developments of Proof \secc(P3P1), using
also the identity $\h_{\l}=\h^{(2)}_\l$,
also this item is verified. \hfill\qed\hskip1em\null
\appendix(ES,Exact symmetries)

The following symmetries will be useful to prove some kernels are
less divergent than what seems from dimensional bounds:
\*

\asub(RE){Reflection.}
Let the ``reflection'' be $\th(k_0,k_1)\defi (-k_0,-k_1)$.
It is easy to verify the interactions $\VV$, $\NN$ and
$\DD$, as well as the free action, are all
invariant under the transformation of the fields
$$\hp^{\s}_{k,\o}\rightarrow i\o\hp^{\s}_{\th k,\o}\;.\Eqa(refl1)$$
In terms of graphs, under reflection the propagator
$\hg^{(j)}_{\m,\o}(k)$ transforms as follows
$$\hg^{(j)}_{\m,\o}(\th k)=-\m\o\hg^{(j)}_{\m,\o}(k)\;;\Eqa(refl2)$$
while the interactions are all invariant, except the ones corresponding
to the interactions $\DD$, which is odd.
Specifically, let any graph contributing to the kernel $\hW^{(j)}_{2,\o,\o}(k)$
be considered: calling $m_2(\o)$ and $m_2(-\o)$ respectively the
number of vertices with interaction linear in $\ps_{\o}\ps_{\o}$
and $\ps_{-\o}\ps_{-\o}$, after the contraction of only the
off-diagonal propagators, they are left $2(l+m_2(\o)-1)$ half
lines of kind $\o$ and $2(l+m_2(-\o))$ half lines of kind $-\o$ to
be contracted with diagonal ({\it odd}) propagators. As the number
of odd vertices is $m_2(\o)+m_2(-\o)$, and the number of odd
propagators is $2l+m_2(\o)+m_2(-\o)-1$, then $\hW^{(j)}_{2,\o,\o}(k)$
is odd. With a similar argument it is possible to prove
$\hW^{(j)}_{2,\o,-\o}(k)$ is even. Therefore
$$\hW^{(j)}_{2,\a,\b}(\th k)=-\a\b\hW^{(j)}_{2,\a,\b}(k)\;,
 \qquad
 \left(\partial_\s\hW^{(j)}_{2,\a,\b}\right)(\th k)
 =
 \a\b\left(\partial_{\s}\hW^{(j)}_{2,\a,\b}\right)(k)\;.
\Eqa(refl)$$
\*

\asub(SR)
{Space reflection.}
Let the ``space reflection'' be
$\th_1(k_0,k_1)\defi (k_0,-k_1)$.
It is easy to verify the interactions $\VV$, $\NN$ 
and $\DD$, as well as the free action, are all
invariant under the transformation of the fields
$$\hp^{\s}_{k,\o}\rightarrow\hp^{\s}_{\th_1k,-\o}\;.$$
In terms of graphs, under space reflection the propagator
$\hg^{(j)}_{\a,\b}(k)$ transforms as follows
$$\hg^{(j)}_{\a,\b}(\th_1k)=\hg^{(j)}_{-\a,-\b}(k)\;;$$
while the vertices are invariant; therefore,
$$
 \hW^{(j)}_{2,\a,\b}(\th_1k)=\hW^{(j)}_{2,-\a,-\b}(k)\;,
 \qquad
 \left(\partial_\s\hW^{(j)}_{2,\a,\b}\right)(\th_1k)
 =
 \left(\partial_{-\s}\hW^{(j)}_{2,-\a,-\b}\right)(k)\;.
\Eqa(srefl)$$
Furthermore, with similar arguments, it is easy to prove
$$
 \hW^{(j)}_{1;2,\a;\b}(\th_1p;\th_1k)=\hW^{(j)}_{1;2,-\a;-\b}(p;k)\;.
\Eqa(srefl1)$$
\*

\asub(Rot){Rotation.}
Let the ``rotation'' of $\p/ 2$ be
$(k_0,k_1)^*\defi (-k_1, k_0)$.
It is easy to verify the interactions $\VV$ and $\NN$, as well as the free
action of the massive Thirring model, are invariant under the
transformation of the fields:
$$\hp^\s_{k,\o}\rightarrow e^{i\o{\p\over 4}}\hp^\s_{k^*,\o}\;.$$
In terms of graphs, under rotation the propagator 
$\hg^{({\rm E},k)}_{\a,\b}(k)$ transforms as follows
$$ \eqalign{
 \hg^{({\rm E},k)}_{\a,\b}(k^*)
&=-i\o \hg^{({\rm E},j)}_{\o,\o}(k)\;,\cr
  \hg^{({\rm E},j)}_{\o,-\o}(k^*)
&=\hg^{({\rm E},j)}_{\o,-\o}(k) \;.}$$
Let $\hW^{({\rm E},j)}_{2,\m,\n}(k)$ be defined as the sum of the
graphs of $\hW^{(j)}_{2,\m,\n}(k)$ which are made only with
propagators $\hg^{({\rm E},j)}_{\m,\m}(k)$ and only with
vertices $\VV$.

Then, each graph of $\hW^{({\rm E},j)}_{2,\o,\o}(k)$ is made of
$l$ diagonal propagators $\hg^{({\rm E},j)}_{\o,\o}$ and $l+1$
diagonal propagators $\hg^{({\rm E},j)}_{-\o,-\o}$; whereas
each graph of $\hW^{({\rm E},j)}_{2,\o,-\o}(k)$ is made of $l$
diagonal propagators $\hg^{({\rm E},j)}_{\o,\o}$ and $l$
diagonal propagators $\hg^{({\rm E},j)}_{-\o,-\o}$ (and also at
least one off-diagonal propagator). Therefore it holds
$$\eqalign{
 \hW^{({\rm E},j)}_{2,\o,\o}(k^*)
&=i\o\hW^{({\rm E},j)}_{2,\o,\o}(k)\;,
 \qquad
 \left(\partial_\s\hW^{({\rm E},j)}_{2,\o,\o}\right)(k^*)
 =\s\o
  \left(\partial_{\s}\hW^{({\rm E},j)}_{2,\o,\o}\right)(k)\;,\cr
 \hW^{({\rm E},j)}_{2,\o,-\o}(k^*)
&=\hW^{({\rm E},j)}_{2,\o,-\o}(k)\;,
 \qquad
 \left(\partial_\s\hW^{({\rm E},j)}_{2,\o,-\o}\right)(k^*)
 =-i\s
  \left(\partial_{\s}\hW^{({\rm E},j)}_{2,\o,-\o}\right)(k)\;,
}\Eqa(rot)$$
and, with similar definitions and arguments:
$$\eqalign{
 \hW^{({\rm E},j)}_{1;2,\m;\n}(p^*;k^*)
&=\m\n\hW^{({\rm E},j)}_{1;2,\m;\n}(p;k)\;. }\Eqa(rot1)$$
\appendix(P,Proof of Theorem {\thm(FC) })

\asub(BaGF)
{Beta and Gamma functions.} Let $x_N\defi(\n_N,\d_N)$,
$\m_h\defi \m \bar Z_h$
and $\D\l_h\defi \l_h-\l$; a conventional
way of writing the relation \equ(ga1), \equ(ga3) and \equ(b1),
\equ(b2) is in terms of the {\it Gamma functions}:
$$\eqalign{
&\log_\g{Z_{h-1}\over Z_h}
 =\G_h(\l_h,x_h;\ldots;\l_N,x_N)\;,\cr
&\log_\g{\bar Z_{h-1}\over \bar Z_h}
 =\bar\G_h(\l_h,\m_h,x_h;\dots;\l_N,\m_N,x_N)\;,\cr
&\log_\g{Z^{(2,\s)}_{h-1}\over Z^{(2,\s)}_h}
 =\G^{(2,\s)}_h(\l_h,x_h;\dots;\l_N,x_N)\;;}\Eqa(d1)$$
and {\it Beta functions}:
$$\eqalign{
&\n_{h-1}-\g\n_h
 =\b^{(\n)}_h(\l_h,x_h;\dots;\l_N,x_N)\;,\cr
&\d_{h-1}-\d_h
 =\b^{(\d)}_h(\l_h,x_h;\dots;\l_N,x_N)\;,\cr
&\D\l_{h-1}-\D\l_h
 =\b^{(\l)}_h(\l_h,x_h;\dots;\l_N,x_N)\;.}\Eqa(d2)$$

Furthermore, such Gamma and Beta function 
are given by convergent graph expansion.
\*
\alemma(abgf){In the domain of the effective parameters
given by \equ(ind2), if \equ(ind1) are satisfied,
the Gamma and Beta function in \equ(d1) and \equ(d2) 
are well defined and analytic in $\{\l_k,\d_k,\n_k\}_{k\le N}$.}
\*
\proof. Like the proof of the convergence of the Schwinger function,
it is a consequence of the Lemmas \lm(SS) and \lm(DBST),
for the set of fixed points, $F_v$, given by only one point.
\hfill\qed\hskip1em\null

\* 
The evolution of the effective parameters 
is determined by the equations \equ(d1) and \equ(d2),
and by fixing the ``initial data''; they are chosen to be:
$$\eqalign{
&\D \l_{-\io}=0\;,\qquad
\d_{-\io}=0\;, \qquad\n_{-\io}=0\;,\cr
&\phantom{***}
\log_\g(Z_0)=0\;,\qquad\log_\g(\bar Z_0)=0\;,\cr
&\phantom{*}
\log_\g(Z^{(2,+)}_0)=0\;,\qquad\log_\g(Z^{(2,-)}_0)=0\;.}\Eqa(indat)$$
Well then, the strategy to find the solution of the evolution problem
is first to skip the flow of the mass, and to find
the solution of the other flows by a fixed point theorem in a suitable 
linear space; then to solve also the flow of the 
mass with the other flow already fixed. 
\*

\asub(cf){Flows of the couplings}.
Let 
$\MMM$ be the linear space of sequences $y$,
$$y\defi
\lft\{\lft(\D \l_k,\d_k,\n_k,
\log_\g(Z_k),
\log_\g\lft(Z^{(2,+)}_k\rgt), 
\log_\g\lft(Z^{(2,-)}_k\rgt)\rgt)\in \RRR^6:k\leq N\rgt\}\;,$$
such that, for any \hbox{$\th<1/16$},
the following properties hold.
\elenco{
\item{\bf i.}
The initial data are as in \equ(indat).
\item{\bf ii.}
The increments of the effective coupling
satisfy \equ(l1), for any $h:h\leq N$.
}
\0Then, let such a space be endowed with the norm $\norm{y}_\th$,
which is the smallest real number such that all the 
following inequalities hold.
\elenco{ 
\item{\bf iii.}
There exist two positive constants,  $c_0$ and $c_1$,
such that, for every $k\leq N$,
$$\eqalign{
\left|\D\l_k\right|\le c_1 \e^2&\g^{-(\th/2)(N-k)}\norm{y}_\th\;,\cr
\left|\d_k\right|\le 2\e\g^{-(\th/2)(N-k)}\norm{y}_\th\;,
\qquad&
\left|\n_k\right|\le 2\e\g^{-(\th/2)(N-k)}\norm{y}_\th\;,\cr
\left|\log_\g(Z_{k-1}/Z_k)\right|&\le c_0\e^2 \norm{y}_\th\;,\cr
\left|\log_\g(Z^{(2,+)}_{k-1}/Z^{(2,+)}_k)\right|\le 2c_0\e^2 \norm{y}_\th\;,
\qquad&
\left|\log_\g(Z^{(2,-)}_{k-1}/Z^{(2,-)}_k)\right|\le 2c_0\e^2 \norm{y}_\th\;.}\Eqa(ind3)$$
}The space $\MMM_{\th}$ is defined 
as $\{y\in \MMM:\norm{y}_\th\leq 1\}$ and 
is clearly complete. 
Let the equation $y=Ty$ read in $\MMM_\th$:
$$\eqalign{
&\D \l_h=-\sum_{k\leq h} \b^{(\l)}_k\;,
 \qquad
 \d_h=-\sum_{k\leq h} \b^{(\d)}_k\;,
 \qquad
 \n_h=-\sum_{k\le h} \g^{-(h-k+1)}\b^{(\n)}_k\;,\cr
&\phantom{*********}
 \log_\g(Z_h)=\sum_{k=0}^h \G_k\;,
 \qquad
 \log_\g(Z^{(2,\s)}_h)=\sum_{k=0}^h \G^{(2,\s)}_k\;,}\Eqa(hh)$$
where, for $h<0$, let $\sum_{k=0}^h\defi -\sum_{k=h}^0$.
\*
\alemma(hh){
There exist $\e>0$, and  $c,c_0,c_1>0$ such that 
there exists a (unique) solution to \equ(hh) in the space
$\MMM_\th$, for $c_0$ and $c_1$ the constants in \equ(ind3),
and $c$ the constant in \equ(l1). Furthermore, 
such a solution is analytic in $\l$.}
\*
\proof. The equation
makes sense since $\norm{y}_\th\le 1$ and $|\l|\leq \e$,
together to the first of \equ(ind3), for $\e$ small enough,
imply \equ(ind1) and \equ(ind2), and hence Lemma \lm(abgf).

The existence of a solution
is consequence of the fact that $T$ is a contraction
from $\MMM_\th$ into itself.
Indeed, because of the following arguments,
if $y\in \MMM_\th$, then $Ty\in\MMM_{\th}$. 
\elenco{
\art
By inductive hypothesis and convergence of the graph
expansion, there exists a constant 
$c_2\geq 0$, such that $|z_{h-1}|\leq c_2\e^2$;
hence, for $\e$ small enough and $c_0\geq 2c_2$,
it holds the statement in \equ(ind3) 
regarding the field strength flow.
\art
For the density strengths,
by definitions \equ(ga3), it is more 
convenient to define two new  strengths,
$\z^{\rm (u)}_k\defi (\z^{(2,+)}_k+ \z^{(2,-)}_k)/ 2$ and
$\z^{\rm (d)}_k\defi (\z^{(2,+)}_k- \z^{(2,-)}_k)/ 2$,
so that the their flows are given by  
$$\eqalign{
&{\z^{\rm (u)}_{h-1}\over \z^{\rm (u)}_h}
 ={Z_h\over Z_{h-1}}
 \left(1+ z^{(2)}_{h-1}+\D z^{(2,+)}_{h-1}+\D z^{(2,-)}_{h-1}\right)\;,\cr
&{\z^{\rm (d)}_{h-1}\over Z^{\rm (d)}_h}
 ={Z_h\over Z_{h-1}}
 \left(1+ z^{(2)}_{h-1}+\D z^{(2,+)}_{h-1}-\D z^{(2,-)}_{h-1}\right)\;.}
$$
Then, an argument similar to the one of the previous item
proves statement in \equ(ind1) regarding the density strengths.
\art
For the flow of the effective coupling,
the argument is more involved: it is based on a cancellation,
the vanishing of the Beta function, 
which exactly holds only in the limit of removed cutoff.
Let $\b^{\rm(T)}_k(\l_k,\ldots,\l_N)$ be
the sum of the graphs of 
$\b_k^{(\l)}$ which are made only with 
diagonal propagators
$\{g^{\rm(E1),k}_{\o,\o}\}_k$ and interactions $\VV$; then, 
setting all the arguments equal, let 
$\b^{\rm(T)}_k(\l_k)\defi \b^{\rm(T)}_k(\l_k,\ldots,\l_k)$.
As proved in \secc(VVBBFF),
there exists a constant $c_2\geq 0$ such that
$| \b^{\rm(T)}_k(\l_k)|\leq c_2\e^2 \g^{-\th(N-k)}$. 
Accordingly, it is convenient to 
expand each coupling $\l_m$ in
the  function  
$\b^{\rm(T)}_k(\l_k,\ldots,\l_N)$ 
as \hbox{$\l_m=\l_k+(\l_m-\l_k)$}, so that 
the following decomposition of the whole 
Beta function holds:
$$\eqalign{
\b^{(\l)}_k=
&\b^{\rm(T)}_k(\l_k)
+\sum_{m=k}^N 
\b^{({\rm T},\l)}_{k,m}(\l_m-\l_k) 
+\sum_{m=k}^N \b^{({\rm R},\l)}_{k,m}
+\sum_{a=\d,\n}\sum_{m=k}^N \b^{(\l,a)}_{k,m}a_m }\;,$$
where $\b^{({\rm T},\l)}_{k,m}$ is the sum of the graphs in 
$\b^{\rm(T)}_k(\l_k,\ldots,\l_N)$, with the replacement 
of the all the couplings $\l_n:k\leq n< m$ with $\l_k$,
and a coupling $\l_m-\l_k$ on scale $m$ put apart from it; 
$\b^{({\rm R},\l,)}_{k,m}$ is the sum of the graphs made with 
interactions $\VV$ and  with at least one propagator 
$g^{({\rm R1,m})}_{\o,\o}$ on scale $m$; 
$\b^{(\l,a)}_{k,m}$ is the sum of
the graphs with at least one coupling $a_m$ on scale $m$
and only diagonal propagators $g^{({\rm E1,m})}_{\o,\o}$
-- if a graph falls in more than one category 
the assignment is  arbitrary.  
By the convergence of power expansion in $\l$,
as stated in \lm(abgf),
and  the short memory property
of the tree ordering, the following bounds holds
for the same constant $c_2$ -- if it is chosen large enough:
$$
 |\b^{({\rm T},\l)}_{k,m}|\leq c_2\e\g^{-\th(m-k)}\;,
 \qquad
 |\b^{({\rm R},\l)}_{k,m}|\leq \g^{-(3/4)(N-m)}c_2\e^2\g^{-\th(m-k)}\;,$$
$$|\b^{(\l,a)}_{k,m}|\leq c_2\e^2\g^{-\th(m-k)}\;.$$
It is straightforward to conclude 
that, to obtain \equ(ind3) and
\equ(l1), as far as the flow $\{\l_h\}_h$ is regarded,
$c_1$ and $c$  have to be chosen 
$c\geq 4c_2(1-\g^{-(\th/2)})^{-1}$
and $c_1\geq c(1-\g^{-(\th/2)})^{-1}$.
\art
Similarly, it is possible to decompose the Beta function
for the couplings $a=\d,\n$:
$$\b_k^{(a)}
\defi
 \sum_{m=k}^N\b_{k,m}^{(a,{\rm R})}
 +
 \sum_{b=\n,\d}
 \sum_{m=k}^N\b_{k,m}^{(a,b)} b_m\;,$$
where $\b_{k,m}^{(a,{\rm R})}$ contains all the graphs made
only with interactions $\VV$ and with at least 
one diagonal 
propagator $g^{({\rm R1,m})}_{\o,\o}$ on scale $m$; whereas
$\b_{k,m}^{(a,b)}$ is made with all the graphs with an interaction $b$ 
on scale $m$ and only diagonal propagators 
$g^{({\rm E1,m})}_{\o,\o}$ -- in ambiguous cases
the assignment is arbitrary.
Again, by convergence of the power expansion in $\l$,
and by the short memory property of 
the tree ordering, 
$$|\b^{(a,2)}_{k,m}|\leq \g^{-\th(N-m)}c_2\e^2\g^{-\th(m-k)}\;,
 \qquad|\b^{(a,b)}_{k,m}|\leq c_2\e^2\g^{-\th(m-k)}\;;$$
and since for $\e$ small enough $5c_2\e^2\lft(1-\g^{-(\th/2)}\rgt)^{-1}\le 2\e$,
then \equ(ind3) holds also for what concerns $\{\d_k\}_k$ and  $\{\n_k\}_k$.
}%

\0Therefore $Ty$ is in $\MMM_\th$ for $\e$ small enough;
and, by Lemma \lm(abgf), if $y$ is analytic 
in \hbox{$\l:|\l|\le \e$}, then also $Ty$ does. 
The next step is to prove that, taken any two 
$y,y'\in \MMM_\th$, it holds $\norm{Ty-Ty'}_\th\leq \r\norm{y-y'}_\th$,
for a constant $\r<1$.
\elenco{
\art
The variation of the Beta function $\b^{(\l)}$
due to the variation of the $y$
is given by:
$$ \eqalign{
 \b^{(\l)}_k-{\b'}^{(\l)}_k
 =&
 \sum_{m=k}^N\D\b^{(\l)}_{k,m}(\l_m-\l'_m)
 +\sum_{m=k}^N\b^{({\rm T},\l)}_{k,m}\big[(\l_m-\l_k)-(\l'_m-\l'_k)\big]\cr
&+\sum_{m=k}^N \D\b^{(\l,Z)}_{k,m}\left({Z_{m-1}\over Z_m}-{Z'_{m-1}\over Z'_m}\right)
 +\sum_{a=\d,\n}\sum_{m=k}^N
 \D\b^{(\l,a)}_{k,m}(a_m-{a'}_m)\;,}$$
where $\D\b^{(\l)}_{k,m}$ corresponds to a variation 
of the coupling $\l_m$ in one of the two previously
defined $\b^{({\rm T})}_{k,m}$ and 
$\b^{({\rm T},\l)}_{k,m}$; 
the term  $\D\b^{(\l,Z)}_{k,m}$
is due to a variation one factor $Z_{m-1}/Z_m$;
and $\D\b^{(\l,a)}_{k,m}$ to a variation of $a_m$.
Since the power series of the 
variation has the same domain of convergence 
of the Beta function itself, and since 
the vanishing of the Beta function holds 
for each order of the power series, 
using also the short memory property,
the following bounds holds
for a suitable constant $c_3\ge 0$:
$$|\D\b^{(\l)}_{k,m}|
\leq \g^{-(\th/2)(N-k)}c_3\e\g^{-\th(m-k)}\;,
\qquad
|\D\b^{(\l,a)}_{k,m}|
\leq c_3\e^2\g^{-\th(m-k)}\;,$$
$$
|\D\b^{(\l,Z)}_{k,m}|
\leq \g^{-(\th/2)(N-k)}c_3\e^2\g^{-\th(m-k)}\;,$$
where the factors $\g^{-(\th/2)(N-k)}$ in the 
first and third bound come from the 
bound on the Beta function on its own, which has been made 
previously.
\art
The variation of the Beta functions $\{\b^{(a)}\}_{a=\n,\d}$
is given by:
$$\eqalign{
 \b_k^{(a)}-{\b'}_k^{(a)}
 \defi &
 \sum_{m=k}^N\D\b_{k,m}^{(a,\l)}(\l_m-\l'_m)
+\sum_{m=k}^N \D\b^{(a,Z)}_{k,m}\left({Z_{m-1}\over Z_m}-{Z'_{m-1}\over Z'_m}\right)\cr 
&+\sum_{b=\n,\d}\sum_{m=k}^N
 \D\b_{h,m}^{(a,b)}\left(b_m-{b'}_m\right)\;,}$$
where $\D\b_{k,m}^{(a,\l)}$ is due to 
the variation of the coupling $\l_m$; 
$\D\b_{k,m}^{(a,Z)}$ to the variation 
of the ratio $Z_{m-1}/Z_m$; 
$\D\b_{k,m}^{(a,\l)}$ to the variation of the coupling
$\b_m$. And they  holds the bounds: 
$$|\D\b^{(a,\l)}_{k,m}|\leq c_3\e \g^{-\th(m-k)}\g^{-(\th/2)(N-k)}\;,\qquad
|\D\b^{(a,b)}_{h,k}|\leq c_3\e^2\g^{-\th(m-k)}\;,$$
$$|\D\b^{(a,Z)}_{k,m}|
\leq \g^{-(\th/2)(N-k)}c_3\e^2\g^{-\th(m-k)}\;.$$
\art
The variation of the Gamma function of the field strength is
$$\eqalign{
 \G_k-{\G'}_k
 \defi &
 \sum_{m=k}^N\D \G_{k,m}^{(\l)}(\l_m-\l'_m)
+\sum_{m=k}^N \D\G^{(Z)}_{k,m}\left({Z_{m-1}\over Z_{m}}-{Z'_{m-1}\over Z'_{m}}\right)\cr 
&+\sum_{b=\n,\d}\sum_{m=k}^N
 \D\G_{h,m}^{(b)}\left(b_m-{b'}_m\right)\;.}$$
with clear justification of the various addends.
Now, by the short memory property,
$$|\D\G^{(\l)}_{k,m}|\leq c_3\e \g^{-\th(m-k)}\;,
\qquad
|\D\G^{(Z)}_{k,m}| 
\leq c_3\e^2 \g^{-\th(m-k)}\;,$$
$$|\D\G^{(b)}_{k,m}| 
\leq c_3\e \g^{-\th(m-k)}\;.$$
\art
Similar arguments hold for the 
field strengths.}
\0By such bounds, the operator $T$ is a contraction with rate
$\r\defi e^2(c_3c_1+2c_2c_1+c_3c_0+2c_3)$: 
for $\e$ small enough, $\r<1$.
The proof of the Lemma
is obtained by the
fixed point theorem with analytic parameterization.\hfill\qed\hskip1em\null

\*
Once the flows $y$ has been found, it is possible 
to consider the flow for the mass:
$$ \log_\g(\bar Z_h)=\sum_{k=0}^h\bar \G_k\;,\Eqa(mm)$$
restricted to the range $0\leq k\le N$. In the remaining 
scales, $h^*\le k<0$, in fact, the flow is determined 
directly, and not by an equation; and since $h^*$,
in its turn, depends on the flow, it is more 
convenient to exclude it from the fixed point theorem. 

As for the other flow, it is defined the linear space 
$\bar{{\MMM}}$ of the sequences 
$$x\defi\lft\{\log_\g(\bar Z_k)\in \RRR : 0\leq k\leq N\rgt\}$$
such that
\elenco{
\item{\bf i.}
the initial datum is as in \equ(indat). 
}
\0Furthermore, such a space is endowed with the 
norm $\norm{x}$, the lowest real number such that
\elenco{
\item{\bf ii.}
for the same constant $c_0$ in \equ(ind3),
and for $0\leq k\le N$,
$$\lft|\log_\g(\bar Z_{k-1}/\bar Z_k)\rgt|\leq 2c_0\e^2\norm{x}\;.\Eqa(ind4)$$
}
The equation $x=Tx$, which is defined to be \equ(mm), 
can be solved in $\bar{{\MMM}}_\th$,
the subspace of $\bar{{\MMM}}$ of the sequences $x$ with 
$\norm{x}\le 1$, with the fixed point theorem.

\*
\alemma(mm){There exists $\e>0$ and the positive constant $c_0$
such that there exists a (unique) solution of 
\equ(mm) in the space $\bar{{\MMM}}_\th$,
for $c_0$ the constant in \equ(ind4).}
\elenco{
\art
If $x\in \bar{{\MMM}}_\th$, then also 
$Tx\in \bar{{\MMM}}_\th$ 
by the following argument.
The local part 
$s_{h-1}$ is the sum  of the graphs
with one antidiagonal propagator $g^{({\rm E1},k)}_{\o,-\o}$
or $g^{({\rm R1},k)}_{\o,-\o}$.
As consequence of the convergence of the graphs expansion 
and of the dimensional bounds of $s_{h-1}$,
calling $s_{h-1,k}$ the sum of all the graphs of  $s_{h-1}$
with $g^{({\rm E1},k)}_{\o,-\o}$ or $g^{({\rm R1},k)}_{\o,-\o}$ on scale $k$
and
divided by $\m_k/ \k\g^k$, 
$$s_{h-1}\defi\sum_{k=h}^N s_{h-1,k}{\m_k\over \k\g^k}\;,
\qquad{\rm with\ } |s_{h-1,k}|\leq \g^{h-1}c_2\e \;.$$
By \equ(ind4),
for $\e$ small enough, 
it holds $(\m_k/\m_h)\leq \g^{2c_0\e(k-h)}<\g^{(1/2)(k-h)}$,
and hence $m_{h-1}=(s_{h-1}/\m_h)\leq c_1(1-\g^{-(1/2)})^{-1}\e$:
since by \equ(ind3) $\g^{-c_0\e^2}(Z_{h-1}/Z_{h})\leq \g^{c_0\e^2}$ and
\hbox{$\log_\g(1+m_{h-1})\leq \left|m_{h-1} \ln(\g)
\int_0^1\!\der t(1+tm_{h-1})^{-1}\right|$},
it is straightforward to obtain that 
\hbox{$\g^{-2c_0\e}\leq (\m_{h-1}/\m_h)\leq\g^{2c_0\e}$}
for $\e$ small enough and $c_0\geq 2c_2(1-\g^{-(1/2)})^{-1}$.
\art
If $x,x'\in \bar{{\MMM}}_\th$, then $\norm{Tx-Tx'}\le \r\norm{x-x'}$,
for $\r<1$. Indeed, under variation of the mass flow,
-- having fixed all the other flows -- 
$$\bar\G_k-{\bar\G'}_k
 \defi 
\sum_{m=k}^N \D\bar\G^{(\m)}_{k,m}\left({\m_m\over \m_k}-{\m'_m\over \m'_k}\right)\;.$$
Now, by the short memory property,
and by \equ(ind4),
$$|\D\bar \G^{(\m)}_{k,m}| 
\leq c_3\e \g^{-\th(m-k)}\;,
 \qquad
\left|{\m_m\over \m_k}-{\m'_m\over \m'_k}\right|\leq 
c_4\g^{(\th/2)(m-k)}
\sup_{n\geq 0}\left|\bar\G_{n}-\bar\G'_{n}\right|\;;$$
-- indeed, $\left|(\m_m/ \m_k)-(\m'_m/ \m'_k)\right|\leq
\max\big\{(\m_m/ \m_k),(\m'_m/ \m'_k)\big\}\ln(\g)\sum_{n=k}^m 
\left|\bar\G_{n}-\bar\G'_{n}\right|$,
which, by \equ(ind4), is less or
equal to ${(4/\th)}\ln(\g)
\g^{\big(2c_0\e+(\th/4)\big)(m-k)}\sup_{n}\left|\bar\G_{n}
-\bar\G'_{n}\right|$. 
Then the assertion follows enlarging $c_0$ chosen for the 
field strength to $c_0\geq c_3c_4(1-\g^{-(\th/2)})^{-1}$.}

This proves the Lemma.
\hfill\qed\hskip1em\null
\*
\asub(FP){Further properties of the Gamma functions.}
In order to complete the proof of the Theorem \thm(FC), 
it is left to prove the 
existence of the critical indexes $\h_\l$, $\h^{(2)}_\l$ and 
$\bar\h_\l$, which only depends on the 
choice of $\l$ and on the graphs that 
can be obtained using the diagonal propagator 
$\{g^{({\rm E1}, h)}_{\o,\o}\}$
and the interaction $\VV$, 
and not from the mass, or from 
the regularization of the model.
Indeed, let it be inductively supposed that 
there exists a positive constant $c_2$
such that, for any $k:h\leq k\leq N$, 
$${Z_{k-1}\over Z_{k}}=\g^{\G^{(0)}_k
+ \G^{(1)}_k}\;,
\qquad{\rm with}\
|\G^{(1)}_k|\leq c_4 \e^2 \g^{-(\th/2)(N-k)}\;,
\Eqa(sindz)$$
while 
$\G^{(0)}_k$ is given in terms of graphs 
made only with the 
diagonal propagator $\{g^{({\rm E1}, h)}_{\o,\o}\}$
and the interaction $\l\VV$, 
and bounded, $|\G^{(0)}_h|\leq c_2 \e^2$. 
Then, 
let the following decomposition be considered:
$$\eqalign{
z_{h-1}=
&z^{(0)}_{h-1}
+\sum_{k=h}^N  \D z^{(\l)}_{h-1,k}
\D\l_k
+\sum_{k=h}^N  \D z^{(Z)}_{h-1,k}
\left({Z_{k-1}\over Z_k}-\g^{\G^{(0)}_k}\right)\cr
&
+
\sum_{k=h}^N \D z^{(2)}_{h-1,k}
+\sum_{a=\d,\n}\sum_{k=h}^N \D z^{(a)}_{h-1,k}a_k\;,}$$
where $z^{(0)}_{h-1}$ is the sum of the graphs contributing to
$z_{h-1}$ which are made only with propagators 
$\{g^{({\rm E1},k)}\}$ and interactions $\VV$, 
with all the coupling $\{\l_k\}_k$ replaced by  
coupling $\l$ and all the ratios $(Z_{k-1}/ Z_k)$
replaced by $\g^{\G_k^{(0)}}$; 
$\D z^{(\l)}_{h-1,k}$ 
is due to the replacement of $\l_k$ with $\D\l_k$;
$\D z^{(Z)}_{h-1,k}\left[(Z_{k-1}/Z_k)-\g^{\G^{(0)}_k}\right]$
is the sum of the same graphs, but with 
at least a factor  $(Z_{k-1}/Z_k)-\g^{\G^{(0)}_k}$
in place of the ratio $(Z_{k-1}/Z_k)$;
$\D z^{(2)}_{h-1}$ is the sum of the graphs  
which do not contain interactions $\NN$ or $\DD$,
and have a propagator $g^{({\rm R1},k)}$ on scale $k$;
$\D z^{(a)}_{h-1,k}$ is the sum of the graphs 
 with an interaction $a=\d,\n$
on scale $k$ -- whenever a graph
falls in more than one of the above categories, the
assignment is made in arbitrary way.
Because of the following bound 
$$|z^{(0)}_{h-1}|\leq c_3\e^2\;,\qquad
|\D z^{(\l)}_{h-1,k}|\leq c_3\e^2\g^{-\th(k-h+1)}\;,\qquad
|\D z^{(Z)}_{h-1,k}|\leq c_3\e^2\g^{-\th(k-h+1)}\;,$$
$$|\D z^{(2)}_{h-1,k}|\leq \g^{-\th(N-k)}c_3\e^2\g^{-\th(k-h+1)}\;,\qquad
|\D z^{(a)}_{h-1,k}|\leq c_3\e^2\g^{-\th(k-h+1)}\;,$$
$$
|\D\l_k|\leq c_1 \e^2\g^{-(\th/2)(N-k)}\;,\qquad
|(Z_{k-1}/Z_k)-\g^{\G_k^{(0)}}|\leq 2c_4 \e^2\g^{-(\th/2)(N-k)}\;,$$
$$|a_{k}|\leq 2\e\g^{-(\th/2)(N-k)}\;,$$
the property \equ(sindz) follows straightforwardly
for $c_4\geq 5c_3(1+c_1)(1-\g^{-(\th/2)})^{-1}$
and 
$$\G^{(0)}_h\defi\log_\g\left(1+z^{(0)}_{h-1}\right)\;.$$
By construction,  $\G^{(0)}_{h}$ 
is the sum of scaling invariant graphs: 
again using  the fixed point theorem theorem
with analytic parameterization, it is possible to
prove the existence of $\h_\l$, limit for $N\to \io$ 
of $\G^{(0)}_{h}$, analytic in $\l$ and such that
there exists a constant $c_5$
for which
$\left|\G^{(0)}_{h}-\h_\l\right|\leq c_5\e^2\g^{-(\th/2)(N-h)}$,
and then the statements in \equ(z1) referring to
the field strength flow holds for
$c_2\geq (c_4+c_5)(1-\g^{-(\th/2)})^{-1}$.
\*

For the Gamma function of the mass a similar argument 
can be applied. Let it be inductively supposed 
for any $k:h\leq k\leq N$ that 
$${\m_k\over \m_{k+1}}=\g^{\bar \G^{(0)}_{k+1} + \bar \G^{(1)}_{k+1}}\;,
\qquad{\rm with}\
|\bar \G^{(0)}_k|\leq c_2 \e\;, 
\quad |\bar \G^{(1)}_k|\leq c_4 \e\g^{-(\th/2)(N-k)}\;,
\Eqa(sind)$$
and $\bar \G^{(0)}_k$ only made with  the propagator 
$\{g^{({\rm E1},k)}\}_k$  and interactions $\l\VV$. It follows that $(\m_k/ \m_h)
=\g^{-\sum_{m=h}^{k-1}\bar \G^{(0)}_m} + \bar\D_{k,h}$
with $|\bar\D_{k,h}|\leq c_6\e\g^{-(\th/2)(N-k)}$, for 
$c_6\geq 2c_42(1-\g^{-\th})^{-1}$ and $\e$ small enough.
Then, with a decomposition similar to the 
case of the field strength:
$$\eqalign{
m_{h-1}=
&m^{(0)}_{h-1}
+\sum_{k=h}^N \D m^{(\l)}_{h-1,k}(\l_k-\l)
+\sum_{k=h}^N \D m^{(Z)}_{h-1,k}\left({Z_{k-1}\over Z_k}-\g^{\G_k^{(0)}}\right)\cr
&
+\sum_{k=h}^N \D m^{(1)}_{h-1,k}\bar\D_{k,h}
+\sum_{k=h}^N m^{(2)}_{h-1,k}+\sum_{a=\d,\n}\sum_{k=h}^N m^{(a)}_{h-1,k}a_k\;;}$$
where  
$m^{(0)}_{h-1}$ is the sum of the graphs 
made only with interactions $\l\VV$, 
all the ratios $\{Z_{m-1}/Z_m\}$ replaced with $\g^{\G^{(0)}_k}$,
all the ratios $\{\m_m/\m_h\}$ replaced with $\g^{-\sum_{n=h}^{m-1}\G^{(0)}_n}$
and all diagonal propagators $g^{({\rm E1}, k)}_{\o,\o}$  
on scale $k\geq h$, except one, which is  antidiagonal, 
$g^{({\rm E1}, k)}_{\o,-\o}$;   
$\D m^{(\l)}_{h-1,k}$
is the sum of the graphs of $m_{h-1}$ 
with all the couplings $\{\l_m\}_m$
replaced, for $m < k$, by $\l$, and at 
a coupling $\l_k$ neglected;
$\D m^{(1)}_{h-1,k}$
is the sum of the graphs in which 
one ratio $\m_k/\m_h$ neglected. 
Then
equation \equ(sind) holds true also in
the case $k=h-1$ for 
$c_4$ large enough and 
$$\bar \G^{(0)}_{h}\defi \G^{(0)}_h+
\log_\g\left(1+ m^{(0)}_{h-1}\right)\;.$$
Finally, since $\bar \G^{(0)}_k$ is given by scale invariant graphs,
using the fixed point theorem with analytic parameterization, 
it would be possible 
to prove the existence of an $\bar\h_\l$ 
analytic in $\l$ and  such that 
$\left|\bar \G^{(0)}_k-\bar\h_\l
\right|\leq c_5\e\g^{-\th(N-k)}$ and the statements about the
mass flow in \equ(z1) holds
for  \hbox{$c_2\geq (c_5+c_4)(1-\g^{-\th})^{-1}$}.

Finally, with similar arguments, it
is straightforward to  prove \equ(zz1).
\appendix(BD,Proof of Lemma {\lm(L2)})

By definition
$$\eqalign{U^{(i,j)}_\o(k,p)
\defi&
 C_\o(k,p)\hg_\o^{(i)}(k)\hg_\o^{(j)}(p)\cr
=&
 f_i(k)\left(1-\c^{-1}_{h,N}(k)\right)
 {f_j(p)\over D_\o(p)}
 -
 f_j(p)\left(1-\c^{-1}_{h,N}(k)\right)
 {f_i(k)\over D_\o(k)}\;.}$$
Setting:
$$
 u_N(k)\defi
 \left\{
 \matrix{
 0&{\rm \ for\  }|k|<\k\g^N\cr
 1-f_N(k)
 &{\rm \ for\  }|k|\geq\k\g^N\;,\cr}
 \right.$$
$$u_h(k)\defi
 \left\{
 \matrix{
 0&{\rm \ for\  }|k|\geq\k\g^h\cr
 1-f_h(k)
 &{\rm \ for\  }|k|<\k\g^h\;,\cr}
 \right.$$
the expansion of $U^{(i,j)}_\o(k,p)$ in terms of
$\Big\{S^{(i,j)}_{\o,\s}(k,p)\Big\}_{\s=\pm}$ 
can be explicitly given in each of the possible case.
\elenco{
\art 
For $i=j=N$,
$$\eqalign{
&U^{(N,N)}_\o(k,p)
 = {u_N(p)f_N(k)\over D_\o(k)}-
 {u_N(k)f_N(p)\over D_\o(p)}\cr
&=
 \sum_{\s=\pm} D_\s(p-k)
 \left[
 \d_{\o,\s}{u_N(k)f_N(p)\over D_\o(p)D_\o(k)}
 +
 {f_N(p)\over D_\o(k)}
 \int_0^1 d\t\  \big(\partial_\s u_N\big)\big(p+\t(k-p)\big)\right.\cr
&\phantom{************************}
 \left.
 -{u_N(p)\over D_\o(k)}
 \int_0^1 d\t\  \big(\partial_\s f_N\big)\big(k+\t(p-k)\big)\right]\cr
&\defi
 \sum_{\s=\pm} D_\s(p-k) S^{(N,N)}_{\o,\s}(k,p)\;.}$$
\art
For $i=N$ and $h<j<N$:
$$U^{(N,j)}_\o(k,p)
 =-{u_N(k)f_j(p)\over D_\o(p)}\;.$$
Being that $u_N(p)f_j(p)\equiv 0$, it holds
$$\eqalign{
 U_\o^{(N,j)}(k,p)
&=\sum_\s D_\s(p-k){f_j(p)\over D_\o(p)}
 \int_0^1d\t\
 \big(\partial_s u_N\big)\big(p+\t(k-p)\big)\cr
&\defi
 \sum_\s D_\s(p-k)S^{(N,j)}_{\o,\s}(k,p)\;.}$$
\art
For $i=N$ and $j=h$
$$U^{(N,h)}_\o(k,p)
 =-{u_N(k)f_h(p)\over D_\o(p)}
 +
 {u_h(p)f_N(k)\over D_\o(k)}\;.$$
The first addend was already studied in point 2. For the second,
the expansion is similar to the first since $u_h(k)f_n(k)\equiv
0$; finally:
$$\eqalign{
 U_\o^{(N,j)}(k,p)
&=\sum_\s D_\s(p-k)
 \left[
 {f_j(p)\over D_\o(p)}
 \int_0^1d\t\
 \big(\partial_s u_N\big)\big(p+\t(k-p)\big)\right.\cr
&\phantom{***********}
 -
 \left.{f_N(k)\over D_\o(k)}
 \int_0^1d\t\
 \big(\partial_s u_N\big)\big(k+\t(p-k)\big)\right]\cr
&\defi
 \sum_\s D_\s(p-k)S^{(N,h)}_{\o,\s}(k,p)\;.}$$
\art
For $h<i<N$ and $j=h$:
$$
 U^{(i,h)}_\o(k,p)
 ={u_h(p)f_i(k)\over D_\o(k)}\;.$$
Being that $u_h(k)f_i(k)\equiv 0$ it holds
$$\eqalign{
 U_\o^{(N,j)}(k,p)
&=\sum_\s D_\s(p-k){f_i(p)\over D_\o(p)}
 \int_0^1d\t\
 \big(\partial_s u_h\big)\big(k+\t(p-k)\big)\cr
&\defi
 \sum_\s D_\s(p-k)S^{(N,j)}_{\o,\s}(k,p)\;.}$$
}
For $i=j=h$, expanding like in point 1
$$\eqalign{
 U^{(h,h)}_\o(k,p)
&=
 \sum_{\s=\pm} D_\s(p-k)
 \left[
 \d_{\o,\s}{u_h(k)f_h(p)\over D_\o(p)D_\o(k)}\right.\cr
&\phantom{***************}
 +
 {f_h(p)\over D_\o(k)}
 \int_0^1 d\t\  \big(\partial_\s u_h\big)\big(p+\t(k-p)\big)\cr
&\phantom{***************}
 \left.
 -{u_h(p)\over D_\o(k)}
 \int_0^1 d\t\  \big(\partial_\s f_h\big)\big(k+\t(p-k)\big)\right]\cr
&\defi
 \sum_{\s=\pm} D_\s(p-k) S^{(h,h)}_{\o,\s}(k,p)\;.}$$
\*

\0By inspection in each case, since for $n=N,h$ it
holds $\Big|\big(\partial_\s
f_n\big)(k)\Big|, \Big|\big(\partial_\s u_n\big)(k)\Big|\leq
c\g^{-n}$, it is simply to get the following bound
$$\Big|\big(\partial_k^{s_i}\partial_p^{s_j} S^{(i,j)}_{\o,\s}\big)(k,p)\Big|
\leq  c\g^{-i(1+s_i)-j(1+s_j)}\;.$$
\appendix(PL,Proof of Theorems {\thm(TWTI)} and {\thm(TT)})

It is natural to introduce the Beta functions
also for the flow of the counterterms 
$\{\n^{(\s)}_N\}_{\s=\pm}$, and the coupling 
$\wt\l^{\m}_{N-1}$, generated
in the multiscale integration of the generating functional
$\WW_{\TT,\m}^{(h)}$:
$$\eqalign{
&\n^{(\s)}_{j-1}-\n^{(\s)}_{j}
 =
 \b^{(\s)}_{j}\left(\l_j,\n_j;\dots,\l_N,\n_N\right)\;,\cr
&\wt\l_{j-1}^{(\m)}-\wt\l_{j}^{(\m)}
 =\wt\b_j^{(\m)}\left(\l_j,\n_j,\wt\l_{j}^{(\m)},\wt z_{j}^{(\m)};
  \ldots,\l_N,\n_N\right)\;.}$$
It has to be remarked that the above Beta function are 
defined for the generating functionals 
$\WW_\AAA^{(h)}$ and  $\WW_{\TT,\m}^{(h)}$ with
infrared cutoff $h=-\io$: this is not restrictive, 
since, by inspection of the 
properties of the kernel $U^{(i,j)}_\o$, the
flows obtained have the property that 
$\wt \l^{(\m)}_k$  and
$\n^{(\s)}_k$,
are, in the range $k: h+1\le k\le N$, 
{\it exactly equal to the effective coupling 
of such  generating functionals with infrared cutoff
on scale $h$ finite.}  
\*

\proof{\bf of Theorem \thm(TWTI).}
Let $\BB_\th$ be the Banach space of all
the finite sequences of vectors
$x\defi\left\{(\n^{(+)}_j,\n^{(-)}_j) :j\leq N\right\}$ s.t.
$$
\norm{x}_{\th}\defi \max_{\s=\pm, j\le N} 
|\n^{(\s)}_j|\g^{(\th/ 2)(N-j)}\leq c_1\e\;.
$$
In this space, it is possible to find a solution for the 
fixed point equation $x=Tx$, which explicitly reads
$$\n^{(\s)}_j=-\sum_{m=-\io}^j \b_m^{(\s)}(x)\Eqa(fpe)$$
(where the argument of the Beta function has been abridged);
such a solution gives a choice of $\{\n^{(\s)}_N\}_{\s\pm}$,
such that  their flows $\{\n^{(\s)}_N\}^{\s\pm}_{h+1\le j\le N}$
have the required decay property. 
Indeed, given  $x,x'\in \BB_\th$:
$$\b_m^{(\s)}(x)
 \defi
 \b_{m,N}^{(\s,0)}
 +
 \sum_{n=m}^N\b_{m,n}^{(\s)}\n^{(\s)}_n\;,\qquad
 \b_m^{(\s)}(x)-\b_m^{(\s)}(x')
 \defi
 \sum_{n=m}^N\b_{m,n}^{(\s)}\left(\n^{(\s)}_n-{\n'}^{(\s)}_n\right)\;,$$
where $\b_{m,N}^{(\s,0)}$ is the localization of the 
sum of the graphs made with no interaction 
$\{\n^{(\s)}_k\AAA_\s\}_k$ and one
propagator connecting the interaction $\AAA_0$ 
contracted on scale  $N$; whereas
$\b^{(\s)}_{m,n}$ is the localization of the sum 
of the graphs made with an interaction $\n^{(\s)}_n\AAA_\s$,
and deprived of  $\n^{(\s)}_n$.
The following bounds hold:
$$
 \left|\b^{(\s,0)}_{m,N}(x)\right|
 \leq c_2\e\g^{-\th(N-m)},
 \qquad
 \left|\b^{(\s)}_{m,n}\right|
 \leq c_2\e\g^{-\th(n-m)},
\Eq(d11)$$
Therefore, if $x\in \BB_\th$, then also $Tx\in \BB_\th$
for $\e$ small enough and if $c_1\ge 2c_2(1-\g^{-(\th/2)})^{-1}$;
and $\norm{x-x'}\le C\e \norm{Tx-Tx'}$ for $C>c_2(1-\g^{-(\th/2)})^{-2}$, so that, 
for $\e $ small enough, $T$ is a contraction in a Banach space; therefore there exists $x\in \BB_\th$,
solution of the fixed point equation, with analytic parameterization 
in $\l:|\l|\le \e$.

Finally, since all the graphs contributing to $\b^{(\s)}_m$,
are scale invariant, by \equ(fpe) for $j=N$ it is easy to 
realize that $\{\n^{(\s)}_N\}_{\s=\pm}$ are constant 
in the scale of the cutoff, $N$: hence 
$$\n^{(\s)}_N=\n^{(\s)}_{N+1}=\n^{(\s)}\;.$$
{The proof of the theorem is completed.\hfill\qed\hskip1em}

\*
\proof {\bf of Theorem \thm(TT).}
The strategy is based on the fixed point theorem  as the previous proof. Let 
$x\defi\left\{\left(\wt\l_j^{(+)}-\a_N^{(+)}\l_j,
\wt\l_j^{(-)}-\a_N^{(-)}\l_j\right):j\le N\right\}$
(with $\l_N^{(\m)}=0$):
the fixed point equation to be solved in $\BB_{\th/2}$
is $x=Tx$, which explicitly reads:
$$ \wt\l_j^{(\m)}-\a_N^{(\m)}\l_j
=-\sum_{m=-\io}^j\left(\wt\b^{(\m)}_m-\a^{(\m)}_N\b_m\right)\;.$$
Given $\a^{(\m)}_N$ and ${\a'}^{(\m)}_N$ such that 
both $\wt\l_j^{(\m)}-\a_N^{(\m)}\l_j$ and $\wt\l_j^{(\m)}-{\a'}_N^{(\m)}\l_j$
are in $\BB_{\th/2}$, it holds:
$$\eqalign{
 \wt\b^{(\m)}_m-\a^{(\m)}_N\b_m
\defi&
 \wt\b^{(\m,o)}_{m,N}-\a^{(\m)}_N\l_N\b^{(\l)}_{m,N} 
 +\sum_{\s=\pm}\sum_{n=m}^N\wt\b^{(\m,\s)}_{m,n}\n^{(\s)}_n{Z_N\over Z_n}\cr
&+\sum_{n=m}^{N-1}\b_{m,n}\left(\wt\l_n^{(\m)}-\a_N^{(\m)}\l_n\right)\;;}$$
while
$$\eqalign{
 \left({\a'}^{(\m)}_N-\a^{(\m)}_N\right)\b_m
\defi&
 \left({\a'}^{(\m)}_N-\a^{(\m)}_N\right)\l_N\b^{(\l)}_{m,N} 
+\sum_{n=m}^{N-1}\b_{m,n}\left({\a'}_N^{(\m)}-\a_N^{(\m)}\right)\l_n\;;}$$
where $\wt\b^{(\m,o)}_{m,N}$ is the sum of the graphs 
made with an interaction $\AAA_o$, contracted on scale $N$;
$\wt\b^{(\m,\s)}_{m,n}$ is the sum of the graphs 
with an interaction $\TT_\s^{(\m)}$ on scale $n$,
deprived of the coupling $\n^{(\s)}_n(Z_N/ Z_n)$;
$b^{(\l)}_{m,N} $ is the sum of the graphs contributing
to he flow of $\a^{(\m)}_N\l_m$ which have an interaction $\BB^{(3)}$
on scale $N$, deprived of the coupling $\a^{(\m)}_N\l_N$;
$\b_{m,n}$ is the sum of the graphs contributing
to the flow of $\a^{(\m)}_N\l_m$ with an interaction $\BB^{(3)}$
on scale $n$, deprived of the coupling $\a^{(\m)}_N\l_n$.
Since the following bounds hold,
$$|\wt\b_{m,N}^{(\m,o)}|,|\b_{m,N}^{(\l)}|\leq c_2\e\g^{-\th(N-m)}\;,
\qquad
 |\wt\b_{m,n}^{(\m,\s)}|\leq c_2\e^2\g^{-\th(n-m)}
\;,\qquad
 |\b_{m,n}|\leq c_2\e\g^{-\th(n-m)}\;,$$
if $x\in \BB_{\th/2}$, also $Tx\in \BB_{\th/2}$, for $\e$ small enough
and $c_1\ge 2c_2(1-\g^{-(\th/4)})^{-1}$; moreover,m
for $C>2c_2(1-\g^{-(\th/4)})^{-2}$, 
$\norm{x-x'}_{\th/2}\leq C\e\norm{Tx-Tx'}_{\th/2}$
so that, for $\e $ small enough, $T$ is a contraction: by the 
fixed point theorem, the solution of such an equation exists and is
in $\BB_{\th/2}$. As consequence, since
$$\wt z_j^{(\m)}-\a_N^{(\m)} z_j
 =\wt z_{j,N}^{(\m,o)}-  \a^{(\m)}_N\l_N z_{j,N}^{(\l)}
 +\sum_{\s=\pm}\sum_{n=j}^N\wt z_{j,n}^{(\m,\s)}\n^{(\s)}_n{Z_N\over Z_n}
 +\sum_{n=j}^{N-1}z_{j,n}\left(\wt\l_n^{(\m)}-\a_N^{(\m)}\l_n\right)\;,$$
where $\wt z^{(\m,o)}_{j,N}$ is the sum of the graphs 
made with an interaction $\AAA_o$, contracted on scale $N$;
$\wt z^{(\m\s)}_{j,n}$ is the sum of the graphs 
with an interaction $\TT_\s^{(\m)}$ on scale $n$,
deprived of the coupling $\n^{(\s)}_n(Z_N/ Z_n)$;
$z^{(\l)}_{j,N} $ is the sum of the graphs contributing
to the flow of $\a^{(\m)}_N z_j$ which have an interaction 
$\BB^{(3)}$ on scale $N$, deprived of the coupling $\a^{(\m)}_N\l_N$;
$z_{m,n}$ is the sum of the graphs contributing
to the flow of $\a^{(\m)}_N z_j$ with an interaction $\BB^{(3)}$
on scale $n$, deprived of the coupling $\a^{(\m)}_N\l_n$.
Since the following bounds hold,
$$|\wt z_{j,N}^{(\m,o)}|,|z_{j,N}^{(\l)}|\leq c_2\e\g^{-\th(N-j)}
\;,\qquad
 |\wt z_{j,n}^{(\m,\s)}|\leq c_2\e^2\g^{-\th(n-j)}
\;,\qquad
 |z_{j,n}|\leq c_2\e\g^{-\th(n-j)}\;,$$
also $\left\{(\wt z_j^{(+)}-\a_N^{(+)} z_j,
\wt z_j^{(-)}-\a_N^{(-)} z_j)\right\}_j\in \BB_{\th/2}$.
Finally, since all the graphs contributing 
to $\{\wt\l^{(\m)}_m\}_m$ and to $\{\l_m\}_m$ are scale invariant,
$$ \a^{(\m)}_N= \a^{(\m)}_{N+1}=\a^{(\m)}\;.$$
{The proof of the theorem is completed.\hfill\qed\hskip1em}
\appendix(a3,Schwinger-Dyson equation)

\asub(FD){Functional derivation.} By decomposing the fermionic
fields $\ps^+_{k,\o}\longrightarrow\ps^+_{k,\o}+\hb_{k,\o}$, it
holds:
$$\eqalign{
 \WW^{(h)}(\jm,\f)=
 &\WW^{(h)}_\BB(\b,\jm,\f)
 +\sum_{\o=\pm}
 \int_{D}\!{\der^2k\over (2\p)^2}\
 \hb_{k,\o}\hf^-_{k,\o}\cr
 &-
 \sum_{\o=\pm}
 \int_{D}\!{\der^2k\over (2\p)^2}\
 \hb_{k,\o}D_\o(k)
 \left[1+Z_N\left(\c^{-1}_{h,N}(k)-1\right)\right]
 {\partial \WW\over \partial \hf^+_{k,\o}}(\jm,\f)
 +{\rm O}(\b^2)\;,}\Eqa(trans)$$
where $\WW^{(h)}_\BB$ is the following functional with the further
source field $\b$:
$$\eqalign{
 e^{\WW^{(h)}_\BB(\b,\jm,\f)}
 \defi
 \int\!\der P^{[h,N]}(\ps)
&\exp\left\{
 -l_N \VV(\ps)
 +Z^{(2)}_N\JJ(\jm,\ps)
 +\FF(\f,\ps)
 \right\}\cr
&\exp\left\{
 -l_N \BB^{(3)}(\b,\ps)+Z^{(2)}_N\BB^{(2)}(\b,\jm,\ps)
 -z_N\BB^{(1)}(\b,\ps)
 \right\}\;,
}$$
with:
$$\eqalign{
 \BB^{(3)}(\b,\ps)
&\defi
 \sum_{\o=\pm}\int_{D}
 {\der^2k\over (2\p)^2}{\der^2p\over (2\p)^2}
 {\der^2q\over (2\p)^2}\
 \hb_{p+k-q,\o}\hp^-_{p,\o}\hp^+_{q,-\o}\hp^-_{k,-\o}\;,\cr
 \BB^{(2)}(\b,\jm,\ps)
&\defi
 \sum_{\o=\pm}\int_{D}{\der^2k\over (2\p)^2}
 {\der^2p\over (2\p)^2}\
 \hb_{k,\o}\hj_{p-k,\o}\hp^-_{p,\o}\;,\cr
 \BB^{(1)}(\b,\ps)
&\defi
 \sum_{\o=\pm}\int\!{\der^2k\over (2\p)^2}\
 \hb_{k,\o} D_\o(k)\hp^-_{k,\o}\;.
}$$
Therefore, extracting the linear part of \equ(trans), for
$k:\g^h\k\leq |k|\leq \g^N\k$ (so that $\c^{-1}_{h,N}(k)-1=0$), it
yield the \sde:
$$\hg^{-1}_\o(k){\partial \WW_\BB\over \partial \hf^+_{k,\o}}
(0,\jm,\f)
 =
 \hf^-_{k,\o}
 +{\partial \WW_\BB\over \partial \hb_{k,\o}}(0,\jm,\f)\;.
 \Eqa(DSE1)
$$
Now, writing the last derivative in terms of the derivative of
$\WW$ -- but loosing in this way the evidence of the
renormalization of composite operators -- and multiplying both
members by $e^{\WW^{(h)}}$ in order to shorten the equations, it
simply holds \eq(DSE2). By derivatives in the sources $\hj$ and
$\hf$, for $\hj=\hf=0$, such an equation generates all the \sde:
for instance, taking a derivative in $\hf^-_{k,\o}$ gives
\equ(DS2).
\appendix(LOC,Lowest Order Computations)

It is interesting  to calculate the 
lowest order expansion of the anomalies.
The computation of the anomaly of the \wti{}
shows {\it a violation of the Adler-Bardeen theorem}:
the correction to the classical identity is not linear
in the coupling, but has at least also a non-vanishing
second order term.
Then, the computation of the anomaly of the \ce{}
-- made in a quite approximate way -- would imply 
{\it the incorrectness of the Johnson solution}.

\*
\asub(WTIA){\wti{} anomaly}
Simplifying the notations, let $\c(k)\defi \c_0(k)$
and $u(k)\defi u_0(k)$. A useful identity is
$$\eqalign{
U_\o(k,k+p)
&=
\left\{u(k+p){\chi(k)\over  D_\o(k)}
-u(k){\chi(k+p)\over  D_\o(k+p)}\right\}\cr
&=D_\o(p)\left\{{u(k+p)\c(k)\over  D_\o(k+p)D_\o(k)}
-\int_0^1\!\der\t\  
{\big(\partial_{\o}\c\big)(k+\t p)\over D_{\o}(k+p)}
\right\}\cr
&-D_{-\o}(p)\int_0^1\!\der\t\  
{\big(\partial_{-\o}\c\big)(k+\t p)\over D_{\o}(k+p)}\;.
}$$
\0To simplify the computations, 
it is performed the following modification to the shape of the cutoff
which, as can be easily checked, it completely harmless 
to the development done in the previous Chapters.
Let $\c(k)\defi\wh\c(|k|)$, and $\wh\c(t)$ is
a Gevrey function 
with compact support $\{t:|t|\leq \k\g^N\g_0\}$,
for $\g_0:1<\g_0<\g$,
and equal to 1 in $\{t:|t|\leq \k\g^N\}$.
\elenco{
\art
{\bf Computation of {$\n^{(-)}$}.}
The lowest order expansion of $\n^{(-)}$ is given by only one 
Feynman graph, which  can be computed exactly:
$$\n^{(-)}=\int\!{\der^2k\over (2\p)^2}
{(\partial_{-\o}\c)(k)\over D_\o(k)}=
-{1\over 4\p}\int_0^\io\!\der t\ 
 \wh\c'(t)={1\over 4\p}\;.$$
where it was used that $(\partial_{-\o}\c)(k)/D_\o(k)=-(1/2|k|)\wh\c'(k)$.
\*
\insertplot{100pt}{90pt}%
{}%
{f7}{\eqg(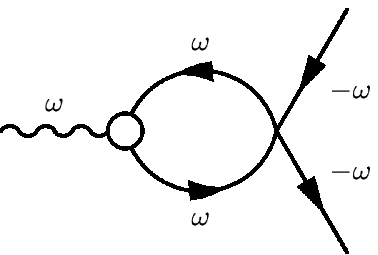)}
\*
\centerline{{\bf Fig \graf(f7.ps)}: Graphical representation of the lowest order contribution to 
$\n^{(-)}$}
\art
{\bf Computation of $\n^{(+)}$.}
Also the lowest order contribution to $\n^{(+)}$ is given by only one Feynman graph:
$$\eqalign{
&\int\!{\der^2 p\over (2\p)^2} 
\left\{{u(p)\c(p)\over  D_\o(p)D_\o(p)}
- 
{\big(\partial_{\o}\c\big)(p)\over D_{\o}(p)}
\right\} 
\int\!{\der^2k\over (2\p)^2} \hat g_{-\o}(k) \hat g_{-\o}(p+k)\cr
&=\int\!{\der^2 p\over (2\p)^2} 
\left\{{u(p)\c(p)-D_\o(p)\big(\partial_{\o}\c\big)(p)\over p^4}
\right\} 
\int\!{\der^2k\over (2\p)^2} {\c(k)\c(p+k)\over k^2 (k+p)^2}
D_{-\o}^2(p)D_\o(k)D_\o(k+p)\;.}\Eqa(nu+)$$
The explicit computation is not so simple as the 
previous; anyway it is possible to prove it is strictly 
non-zero.
Since  $-D_\o(p)\big(\partial_{\o}\c\big)(p)=-(|p|/2)\wh\c'(|p|)\ge 0$,
as well as $u(p)\c(p)\ge 0$,
while, calling $\th$ the angle between $p$ and $k$
and $\x\defi (|k|/|p|)$,
$$D_{-\o}^2(p)D_\o(k)D_\o(k+p)=|k||p|^3
\Big[\cos(\th)+ \x\cos(2\th)\Big]
\defi |k||p|^3J_\x(\th)\;,$$
up to a pure imaginary  contribution which integrated gives zero 
by symmetries.
Now, since by support of the 
cutoff functions $|k|\le \g_0$ and $1\le |p|\le \g_0$,
then  $\cos(\th)<1/2$ if $\g_0$ is chosen $\le 3/2$.
Hence, $J_1(\th)=\big[\cos(\th)-(1/2)\big]\big[\cos(\th)+1\big]<0$, except 
for $\th=\pm (\p/3),\p$, where it vanishes.
Then, the integral over $\th$ of $J_\x(\th)$
is continuous in $\x$, and strictly negative for $\x=1$;
 therefore  it remains strictly negative also for $\x=|k|/|p|$,
if $\g_0-1\geq |k|/|p|-1$ is small enough.
Therefore, for such values of $\g_0$, the 
lowest order contribution to $\n^{(+)}$ is
strictly negative.
\*
\vbox{
\insertplot{150pt}{90pt}%
{}%
{f8}{\eqg(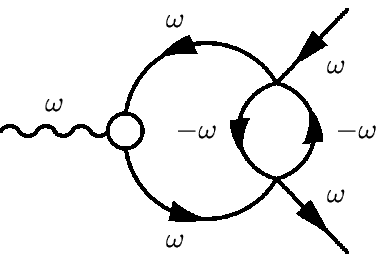)}
\*
\centerline{{\bf Fig \graf(f8.ps)}: Graphical representation of the lowest order contribution 
to $\n^{(+)}$}}
}

\*
\asub(CEA){\ce{} anomaly.}
From \equ(A1), and since $a-\bar a={\rm O}(\l)$,
while $a+\bar a=1+{\rm O}(\l^2)$, 
the contribution ${\rm O}(\l)$
to $A$ is proportional to
the terms ${\rm O}(1)$  of $\a^{(-)}-\s^{(-)}$.
\elenco{
\art 
The $0$-th order of $\a^{(-)}$ is given by two graphs with 
values cancelling each other.
\art
There is no possible graph for $\s^{(-)}$
at the $0$-th order, since there are no possible
tadpoles.
}
\*
\vbox{
\insertplot{250pt}{70pt}%
{}%
{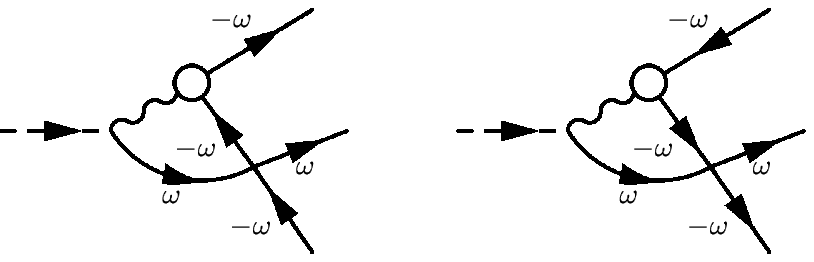}{\eqg(f20)}
\*
\centerline{{\bf Fig \graf(f20)}: Graphical of item 1}}
\*
\0Well then, $A=1+{\rm O}(\l^2)$. Then, the quadratic 
order in $\l$ comes from the linear order 
of $\a^{(-)}-\s^{(-)}$, and the O$(1)$ order 
of $\a^{(+)}-\s^{(+)}$.
\elenco{
\art
There are more than one Feynman graphs contributing to 
the linear order of $\a^{(-)}$.
\subelenco{
\item{\bf $\bullet$}
{\bf First graph.}
A first contribution are the two graphs with 
all and three external leg of $T$ involved: they are
two, with the same value. Furthermore, the factor $1/2!$ of the 
expansion of the interaction is compensated by multiplicity obtained
by exchanging the labels to the two vertices $\VV$ of each graph.
Therefore the sum of them gives the {\it first graph}:
$$\eqalign{
&2\int\!{\der^2k\over (2\p)^2}{\der^2p\over (2\p)^2}
{U_{-\o}(k,k+p)\over D_{-\o}(p)}g_\o(p+k)g_\o(k)\cr
=&
-2\int\!{\der^2p\over (2\p)^2}{\c(p)\over p^2}
\int\!{\der^2k\over (2\p)^2}
{u(k)\c^2(k+p)\over (p+k)^2}\cr
&
-2\int\!{\der^2p\over (2\p)^2}{\c(p)\over p^2}
\int\!{\der^2k\over (2\p)^2}
{u(k+p)\c(k+p)\c(k)\over D_{-\o}(k) D_\o(p+k)}\;.}$$
The latter addend is vanishing in the limit $\g_0\to 1$.
The former is convergent. Indeed:
$$\eqalign{
&\int_{|p|\le 1/2}\!{\der^2p\over (2\p)^2}{\c(p)\over p^2}
\int\!{\der^2k\over (2\p)^2}
{u(k)\c^2(k+p)\over (p+k)^2}\cr
&=
\int_{|p|\le 1/2}\!{\der^2p\over (2\p)^2}{\c(p)\over p^2}
\int\!{\der^2k\over (2\p)^2}
{u(k)(\c^2(k+p)-\c^2(k))\over (p+k)^2}\;,}$$
and $|p+k|\geq |k|-|p|\geq 1/2$; while
$$\eqalign{
&\int_{|p|> 1/2}\!{\der^2p\over (2\p)^2}{\c(p)\over p^2}
\int\!{\der^2k\over (2\p)^2}
{u(k)\c^2(k+p)\over (p+k)^2}\cr
&=
\int_{|p|> 1/2}\!{\der^2p\over (2\p)^2}
{(\c(p)-\c(k))\over p^2}
\int\!{\der^2k\over (2\p)^2}
{u(k)\c^2(k+p)\over (p+k)^2}\;.}$$
\*
\vbox{
\insertplot{100pt}{70pt}%
{}%
{f9}{\eqg(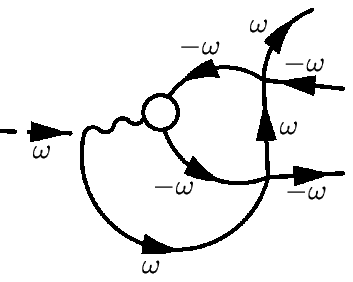)}
\*
\centerline{{\bf Fig \graf(f9.ps)}: First graph}}
\*
\item{\bf $\bullet$}
{\bf Second graphs.}
The second contribution is given by the graph
$$\eqalign{
&-\int\!{\der^2p\over (2\p)^2}{\der^2k\over (2\p)^2}\
g_\o(p) g_\o(p){U_{-\o}(k,k+p)\over D_{-\o}(p)}\cr
& =\int\!{\der^2p\over (2\p)^2}{\der^2k\over (2\p)^2}\
{\c^2(p)\over D_\o(p) D_\o(p)}
\left\{-{u(k+p)\c(k)\over  D_{-\o}(k+p)D_{-\o}(k)}+
\int_0^1\!\der\t\  
{\big(\partial_{-\o}\c\big)(k+\t p)\over D_{-\o}(k+p)}
\right\}\cr
&+\int\!{\der^2p\over (2\p)^2}{\der^2k\over (2\p)^2}\
{\c^2(p)\over  D_\o(p) D_{-\o}(p)}\int_0^1\!\der\t\  
{\big(\partial_{\o}\c\big)(k+\t p)\over D_{-\o}(k+p)}\;;}$$
and, 
subtracting the graph containing the counterterm $\n_N^{(-)}$,
$$ \int\!{\der^2p\over (2\p)^2}\
{\c^2(p)\over D_{\o}(p) D_{-\o}(p)}\int\!{\der^2k\over (2\p)^2}\  
{\big(\partial_{\o}\c\big)(k)\over D_{-\o}(k)}$$
the last addend is convergent;
while the first two terms are convergent automatically:
$$\eqalign{
&\int\!{\der^2p\over (2\p)^2}{\der^2k\over (2\p)^2}\
{\c^2(p)\over D_\o(p) D_\o(p)}{u(k+p)\c(k)\over  D_{-\o}(k+p)D_{-\o}(k)}\cr
&=\int\!{\der^2p\over (2\p)^2}{\der^2k\over (2\p)^2}\
{\c^2(p)\over D_\o(p) D_\o(p)}\left\{{u(k+p)\c(k)\over  D_{-\o}(k+p)D_{-\o}(k)}
-{u(k)\c(k)\over  D_{-\o}(k)D_{-\o}(k)}\right\}}$$
$$\eqalign{
&\int\!{\der^2p\over (2\p)^2}{\der^2k\over (2\p)^2}\
{\c^2(p)\over D_\o(p) D_\o(p)}\int_0^1\!\der\t\  
{\big(\partial_{-\o}\c\big)(k+\t p)\over D_{-\o}(k+p)}\cr
&=\int\!{\der^2p\over (2\p)^2}{\der^2k\over (2\p)^2}\
{\c^2(p)\over D_\o(p) D_\o(p)}\int_0^1\!\der\t\  
\left\{{\big(\partial_{-\o}\c\big)(k+\t p)\over D_{-\o}(k+p)}
-{{\big(\partial_{-\o}\c\big)(k)\over D_{-\o}(k)}}\right\}}$$
since the subtracted terms are zero by transformation under rotation.
\*
\insertplot{200pt}{100pt}%
{}%
{f10}{\eqg(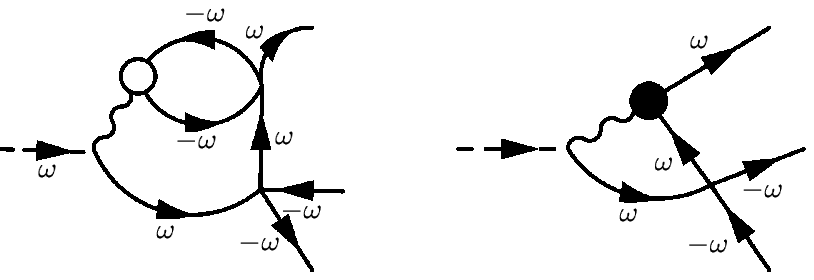)}
\*
\centerline{{\bf Fig \graf(f10.ps)}: Second graphs}
\item{$\bullet$} 
{\bf Vanishing graphs.}
There are four graphs subleading in the limit $\g_0\to 1$:
their total value is the double of the two 
{\it vanishing graphs}
\*
\vbox{
\insertplot{200pt}{70pt}%
{}%
{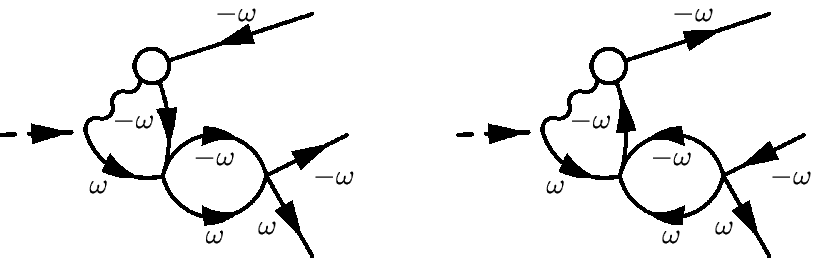}{\eqg(f21)}
\*
\insertplot{200pt}{90pt}%
{}%
{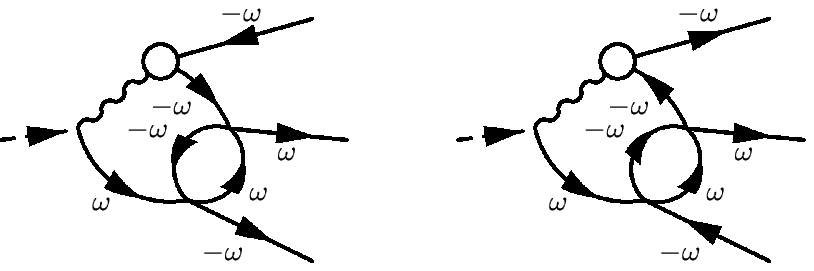}{\eqg(f22)}
\*
\centerline{{\bf Fig \graf(f21)}: Vanishing  graphs}}
}
\art 
The linear order of  $\s^{(-)}$ is given by only one graph.
\subelenco{
\item{$\bullet$}
{\bf Third graph.}
Such graph is very similar to the previous:
it is given by the the second graph, with 
the replacement of $g^2_\o(p)$ with 
$(\partial_\o g_\o)(p)$:
$$\eqalign{
&-\int\!{\der^2p\over (2\p)^2}{\der^2k\over (2\p)^2}\
(\partial_\o g_\o)(p)
{U_{-\o}(k,k+p)\over D_{-\o}(p)}\cr
& =\int\!{\der^2p\over (2\p)^2}{\der^2k\over (2\p)^2}\
\left[{(\partial_\o\c)(p)\over D_\o(p)}-{\c(p)\over D_\o(p) D_\o(p)}\right]
\left\{-{u(k+p)\c(k)\over  D_{-\o}(k+p)D_{-\o}(k)}\rgt.\cr
&\phantom{*********************************}
+\lft.\int_0^1\!\der\t\  
{\big(\partial_{-\o}\c\big)(k+\t p)\over D_{-\o}(k+p)}
\right\}\cr
&+\int\!{\der^2p\over (2\p)^2}{\der^2k\over (2\p)^2}\
\left[{(\partial_\o\c)(p)\over D_{-\o}(p)}-{\c(p)\over D_\o(p) D_{-\o}(p)}\right]
\int_0^1\!\der\t\  
{\big(\partial_{\o}\c\big)(k+\t p)\over D_{-\o}(k+p)}\;;}$$
and, 
subtracting the graph containing the counterterm $\n^{(-)}$,
$$ \int\!{\der^2p\over (2\p)^2}\
\left[{(\partial_\o\c)(p)\over D_{-\o}(p)}-{\c(p)\over D_\o(p) D_{-\o}(p)}\right]
\int\!{\der^2k\over (2\p)^2}\  
{\big(\partial_{\o}\c\big)(k)\over D_{-\o}(k)}$$
\*
\insertplot{250pt}{50pt}%
{}%
{f11}{\eqg(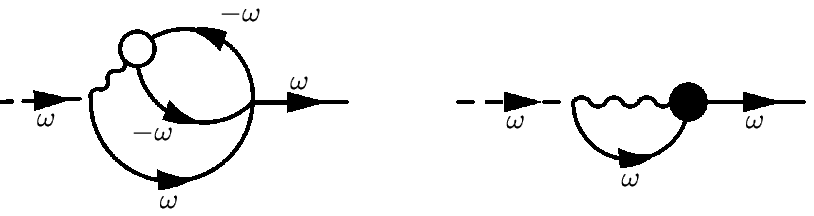)}
\*
\centerline{{\bf Fig \graf(f11.ps)}: Third graph}}\*
the last addend is convergent.
\art
The $0$-th order of $\a^{(+)}$ is given by only one graph,
which is subleading: it vanishes in the limit $\g_0\to 1$.
\*
\insertplot{100pt}{60pt}%
{}%
{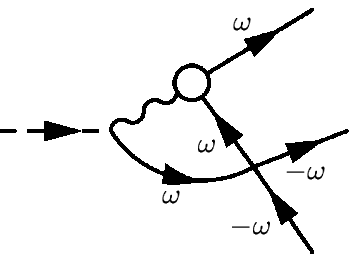}{\eqg(f20b)}
\*
\centerline{{\bf Fig \graf(f20b)}: Graph in item 3.}\*
\art
The $0$-th  order of $\s^{(+)}$ is only given by a tadpole. 
\item{$\bullet$}
{\bf Fourth graph.}
It derives from the tadpole of $T_0^{(+)}$: for any $N'\geq 2$
$$\int\!{\der^2q\over (2\p)^2}{u_N(q)\c_{N+N'}(k-q)\over D_{-\o}(k-q)}\;;$$
the localization of this term is the extraction of the 
zeroth and first order Taylor expansion in the external momentum $k$:
the former is clearly summable and zero by symmetries;
the latter is:
$$\int\!{\der^2q\over (2\p)^2}{u(q)\big(\partial_\o\c_{N'}\big)(q)\over D_{-\o}(q)}
=\int\!{\der^2q\over (2\p)^2}{u_{-N'}(q)(\partial_\o\c)(q)\over D_{-\o}(q)}\;.$$
\*
\vbox{
\insertplot{100pt}{50pt}%
{}%
{f12}{\eqg(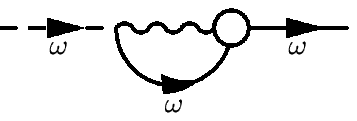)}
\*
\centerline{{\bf Fig \graf(f12.ps)}: Fourth graph}}}
\*
\asub(EC){Explicit computation.}
To make the computation easier, the cutoff is
chosen to be a distribution
$$\c(k)\defi f(k_0)f(k_1)\;,\qquad{\rm for}\ f(x)\defi \th(x+1)-\th(x-1)\;.$$
Then $f'(x)=\d(x+1)-\d(x-1)$.
Since, by definition of $D_\o(k)$,
it holds $k_0=(i/2)\big[D_\o(k)+D_{-\o}(k)\big]$
while  $k_1=(\o/2)\big[D_\o(k)-D_{-\o}(k)\big]$,
then:
$$\Big(\partial_{\o}\c\Big)(k)=
{i\over2}f'(k_0)f(k_1)+{\o\over2}f(k_0)f'(k_1)\;.$$
It is suitable to remark that the above choice of the 
cutoff function, in contrast with what done for the 
anomaly of the \wti, {\it is not allowed} 
in the developments of the previous Chapter. Furthermore
the computation of the following integrals is not exact, 
but rather is performed with a simple
Montecarlo simulation. That is way the 
incorrectness is not proved,
but it has to be considered as a {\it conjecture},
enforced by such a calculation.
\subelenco{
\item{$\bullet$}
{\bf F.} For the first graph it holds:
$$\eqalign{
{\bf F}\defi
&-2\int\!{\der^2p\over (2\p)^2}{\c(p)\over p^2}
\int\!{\der^2k\over (2\p)^2}
{u(k)\c^2(k+p)\over (p+k)^2}\cr
=&-{2\over (2\p)^4}\int_{-1}^1\!\der p_0\der p_1\ 
{1\over p^2}
\int_{-1}^1\!\der k_0\der k_1
{1-f(k_0-p_0)f(k_1-p_1)\over k^2}={ 52.64}{1\over (2\p)^4}\;.}$$
\item{$\bullet$}
\0{\bf S.} Calling $p^\t=(1-\t)p$, for the second graph it holds:
$$\eqalign{
{\bf Sa}\defi
&\int\!{\der^2p\over (2\p)^2}{\der^2k\over (2\p)^2}\
{\c^2(p)\over D_\o(p) D_\o(p)}{u(k+p)\c(k)\over  D_{-\o}(k+p)D_{-\o}(k)}\cr
=&{2\over (2\p)^4}\int_{-1}^1\!\der p_0\der p_1\
{(p_0^2-p_1^2)\over p^4}
\int_{-1}^{1}\!\der k_0 \der k_1\ 
{1-f(k_0+p_0)f(k_1+p_1)\over (k+p)^2}(k_0+p_0){k_0\over k^2}\cr
+&{4\over (2\p)^4}\int_{-1}^1\!\der p_0\der p_1\
{p_0p_1\over p^4}
\int_{-1}^1\!\der k_0\der k_1\ 
{1-f(k_0+p_0)f(k_1+p_1)\over (k+p)^2}(k_1+p_1){k_0\over k^2}\;.}$$
Finally:
$$\eqalign{
{\bf Sa1}\defi
&{2\over (2\p)^4}\int_{-1}^1\!\der p_0\der p_1\
{(p_0^2-p_1^2)\over p^4}
\int_{-1}^{1}\!\der k_0 \der k_1\ 
{1-f(k_0+p_0)f(k_1+p_1)\over (k+p)^2}(k_0+p_0){k_0\over k^2}\cr
&=2.69{1\over (2\p)^4}\;,\cr
{\bf Sa2}\defi
&{4\over (2\p)^4}\int_{-1}^1\!\der p_0\der p_1\
{p_0p_1\over p^4}
\int_{-1}^1\!\der k_0\der k_1\ 
{1-f(k_0+p_0)f(k_1+p_1)\over (k+p)^2}(k_1+p_1){k_0\over k^2}\cr
&=0.29{1\over (2\p)^4}\;.}$$
The second addend of the second graph is:
$$\eqalign{
{\bf Sb}\defi
&\int\!{\der^2p\over (2\p)^2}{\der^2k\over (2\p)^2}\
{\c^2(p)\over D_\o(p) D_\o(p)}\int_0^1\!\der\t\  
{\big(\partial_{-\o}\c\big)(k+\t p)\over D_{-\o}(k+p)}\cr
=&-\int\!{\der^2p\over (2\p)^2}\ 
{\c^2(p)\over p^4}{D^2_{-\o}(p)}\int_0^1\!\der\t\ 
\int\!{\der^2k\over (2\p)^2}\  
{\big(\partial_{-\o}\c\big)(k)\over \left(k+p^\t\right)^2} D_{\o}\left(k+p^\t\right)\cr
=&\int\!{\der^2p\over (2\p)^2}\ 
{\c^2(p)\over p^4}(p_0^2-p_1^2)\int_0^1\!\der\t\ 
\int\!{\der^2k\over (2\p)^2}\  
{(k_0+p_0^\t)f'(k_0)f(k_1)\over \left(k+p^\t\right)^2}\cr
&+2\int\!{\der^2p\over (2\p)^2}\ 
{\c^2(p)\over p^4}p_0p_1\int_0^1\!\der\t\ 
\int\!{\der^2k\over (2\p)^2}\  
{(k_1+p_1^\t)f'(k_0)f(k_1)\over \left(k+p^\t\right)^2}\cr
=
&{2\over (2\p)^4}\int_{-1}^1\!\der p_0\der p_1\ 
{p_0^2-p_1^2\over p^4}\int_0^1\!\der\t\ 
\int_{-1}^{1}\!\der k_1\  
{p_0^\t-1\over \left(p_0^\t-1\right)^2+\left(p_1^\t+k_1\right)^2}\cr
&+{4\over (2\p)^4}\int_{-1}^1\!\der p_0\der p_1\ 
{p_0p_1\over p^4}\int_0^1\!\der\t\ 
\int_{-1}^1\!\der k_1\
{p_1^\t+k_1\over \left(p_0^\t-1\right)^2+\left(p_1^\t+k_1\right)^2}\;.}$$
The third addend of the second graph is
$$\eqalign{
{\bf Sc}
\defi&\int\!{\der^2p\over (2\p)^2}{\der^2k\over (2\p)^2}\
{\c^2(p)\over  D_\o(p) D_{-\o}(p)}\int_0^1\!\der\t\  
{\big(\partial_{\o}\c\big)(k+\t p)\over D_{-\o}(k+p)}\cr
=&\int\!{\der^2p\over (2\p)^2}\
{\c^2(p)\over p^2}\int_0^1\!\der\t\  
\int\!{\der^2k\over (2\p)^2}\ 
{\big(\partial_{\o}\c\big)(k)\over (p^\t+k)^2}D_\o(p^\t+k)\cr
=&\int\!{\der^2p\over (2\p)^2}\
{\c^2(p)\over p^2}\int_0^1\!\der\t\  
\int\!{\der^2k\over (2\p)^2}\ 
{(p_0^\t+k_0)f'(k_0)f(k_1)\over (p^\t+k)^2}\cr
=&{2\over (2\p)^4}\int_{-1}^1\!\der p_0\der p_1\
{1\over p^2}\int_0^1\!\der\t\  
\int_{-1}^1\!\der k_1\ 
{(p_0^\t-1)\over \left(p_0^\t-1\right)^2+\left(p_1^\t+k_1\right)^2}\;;}$$
and its regularization is obtained by subtracting the $\io$ term
$$-{2\over (2\p)^4}\int_{-1}^1\!\der p_0\der p_1\
{1\over p^2} 
\int_{-1}^1\!\der k_1\ 
{1\over 1+k_1^2}\;.$$
Therefore:
$$\eqalign{
{\bf Sc}
\defi
&{2\over (2\p)^4}\int_{-1}^1\!\der p_0\der p_1\
{1\over p^2}\int_0^1\!\der\t\  
\int_{-1}^1\!\der k_1\ 
\left[
{(p_0^\t-1)\over \left(p_0^\t-1\right)^2+\left(p_1^\t+k_1\right)^2}+
{1\over 1+k_1^2}\right]\;.}$$
Setting ${\bf Sd}\defi{\bf Sb}+{\bf Sc}$
finally:
$$\eqalign{
{\bf Sd1}\defi
&-{4\over (2\p)^4}\int_{-1}^1\!\der p_0\der p_1\ 
{p_1^2\over p^4}\int_0^1\!\der\t\ 
\int_{-1}^{1}\!\der k_1\  
\left[{p_0^\t-1\over \left(p_0^\t-1\right)^2+\left(p_1^\t+k_1\right)^2}
+{1\over 1+k_1^2}\right]\cr
&=- 0.49{1\over (2\p)^4}\;,\cr
{\bf Sd2}\defi
&{4\over (2\p)^4}\int_{-1}^1\!\der p_0\der p_1\ 
{p_0p_1\over p^4}\int_0^1\!\der\t\ 
\int_{-1}^1\!\der k_1\
\left[{p_1^\t+k_1\over \left(p_0^\t-1\right)^2+\left(p_1^\t+k_1\right)^2}
-{k_1\over 1+k_1^2}\right]\cr
&=0.0056{1\over (2\p)^4}\;.}$$
\item{$\bullet$}
{\bf T.} For the third graph it holds:
$$\eqalign{
{\bf Ta}\defi
&\int\!{\der^2p\over (2\p)^2}{\der^2k\over (2\p)^2}\
{(\partial_\o\c)(p)\over  D_\o(p)}{u(k+p)\c(k)\over  D_{-\o}(k+p)D_{-\o}(k)}\cr
=&{2\over (2\p)^4}\int_{-1}^1\!\der p_0\
{1\over p^2_0+1}
\int_{-1}^{1}\!\der k_0 \der k_1\ 
{1-f(k_0+p_0)f(k_1-1)\over (k_0+p_0)^2+(k_1-1)^2}
(k_0+p_0){k_0\over k^2}\cr
-&{2\over (2\p)^4}\int_{-1}^1\!\der p_1\
{1\over 1+p_1^2}
\int_{-1}^{1}\!\der k_0 \der k_1\ 
{1-f(k_0-1)f(k_1+p_1)\over (k_0-1)^2+(k_1+p_1)^2}(k_0-1){k_0\over k^2} \cr
+&{2\over (2\p)^4}\int_{-1}^1\!\der p_1\
{p_1\over 1+p^2_1}
\int_{-1}^1\!\der k_0\der k_1\ 
{1-f(k_0-1)f(k_1+p_1)\over (k_0-1)^2+(k_1+p_1)^2}(k_1+p_1)
{k_0\over k^2}\cr
+&{2\over (2\p)^4}\int_{-1}^1\!\der p_0\
{p_0\over p^2_0+1}
\int_{-1}^1\!\der k_0\der k_1\ 
{1-f(k_0+p_0)f(k_1-1)\over (k_0+p_0)^2+(k_1-1)^2}(k_1-1){k_0\over k^2}\;.}$$
Finally
$$\eqalign{
{\bf Ta1}\defi
&{2\over (2\p)^4}\int_{-1}^1\!\der p_0\
{1\over p^2_0+1}
\int_{-1}^{1}\!\der k_0 \der k_1\ 
{1-f(k_0+p_0)f(k_1-1)\over (k_0+p_0)^2+(k_1-1)^2}
(k_0+p_0){k_0\over k^2}\cr
&=1.96{1\over (2\p)^4}\;,\cr
{\bf Ta2}\defi
-&{2\over (2\p)^4}\int_{-1}^1\!\der p_1\
{1\over 1+p_1^2}
\int_{-1}^{1}\!\der k_0 \der k_1\ 
{1-f(k_0-1)f(k_1+p_1)\over (k_0-1)^2+(k_1+p_1)^2}(k_0-1){k_0\over k^2}\cr
&=- 4.1{1\over (2\p)^4}\;,\cr
{\bf Ta3}\defi
+&{2\over (2\p)^4}\int_{-1}^1\!\der p_1\
{p_1\over 1+p^2_1}
\int_{-1}^1\!\der k_0\der k_1\ 
{1-f(k_0-1)f(k_1+p_1)\over (k_0-1)^2+(k_1+p_1)^2}(k_1+p_1)
{k_0\over k^2}\cr
&=- 0.28{1\over (2\p)^4}\;,\cr
{\bf Ta4}\defi
+&{2\over (2\p)^4}\int_{-1}^1\!\der p_0\
{p_0\over p^2_0+1}
\int_{-1}^1\!\der k_0\der k_1\ 
{1-f(k_0+p_0)f(k_1-1)\over (k_0+p_0)^2+(k_1-1)^2}(k_1-1){k_0\over k^2}\cr
&=0.11{1\over (2\p)^4}\;.}$$
The second addend of the third graph is
$$\eqalign{
{\bf Tb}\defi
&\int\!{\der^2p\over (2\p)^2}{\der^2k\over (2\p)^2}\
{(\partial_\o\c)(p)\over D_{\o}(p)}\int_0^1\!\der\t\  
{\big(\partial_{-\o}\c\big)(k+\t p)\over D_{-\o}(k+p)}\cr
=&\int\!{\der^2p\over (2\p)^2}\ 
{(\partial_\o\c)(p)\over p^2}{D_{-\o}(p)}\int_0^1\!\der\t\ 
\int\!{\der^2k\over (2\p)^2}\  
{\big(\partial_{-\o}\c\big)(k)\over \left(k+p^\t\right)^2} D_{\o}\left(k+p^\t\right)\cr
=&{1\over 2}\int\!{\der^2p\over (2\p)^2}\ 
{f'(p_0)f(p_1)p_0-f(p_0)f'(p_1)p_1\over p^2}\int_0^1\!\der\t\ 
\int\!{\der^2k\over (2\p)^2}\  
{(k_0+p_0^\t)f'(k_0)f(k_1)\over \left(k+p^\t\right)^2}\cr
+&{1\over 2}\int\!{\der^2p\over (2\p)^2}\ 
{f'(p_0)f(p_1)p_1+f(p_0)f'(p_1)p_0\over p^2}\int_0^1\!\der\t\ 
\int\!{\der^2k\over (2\p)^2}\  
{(k_1+p_1^\t)f'(k_0)f(k_1)\over \left(k+p^\t\right)^2}\;;}$$
$$\eqalign{
{\bf Tc}\defi
&\int\!{\der^2p\over (2\p)^2}{\der^2k\over (2\p)^2}\
{(\partial_\o\c)(p)\over  D_{-\o}(p)}\int_0^1\!\der\t\  
{\big(\partial_{\o}\c\big)(k+\t p)\over D_{-\o}(k+p)}\cr
=&\int\!{\der^2p\over (2\p)^2}\
{(\partial_\o\c)(p)D_{\o}(p)\over p^2}\int_0^1\!\der\t\  
\int\!{\der^2k\over (2\p)^2}\ 
{\big(\partial_{\o}\c\big)(k)\over (p^\t+k)^2}D_\o(p^\t+k)\cr
=&{1\over 2}
\int\!{\der^2p\over (2\p)^2}\
{f'(p_0)f(p_1)p_0+f(p_0)f'(p_1)p_1\over p^2}\int_0^1\!\der\t\  
\int\!{\der^2k\over (2\p)^2}\ 
{(p_0^\t+k_0)f'(k_0)f(k_1)\over (p^\t+k)^2}\cr
&+{1\over 2}
\int\!{\der^2p\over (2\p)^2}\
{f'(p_0)f(p_1)p_1-f(p_0)f'(p_1)p_0\over p^2}\int_0^1\!\der\t\  
\int\!{\der^2k\over (2\p)^2}\ 
{(p_0^\t+k_0)f(k_0)f'(k_1)\over (p^\t+k)^2}\;.}$$
Setting {\bf Td$\defi$ Tb + Tc}, some cancellation occurs:
$$\eqalign{
{\bf Td}\defi
&\int\!{\der^2p\over (2\p)^2}{\der^2k\over (2\p)^2}\
{(\partial_\o\c)(p)\over  D_{-\o}(p)}\int_0^1\!\der\t\  
{\big(\partial_{\o}\c\big)(k+\t p)\over D_{-\o}(k+p)}\cr
=&\int\!{\der^2p\over (2\p)^2}\
{(\partial_\o\c)(p)D_{\o}(p)\over p^2}\int_0^1\!\der\t\  
\int\!{\der^2k\over (2\p)^2}\ 
{\big(\partial_{\o}\c\big)(k)\over (p^\t+k)^2}D_\o(p^\t+k)\cr
=&
\int\!{\der^2p\over (2\p)^2}\
{f'(p_0)f(p_1)p_0\over p^2}\int_0^1\!\der\t\  
\int\!{\der^2k\over (2\p)^2}\ 
{(p_0^\t+k_0)f'(k_0)f(k_1)\over (p^\t+k)^2}\cr
&+
\int\!{\der^2p\over (2\p)^2}\
{f(p_0)f'(p_1)p_0\over p^2}\int_0^1\!\der\t\  
\int\!{\der^2k\over (2\p)^2}\ 
{(p_1^\t+k_1)f'(k_0)f(k_1)\over (p^\t+k)^2}}$$
Therefore:
$$\eqalign{
{\bf Td1}\defi
&{2\over (2\p)^4}\int_{-1}^1\!\der p_1\ 
{1\over 1+p^2_1}\int_0^1\!\der\t\ 
\int_{-1}^{1}\!\der k_1\  
\left[{\t\over \t^2+\left(p_1^\t+k_1\right)^2}
\right.\cr&\phantom{******************}\left.
 -{\t-2\over \left(\t-2\right)^2+\left(p_1^\t+k_1\right)^2}
-{2\over 1+k_1^2}\right]
=0.86{1\over (2\p)^4}\;,\cr
{\bf Td2}\defi
&{4\over (2\p)^4}\int_{-1}^1\!\der p_0\ 
{p_0\over p^2_0+1}\int_0^1\!\der\t\ 
\int_{-1}^1\!\der k_1\
 \left[{1-\t+k_1\over \left(p_0^\t+1\right)^2+\left(1-\t+k_1\right)^2}
-{k_1\over 1+k_1^2}\right]\cr
&=-0.62{1\over (2\p)^4}\;.}$$
\item{$\bullet$}
{\bf Q.}Regarding the fourth graph,
since $(\partial_\o\c_0)(q)=-\big(1/2|q|\big)\c'_0(q)D_{-\o}(q)$, and since
when $\c'_0(q)\neq 0$, $u_{-N'}(q)\equiv1$,
the last integral is equal to 
$$-{1\over 2}\int\!{\der^2q\over (2\p)^2}{\c_0'(q)\over |q|}
=-{1\over 2}\left.{1\over 4\p}\c_0(q)\right|^{q=\g}_{q=1}={1\over 8\p}\;,$$
{\it independently} on the scale $N'$ and on the shape of the 
function $\c$.
Such a contribution has to be multiplied 
times $(a-\bar a)/2=\n^{(-)}+{\rm O}(\l^2)={\p\over (2\p)^2}$,
obtaining
$$6.18{1\over (2\p)^2}\;.$$
}
\0In the end,
the quadratic coefficient of the second anomaly,
$A$, is non zero, and  in particular $\geq 18/(2\p)^4$ .

\intro(Ref,References)
\*
\halign{\hbox to 1.8truecm {#\hss} & \vtop{\advance\hsize by -1.8
truecm \0#}\cr
[AAR]& {{\cs Abdalla E., Abdalla M.C.B., Rothe, D.K.}:
Non-perturbative methods in 2 dimensional quantum field theory
{\book World Scientific} (2001). }\cr\cr
[A69]& {{\cs Adler S. L.}: Axial-Vector Vertex in Spinor Electrodynamics.
{\journal Phys. Rev.}, {\bf 177},
{\pagine 2426-2438} 1969.}\cr\cr
[BM01]& {{\cs Benfatto G., Mastropietro V.}:
Renormalization group, hidden
symmetries and approximate Ward identities in the $XYZ$ model.
{\journal Rev. Math. Phys.}, {\bf 13},
{\pagine 1323-1435}, 2001.}\cr\cr
[BM02]& {{\cs Benfatto G., Mastropietro V.}:
On the density-density
critical indices in interacting Fermi systems.
{\journal Comm. Math. Phys.}, {\bf 231},
{\pagine 97-134}, 2002. }\cr\cr
[BM04]& {{\cs Benfatto G., Mastropietro V.}:
Ward identities and
vanishing of the Beta function for $d=1$ interacting Fermi systems.
{\journal J. Stat. Phys.}, {\bf 115},
{\pagine 143-184}, 2004.}\cr\cr
[BM05]& {{\cs Benfatto G., Mastropietro V.}:
Ward identities and chiral anomaly in the Luttinger liquid.
{\journal Comm. Math. Phys.}, {\bf 258},
{\pagine 609-655}, 2005. }\cr\cr
[BoM97]& {{\cs Bonetto F., Mastropietro V.}: Critical indices for
the Yukawa$_2$ quantum field theory. {\journal Nucl. Phys. B},
{\bf258}, {\pagine 541-554}, 1997. }\cr\cr
[FMRS85]& {{\cs Feldman J., Magnen J., Rivasseau, V. S\'en\'eorR.}:
Gross-Neveu model: a rigorous perturbative construction.
{\journal Phys. Rev. Lett}, {\bf 54},
{\pagine 1479-1481}, 1985.}\cr\cr
[FW]& {{\cs Fetter L., Walecka J.D.}:
Quantum theory of many particle systems.
{\book McGraw-Hill}, 1971. }\cr\cr
[F79]& {{\cs Fujikawa K.}:
Path Integral Measure for Gauge Invariant Fermion Theories.
{\journal Phys. Rev. Lett.}, {\bf 42}
{\pagine 1195-1198}, 1979. }\cr\cr
[GK85]& {{\cs Gawedzki K, Kupiainen, A.}:
Gross-Neveu model through Convergent Perturbation Expansion.
{\journal Comm. Math.Phys.}, {\bf 102},
{\pagine 1-30}, 1985. }\cr\cr
[G58]& {{\cs Glaser V.}:
An explicit solution of the Thirring model.
{\journal Nouvo Cimento }, {\bf 9},
{\pagine 990-1006}, 1958. }\cr\cr
[GL72]& {{\cs Gomes M., Lowenstein J.H.}:
Asymptotic scale invariance in a massive Thirring model.
{\journal Nucl. Phys. B}, {\bf 45},
{\pagine 252-266}, 1972. }\cr\cr
[H89]& {{\cs Hurd T.R.}:
Soft breaking of gauge invariance in regularized electrodynamics.
{\journal Comm. Math. Phys.}, {\bf 125},
{\pagine 515-526}, 1989. }\cr\cr
[J61]& {{\cs Johnson K.}:
Solution of the Equations for the Green's Functions of a two
Dimensional Relativistic Field Theory.
{\journal Nuovo Cimento}, {\bf 20},
{\pagine 773-790}, 1961. }\cr\cr
[JZ59]& {{\cs Johnson K., Zumino,B.}: Gauge Dependence of the
Wave-Function Renormalization Constant in Quantum Electrodynamics.
{\journal Phys. Rev. Lett.}, {\bf 3}, {\pagine 351-352}, 1959.}\cr\cr
[K68]& {{\cs Klaiber B.}: Lectures in theoretical physics. 
ed: A.O. Barut and W.E. Brittin. {\it
Gordon and Breach.},  1968.}\cr\cr
[MS76]& {{\cs Magnen J., S\'en\'eor R.}:
The Wightman Axioms for the Weakly Coupled Yukawa Model in Two
Dimensions.
{\journal Comm. Math. Phys.}, {\bf 51},
{\pagine 297-313}, 1976. }\cr\cr
[M]& {{\cs Marcushevich A.I.}:
The theory of analytic functions: a brief course.
{\book MIR} (1983). }\cr\cr
[M93]& {{\cs Mastropietro V.}:
Schwinger function in Thirring and Luttinger models.
{\journal Nuovo  Cimento}, {\bf 108},
{\pagine 1095-1107}, (1993). }\cr\cr
[ML65]& {{\cs Mattis D., Lieb E.}: Exact Solution of a Many
Fermion System and its Associated Boson Field {\journal J. Math.
Phys.}, {\bf 6}, {\pagine 304-312}, (1965). }\cr\cr
[MM]& {{\cs Montvay I., M\"unster G.}: Quantum Fields on a
Lattice. {\book Cambridge University Press}, (1994). }\cr\cr
[OS73]& {{\cs Osterwalder K., Schrader R.}: Axioms for Euclidean
Green's Functions. {\journal Comm. Math. Phys.}, {\bf 31},
{\pagine 83-112}, (1973). }\cr\cr
[OS77]& {{\cs Osterwalder, K., Seiler E.}: Gauge Field Theories on
a Lattice. {\journal Annals of Physics}, {\bf 110}, {\pagine }, (1977). }\cr\cr
[S75]& {{\cs Seiler E.}:  Schwinger functions for the Yukawa model in 
two dimensions
with space-time cutoff  {\journal Comm. Math. Phys}, {\bf 42},
{\pagine163-182}, (1975). }\cr\cr
[T58]& {{\cs Thirring W.}: A soluble relativistic field theory.
{\journal Annals of Physics}, {\bf 3}, {\pagine}, (1958).
}\cr\cr
[W69]& {{\cs Wilson K. G.}: 
 9. Non-Lagrangian Models of Current Algebra
{\journal Phys. Rev.}, {\bf 179},
{\pagine 1499-1512}, (1969). }\cr\cr
[W76]& {{\cs Wilson W.} {\book Cargese lectures}, (1976). }\cr\cr
[Z70]& {{\cs Zimmermann W.}: Lectures on elementary particles and
quantum field theory, vol.1 {\book M.I.T. press }, (1970).
}\cr\cr}

\closeout1
\bye